\PassOptionsToPackage{unicode}{hyperref}
\PassOptionsToPackage{hyphens}{url}
\PassOptionsToPackage{dvipsnames,svgnames,x11names}{xcolor}
\documentclass[12pt]{article}

\usepackage{amsmath,amssymb}
\usepackage{iftex}
\usepackage[T1]{fontenc}
\usepackage[utf8]{inputenc}
\usepackage{textcomp}
\usepackage{lmodern}
\IfFileExists{upquote.sty}{\usepackage{upquote}}{}
\IfFileExists{microtype.sty}{%
  \usepackage[]{microtype}
  \UseMicrotypeSet[protrusion]{basicmath}
}{}
\makeatletter
\@ifundefined{KOMAClassName}{%
  \IfFileExists{parskip.sty}{%
    \usepackage{parskip}
  }{%
    \setlength{\parindent}{0pt}
    \setlength{\parskip}{6pt plus 2pt minus 1pt}}
}{%
  \KOMAoptions{parskip=half}}
\makeatother
\usepackage{xcolor}
\makeatletter
\ifx\paragraph\undefined\else
  \let\oldparagraph\paragraph
  \renewcommand{\paragraph}{
    \@ifstar
      \xxxParagraphStar
      \xxxParagraphNoStar
  }
  \newcommand{\xxxParagraphStar}[1]{\oldparagraph*{#1}\mbox{}}
  \newcommand{\xxxParagraphNoStar}[1]{\oldparagraph{#1}\mbox{}}
\fi
\ifx\subparagraph\undefined\else
  \let\oldsubparagraph\subparagraph
  \renewcommand{\subparagraph}{
    \@ifstar
      \xxxSubParagraphStar
      \xxxSubParagraphNoStar
  }
  \newcommand{\xxxSubParagraphStar}[1]{\oldsubparagraph*{#1}\mbox{}}
  \newcommand{\xxxSubParagraphNoStar}[1]{\oldsubparagraph{#1}\mbox{}}
\fi
\makeatother

\usepackage{longtable,booktabs,array}
\usepackage{calc}
\usepackage{etoolbox}
\makeatletter
\patchcmd\longtable{\par}{\if@noskipsec\mbox{}\fi\par}{}{}
\makeatother
\IfFileExists{footnotehyper.sty}{\usepackage{footnotehyper}}{\usepackage{footnote}}
\makesavenoteenv{longtable}
\usepackage{graphicx}
\makeatletter
\def\maxwidth{\ifdim\Gin@nat@width>\linewidth\linewidth\else\Gin@nat@width\fi}
\def\maxheight{\ifdim\Gin@nat@height>\textheight\textheight\else\Gin@nat@height\fi}
\makeatother
\setkeys{Gin}{width=\maxwidth,height=\maxheight,keepaspectratio}
\makeatletter
\def\fps@figure{htbp}
\makeatother

\makeatletter
\@ifpackageloaded{caption}{}{\usepackage{caption}}
\AtBeginDocument{%
\ifdefined\contentsname
  \renewcommand*\contentsname{Table of contents}
\else
  \newcommand\contentsname{Table of contents}
\fi
\ifdefined\listfigurename
  \renewcommand*\listfigurename{List of Figures}
\else
  \newcommand\listfigurename{List of Figures}
\fi
\ifdefined\listtablename
  \renewcommand*\listtablename{List of Tables}
\else
  \newcommand\listtablename{List of Tables}
\fi
\ifdefined\figurename
  \renewcommand*\figurename{Figure}
\else
  \newcommand\figurename{Figure}
\fi
\ifdefined\tablename
  \renewcommand*\tablename{Table}
\else
  \newcommand\tablename{Table}
\fi
}
\@ifpackageloaded{float}{}{\usepackage{float}}
\floatstyle{ruled}
\@ifundefined{c@chapter}{\newfloat{codelisting}{h}{lop}}{\newfloat{codelisting}{h}{lop}[chapter]}
\floatname{codelisting}{Listing}

\makeatother
\makeatletter
\@ifpackageloaded{caption}{}{\usepackage{caption}}
\@ifpackageloaded{subcaption}{}{\usepackage{subcaption}}
\makeatother

\usepackage[]{natbib}
\usepackage{bookmark}

\IfFileExists{xurl.sty}{\usepackage{xurl}}{}
\hypersetup{
  pdftitle={Bayesian variable and hazard structure selection in the General Hazard model},
  pdfauthor={Yulong Chen, Jim Griffin, Francisco Javier Rubio},
  pdfkeywords={Bayesian variable selection, Survival analysis, Censoring, Hazard structure, Model misspecification},
  colorlinks=true,
  linkcolor={blue},
  filecolor={Maroon},
  citecolor={Blue},
  urlcolor={Blue},
  pdfcreator={LaTeX via pandoc}}

\newcommand{\anon}{1}

\newcommand{\bxi}{\bm{\xi}}
\newcommand{\bPsi}{\bm{\Psi}}
\newcommand{\bgamma}{\bm{\gamma}}
\newcommand{\btheta}{\bm{\theta}}

\newcommand{\bmeta}{\bm{\eta}}
\newcommand{\bbeta}{\bm{\beta}}

\newcommand{\balpha}{\bm{\alpha}}

\newcommand{\bt}{{\bf t}}

\newcommand{\bv}{{\bf v}}

\newcommand{\bx}{{\bf x}}

\newcommand{\bz}{{\bf z}}
\newcommand{\tbx}{\tilde{{\bf x}}}
\newcommand{\bX}{{\bf X}}
\newcommand{\tbX}{\tilde{{\bf X}}}

\newcommand{\hz}{{\text{Hz}}}

\DeclareFontFamily{OT1}{pzc}{}
\DeclareFontShape{OT1}{pzc}{m}{it}{<-> s * [1.10] pzcmi7t}{}
\DeclareMathAlphabet{\mathpzc}{OT1}{pzc}{m}{it}
\DeclareMathOperator{\lOp}{\mathcal{O}_p}

\DeclareMathOperator{\lO}{\mathcal{O}}

\usepackage[utf8]{inputenc}
\usepackage[T1]{fontenc}
\usepackage{hyperref}
\usepackage{url}
\usepackage{booktabs}
\usepackage{amsfonts}
\usepackage{nicefrac}
\usepackage{microtype}
\usepackage{graphicx}
\usepackage{doi}
\usepackage{circuitikz}

\usepackage{enumerate}
\usepackage{times,epsfig,makeidx,amsfonts,amsmath,dcolumn,multirow,graphicx,amssymb,colortbl,bm,url}
\usepackage{algorithm}
\usepackage[]{algorithmic}
\usepackage{mathtools,upgreek,amsfonts,eucal}
\usepackage{afterpage,lscape}
\usepackage{paralist}
\usepackage{xr}
\usepackage{color}
\usepackage{booktabs}
\usepackage{multirow}
\usepackage{subcaption}
\usepackage{capt-of}
\usepackage{soul}

\newtheorem{proposition}{Proposition}

\newtheorem{corollary}{Corollary}

\begin{document}

\setstcolor{red}

\def\spacingset#1{\renewcommand{\baselinestretch}%
{#1}\small\normalsize} \spacingset{1}

\if1\anon
{
  \title{\bf Bayesian variable and hazard structure selection in the General Hazard model}
  \author{Yulong Chen\hspace{.2cm}\\
    Department of Statistical Science,
	University College London\\
    and \\
    Jim Griffin\hspace{.2cm}\\
    Department of Statistical Science,
	University College London\\
    and \\
    Francisco Javier Rubio\hspace{.2cm}\\
    Department of Statistical Science,
	University College London}
  \maketitle
} \fi

\if0\anon
{
  \bigskip
  \bigskip
  \bigskip
  \begin{center}
    {\LARGE\bf Bayesian variable and hazard structure selection in the General Hazard model}
\end{center}
  \medskip
} \fi

\bigskip
\begin{abstract}
The proportional hazards (PH) and accelerated failure time (AFT) models are the most widely used hazard structures for analysing time-to-event data. When the goal is to identify variables associated with event times, variable selection is typically performed within a single hazard structure, imposing strong assumptions on how covariates affect the hazard function.
To allow simultaneous selection of relevant variables and the hazard structure itself, we develop a Bayesian variable selection approach within the general hazard (GH) model, which includes the PH, AFT, and other structures as special cases.
We propose two types of $g$-priors for the regression coefficients that enable tractable computation and show that both lead to consistent model selection. We also introduce a hierarchical prior on the model space that accounts for multiplicity in the number of variables, effects and hazard structures. To efficiently explore the GH model space, we extend the Add–Delete–Swap algorithm to jointly sample variable inclusion indicators and hazard structures.
Simulation studies show accurate recovery of both the true hazard structure and active variables across different sample sizes and censoring levels. Two real-data applications are presented to illustrate the use of the proposed methodology and to compare it with existing variable selection methods.
\end{abstract}

\noindent%
\vfill

\newpage
\spacingset{1.8}

\section{Introduction}\label{sec:intro}

Survival regression models are central to the analysis of time-to-event data across scientific areas such as medicine, biology, engineering, and related fields.
These models are typically specified by linking the hazard function to a set of covariates $\bx \in \mathbb{R}^p$, where the specific manner in which covariates enter the hazard function determines the \textit{hazard structure}.
Popular formulations include the proportional hazards (PH; \citealp{cox:1972}), accelerated failure time (AFT; \citealp{cox:2018}), and accelerated hazards (AH; \citealp{chen:2001}) models, as well as more general hazard structures (GH; \citealp{etezadi:1987,chen:2001}) that encompass these as special cases. These structures differ in whether covariates act multiplicatively on the hazard, on the time scale, or on both.
Recent advances in data linkage and data collection have led to the availability of a large number of candidate covariates, only some of which are truly associated with the outcome. Identifying these covariates has become a key part of analysis leading to parsimony and better interpretability. The literature offers two main perspectives on how to approach the problem of selecting the variables that best explain the times-to-event and the hazard structure that most appropriately represents the data-generating process. One stream focuses on selecting the hazard structure under the assumption that all relevant covariates are included. In the Bayesian setting, \cite{zhang:2018} studied model selection across six parametric survival models (including PH, AFT, and proportional odds), and \cite{zhou:2018} extended this idea to a semiparametric framework.
The second stream of the literature starts from a fixed hazard structure and develops Bayesian variable-selection methods for sparse settings. Recent work includes spike-and-slab priors for AFT models \citep{sha:2006}, generalised $g$-priors that incorporate censoring for PH and AFT models \citep{castellanos:2021,Donato:2023}, and non-local priors for PH and AFT models \citep{niko:2017,rossell:2023}. While these methods address covariate sparsity, they rely on a pre-specified hazard structure, which, if misspecified, can affect both variable selection and inference \citep{rossell:2023}.
The limitations of these two approaches motivate joint selection of variables and the hazard structure. The GH family provides an appealing framework for this task, as effects are easily interpretable within a single, encompassing model. The only previous work in this direction is \cite{tong:2013} who proposed penalised likelihood methods for variable selection
in the GH model.

Beyond identifying which covariates are associated with survival, a question of primary scientific interest is \emph{how} they act: whether a covariate scales the instantaneous risk (a hazard-level effect, as in proportional hazards) or rescales the time evolution of the event process (a time-level effect, as in accelerated failure time). This distinction is not a modelling choice but a statement about the effect of the covariate, with direct consequences for how risk evolves over follow-up and how the covariate should be interpreted and used. It is biologically natural for different covariates to act through different mechanisms, however, standard analyses impose a single structure on all of them, and the consequences of fixing this structure in advance can be substantive rather than merely technical.
As a practical example, \cite{Kwon2014} analyse
 a phase III ipilimumab immunotherapy trial (NCT00861614) of $799$ patients with metastatic castration-resistant prostate cancer who were randomised to the immune checkpoint inhibitor ipilimumab or placebo. The reported Cox PH analysis of overall survival is not significant (hazard ratio $0.85$, $p = 0.053$) but there is clear evidence of a treatment effect with
the hazard ratio only exceeding one during the first five to ten months, and an excess of long-term survivors in the ipilimumab arm \citep{Fizazi2020}. The PH model fails to detect the treatment effect since it does not allow for crossing survival curves, which are common in areas such as immuno-oncology trials \citep{Plana22}.
A detailed analysis of this trial, presented in the Supplementary Material,  highlights the dangers of variable selection under a misspecified hazard structure. 

In this paper, we develop the first Bayesian treatment of variable and hazard selection within the GH model. This is challenging since the model priors on the two sets of model coefficients and the model space must be carefully designed to accommodate the differing roles that covariates may play, to reflect the partitioning of the model space into distinct hazard structures, and to adjust for multiplicity in both variable and hazard structure selection.
Our approach enables identification of the variables that predict survival and whether those variables act on the hazard and time scales, as well as offering principled model uncertainty quantification and providing theoretical guarantees for model selection consistency. Our main contributions are as follows.
We introduce two $g$-prior formulations for the regression coefficients, a likelihood-curvature-matching prior and a product $g$-prior, that align with the GH model's two-component structure and lead to consistent model selection under suitable regularity conditions.
We propose a prior on the model space that generalises the Beta–Binomial model-multiplicity correction \citep{scott:2010} to situations where covariates may appear in multiple roles (hazard level and/or time scale), a key feature of GH models.
We develop computational tools that enable efficient exploration of the model space via a tailored Markov chain Monte Carlo (MCMC) strategy that accommodates moves both within and across hazard structures. We also implement fast approximations to the marginal likelihood, which is required for computing posterior model probabilities and for evaluating MCMC moves, using Laplace and integrated Laplace approximations.

The remainder of the paper is organised as follows. Section \ref{sec:hazregsel} introduces the key notation and the GH model, and formalises the model selection problem within this class of models. Section \ref{sec:priors} outlines prior specifications for structure and variable selection. Section \ref{sec:mcmc} introduces the approximation and MCMC methods used to compute posterior model probabilities. Section \ref{sec:theory} presents theoretical results on the model selection consistency of the proposed approach. Section \ref{sec:simulation} evaluates its performance through a series of simulation studies. Section \ref{sec:realapp} illustrates the use of the proposed methodology using two real-data applications. Finally, Section \ref{sec:discuss} concludes with a discussion and possible extensions of this work.

\section{Model and Variable Selection for the GH Structure}\label{sec:hazregsel}

\subsection*{Hazard-based regression models}
Let $\{o_1,\dots,o_n\}$ be a sample of times-to-event, and $\{c_1,\dots,c_n\}$ be right-censoring times. Suppose that we observe $\bt = \{t_1,\dots,t_n \}$, where $t_i = \min\{o_i,c_i\}$, $i=1,\dots,n$. Let $\delta_i = I(o_i < c_i)$ be the corresponding right-censoring indicators ($\delta_i=1$ observed, $\delta_i=0$ right-censored); and $\bz_i\in {\mathbb R}^d$ be the vector of available covariates associated to the $i$th individual. Throughout, we assume that the times-to-event are independent of the censoring times conditional on the covariates (non-informative censoring).
We adopt the GH structure to model the time-to-event outcomes of interest. The GH model is defined through the hazard function:
\begin{eqnarray}
h(t\mid \balpha,\bbeta,\bxi) = h_0\left( t \exp \left\{ \tbx_i^{\top}\balpha \right\}  \mid \bxi \right)\exp \left\{ \bx_i^{\top}\bbeta \right\},
\label{eq:GH}
\end{eqnarray}
where $h_0(\cdot \mid \bxi)$ is a parametric baseline hazard function, with parameter $\bxi\in{\mathbb R}^r$, $\bx_i\in{\mathbb R}^p$ are the covariates included as hazard-level effects, and $\tbx_i\in{\mathbb R}^q$ are the covariates included as time-level effects, with $\bx_i \subseteq \bz_i$ and $\tbx_i \subseteq \bz_i$. Then, the cumulative hazard function can be written as:
\begin{eqnarray*}
H(t\mid \balpha,\bbeta,\bxi) = H_0\left( t \exp \left\{ \tbx_i^{\top}\balpha \right\}  \mid \bxi \right)\exp \left\{ \bx_i^{\top}\bbeta - \tbx_i^{\top}\balpha \right\},
\end{eqnarray*}
thus avoiding the need for numerical integration whenever $H_0(\cdot \mid \bxi)$ is available in closed-form.
Such models have been used in several practical applications \citep{rubio:2019}, as they provide a simple way to incorporate time-level effects (through the coefficients $\balpha$) and hazard-level effects (through the coefficients $\bbeta$) while retaining tractable estimation. The GH family includes the proportional hazards (PH) model ($\balpha = 0$), the accelerated failure time (AFT) model ($\balpha = \bbeta$ and $\bx_i = \tbx_i$), and the accelerated hazards (AH) model ($\bbeta = 0$) as special cases (see \citealp{rubio:2019} for a more detailed discussion of this hazard structure).
 The GH model \eqref{eq:GH} is identifiable if and only if the baseline hazard is not Weibull \citep{chen:2001} and the covariates are not collinear. If a Weibull baseline hazard is employed, which is proportional to $t^{\nu-1}$ where $\nu>0$ is the shape parameter, the PH, AFT, and AH models coincide, corresponding to a situation in which a simpler model structure suffices.
This theoretical identifiability should be distinguished from  \emph{practical} non-identifiability that can arise in finite samples: even when a model is theoretically identifiable, a fitted baseline hazard that is close to monotonic may be difficult to separate from a Weibull form, and high censoring rates, small sample sizes, or strongly correlated covariates can each flatten the likelihood and blur the distinction between the PH, AFT, and GH structures \citep{rubio:2023}.
We focus on the case where the baseline hazard $h_0$ belongs to the log-location-scale family. That is, the corresponding probability density function (pdf) can be expressed as:
\begin{eqnarray}\label{eq:LLS}
f_0(t\mid \bxi) = f_0(t\mid \mu,\sigma) = \dfrac{1}{\sigma t}f\left(\dfrac{\log(t)-\mu}{\sigma}\right),
\end{eqnarray}
where $\mu \in \mathbb{R}$ is a location parameter, $\sigma > 0$ is a scale parameter, and $f$ is a fully specified pdf with support on $\mathbb{R}$ that is symmetric about $0$. The corresponding hazard function is defined in the usual way $h_0(t\mid \mu,\sigma) = \frac{f_0(t\mid \mu,\sigma)}{1-F_0(t\mid \mu,\sigma)}$.
This family includes distributions commonly used in survival analysis, such as the log-normal, log-logistic, and log-Laplace \citep{vallejos:2015}. Members of this family can accommodate a variety of hazard shapes, including unimodal (log-normal) and monotone decreasing (log-logistic) forms. This interpretability of the parameters also facilitates prior elicitation. Within the GH structure \eqref{eq:GH} with baseline given by \eqref{eq:LLS}, the parameter $\mu$ acts as a log-location parameter on the same scale as the linear predictor $\tbx_i^{\top}\balpha$ and can be interpreted as an intercept.

We consider the reparametrisation $\exp(\nu) = 1/\sigma$, $\theta_0 =  \mu/\sigma$, $\btheta = -{\balpha}/\sigma$, and $\bmeta = -{\bbeta}$, which transforms all parameters to the real line and absorbs the original scale into the coefficient $\btheta$. After some algebra, the log-likelihood function can then be expressed as follows:
\begin{eqnarray}\label{eqn:llh}
\ell_n(\bPsi) &=&  n_o \nu - \sum_{i=1}^{n} \delta_i \log(t_i) + \sum_{i=1}^{n} {\delta_i}\left( \tilde{\bx}_i^{\top}\btheta  e^{-\nu} - \bx_i^{\top}\bmeta \right) \nonumber + \sum_{i=1}^{n} {\delta_i} \log f\left(  e^{\nu}\log(t_i) - \tilde{\bx}_i^{\top}\btheta -\theta_0 \right)\nonumber\\
&+& \sum_{i=1}^{n} \log F\left( -e^{\nu}\log(t_i) + \tilde{\bx}_i^{\top}\btheta +\theta_0 \right) \left[\exp\left(  { \tilde{\bx}_i^{\top}\btheta e^{-\nu} - \bx_i^{\top}\bmeta }\right) -\delta_i\right],
\end{eqnarray}
where $\bPsi= (\nu,\theta_0,\btheta^{\top}, \bmeta^{\top})^{\top}$.
Thus, the likelihood function for the GH structure is as tractable as that of standard parametric AFT and PH models.
The gradient and Hessian of the log-likelihood function are provided in the Supplementary Material.

\subsection{Hazard structure and variable selection}\label{subsec:modsel}
We now formulate the model selection problem, which defines the proposed methodology for performing Bayesian variable and hazard-structure selection within the GH family.
Suppose we begin with the full vector of available covariates, $\bx_i = \tilde{\bx}_i = \bz_i$, of dimension $p$.
Since our goal is to perform simultaneous variable and hazard structure selection, we formalise this task by introducing the following notations for individual variables and hazard structure:
\begin{equation*}
\gamma_j=\left\{
\begin{aligned}
&0, \mbox{ if } \theta_j= 0, \eta_j=0, \quad \text{(No effect)} \\
&1, \mbox{ if } \theta_j \neq 0, \eta_j = 0 , \quad \text{(Time-level effect)}\\
&2, \mbox{ if } \theta_j=  0, \eta_j \neq 0 ,\quad \text{(Hazard-level  effect)}\\
&3, \mbox{ if } \theta_j \neq 0, \eta_j\neq 0  \text{ and }e^{-\nu}\theta_j \neq \eta_j,\quad \text{(Both effects at different level)} \\
&4, \mbox{ if } \theta_j \neq 0, \eta_j\neq 0  \text{ and }e^{-\nu}\theta_j = \eta_j, \quad \text{(Both effects at same level)}
\end{aligned}, \right.
\end{equation*}
corresponding to no effect, a time-level effect only, a hazard-level effect only, and both effects with different/same magnitudes of each covariate $j=1,\ldots,p$.
That is, $\bgamma=(\gamma_1,\ldots,\gamma_p)^{\top}$ determines what covariates enter the model and their role in the model.

The different hazard structures can then be characterised in terms of $\bgamma$: let $\varnothing$ denote the null model and $\text{Hz}(\bgamma)$ denotes the hazard structure of $\bgamma$. These four types of hazard structure are characterised as
\begin{equation*}
\text{Hz}(\bgamma)=\left\{ \begin{aligned}
\varnothing, &\mbox{ if }\bgamma=\boldsymbol{0},\\
\text{AH}, &\mbox{ if }\bgamma \neq \boldsymbol{0}  \text{ and }\forall j\in \{1,...,p\},\gamma_j\in \{0,1\},\\
\text{PH}, &\mbox{ if }\bgamma \neq \boldsymbol{0}  \text{ and }\forall j\in \{1,...,p\},\gamma_j\in \{0,2\},\\
\text{GH}, &\mbox{ if }\bgamma \neq \boldsymbol{0}  \text{ and }\forall j\in \{1,...,p\},\gamma_j\in \{0,1,2,3\}\text{ and } \text{Hz}(\bgamma)\neq \text{AH} \text{ or }\text{PH},\\
\text{AFT}, &\mbox{ if }\bgamma \neq \boldsymbol{0}  \text{ and }\forall j\in \{1,...,p\},\gamma_j\in \{0,4\}.\\
\end{aligned}\right.
\end{equation*}
In other words, for the rest of this paper, we will use GH to denote any GH model defined in \eqref{eq:GH} that is not one of AH, PH, and AFT for convenience. This formulation leads to $4^p-1$ PH, AH, and GH models and $2^p-1$ AFT models, which, including the null model, come down to $4^p+2^p-1$ total models to consider. One implication of this formulation is that we do not consider variables to have both effects at the exact same level for GH models to reduce the number of trivial or redundant model configurations.

Let ${\bPsi}_{\bgamma}= (\nu,\theta_0,\btheta^{\top}_{\bgamma}, \bmeta^{\top}_{\bgamma})^{\top}$ denote the parameters associated to the model $\bgamma$, and $\bX_{o,\bgamma}$ and $\bX_{c,\bgamma}$ denote the submatrices of the design matrices for observed times $\bX_o$ and censored times $\bX_c$, respectively.
The main quantities of interest are the posterior model probabilities, which are given by
\begin{equation*}\label{eq:modelposterior}
\pi(\bgamma \mid \bt) = \frac{p(\bt \mid \bgamma)\pi(\bgamma)}{\sum_{\bgamma} p(\bt \mid \bgamma)\pi(\bgamma)}
= \left( 1 + \sum_{\bgamma' \neq \bgamma}  B_{\bgamma', \bgamma} \frac{\pi(\bgamma')}{\pi(\bgamma)}  \right)^{-1},
\end{equation*}
where $\pi(\bgamma)$ is the model prior probability,
$B_{\bgamma',\bgamma}= p(\bt \mid \bgamma')/p(\bt \mid \bgamma)$ is the Bayes factor between $(\bgamma',\bgamma)$, and $p(\bt \mid \bgamma)= \int \exp\left( \ell(\bPsi_{\bgamma})\right) \pi(\bPsi_{\bgamma} \mid \bgamma) d\bPsi_{\bgamma}$
is the integrated likelihood $\exp\left( \ell(\bPsi_{\bgamma})\right)$ with respect to a prior density $\pi(\bPsi_{\bgamma} \mid \bgamma)$.
Note that the parameters $\btheta_{\bgamma}$ and $\bmeta_{\bgamma}$ are specific to each model, while $(\nu,\theta_0)$ are common to all models.
Next, we discuss the choices for these two parts and the prior on the model space $\pi(\bgamma)$, along with the computational tools used to calculate the model posterior probability $\pi(\bgamma \mid \bt)$.

\section{Priors}\label{sec:priors}

We assign the following vague prior distributions to the common parameters $(\nu,\theta_0)$:
\begin{equation}\label{eqn:commonprior}
    \pi(\nu) = \frac{\beta_{\nu}^{\alpha_{\nu}} e^{\alpha_{\nu}\nu} e^{-\beta_{\nu} e^{\nu}}}{\Gamma(\alpha_{\nu})}, \quad\pi(\theta_0) = N(0,K),
\end{equation}
where $\alpha_{\nu}$ and $\beta_{\nu}$ are hyper-parameters with small values and $K$ is a large constant. The prior on $\nu$ corresponds to a $Gamma(\alpha_\nu, \beta_\nu)$ prior on $1/\sigma$, which is, with small values of $\alpha_{\nu}$ and $\beta_{\nu}$, a common vague prior for an inverse scale parameter in the literature. The parameter $\theta_0$ is the scaled intercept, and we assign it a vague normal prior with zero mean and large variance. The default values we choose are $\alpha_{\nu}=\beta_{\nu}=0.01$ and $K=10^6$.
In Supplementary Material Section 8 we establish that the improper prior $\pi(\mu,\sigma) \propto \sigma^{-k}, k\geq1$ (with the prior on the model-specific coefficients in the following section) leads to a proper posterior, and that every Bayes factor is well defined (under certain conditions on the baseline distribution). This justifies our choice of a vague proper prior resembling the shape of $\sigma^{-2}$.

\subsection{Likelihood curvature matching prior for model-specific coefficients}
The prior on model-specific coefficients, $\pi (\btheta_{\bgamma}, \bmeta_{\bgamma})$, plays a crucial role in Bayesian variable selection as it determines the asymptotic behaviour of the Bayes factor and model selection consistency rates. In the context of Bayesian survival analysis, there are two common types of priors on model coefficients: the $g$-prior \citep{castellanos:2021,zhang:2018,Donato:2023} and non-local priors \citep{rossell:2023,niko:2017}.
The calibration of hyperparameters in non-local priors is essential, as they directly control the threshold beyond which an effect is detectable. However, such calibration could be difficult under the GH model, as this requires simultaneously calibrating the time level and the hazard level, and the same values of the coefficients in time and hazard levels do not usually imply an equal influence on the survival function.
On the other hand, the $g$-prior \citep{Zellner:1986}, having the form $N(\boldsymbol{0}, g\mathcal{I}^{-1})$, incorporates the expected Fisher information matrix $\mathcal{I}$ into the prior covariance matrix, which intuitively reflects the curvature of the log-likelihood. The single hyperparameter $g$ can be calibrated based on, for example, the unit information principle \citep{li:2018}. \cite{castellanos:2021} and \cite{Donato:2023} applied this $g$-prior construction, respectively, to AFT and PH models, demonstrating that in the survival context it can automatically adjust the contribution of each observation to the prior covariance according to its censoring status. We extend the $g$-prior construction to the model coefficients $[\btheta_{\bgamma}, \bmeta_{\bgamma}]$ of the GH model, following the reasoning outlined above. Since the $g$-prior is defined using the expected Fisher information matrix, we first analyse the structure of this matrix in the following result.
\begin{proposition}\label{prop:ef}
    The $[\btheta_{\bgamma},\bmeta_{\bgamma}]$ block of the expected Fisher information could be decomposed into
\begin{equation*}
    \begin{aligned}
\mathcal{I}_{\btheta_{\bgamma},\bmeta_{\bgamma}}(\bPsi_{\bgamma})
&= \sum_{k=1}^n \mathbb E \left[ -\nabla^2_{\btheta_{\bgamma},\bmeta_{\bgamma}} \ell({\bPsi_{\bgamma}}) \mid {\bPsi_{\bgamma}},\delta_k=0 \right]_k \cdot p(\delta_k=0\mid {\bPsi_{\bgamma}}) \\
&+ \sum_{k=1}^n \mathbb E \left[ -\nabla^2_{\btheta_{\bgamma},\bmeta_{\bgamma}} \ell({\bPsi_{\bgamma}})\mid {\bPsi_{\bgamma}},\delta_k=1 \right]_k \cdot p(\delta_k=1\mid {\bPsi_{\bgamma}}),
\end{aligned}
\end{equation*}
 where $\nabla^2_{\btheta_{\bgamma},\bmeta_{\bgamma}} \ell({\bPsi_{\bgamma}})$ is the Hessian matrix of $\ell({\bPsi_{\bgamma}})$ with respect to $[\btheta_{\bgamma},\bmeta_{\bgamma}]$, and
$v_k = \exp\left({ e^{-\nu}\tilde{\bx}_{\bgamma,k}^{\top}\btheta_{\bgamma} - \bx_{\bgamma,k}^{\top}\bmeta_{\bgamma} } \right)$.
Then, $\mathbb E \left[ -\nabla^2_{\btheta_{\bgamma},\bmeta_{\bgamma}} \ell({\bPsi_{\bgamma}}) \mid {\bPsi_{\bgamma}},\delta_k \right]$ has the form $\begin{pmatrix}
  \tilde \bX_{\bgamma}^{\top} \boldsymbol{M}^{\delta_k}_{\btheta_{\bgamma}} \tilde \bX_{\bgamma} & \tilde \bX_{\bgamma}^{\top} \boldsymbol{M}^{\delta_k}_{\btheta_{\bgamma}\bmeta_{\bgamma}} \bX_{\bgamma}\\
  \bX_{\bgamma}^{\top} \boldsymbol{M}^{\delta_k}_{\btheta_{\bgamma}\bmeta_{\bgamma}} \tilde \bX_{\bgamma} & \bX_{\bgamma}^{\top} \boldsymbol{M}^{\delta_k}_{\bmeta_{\bgamma}} \bX_{\bgamma}
\end{pmatrix}$, where $\boldsymbol{M}^{\delta_k}_{\cdot}$ are $n\times n$ diagonal matrices (with no closed-form expressions).
\end{proposition}
The derivation and explicit expressions are provided in the Supplementary Material, including the particular case where $f$ is standard normal. These expressions show that many components of the information matrix depend on integrals of the baseline hazard and its derivative with respect to the lifetime distribution. These integrals have closed-form solutions for PH models and, assuming the censoring times are known, also for AFT models. However, for the AH and GH models, closed-form solutions are not available. Moreover, the expected Fisher information depends on the value of the regression coefficients, which are unknown, resulting in a non-standard distribution. In summary, there are two challenges that prevent us from directly using the expected Fisher information matrix in the prior covariance: (i) it does not have a closed-form expression, and (ii) it depends on the unknown regression coefficients themselves.

The first challenge on tractability of the expression of the expected Fisher information matrix could be solved by using the sample average of the observed Fisher information matrix as an estimate of the expected Fisher information, which reflects the curvature of the finite-sample likelihood function.
For the second challenge, there are two common solutions: either evaluate the parameters at $0$ \citep{castellanos:2021} or at the model-specific maximum likelihood estimate (MLE) \citep{WangGeorge:2007}. The former choice
follows from arguing that only the null model, which assigns positive probabilities of them being $0$, provides information about the model-specific parameters  \citep{castellanos:2021}. The latter choice, our preferred choice, is better at capturing the large sample covariance structures of the model coefficients and the local geometry under the current model \citep{li:2018}. This feature is particularly important in our setting,
as it captures the structure of the coefficients both between and within the time and hazard levels.
Combining these two solutions leads to the following likelihood-curvature-matching (LCM) $g$-prior:
\begin{equation}\label{eq:ebprior}
    \left(\boldsymbol{\theta}_{\bgamma}, \boldsymbol{\eta}_{\bgamma} \right)^{\top} \mid g_E \sim N(\boldsymbol 0, ng_E\mathcal J_{\boldsymbol{\theta_{\bgamma}},\boldsymbol{\eta_{\bgamma}}}(\widehat\bPsi_{\bgamma})^{-1}),
\end{equation}
where $\mathcal J_{\boldsymbol{\theta_{\bgamma}},\boldsymbol{\eta_{\bgamma}}}(\widehat\bPsi_{\bgamma})$ is the $\btheta_{\bgamma}$ and $\bmeta_{\bgamma}$ blocks of the observed Fisher information matrix with all parameters evaluated at the MLE $\widehat{\bPsi}_{\bgamma}$, and $g_E>0$ is a hyperparameter. The single hyperparameter $g_E$ plays the same role as in the standard $g$-prior, and we set its default value to $g_E = 1$ based on the unit information principle. For AFT models, where $\bmeta_{\bgamma}=e^{-\nu}\btheta_{\bgamma}$, the prior \eqref{eq:ebprior} is placed on $\btheta_{\bgamma}$ alone, using the corresponding block of $\mathcal J$.
In Section 14 of the Supplementary Material, we consider its properties
 under extreme censoring, and empirically find that
 it encodes a preference for sparsity. We  also discuss an alternative
choice based on predictive matching \citep{Bayarri2012CriteriaSelection,castellanos:2021} which avoids this preference.

A natural extension of \eqref{eq:ebprior} consists of placing a prior on the hyperparameter $g_E$. Although many options exist \citep{Liang2008}, we select a single specification for illustration, as benchmarking these priors is not the central objective of this paper and, as we show in Section \ref{sec:mcmc}, the computational methods and costs are comparable across different choices. While our main analysis considers fixed $g_E$, we also examine the robust prior of \cite{Bayarri2012CriteriaSelection} due to its appealing theoretical properties. The robust prior has the following form \citep{castellanos:2021}:
\begin{equation}\label{eqn:robust_g}
    \pi(g_{E}) =\frac{1}{2} \sqrt{\frac{1+n}{n(d_{\bgamma}+1)}}\left(g_{E}+\frac{1}{n}\right)^{-3 / 2},\quad g_{E}>\frac{1+n}{n(d_{\bgamma}+1)}-\frac{1}{n},
\end{equation}
where $d_{\bgamma}=p_{\bgamma}+q_{\bgamma}$ is the model size, with $p_{\bgamma}$ and $q_{\bgamma}$ the numbers of hazard-level and time-level coefficients in model $\bgamma$ (for AFT models, $p_{\bgamma}=0$ and $q_{\bgamma}$ is the number of included variables), and $n$ is the sample size.

\subsection{Product prior for model-specific coefficients}\label{ssec:fbprior}

Although the prior in \eqref{eq:ebprior} offers practical benefits, its reliance on the observed survival times and censoring indicators might be less appealing to some users. Consequently, we now seek an alternative specification of the
$g$-prior for the GH model.
Proposition~\ref{prop:ef} shows that the expected Fisher information matrix for $\btheta_{\bgamma}$ and $\bmeta_{\bgamma}$ can be written in the form of $\tilde \bX_{\bgamma}^{\top} \boldsymbol{M}_{\btheta_{\bgamma}}\tilde \bX_{\bgamma}$ and $\bX_{\bgamma}^{\top} \boldsymbol{M}_{\bmeta_{\bgamma}}\bX_{\bgamma}$, where $\boldsymbol{M}_{\btheta_{\bgamma}}$ and $\boldsymbol{M}_{\bmeta_{\bgamma}}$ are diagonal matrices.
Motivated by this structure, we propose a simpler and intuitive Product $g$-prior specification for AH, PH, AFT and GH models, and their corresponding coefficients:
\begin{eqnarray}\label{eqn:fbprior}
{\pi_{P}(\btheta_{\bgamma},\bmeta_{\bgamma} \mid g_{C_t}, g_{C_h})}
  &=&  N\left(\btheta_{\bgamma};\boldsymbol{0},  g_{C_t} n \left(\tbX_{\bgamma}^\top \tbX_{\bgamma}\right)^{-1}\right) \cdot
       N\left(\bmeta_{\bgamma};\boldsymbol{0}, g_{C_h} n\left(\bX_{\bgamma}^\top \bX_{\bgamma}\right)^{-1}\right),
\end{eqnarray}
where $g_{C_t}>0$ and $g_{C_h}>0$ are hyperparameters. For AFT models, the prior becomes $N\left(\btheta_{\bgamma};\boldsymbol{0},  g_{C_t} n \left(\tbX_{\bgamma}^\top \tbX_{\bgamma}\right)^{-1}\right)$, as $e^{-\nu}\btheta_{\bgamma} =\bmeta_{\bgamma}$.
In other words, the formulation in \eqref{eqn:fbprior} assumes that all samples contribute equally to the prior covariance.
Consequently, we expect the prior covariance for $\btheta_{\bgamma}$ to deviate slightly from the inverse of the expected Fisher information matrix in the AH, AFT, and $\btheta_{\bgamma}$ component of the GH models, while it remains aligned with the expected information for the PH model and the $\bmeta_{\bgamma}$ component of the GH model, though it may be sensitive to the calibration of $g_{C_t}$ and $g_{C_h}$ depending on the quality of the approximation.
The default value we set is $g_{C_t}=g_{C_h}=1$ for the sake of comparison with \eqref{eq:ebprior}. However, since the censoring contribution is, to some extent, absorbed into $g_{C_t}$ and $g_{C_h}$, a more thoughtful choice that reasonably accounts for this information could be considered.

\subsection{Prior on the model space}\label{ssec:priormod}
For the prior on the model space, $\pi(\bgamma)$, we aim to specify a non-informative prior that assigns equal weight across model sizes and hazard structures. A naive uniform prior, $\pi(\bgamma) = (4^p + 2^p - 1)^{-1}$, is not appropriate because, for example, the number of GH models far exceeds that of PH, AH, and AFT models, thereby allocating most of the prior mass to GH models. To correct for this kind of imbalance, we adopt a multiplicity-adjusted formulation that extends the widely used Beta–Binomial prior \citep{scott:2010}.

We identify three sources of multiplicity that we would like to control in our setting: (i) variation in the number of variables included, (ii) variation in the hazard structure, and (iii) variation in the number of effects. We distinguish here between the number of variables and the number of effects: there are $p$ variables in total, and each variable can contribute $0$, $1$, or $2$ effects. A variable having $0$ effects corresponds to not being included in the model; $1$ effect corresponds to having a time level OR hazard level effect, or both effects at the same scale; $2$ effects corresponds to having both effects at different scales. This three-layered multiplicity naturally suggests a hierarchical prior structure, which sequentially accounts for these sources. Define $\boldsymbol{\Lambda}$ as a binary vector of length $p$ with $\boldsymbol{\Lambda}_j=1$ indicating variable $j$ is included. Let $p_i$ denote the number of variables with $\gamma_j=i$ for $i\in\{1,2,3,4\}$, so that the number of included variables is $|\boldsymbol{\Lambda}|=\sum_{i=1}^4 p_i$, and $p_0=p-|\boldsymbol{\Lambda}|$ counts the not-included variables. Denote $\omega=|\boldsymbol{\Lambda}|+p_3$ as the number of effects. We set $\pi(\bgamma=\mathbf{0})=\pi(\boldsymbol{\Lambda}=\mathbf{0})$ for the null model and, for $|\boldsymbol{\Lambda}|\ge 1$, consider the factorisation
\begin{equation}\label{eq:gamfact}
\pi(\bgamma) = \pi(\boldsymbol{\Lambda})\,\pi(\hz(\bgamma)\mid \boldsymbol{\Lambda})\left[\pi(\omega \mid \text{GH},\boldsymbol{\Lambda})\,\pi(\bgamma\mid\omega,\text{GH},\boldsymbol{\Lambda})\right]^{\mathbf{1}_{\hz(\bgamma)=\text{GH}}},
\end{equation}
in which every factor is a proper distribution.
For $\pi(\boldsymbol{\Lambda})$, we consider only the number of included variables (not their individual effects) and adopt the Beta–Binomial prior of \cite{scott:2010}. Marginalising a prior $\mathrm{Beta}(a_\Lambda, b_\Lambda)$ on the inclusion probability yields the Beta–Binomial prior $\pi(\boldsymbol{\Lambda}) \propto \mathrm{B}\left(a_\Lambda+|\boldsymbol{\Lambda}|, b_\Lambda+p_0\right),$
where $\mathrm{B}(\cdot,\cdot)$ is the Beta function. The hyperparameters $a_\Lambda$ and $b_\Lambda$ control prior sparsity: for a fixed value of $a_\Lambda$, a larger $b_\Lambda$ favours fewer included variables. A simple and intuitive approach to calibrate them is by setting $a_\Lambda=1$ and interpreting $b_\Lambda$ as the prior odds between $p_0$ and $|\boldsymbol{\Lambda}|$. The default choice we recommend is setting $a_\Lambda=b_\Lambda=1$, which corresponds to a $\mathrm{Beta}(1,1)$ (uniform) prior on the inclusion probability and hence a non-preferential prior for the number of included variables. Apart from the Beta-Binomial prior, there are plenty of choices for $\pi(\boldsymbol{\Lambda})$: for example, for larger $p$ settings, the priors considered by \cite{LeySteel:2007} might be of interest.

Next, for the conditional $\pi(\hz(\bgamma)\mid \boldsymbol{\Lambda})$, we consider a categorical distribution:
\begin{equation}\label{eq:priorgam_2}
\pi(\hz(\bgamma)=l \mid \boldsymbol{\Lambda}) = \frac{h_l}{h_{AH}+h_{PH}+h_{AFT}+h_{GH}}, \qquad l\in\{\text{AH},\text{PH},\text{AFT},\text{GH}\},
\end{equation}
where $h_{GH},h_{AH},h_{PH},h_{AFT} >0$ are hyperparameters. These four hyperparameters, or more precisely the ratios between them, encode the prior beliefs on the four hazard structures. The default choice is setting all to $1$, leading to a non-preferential prior on hazard structures with each structure having prior probability $1/4$ given $\boldsymbol{\Lambda}$.

While the marginal model space prior for AH, PH, and AFT models are Beta-Binomial, it is not for GH models as they have different multiplicity than the other structures at each given level of $|\boldsymbol{\Lambda}|>1$ (for $|\boldsymbol{\Lambda}|=1$ the only valid GH model has a single active variable with $\gamma_j = 3$ hence no correction is needed).
For a given $|\boldsymbol{\Lambda}|$ and $\hz(\bgamma) = \text{GH}$, the possible number of effects, $\omega$, takes values from $|\boldsymbol{\Lambda}|$ to $2|\boldsymbol{\Lambda}|$. In general, $N_\omega$, the number of GH models for a fixed $|\boldsymbol{\Lambda}|$ that has $\omega$ effects, is
\begin{equation*}
N_\omega \mid |\boldsymbol{\Lambda}|>1=\binom{|\boldsymbol{\Lambda}|}{\omega-|\boldsymbol{\Lambda}|} \cdot 2^{2\cdot|\boldsymbol{\Lambda}|-\omega}-2\cdot \mathbf{1}_{\omega=|\boldsymbol{\Lambda}|} .
\end{equation*}
In other words, if all GH models with a given $\boldsymbol{\Lambda}$ were equally likely, $\omega-|\boldsymbol{\Lambda}|$ would follow a $\text{Bin}(|\boldsymbol{\Lambda}|, 1/3)$ distribution (\textit{i.e.} each included variable has both effects with probability $1/3$) with a correction term when $\omega = |\boldsymbol{\Lambda}|$. Without adjusting for this source of multiplicity, the prior probability of GH models would concentrate around models with $\frac43|\boldsymbol{\Lambda}|$ effects. To address this, we weight each GH model inversely to the binomial probability of its number of effects, in spirit of the multiplicity correction of Beta-Binomial prior:
\begin{equation}\label{eq:priorgam_3}
\pi(\omega \mid \text{GH},\boldsymbol{\Lambda}) = \frac{N_\omega w_\omega}{\sum_{\omega'=|\boldsymbol{\Lambda}|}^{2|\boldsymbol{\Lambda}|} N_{\omega'} w_{\omega'}}, \qquad \pi(\bgamma\mid\omega,\text{GH},\boldsymbol{\Lambda}) = \frac{1}{N_\omega},
\end{equation}
where $w_\omega=\left[P_{Bin}(\omega-|\boldsymbol{\Lambda}|\mid |\boldsymbol{\Lambda}|,q)\, c_\omega\right]^{-1}$, $c_\omega=1-2^{1-|\boldsymbol{\Lambda}|}$ if $\omega=|\boldsymbol{\Lambda}|$ and $c_\omega=1$ otherwise, and $P_{Bin}(x \mid a,b)$ is the probability mass function of the Binomial distribution with parameter $n=a$ and $p=b$ evaluated at $x$. $\pi(\omega \mid \text{GH},\boldsymbol{\Lambda})$ controls the prior spread over the number of effects, and $\pi(\bgamma\mid\omega,\text{GH},\boldsymbol{\Lambda})$ adjusts the weights for each $\omega$ level such that the total weights of GH models is the same as other structures at each level of $|\boldsymbol{\Lambda}|$. The hyper-parameter $q\in [0,1]$ controls the number of effects in GH models: the closer it is to $1$, the more heavily we penalise the number of effects. The default choice is $q=1/3$, so that $\pi(\omega\mid\text{GH},\boldsymbol{\Lambda})=1/(|\boldsymbol{\Lambda}|+1)$ is uniform.

With these components, the marginal prior of $|\boldsymbol{\Lambda}|$ is exactly $\text{Beta-Binomial}(a_\Lambda,b_\Lambda)$ and, given $|\boldsymbol{\Lambda}|\ge 1$, hazard structure $l$ has prior probability $h_l/\sum_{l'}h_{l'}$. The full expression of $\pi(\bgamma)$ is given in the Supplementary Material.
This proposal builds on the widely used Beta-Binomial prior for the model space and provides multiplicity control over the number of variables, the number of effects, and the hazard structure (see Section~\ref{sec:discuss} for a further source of multiplicity).
Moreover, it constitutes a fully Bayesian prior with a set of interpretable hyperparameters, allowing convenient multi-level adjustment of sparsity relative to the given non-preferential baselines.

\section{Computations}\label{sec:mcmc}

\subsection{Marginal Likelihood Approximations}
We begin by describing the computations required to implement the Bayesian variable selection strategy under the LCM prior for model-specific parameters in \eqref{eq:ebprior}. \cite{WangGeorge:2007} proposed an alternative to the Laplace approximation (LA), namely the integrated Laplace approximation (ILA), to approximate the marginal likelihood under this class of priors. Briefly, the key idea of ILA is that, instead of expanding the log-posterior at the maximum a posteriori (MAP) estimate, as is typical in the standard LA \citep{rossell:2023}, ILA expands the log-likelihood function at the MLE and evaluates the marginal likelihood integral using this approximated likelihood. The integral admits a closed-form if the prior on the model parameters is defined as a multivariate normal. The resulting procedure is much faster than the usual LA, as we need to conduct only one optimisation scheme to calculate the MLE for each model, instead of two optimisation steps required by the LA, one to obtain the MLE required in the specification of the LCM prior and one to obtain the MAP.

A direct application of ILA does not yield a closed-form expression for the marginal likelihood, since $\pi(\nu)$ follows a log-Gamma distribution, whereas ILA requires all prior densities to be Gaussian or uniform. To address this, we approximate $\pi(\nu)$ with a normal distribution. In particular, for the choice $Gamma(0.01,0.01)$, which is the vague prior we suggested for $1/\sigma$ in the previous section, a normal prior with mean $9.34$ and standard deviation $41.15$ provides an accurate approximation, with $<1\%$ difference in probability density for $0.32 < \sigma < 8\times 10^5$.

Let $\mathcal J_{aa}$ denote the block of the observed Fisher information matrix of the parameter $a$, and $\mathcal J_{a,b}$ denotes the crossing block of parameters $a,b$. With the approximation for $\pi(\nu)$, the marginal likelihood $p(\bt \mid \bgamma)$ has the following closed-form under the ILA:
\begin{equation}\label{eq:ebmarglik}
    \widehat p(\bt \mid \bgamma)= \exp[\ell(\widehat{\bPsi}_{\bgamma})](1+ng_E)^{-d_{\bgamma}/2} |\mathbf{P}_{\bgamma}|^{-\frac{1}{2}} \exp \left\{\frac{1}{2} \mathbf{h}^{\top}_{\bgamma} \mathbf{P}_{\bgamma}^{-1} \mathbf{h}_{\bgamma}+C_{\bgamma} \right\},
\end{equation}
where $\bf h_{\bgamma}$ is a vector of length $2$, $\bf P_{\bgamma}$ is a $2\times2$ symmetric matrix, and $C_{\bgamma}$ is a scalar. Each of these terms is composed of entries from the observed Fisher information matrix evaluated at the MLE $\widehat{\bPsi}_{\bgamma}$. The full expression for those terms and the detailed derivation are provided in Section 6 of the Supplementary Material. The MLE can be computed numerically using gradient descent or any general-purpose optimisation method. We employ the \verb|nlminb| function in \verb|R|, which works well in our simulations and real-data applications.

If the robust prior on $g_E$ in \eqref{eqn:robust_g} is used, we could simply perform a one-dimensional integration:
\begin{equation*}
\widehat p(\bt \mid \bgamma) = \int_0^{\infty} \widehat p(\bt \mid \bgamma,g_E)\pi(g_E) dg_E .
\end{equation*}
Evaluating this integral through numerical integration is computationally inexpensive because we do not need to re-evaluate the MLE and the observed information matrix at each iteration of the integration. Moreover, this method is applicable for all types of properly defined priors on $g_E$. In our implementation, we use the \texttt{integrate} function within the \texttt{stats} package of base \texttt{R} to perform this numerical integration.

For the Product prior \eqref{eqn:fbprior}, we adopt the Laplace Approximation (LA) to obtain the approximated marginal likelihood:
\begin{equation}\label{eqn:fbmarglik}
\tilde{p}(\bt \mid \bgamma)= \exp\{\ell(\tilde{\bPsi}_{\bgamma}) + \log \pi_{P}(\tilde{\bPsi}_{\bgamma}) \}
(2\pi)^{(d_{\bgamma}+2)/2} \left|H(\tilde{\bPsi}_{\bgamma}) + \nabla^2 \log\pi_{P}(\tilde{\bPsi}_{\bgamma}) \right|^{-1/2},
\end{equation}
where $\tilde{\bPsi}_{\bgamma}= \arg\max_{\bPsi_{\bgamma}} \left\{ \ell(\bPsi_{\bgamma}) + \log \pi_{P}(\bPsi_{\bgamma})\right\}$ is the maximum a posteriori (MAP) under the Product prior and $\pi_{P}(\bPsi_{\bgamma}) = \pi_{P}(\btheta_{\bgamma},\bmeta_{\bgamma} \mid g_{C_t}, g_{C_h})\pi(\nu)\pi(\theta_0)$. We use the \verb|nlminb| function to obtain the MAP in our implementation. For both proposed priors, ach evaluated model requires one optimisation, and the models whose observed information or negative Hessian at the MAP is not positive definite receive zero posterior probability. 

\subsection{Model Space Exploration}
Given the marginal likelihood (or an approximation thereof, under either prior structure) and the prior on model probabilities, we can now obtain samples from the model posterior. For low-dimensional scenarios, it is possible to compute the posterior probabilities for all models. However, since the number of models is $4^p + 2^p - 1$, this becomes infeasible even for relatively small values of $p$. This implies the need for using Markov chain Monte Carlo methods to explore high-probability regions of the model space. To sample from the posterior of the models, $\pi(\bgamma \mid \bt)$, we consider the Add-Delete-Swap (ADS) algorithm \citep{Madigan1995BayesianData,tadesse:2021} due to its extendable design. The original ADS algorithm has three possible moves: the Add move (\textbf{A}) and Delete move (\textbf{D}) pick one variable at random and include or exclude it from the current model; the Swap move (\textbf{S}) draws a pair of variables: one from the current model and one not in the current model and swaps them. Direct application of the ADS algorithm on $\bgamma$ would be problematic in two ways: (i) it is not compatible with the setup of $\hz(\bgamma)$ in Section \ref{sec:hazregsel}, as an AFT model is almost impossible to reach from a GH/AH/PH model with these three moves; (ii) as hinted in Section \ref{ssec:priormod}, the ratio of the number of GH models to that of the other three types grows exponentially with the number of included variables. For example, if the true model is a PH model with a few active variables, then the sampler may take a large number of iterations to converge and could easily get stuck in local modes. Therefore, it is necessary to extend the ADS sampler to accommodate the need to move between the four types of hazard structures by introducing corresponding moves.

\begin{figure}[ht]
  \begin{minipage}[t][][b]{.3\linewidth}
    \centering
    \resizebox{\textwidth}{!}{%
        \begin{circuitikz}
            \tikzstyle{every node}=[font=\small]
            \draw  (7.75,17) circle (1cm) node {\LARGE GH} ;
            \draw  (7.75,20.5) circle (1cm) node {\LARGE PH} ;
            \draw  (4.5,15.25) circle (1cm) node {\LARGE AH} ;
            \draw  (11,15.25) circle (1cm) node {\LARGE AFT} ;
            \draw [<->, >=Stealth] (8.75,19.25) -- (10.5,16.5)node[pos=0.5, fill=white]{$\textbf{C}^A$};
            \draw [<->, >=Stealth] (6.75,19.25) -- (5,16.5)node[pos=0.5, fill=white]{$\textbf{C}^A$};
            \draw [<->, >=Stealth] (5.75,15.25) -- (9.75,15.25)node[pos=0.5, fill=white]{$\textbf{C}^A$};
            \node [font=\small] at (7.75,20) {\textbf{A/D/S}};
            \node [font=\small] at (4.5,14.75) {\textbf{A/D/S}};
            \node [font=\small] at (7.75,16.5) {\textbf{A/D/S/C}};
            \node [font=\small] at (11,14.75) {\textbf{A/D/S}};
            \draw [<->, >=Stealth] (7.75,19.25) -- (7.75,18.25)node[pos=0.5, fill=white]{\textbf{C}};
            \draw [<->, >=Stealth] (5.5,15.75) -- (6.75,16.5)node[pos=0.5, fill=white]{\textbf{C}};
            \draw [<->, >=Stealth] (8.75,16.5) -- (10,15.75)node[pos=0.5, fill=white]{$\textbf{C}^A$};
        \end{circuitikz}
    }%
    \caption{Illustration of the extended ADS sampler. The bold letters represent different moves.}
    \label{fig:mcmc}
  \end{minipage}\hfill
  \begin{minipage}[t][][b]{.6\linewidth}
    \centering
    \resizebox{\textwidth}{!}{%
        \begin{tabular}{@{}cccc@{}}
        \toprule
        \multicolumn{2}{c}{Move} &
          Usage &
          Detail \\ \midrule
        \multirow{4}{*}{\begin{tabular}[c]{@{}c@{}}Within\\ hazard\end{tabular}} &
          \textbf{A} &
          All &
          Add one variable to the current model \\ \cmidrule(l){2-4}
         &
          \textbf{D} &
          All &
          Remove one variable from the current model \\ \cmidrule(l){2-4}
         &
          \textbf{S} &
          All &
          \begin{tabular}[c]{@{}c@{}}Swap the value of $\gamma$ between two variables with different $\gamma$\end{tabular} \\ \cmidrule(l){2-4}
         &
          \multirow{2}{*}{\textbf{C}} &
          GH &
          \multirow{2}{*}{\begin{tabular}[c]{@{}c@{}}Change $\gamma$ of an included variable to a different value in $\{1,2,3\}$\end{tabular}} \\ \cmidrule(r){1-1} \cmidrule(lr){3-3}
        \multirow{2}{*}{\begin{tabular}[c]{@{}c@{}}Between\\ hazard\end{tabular}} &
           &
          AH, PH, GH &
           \\ \cmidrule(l){2-4}
         &
          $\textbf{C}^A$ &
          All &
          \begin{tabular}[c]{@{}c@{}}When all included variables have the same values of $\gamma \in \{1,2,3,4\}$,\\ change all to another; for GH changes only allowed between/from AFT \end{tabular} \\ \bottomrule
        \end{tabular}%
    }
    \captionof{table}
      {%
        Detailed description of each of the moves in the left figure. Note that \textbf{C} is used for both within and between structure sampling. %
        \label{tab:MCMC}%
      }
  \end{minipage}
\end{figure}
Figure \ref{fig:mcmc} provides a general picture of the algorithm. In our setting, we keep the three moves from the original ADS algorithm for moving within the model space of every single structure. One modification we made to the \textbf{S} move is that, for GH models, we only swap one of the time-level or hazard-level effects (at random) for the chosen pair of variables, which prevents the jumps from being too large. For moving between the model structures, we design a pair of moves: Change (\textbf{C}) and Change All ($\textbf{C}^A$), as shown in Table~\ref{tab:MCMC}. The \textbf{C} move randomly chooses a variable $j$ from the current model and randomly changes $\gamma_j$ to one of the other two values in $\{1,2,3\}$. Note that \textbf{C} move is not only used for between-hazard transition, but also for within-hazard sampling of GH models. The $\textbf{C}^A$ move, on the other hand, directly changes the hazard structure by changing all variables in the current model simultaneously. The transition from GH to AFT requires all variables in the model to have both effects, and this proposal-generating step is the same regardless of the prior specification.

Having the new candidate model, we use the standard Metropolis-Hastings update to decide if the candidate model will be accepted. Let $\bgamma'$ denote the candidate model and $Q$ be the Hastings ratio, the acceptance probabilities under the two proposed priors are given by: $\text{acc}_{LCM} = \min \left\{ 1,\dfrac{\widehat p(\bt \mid \bgamma')p(\bgamma')}{\widehat p(\bt \mid \bgamma)p(\bgamma)}\cdot Q \right\}$, $\text{acc}_{P} = \min \left\{ 1,\dfrac{\tilde p(\bt \mid \bgamma')p(\bgamma')}{\tilde p(\bt \mid \bgamma)p(\bgamma)}\cdot Q \right\}$.
The detailed expression for $Q$ depends on the selected move in the extended ADS sampler. Details of the extended ADS sampler and methods of summarising the MCMC samples are provided in the Supplementary Material.

\section{Asymptotic Theory}\label{sec:theory}
In this section, we characterise the asymptotic Bayes factor rates and model selection consistency for both prior types (LCM and product priors), using their corresponding approximations (ILA and LA). Throughout, we denote by $\varphi_0(o_i, c_i, \bz_i)$ the true generating mechanism, which may depend on additional unavailable covariates (that is, $\tbx_i\subseteq \bz_i$ and $\bx_i\subseteq \bz_i$). Let us denote $\bv = (t_1, o_1, c_1, \delta_1)^{\top}$, and define the functions
\begin{equation*}
\begin{split}
m(\bPsi_{\bgamma}, \bv) &= \delta_1 \log h(o_1 \mid \bPsi_{\bgamma}) -  H(t_1 \mid \bPsi_{\bgamma}),\\
M(\bPsi_{\bgamma}) &= E_{\varphi_0}\left[m(\bPsi_{\bgamma}) \right],
\end{split}
\end{equation*}
which represent the contribution of a single observation to the log-likelihood function, and the expected log-likelihood function, respectively. Define ${\bPsi}^*_{\bgamma} =\operatorname{argmax}_{\Gamma_{\bgamma}} M({\bPsi}_{\bgamma})$, the value that maximises the expected log-likelihood. Let also $\bgamma^*$ be the indicator of the non-zero variables (most parsimonious model) that maximise $M({\bPsi})$, that is, $\bPsi^*_{\bgamma^*}=\operatorname{argmax}_{\Gamma_{\bgamma}} M({\bPsi}_{\bgamma}^*)$. Let us also denote by $\bX_{\bgamma}$ and $\tbX_{\bgamma}$, the design matrices associated with model $\bgamma$, and $\bX_{\bgamma,O}$ and $\tbX_{\bgamma,O}$, the design matrices associated with model $\bgamma$ and the uncensored observations.
 The results depend on
the standard regularity assumptions for $M$-estimation under possible misspecification with a
\emph{fixed} model dimension $d_{\bgamma}$
\citep{rossell:2018,rossell:2023}, tailored to the GH likelihood, and are listed in the Supplementary Material as Conditions C1--C7.
 The compactness in C1 is a technical convenience: $\Gamma_{\bgamma}$ may be taken arbitrarily large, and the results extend to locally compact parameter spaces \citep{white:1981}. The positive-definiteness in C2 is a full-column-rank (non-collinearity) requirement, which needs at least $n_o\ge d_{\bgamma}+2$ uncensored observations to identify a model with $d_{\bgamma}+2$ parameters (see the propriety analysis in the Supplementary Material).
 The non-singular information in C7 is the local-identifiability condition, whose finite-sample counterpart is the practical near-identifiability discussed in Section~\ref{sec:hazregsel}.These conditions do \emph{not} cover the high-dimensional regime in which $p$ grows with $n$ (or $p\gg n$) where C2 eventually fails for the full model, and the fixed-$d_{\bgamma}$ Laplace and integrated-Laplace expansions underlying Propositions~\ref{prop:BFrates_EB} and~\ref{prop:BFrates} no longer apply.

We first provide a characterisation of the Bayes factor rates for the LCM prior obtained via the integrated Laplace approximation.
We say that the model indicator vector $\boldsymbol{\gamma}$ contains $\boldsymbol{\gamma}^*$, denoted by $\boldsymbol{\gamma}^* \subset \boldsymbol{\gamma}$, if every covariate that is active in the  model $\boldsymbol{\gamma}^*$ is also active in the model $\boldsymbol{\gamma}$. Formally, for all $j = 1, \ldots, p$, $\gamma_j^* \neq 0 \;\Rightarrow\; \gamma_j \neq 0$ and $\gamma_j \in \{\gamma_j^*,\, 3\}$, while $\gamma_k$ is unrestricted for any $k$ such that $\gamma_k^* = 0$.

\begin{proposition}\label{prop:BFrates_EB}
Consider the GH model \eqref{eq:GH} with log-location–scale baseline hazard \eqref{eq:LLS}, together with the prior \eqref{eqn:commonprior} on the common parameters and \eqref{eq:ebprior} on the model-specific parameters, where $g_E$ is assumed to be non-decreasing in $n$. Let $\widehat{B}_{\bgamma,\bgamma^*}= \frac{\widehat{p}(\bt \mid \bgamma)}{\widehat{p}(\bt \mid \bgamma^*)}$ denote the Bayes factor obtained via an integrated Laplace approximation, where $\bgamma^*$ is the model maximising $M(\bPsi_{\bgamma})$, and $\bgamma \neq \bgamma^*$ is another regression model of type \eqref{eq:GH}.
Assume that both models associated with $\bgamma^*$ and $\bgamma$ satisfy Conditions C1–C7 in Section 7 of the Supplementary Material. Then,
\begin{enumerate}
\item Overfitted models. If $\bgamma^* \subset \bgamma$, then
$$
\log (\widehat{B}_{\bgamma, \bgamma^*})= \frac{d_{\bgamma^*} - d_{\bgamma}}{2}\log (1+ng_E) + \lO_p(1),
$$
\item Non-overfitted models. If $\bgamma^* \not\subset \bgamma$, then
$$
\log(\widehat{B}_{\bgamma, \bgamma^*})= -n[M(\bPsi^*_{\bgamma^*}) - M(\bPsi^*_{\bgamma})]
+ \frac{d_{\bgamma^*} - d_{\bgamma}}{2}\log (1+ng_E) + \lOp(1).
$$
\end{enumerate}
\end{proposition}
Based on the results of Proposition \ref{prop:BFrates_EB}, we can establish model selection consistency for the LCM prior, as described in the following corollary.
\begin{corollary}\label{cor:EB}
Assuming the prior probability assigned to $\bgamma^*$ is not ${\bf 0}$, \textit{i.e.} $\pi (\bgamma^*)\not = 0$, and the Conditions C1-C7 hold, then $\widehat{\pi}\left(\boldsymbol{\bgamma}^* \mid \boldsymbol{t}, \boldsymbol{\delta}\right) \xrightarrow{\Pr} 1$ as $n\to\infty$.
\end{corollary}

The following result provides a characterisation of the Bayes factor rates obtained with the Product prior \eqref{eqn:fbprior}, and approximated with the Laplace approximation.

\begin{proposition}\label{prop:BFrates}
Consider the GH model \eqref{eq:GH} with log-location–scale baseline hazard \eqref{eq:LLS}, together with the prior \eqref{eqn:commonprior} on the common parameters and \eqref{eqn:fbprior} on the model-specific parameters, where $g_{C_t}$ and $g_{C_h}$ are assumed to be non-decreasing in $n$. Let $\tilde{B}_{\bgamma,\bgamma^*} = \frac{\tilde{p}(\bt \mid \bgamma)}{\tilde{p}(\bt \mid \bgamma^*)}$ denote the Bayes factor obtained via a Laplace approximation, where $\bgamma^*$ is the model maximising $M(\bPsi_{\bgamma})$, and $\bgamma \neq \bgamma^*$ is another hazard-based regression model of type \eqref{eq:GH}.
Assume that both models associated with $\bgamma^*$ and $\bgamma$ satisfy Conditions C1–C7 in Section 7 of the Supplementary Material. Then,

\begin{enumerate}

\item  Overfitted models. If $\bgamma^* \subset \bgamma$, then
$$
\log (\tilde{B}_{\bgamma, \bgamma^*}) = {\frac{p_{\bgamma^*} - p_{\bgamma}}{2}}\log (n g_{C_h}) +  {\frac{q_{\bgamma^*} - q_{\bgamma}}{2}}\log (ng_{C_t}) + \lO_p(1).
$$
\item Non-overfitted models. If $\bgamma^* \not\subset \bgamma$, then
$$
\log(\tilde{B}_{\bgamma, \bgamma^*})= -n[M(\bPsi^*_{\bgamma^*}) - M(\bPsi^*_{\bgamma})]
+ {\frac{p_{\bgamma^*} - p_{\bgamma}}{2}}\log (n g_{C_h}) +  {\frac{q_{\bgamma^*} - q_{\bgamma}}{2}}\log (ng_{C_t}) + \lOp(1).
$$
\end{enumerate}
\end{proposition}
Finally, the corollary below shows the model selection consistency for the Product prior.
\begin{corollary}\label{cor:FB}
Assuming the prior probability assigned to $\bgamma^*$ is not ${\bf 0}$, \textit{i.e.} $\pi (\bgamma^*)\not = 0$, and the Conditions C1-C7 hold, then $\tilde{\pi}\left(\boldsymbol{\bgamma}^* \mid \boldsymbol{t}, \boldsymbol{\delta}\right) \xrightarrow{\Pr} 1$ as $n\to\infty$.
\end{corollary}
For both priors, in overfitted models, the asymptotic Bayes factor rates do not depend on censoring or model misspecification, but only on the number of spurious variables at each level and the values of the hyperparameters. For non-overfitted models, the term $M(\bPsi^*_{\bgamma^*}) - M(\bPsi^*_{\bgamma})$ depends on censoring and potential model misspecification through the expected log-likelihood. Although the GH model encompasses a range of simpler models, when the true model structure is, for example, a proportional odds or cure model, the value of this term depends on how the true generating model projects onto the GH family.
Comparing Propositions \ref{prop:BFrates_EB} and \ref{prop:BFrates}, we see that when $g_E = g_{C_t} = g_{C_h}$, the Bayes factor rates are similar despite differences in the prior covariance matrices, with the LCM prior exhibiting a marginally better rate due to the additional “plus-one” term.
In other words, the Bayes factor rates coincide when $g_{C_t} = g_{C_h} = g_E + 1/n$.
Since a larger $g$ implies stronger sparsity, the LCM prior, under the default hyperparameter setting, has slightly greater ability to rule out spurious variables than the Product prior.

The quantity $\bgamma^*$ is, by definition, the most parsimonious model maximising the expected log-likelihood $M(\bPsi_{\bgamma})$. Equivalently, it indexes the projection of the true generating mechanism $\varphi_0$ onto the GH family in the generalised Kullback--Leibler sense that accounts for the presence and type of censoring \citep{rossell:2023}. Corollaries~\ref{cor:EB} and~\ref{cor:FB} therefore state that the posterior concentrates on this KL-optimal model, \emph{not} necessarily on a true data-generating model. Under misspecification, for instance, when the truth is a proportional-odds, or when the baseline lies outside the assumed family, $\bgamma^*$ and $\bPsi^*_{\bgamma^*}$ describe the best GH approximation, and the term
$M(\bPsi^*_{\bgamma^*})-M(\bPsi^*_{\bgamma})$ governing the non-overfitted Bayes factor rate is exactly the excess discrepancy of $\bgamma$ relative to that approximation. When the model is correctly specified, however, the KL-optimal model coincides with the data-generating one: as made explicit in Corollary~\ref{cor:recovery} below, $\bgamma^*=\bgamma_0$, so that posterior consistency for $\bgamma^*$ delivers recovery of both the true active set and the true hazard structure.

\begin{corollary}\label{cor:recovery}
Suppose the conditions of Corollary~\ref{cor:EB} (resp.\ Corollary~\ref{cor:FB}) hold, and let $\widehat\pi$ (resp.\ $\tilde\pi$) denote the corresponding posterior. Then, for any deterministic
function $g$ of the model indicator, $\widehat\pi\big(g(\bgamma^*)\mid\bt,\boldsymbol{\delta}\big)\xrightarrow{\Pr}1$, as $n\to\infty$. In particular, writing $\mathcal{S}(\bgamma)=\{j:\gamma_j\neq 0\}$ for the active set and $\mathrm{Hz}(\bgamma)$ for the hazard structure,
\begin{equation*}
\widehat\pi\big(\mathcal{S}(\bgamma^*)\mid\bt,\boldsymbol{\delta}\big)\xrightarrow{\Pr}1,
\qquad
\widehat\pi\big(\mathrm{Hz}(\bgamma^*)\mid\bt,\boldsymbol{\delta}\big)\xrightarrow{\Pr}1,
\end{equation*}
and the marginal inclusion probabilities satisfy $\widehat\pi(\gamma_j\neq 0\mid\bt,\boldsymbol{\delta})\xrightarrow{\Pr}1$ for $j\in\mathcal{S}(\bgamma^*)$ and $\widehat\pi(\gamma_j\neq 0\mid\bt,\boldsymbol{\delta})\xrightarrow{\Pr}0$ otherwise, as $n\to\infty$. If, moreover, the model is correctly specified, that is, $\varphi_0$ follows a GH model \eqref{eq:GH} with a baseline
in the log-location-scale family \eqref{eq:LLS}, and with data-generating indicator $\bgamma_0$, then $\bPsi^*_{\bgamma_0}=\bPsi_0$ and $\bgamma^*=\bgamma_0$, so $\mathcal{S}(\bgamma^*)$ and $\mathrm{Hz}(\bgamma^*)$ are the true active set and the true hazard structure, respectively.
\end{corollary}

\section{Simulation Study}\label{sec:simulation}

In this section, we assess how well the proposed methods recover the active variables and the hazard structure, including under a misspecified baseline hazard, and whether selecting the structure affects detection power in variable selection. We consider $n \in \{250, 500, 1000\}$, $p \in \{10, 50, 100\}$, the censoring rate $(25\%, 50\%)$, the true hazard structure (PH, AFT, GH) and the true baseline hazard: log-normal, or power generalised Weibull (PGW), whose hazard decreases monotonically. All analyses assume a log-normal baseline, so the PGW scenarios are misspecified. GH data are generated with the log-normal baseline only, giving 90 scenarios, each replicated 250 times. Covariates follow $N(\mathbf{0}, \boldsymbol{\Sigma})$ with $\Sigma_{ij} = 0.7^{|i-j|}$ and $\tilde{\mathbf{X}} = \mathbf{X}$. Four of them are active, with large and small effects of both signs. Data generation, MCMC settings, initialisation, and convergence diagnosis are given in Supplementary Material Section 10.

\begin{figure}[ht]
    \centering
    \begin{subfigure}[b]{0.3\textwidth}
        \centering
        \includegraphics[width=\textwidth]{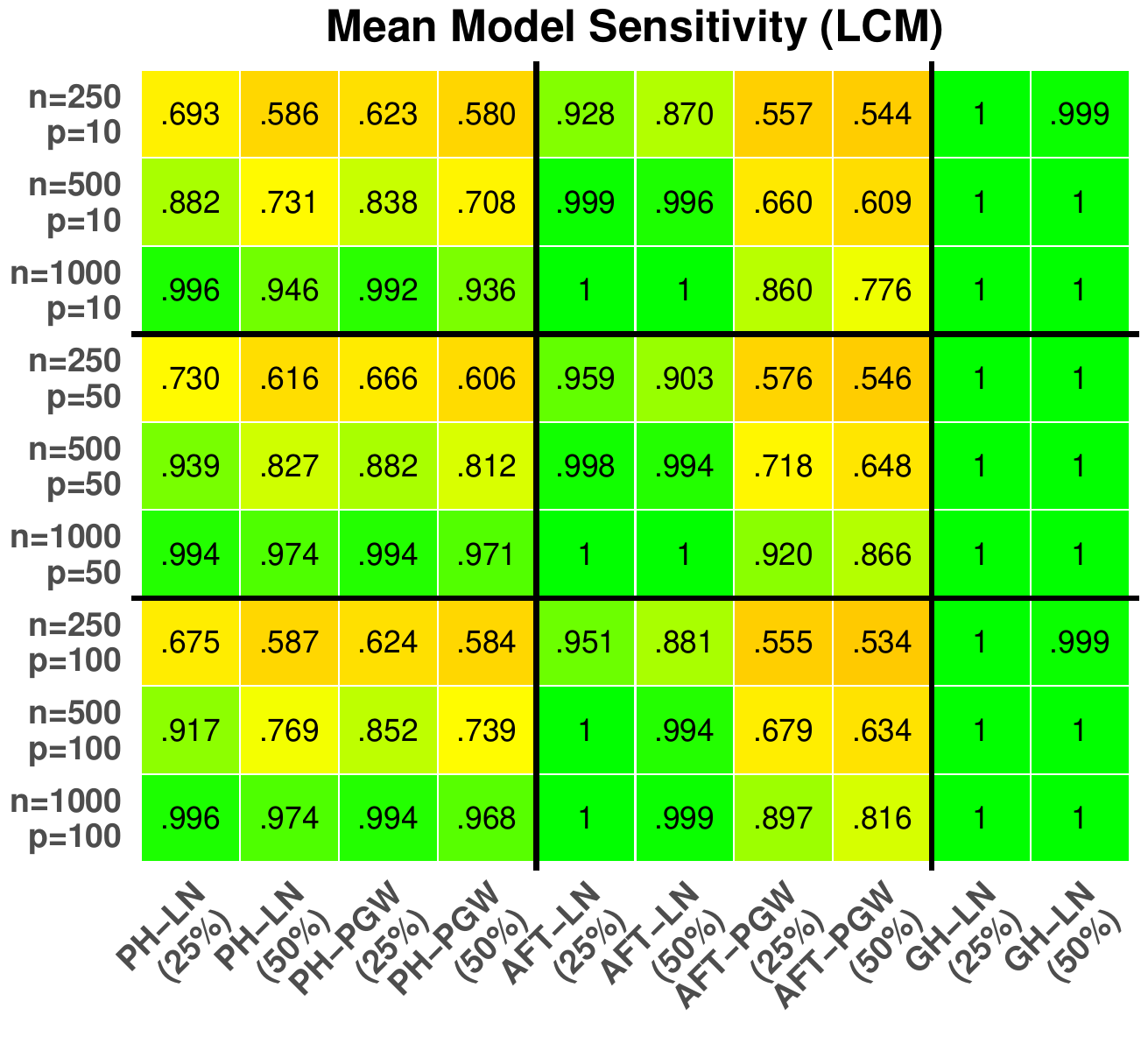}
        \caption{}
    \end{subfigure}
    \hfill
    \begin{subfigure}[b]{0.3\textwidth}
        \centering
        \includegraphics[width=\textwidth]{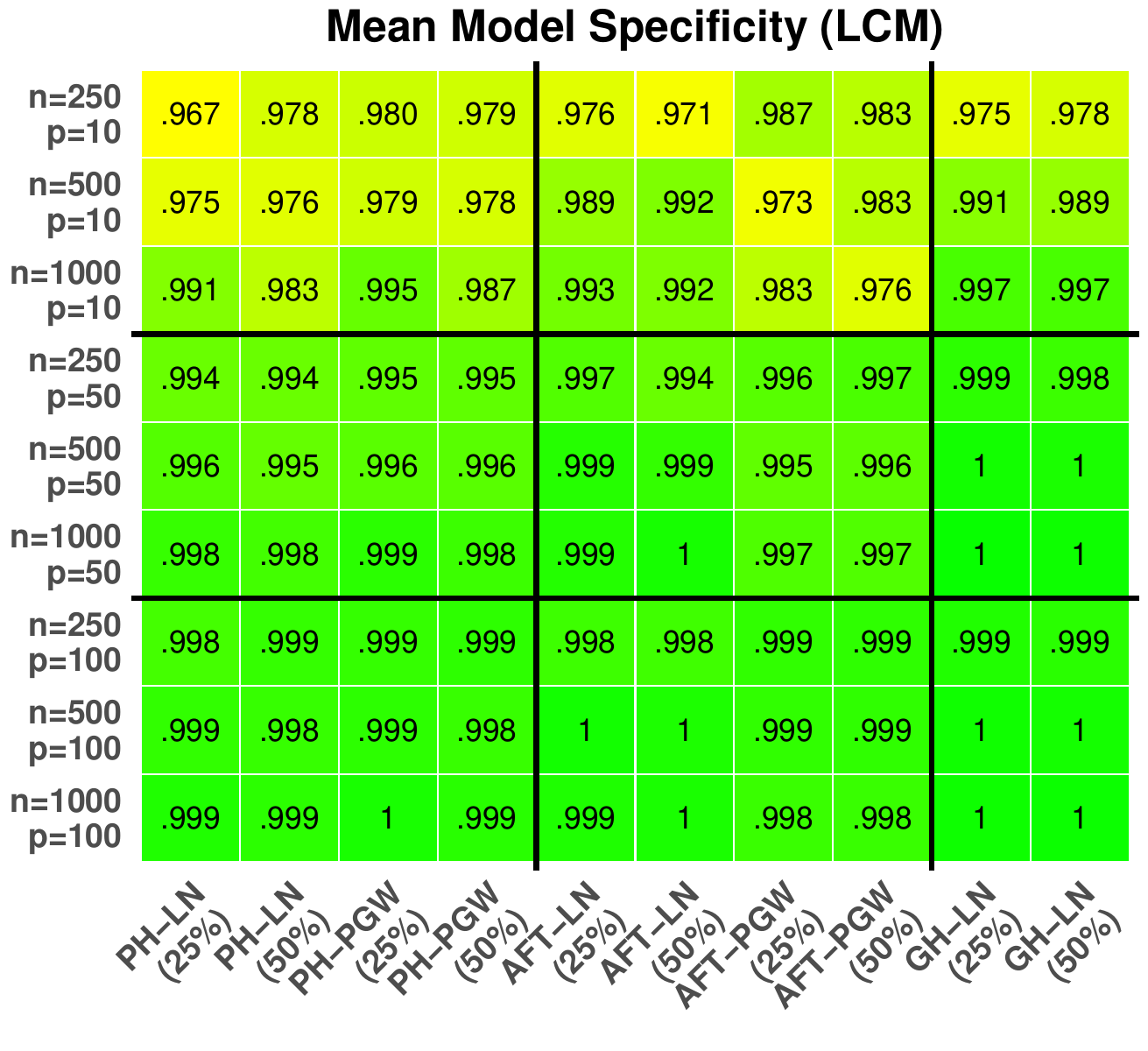}
        \caption{}
    \end{subfigure}
    \hfill
    \begin{subfigure}[b]{0.3\textwidth}
        \centering
        \includegraphics[width=\textwidth]{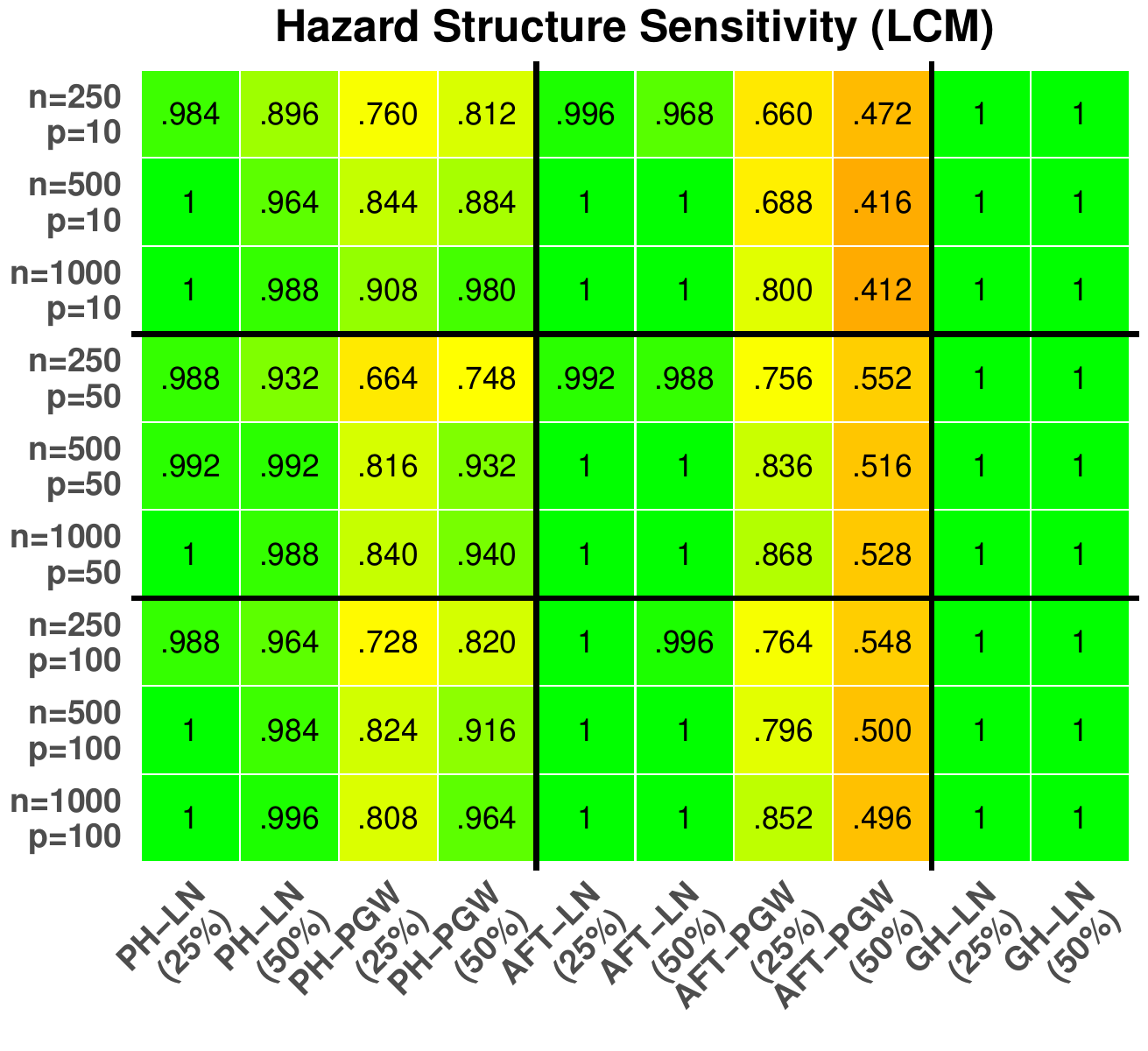}
        \caption{}
    \end{subfigure}
    \vspace{1em}
    \begin{subfigure}[b]{0.3\textwidth}
        \centering
        \includegraphics[width=\textwidth]{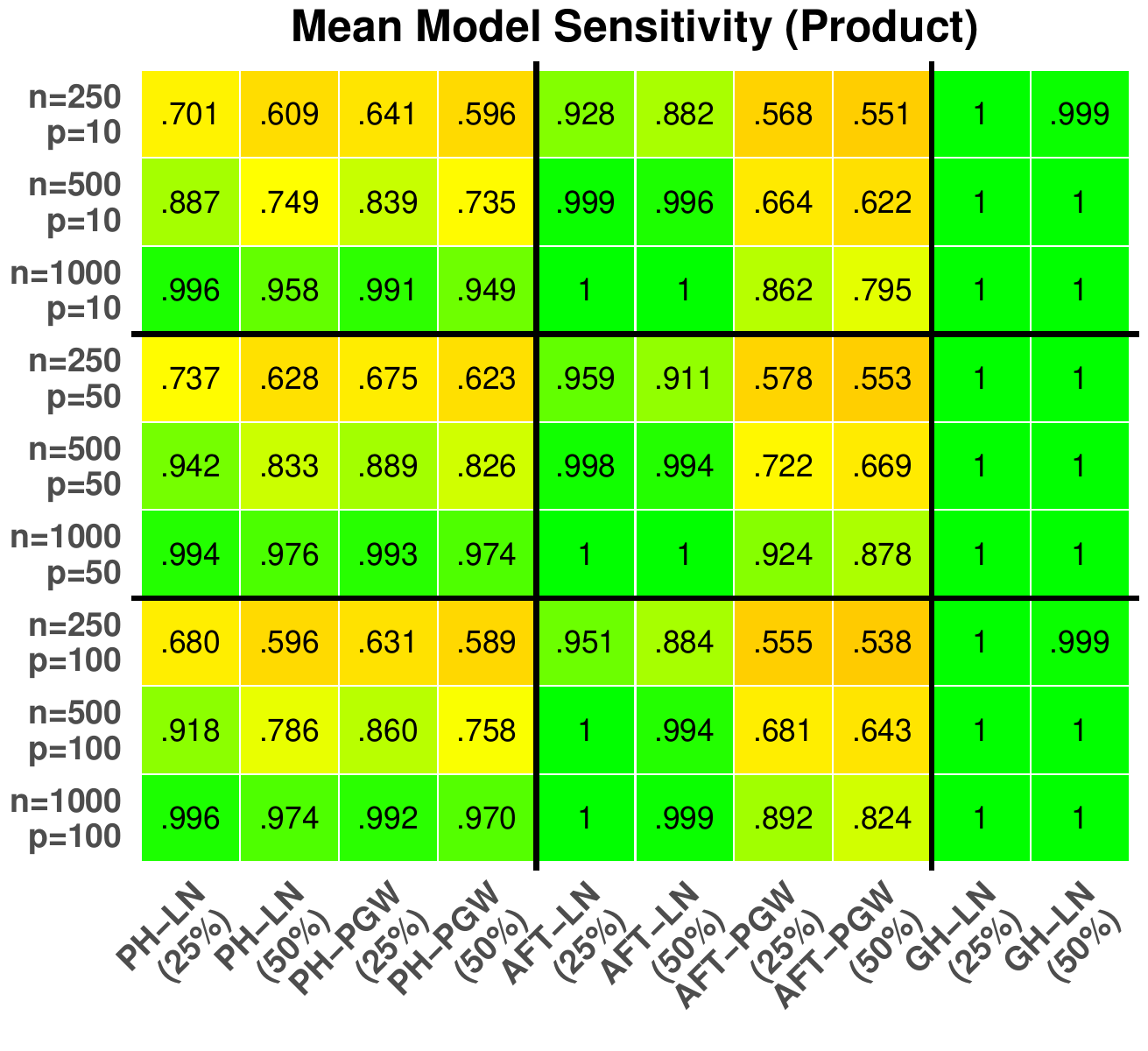}
        \caption{}
    \end{subfigure}
    \hfill
    \begin{subfigure}[b]{0.3\textwidth}
        \centering
        \includegraphics[width=\textwidth]{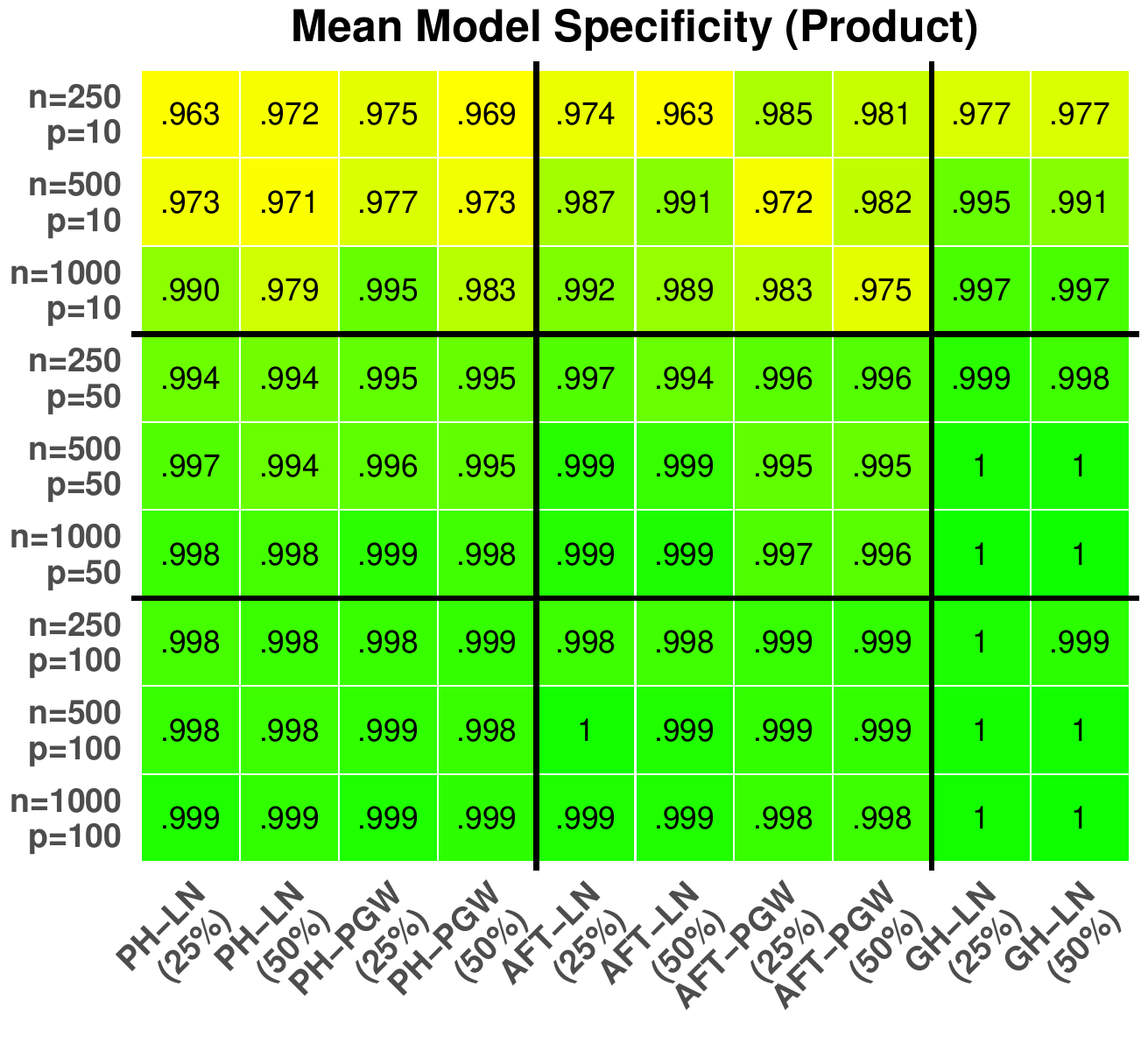}
        \caption{}
    \end{subfigure}
    \hfill
    \begin{subfigure}[b]{0.3\textwidth}
        \centering
        \includegraphics[width=\textwidth]{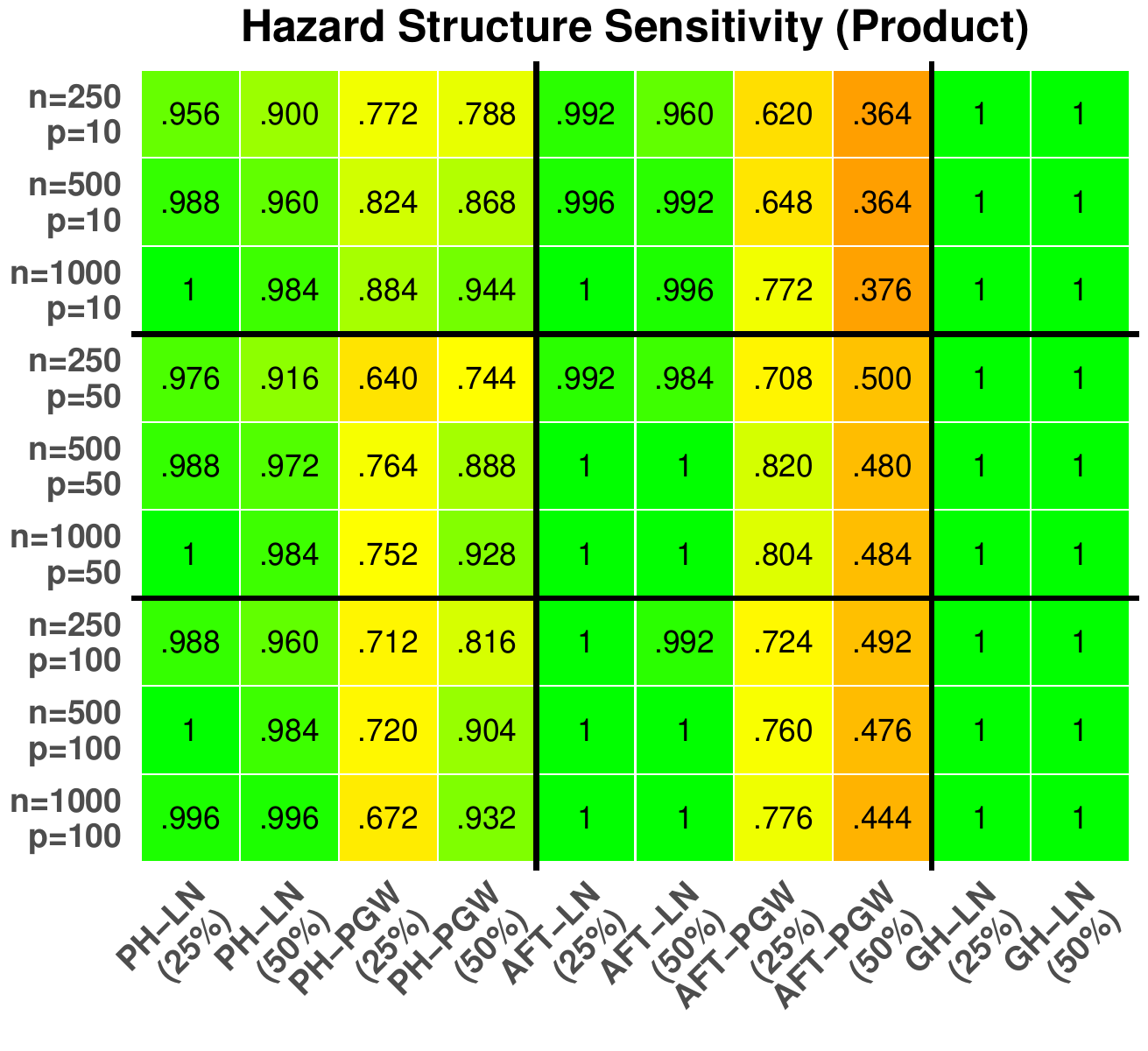}
        \caption{}
    \end{subfigure}
    \caption{
    Means over 250 replicates under the LCM (a–c) and Product (d–f) priors: sensitivity (a, d) and specificity (b, e) of the highest posterior probability model, and hazard-structure selection accuracy (c, f). Columns: true structure, baseline (LN: log-normal; PGW: power generalised Weibull) and censoring rate; rows: $n$ and $p$.
    }
    \label{fig:sim_varsel}
\end{figure}

Figure~\ref{fig:sim_varsel} reports the sensitivity and specificity of the highest posterior probability model, averaged over 250 replicates. Specificity is generally good: it exceeds 0.95 in every scenario, and the expected number of false positives stays around 0.1 for all levels of $p$. The active variables under GH are detected in essentially every replicate. For PH and AFT truths, sensitivity is largely driven by $n$: at $n = 250$ only the two largest effects are detected consistently, as shown in Figure S2 in Supplementary Material, whereas at $n = 1000$ sensitivity is at least 0.94 under the correct baseline. A misspecified baseline mildly affects performance under a true PH model but lowers it by around $0.3$ on average for a true AFT model, as under a monotonic baseline, time-level and hazard-level effects are hard to separate. Higher censoring rates also lower sensitivity, especially in increasing the difficulty of detecting the small effects, as shown in Figure S2. Lastly, the two priors perform quite similarly. The LCM prior is marginally more parsimonious for PH and AFT truths, indicated by its slightly lower sensitivity but higher specificity. In the GH scenarios, the specificity of the two priors differs by at most 0.004.

Next, we examine hazard selection performance. In each replicate we select the structure with the highest posterior probability summed over its models. With the correct baseline, the true structure is selected in at least $89\%$ of replicates in every scenario. Under the PGW baseline, accuracy is $0.64$--$0.98$ for a true PH model and $0.62$--$0.87$ for a true AFT model with $25\%$ censoring, and in both cases it generally increases with $n$. The exception is AFT with the PGW baseline and $50\%$ censoring. Accuracy there is $0.36$--$0.55$ and does not improve with increasing $n$, as the misclassified replicates go almost entirely to PH, as shown in Figure S2. A potential reason for this is that the two structures coincide under a Weibull baseline, which is monotonic, and high censoring removes the late follow-up that informs time-level effects.
Indeed, increasing the length of follow-up, and thus reducing censoring, markedly improves the identification of the correct hazard structure, as shown in Figure \ref{fig:sim_varsel}.
Lastly, with the correct baseline the accuracy is almost identical under the two priors, so structure selection is governed by the shared model-space prior rather than by the prior on the coefficients; under the PGW baseline, the LCM prior is slightly more accurate, by $0.04$ on average.

For $p = 10$ and $n = 250, 500$, we repeated the PH and AFT scenarios with the model space restricted to the true structure, \textit{i.e.} we conduct variable selection within a fixed PH or AFT model under the same priors (details in Supplementary Section 10.2). When the structure is correctly specified, the posterior inclusion probabilities closely match those obtained with selection under the GH model on the full model space. Thus, selecting the structure has a low cost on variable selection power. Only in the misspecified case described above, the inclusion probabilities of small effects are more clearly affected. Supplementary Material Section 10.3 also contains comparison with BIC based approaches.

\section{Real Data Applications}\label{sec:realapp}

\subsection{Serum free light chain data set (\texttt{FLC})}

We analyse the \texttt{FLC} data set, which is publicly available in the \texttt{survival} R package. This data set has been used to study the prevalence of monoclonal gammopathy of undetermined significance (MGUS) in Olmsted County, Minnesota, based on approximately two-thirds of residents aged 50 years or older. \cite{Dispenzieri2012-flc} assayed FLC levels in these samples and found that elevated levels were associated with increased mortality.
The \texttt{FLC} data set contains a stratified random sample (by age and sex) comprising $1/2$ of the originally studied subjects, with $n=7,874$ samples. Details on the $8$ variables relevant here are provided in Supplementary Material Section 11. The longest follow-up time in the study is approximately $14$ years. We consider the following $6$ variables, matching the multivariate analysis in \cite{Dispenzieri2012-flc}: two binary variables (Sex and Mgus) and four numerical variables (Age, Kappa, Lambda, and Creatinine). Retaining only complete cases for these $p=6$ variables produces $n=6,521$ observations, of which $70\%$ are censored. All numerical variables are then centred and scaled.

For each prior (LCM and Product), we run the MCMC sampler for $50{,}000$ iterations with $10{,}000$ burn-in and thinning of $10$, assuming a log-normal baseline hazard and the default hyperparameters of Section~\ref{ssec:priormod}. Table~\ref{tab:flc-hpm} reports the $90\%$ highest posterior probability credible sets \citep{griffin:2025}. All selected models are GH models, and the four selected variables (Age, Kappa, Lambda, and Sex) are broadly consistent with those identified as significant in the Cox PH analysis of \cite{Dispenzieri2012-flc} (Age, Sex, and the sum of Kappa and Lambda).

\begin{table}[h]
\centering
\caption{The $90\%$ highest posterior probability model credible set for each prior structure for the FLC data set. $haz$ in the superscript within the brackets stands for hazard-level effect only, and no superscript means the variable has both time-level and hazard-level effects but at a different level (\textit{i.e.}~$\gamma_j=3$). The superscript on the right bracket indicates the hazard type.}
\label{tab:flc-hpm}
\begin{tabular}{@{}cccc@{}}
\toprule
Model \textbackslash{} Posterior probability         & LCM  ($g_E=1$) & Product ($g_{C_t}=g_{C_h}=1$) & LCM (Robust $g$) \\ \midrule
\{age, lambda, sex$\}^{GH}$                       & 0.511          & 0.328                         & 0.193            \\
\{age, kappa, $\text{lambda}^{haz}$, sex$\}^{GH}$ & 0.273          & 0.265                         & 0.275            \\
\{age, kappa, lambda, sex$\}^{GH}$                & 0.152          & 0.335                         & 0.452            \\ \bottomrule
\end{tabular}
\end{table}
All three priors select the same model set but distribute the posterior mass differently. Under the top model (LCM prior), the MAP estimates of $(\balpha,\bbeta)$ are $(-0.619, 0.967)$ for Age, $(0.489, 0.357)$ for Lambda, and $(-1.117, 0.202)$ for Sex, with similar estimates under the Product prior. Age and being male thus show a proportional increase in hazard together with a later, slower-decaying hazard peak, while Lambda's hazard- and time-level effects are smaller and comparable in magnitude; equal coefficients at the two levels need not imply equal influence on the baseline hazard, which depends on study duration and baseline form. Under the fixed-$g$ LCM prior, posterior mass decreases monotonically as the number of effects grows from $6$ to $8$, whereas the Product prior gives similar mass to the $6$- and $8$-effect models; the LCM prior thus penalises additional effects more strongly, consistent with the simulation results.
The hyper-$g$ LCM prior instead favours the largest model in the set, $\{\text{age, kappa, lambda, sex}\}^{GH}$ with $8$ effects (posterior probability $0.452$), adding Kappa with MAP estimates $(0.382, 0.164)$; the remaining coefficients change little except Lambda, whose estimate shifts once Kappa is included, reflecting their $0.83$ correlation. Compared with the fixed-$g$ LCM prior, which concentrates on the smallest model, the hyper-$g$ prior shifts posterior weight towards the larger models.
To assess the role of this strong correlation, we replace Kappa and Lambda with their sum (``kap+lam'', an overall FLC level) and ratio (``kap/lam'', used clinically to detect monoclonal light chains), both used by \cite{Dispenzieri2012-flc}, reducing their correlation to $0.01$. Re-running all three priors on this transformed set, we find they now agree on a single top model, $\{\text{Age, kap+lam, Sex}\}^{GH}$, with posterior probability $0.936$, $0.875$, and $0.913$ under the fixed-$g$ LCM, fixed-$g$ Product, and hyper-$g$ LCM priors respectively.

Beyond recovering variables already known to be relevant, our framework identifies their \emph{role}: the selected GH structure attributes to Age, Sex, and kap+lam both hazard-level and time-level effects, rather than the single proportional effect assumed by a Cox PH analysis. For kap+lam, an established adverse prognostic marker \citep{Dispenzieri2012-flc}, this suggests that higher free light chain burden is associated not only with elevated instantaneous risk but also with a faster survival timescale. This temporal decomposition is a model-based finding that a proportional-hazards summary cannot express, rather than an independently established mechanism, which would require
confirmation by a dedicated clinical investigation.

\subsection{Metastasis-free survival of breast cancer patients (\texttt{NKI70})}

We now analyse the \texttt{NKI70} data set from the Netherlands Cancer Institute, publicly available in the \verb|penalized| R package (the variable dictionary is in Supplementary Material Section 11). It contains metastasis-free survival times for $n=144$ lymph node--positive patients, $5$ clinical risk factors, and expression measurements for $70$ genes (longest follow-up approximately $17$ years), giving $p=75$ covariates, $4$ of which are categorical. The censoring rate is $66.7\%$, there are no missing values and the numerical variables are centred and scaled. The small sample size relative to $p$, combined with the high censoring rate, makes this a challenging setting for variable and hazard structure selection.

Alongside our two proposed priors, we benchmark against two reference methods: a Zellner-type fixed-$g$ prior for the AFT model \citep{rossell:2023}, via the \verb|modelSelection| function in the \verb|mombf| package with $g=144$ (a unit information prior; for AFT models \texttt{mombf} uses the block-diagonal version of Zellner's prior, i.e. independent normal priors with variance $g\sigma^2/\mathbf{x}_j^{\top}\mathbf{x}_j$) and a default $\text{Beta-Binomial(1,1)}$ model-space prior; and the non-local piMOM prior for the Cox partial likelihood \citep{niko:2017}, via the \verb|bvs| function in the \verb|BVSNLP| package with its default hyperparameters and the same model-space prior. Full MCMC specifications are given in Supplementary Material Section 12.
Table \ref{tab:nki} reports the model selection results. Consistent with the simulation study, the LCM prior concentrates more posterior mass on smaller models than the Product prior.

\begin{table}[ht]
\centering
\caption{Rank of the top models, in terms of estimated posterior probability, for the proposed and reference methods for NKI70 data set. The superscript on the right bracket indicates model type.}
\label{tab:nki}
\resizebox{\textwidth}{!}{%
\begin{tabular}{@{}cccc|c@{}}
\toprule
Model \textbackslash{} Rank (Posterior probability) &
  LCM  ($g_E=1$) &
  Product ($g_{C_t}=g_{C_h}=1$) &
  AFT (Zellner-$g$) &
  Partial-Cox (piMOM) \\ \midrule
$\{\text{PRC1}\}^{AFT}$ &
  1 (0.138) &
  1 (0.066) &
  1 (0.065) &
  1 (0.129): \{PRC1\} \\
$\{\text{PRC1}$, $\text{KNTC2}\}^{AFT}$ &
  2 (0.070) &
  2 (0.049) &
  2 (0.051) &
  2 (0.039): \{PRC1, N\} \\
$\{\text{PRC1}$, $\text{Age}\}^{AFT}$ &
  3 (0.045) &
  4 (0.026) &
  3 (0.028) &
  3 (0.036): \{PRC1, KNTC2\} \\
\{PRC1$\}^{GH}$ &
  4 (0.036) &
  3 (0.040) &
  - &
  4 (0.025): \{PRC1, IGFBP5.1\} \\
$\{\text{PRC1}$, $\text{IGFBP5.1}\}^{AFT}$ &
  5 (0.028) &
  7 (0.015) &
  5 (0.018) &
  5 (0.024): \{PRC1, KNTC2, IGFBP5.1\} \\
$\{\text{PRC1}$, KNTC2, $\text{IGFBP5.1}\}^{AFT}$ &
  6 (0.024) &
  5 (0.019) &
  4 (0.021) &
  6 (0.022): \{PRC1, IGFBP5\} \\ \bottomrule
\end{tabular}%
}
\end{table}
The gene PRC1 appears in every listed model, and the single-covariate model $\{\text{PRC1}\}$ has the highest posterior probability under all methods. The model $\{\text{PRC1}\}^{GH}$ receives posterior probability $0.036$ and $0.040$ under the LCM and Product priors, respectively, suggesting that a difference between PRC1's hazard- and time-level effect sizes may become identifiable with larger sample. The remaining models mostly add one or two of $\{\text{KNTC2, Age, IGFBP5.1}\}$ to PRC1.

The posterior distribution over hazard structures (AH, PH, AFT, GH) is $(0, 0.060, 0.878, 0.061)$ under the LCM prior and $(0, 0.105, 0.661, 0.234)$ under the Product prior: both proposed methods favour AFT over PH, which explains why the Cox-piMOM ranking, fixed at PH, ranks $\{\text{PRC1, N}\}$ second and contains no model with Age. This highlights the value of hazard structure selection.
The AFT-Zellner-$g$ results closely match the Product prior within the AFT structure: the four most probable AFT models coincide, in the same order with comparable posterior probabilities ($0.065$, $0.051$, $0.028$, $0.021$ against $0.066$, $0.049$, $0.026$, $0.019$), consistent with the two methods' similar prior and likelihood conditional on the AFT structure, which receives most posterior mass under the Product prior.

The preference for an AFT structure over PH is consistent with both empirical and biological evidence for this cohort. PRC1, the gene most robustly selected by our methods, is a regulator of cytokinesis whose overexpression promotes proliferation through the cell cycle and is an adverse prognostic marker across several solid tumours, including breast cancer \citep{li:2023}. As a determinant of proliferation rate, PRC1 plausibly accelerates the time scale of progression to recurrence rather than inducing a constant relative hazard. We therefore interpret the AFT preference as indicating that, in this cohort, the prognostic effect concentrates early and acts by rescaling time to recurrence, which is a model-based and biologically plausible interpretation rather than an established mechanism.

\section{Discussion}\label{sec:discuss}
We have developed a Bayesian methodology for simultaneously performing variable selection and hazard-structure selection for time-to-event data within the GH modelling framework. This model encompasses hazard structures of practical interest, including the PH and AFT models, enabling users not only to identify relevant variables but also to determine the level at which each covariate acts.
We provide a principled approach for specifying covariate effects based on $g$-prior and propose a new model-space prior that extends the Beta-Binomial prior.

Through the simulation study, we found that the two proposed priors behave very similarly, with the LCM prior being slightly more parsimonious. Specificity remains high and the number of false positives does not grow with $p$, while the sample size governs the detection of small effects. With a correctly specified baseline, the true hazard structure recovery rate is at least $89\%$ in all settings. Time-level effects are the hardest to detect and attribute: combined with high censoring rates, a misspecified monotone baseline can make an AFT structure indistinguishable from PH, however, correct hazard structure detection can be markedly improved with a longer follow-up. Overall, allowing selection among hazard structures has little effect on variable-selection performance.

The results from the \texttt{FLC} dataset confirm the ability of our proposed methodology to identify important variables and correctly characterise their roles in a real-data setting.
The comparison of results using fixed $g$ priors with those obtained under the LCM prior combined with a robust prior on $g$ indicates that the mixture-$g$ prior can be sensitive to high collinearity among active variables.
The results from the \texttt{NKI70} dataset, by contrast, evaluate the performance of the proposed methods relative to existing approaches in a more challenging scenario with moderately large $p$, small $n$, and a high censoring rate, leading to higher uncertainty in model selection.

There are several promising avenues for future work.
The ideas used to construct the prior on $\bgamma$ can be readily extended to other models in which covariates play multiple roles, such as distributional regression, where covariates may influence more than two linear or additive predictors, as well as sample selection models, cure models, and related frameworks.
 The factorisation in \eqref{eq:gamfact} is one of several possible ways of controlling multiplicity, and alternative structures could encode different notions of a non-preferential prior. More broadly, formulating model-space priors for models in which covariates play multiple roles remains an open research direction, as well as a more detailed treatment of categorical covariates \citep{García-Donato2022}.
Finally, based on the computing-time plots in the Supplementary Material Section 10.1 for the simulation study, we observe that the run time grows roughly linearly with $p$, and the current implementation requires approximately $3$ to $19$ minutes, depending on the setting, to run the sampler for $20{,}000$ iterations in scenarios where $n$ is on the order of $1{,}000$s and $p$ is on the order of $100$s, which aligns with the applications of interest in most practical settings. Although the current implementation has substantial scope for efficiency improvements, for example, by rewriting the code in C++ instead of R, a more efficient MCMC scheme remains the key component for scaling to higher-dimensional problems, such as those arising in genomics (where $p$ can be on the order of $1{,}000$s). Recent advances in MCMC algorithms for variable selection \citep{liang:2023} suggest several opportunities to enhance the efficiency of our extended ADS algorithm.

\section{Disclosure statement}\label{disclosure-statement}

The authors have no conflicts of interest to declare.

\section{Data Availability Statement}\label{data-availability-statement}

Deidentified data have been made available at the following URL: \url{https://github.com/Eric-YChen/GHBVS}

\bibliography{bibliography.bib}

\end{document}


\def\spacingset#1{\renewcommand{\baselinestretch}%
{#1}\small\normalsize} \spacingset{1}

\if1\anon
{
  \title{\bf Supplementary material \\for\\ Bayesian variable and hazard structure selection in the General Hazard model}
  \author{Yulong Chen\hspace{.2cm}\\
    Department of Statistical Science,
	University College London\\
    and \\
    Jim Griffin\hspace{.2cm}\\
    Department of Statistical Science,
	University College London\\
    and \\
    Francisco Javier Rubio\hspace{.2cm}\\
    Department of Statistical Science,
	University College London}
  \maketitle
} \fi

\if0\anon
{
  \bigskip
  \bigskip
  \bigskip
  \begin{center}
    {\LARGE\bf Supplementary material for Bayesian variable and hazard structure selection in the General Hazard model}
\end{center}
  \medskip
} \fi

\newpage
\spacingset{1.8}

\tableofcontents
\newpage

\section{Details and additional study on the ipilimumab immunotherapy trial}

{
The ipilimumab immunotherapy trial (NCT00861614) dataset, reconstructed from \cite{Kwon2014} and available in the KMDATA database \citep{Fell2021}, includes information on $n = 799$ individuals enrolled in a phase 3 trial investigating immunotherapy for advanced prostate cancer. It contains time-to-event data (with the longest follow-up of 36.1 months), right-censoring indicators ($\sim 30\%$ censoring rate), and treatment arm assignments (binary, ipilimumab \textit{vs.} placebo) \citep{ChrisRubio26}. The methods we use are:
\begin{enumerate}
    \item Kaplan–Meier estimates by arm.
    \item A semi-parametric Cox PH model (`coxph` in the `survival` R package).
    \item A log-normal AFT model (`survreg`, also in the `survival` R package).
    \item The proposed methodology under LCM prior, with
    \begin{enumerate}
        \item PH models only, obtained by setting $h_{AH} = h_{GH} = h_{AFT} = 0$ in equation (9);
        \item AFT models only, obtained analogously;
        \item Whole GH space (joint selection, with default hyperparameters on model space prior).
    \end{enumerate}
\end{enumerate}

\begin{figure}[h]
    \centering
    \includegraphics[width=0.8\linewidth]{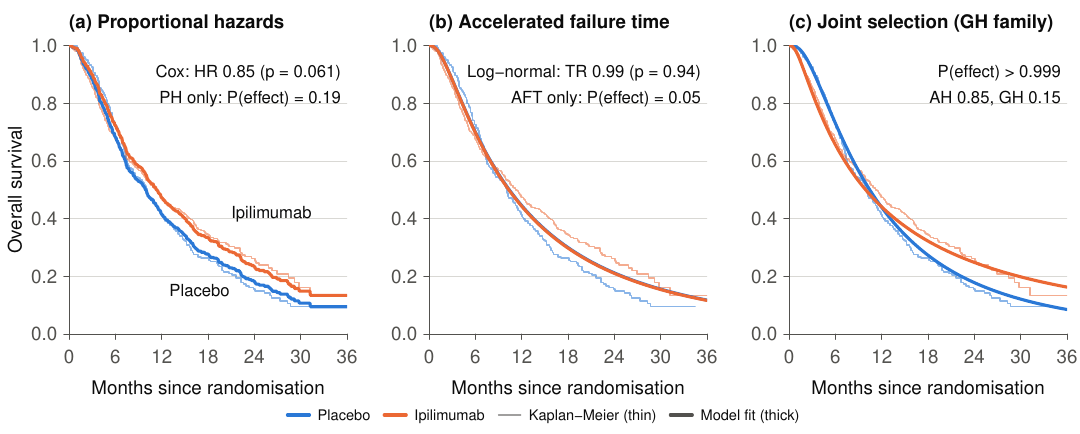}
    \caption{Overall survival in the CA184-043 trial. Thin lines are Kaplan–Meier estimates and thick lines are fitted survival functions for placebo (blue) and ipilimumab (orange). (a) Cox PH model (HR: hazard ratio); (b) Log-normal AFT model (TR: time ratio); (c) Posterior model average from the proposed method with the hazard structure selected jointly (log-normal baseline, LCM prior with $g_E = 1$). ``P(effect)'' is the posterior probability that treatment enters the model when the proposed method is restricted to (a) PH models, (b) restricted to AFT models, or (c) the whole GH family. The numbers in the top right corner of (c) are the posterior probabilities of the structures.}
    \label{fig:ipil}
\end{figure}

When looking at the fitted survival curves of the averaged model and the Kaplan-Meier estimates in Figure~\ref{fig:ipil} (c), there exists a crossing between the baseline survival and the treatment survival at around $t=10$. This is known as crossing survival, and it is common for therapeutic variables, particularly in trials of immune checkpoint inhibitors \citep{Plana22}, where the survival probability after taking the treatment is lower/higher than the baseline at the beginning but becomes higher/lower after a certain time, \textit{i.e.} the effect varies with time. \cite{Kwon2014} notice a similar pattern through an exploratory piecewise hazard model, where they suspect that ``with ipilimumab seeming to be associated with better survival than placebo at later timepoints''.
A PH model assumes that the hazards for any covariates levels are proportional and cannot capture the variation at the time level. This is clearly visible in Figure~\ref{fig:ipil} (a) which compares the semi-parametric PH fit with the Kaplan-Meier curve.  The fitted PH survival curves do not cross, so they cannot reproduce the crossing of the Kaplan--Meier curves. The AFT model
is also unable to represent crossing survival curves, since it assumes  equal time and hazard level effects. The  MAP estimate of the treatment effects under the GH model is $(0.953,-0.140)$, which is far from equality. After projecting this estimate to the AFT space, it is likely that the two effects  nearly cancel out, as shown by the nearly overlapped fitted curves in Figure~\ref{fig:ipil} (b). Lastly, while the average model under joint selection is able to discover the crossing, it happens slightly later than the crossing indicated by the Kaplan-Meier curve. This is likely due to the log-normal assumption of the baseline hazard -- under a more flexible baseline hazard, which is regarded as one of the most important future works, the location of the crossing will be identified more precisely.

\subsection{Additional study with one extra noise covariate}
With a single covariate, selecting the treatment is equivalent to testing whether it has an effect. To show that the conclusions of the previous study also hold when the treatment must be selected from a set of candidate covariates, we add a spurious variable generated from a standard normal distribution. The same set of methods are used to fit this new data.

Table~\ref{tab:motiv_hpm} shows that, under settings 4(a) and 4(b), the highest probability model is the null model, both with over $80\%$ posterior probability. Similarly, the results for the two reference methods in Table~\ref{tab:motiv_standard} suggest that, at a significance level of $0.05$, neither of the two methods identifies the effect of the treatment as significant, while the noise is correctly identified as insignificant. However, under a significance level of $0.1$, the PH model identifies both the treatment and the noise as significant, while the AFT model identifies only the noise as significant.
On the other hand, under setting 4(c) the treatment variable has PIP of nearly $1$, and the posterior distribution of the hazard structure is $(0.829, 0, 0, 0.170)$ for (AH, PH, AFT, GH), which is almost identical to the previous treatment-only study.

\begin{table}[h]
\centering
\caption{$90\%$ highest probability models sets and posterior inclusion probabilities of the variables under each hazard structure assumption using the proposed methodology under LCM prior. }
\label{tab:motiv_hpm}
\begin{tabular}{@{}cccc@{}}
\toprule
Summary                     & GH                                 & PH              & AFT             \\ \midrule
\multirow{2}{*}{90\% HPM set} & 1: \{\textbf{trt}\}$^{AH}$ (0.767) & 1: Null (0.827) & 1: Null (0.872) \\
                              & 2: \{\textbf{trt}\}$^{GH}$ (0.134) & 2: \{\textbf{trt}\} (0.095)  & 2: \{spur\} (0.096) \\  \midrule
$(\text{PIP}_\text{trt},\text{PIP}_\text{spur})$  & $(0.999,0.098)$  & $(0.122,0.079)$ & $(0.031,0.106)$ \\ \bottomrule
\end{tabular}
\end{table}

\begin{table}[h]
\centering
\caption{Fitted coefficients and p-values under semi-parametric PH and log-normal AFT model.}
\label{tab:motiv_standard}
\begin{tabular}{@{}ccccc@{}}
\toprule
Model       & coef$_\text{trt}$ & p-value & coef$_\text{spur}$ & p-value \\ \midrule
Cox PH & -0.166    & 0.051   & 0.068      & 0.098   \\
AFT    & 0.003     & 0.965   & -0.075     & 0.055   \\ \bottomrule
\end{tabular}
\end{table}

From an empirical perspective, merely performing standard PH or AFT analyses can lead to potentially misleading conclusions. One may claim the treatment effect is significant based on the PH model fit and hence conclude that ipilimumab reduces the hazard by the same proportion throughout follow-up, or they may even suggest that the noise has significant effect based on the AFT model fit. However, performing this joint selection immediately reveals that the ipilimumab therapy is much more important than the noise and is likely to be harmful only at the beginning but gradually become more effective than the placebo as time progresses. While the Kaplan-Meier estimator is also capable of discovering the time-dependency and importance of the treatment, as there is only one binary variable of interest in the example, when the dimension and cardinality increase, it is impossible to plot out every combination of variables' values to identify their effects.

}

\clearpage

\section{Log-Likelihood, Gradient and Hessian functions}
\subsection{Log-likelihood function}
Recall that the hazard and cumulative hazard functions can be written as
\begin{equation*}
h(t\mid \mu,\sigma) = \dfrac{\frac{1}{\sigma t}f\left(\dfrac{\log(t)-\mu}{\sigma}\right)}{1-F\left(\dfrac{\log(t)-\mu}{\sigma}\right)}, \,\,\,\,\, H(t\mid \mu,\sigma) = -\log\left[1-F\left(\dfrac{\log(t)-\mu}{\sigma}\right)\right].
\end{equation*}
The likelihood function of $(\balpha^{\top},\bbeta^{\top},\mu, \sigma)$ is
\begin{eqnarray*}
L(\balpha^{\top},\bbeta^{\top},\mu, \sigma) &=&  \prod_{i=1}^{n}   h(t_i\mid \balpha,\bbeta,\mu, \sigma)^{\delta_i} \exp\left[-H(t_i\mid \balpha,\bbeta,\mu, \sigma)\right].
\end{eqnarray*}
Thus, the log likelihood function can be rewritten as
\begin{eqnarray*}
\ell(\balpha^{\top},\bbeta^{\top},\mu, \sigma) &=&  \sum_{i=1}^{n} {\delta_i} \log h(t_i\mid \balpha,\bbeta,\mu, \sigma) - \sum_{i=1}^{n}H(t_i\mid \balpha,\bbeta,\mu, \sigma)\\
 &=& -n_o\log(\sigma) -\sum_{i=1}^{n} \delta_i \log(t_i) - \sum_{i=1}^{n} {\delta_i}\tilde{\bx}_i^{\top}\balpha + \sum_{i=1}^{n} {\delta_i}\bx_i^{\top}\bbeta \\
 &+& \sum_{i=1}^{n} {\delta_i} \log f\left(\dfrac{\log(t_i) + \tilde{\bx}_i^{\top}\balpha -\mu}{\sigma}\right) - \sum_{i=1}^{n} \delta_i \log\left[ 1- F\left(\dfrac{\log(t_i) + \tilde{\bx}_i^{\top}\balpha -\mu}{\sigma}\right) \right]\\
 &+& \sum_{i=1}^{n} \log\left[ 1- F\left(\dfrac{\log(t_i) + \tilde{\bx}_i^{\top}\balpha -\mu}{\sigma}\right) \right]\exp\left(   \bx_i^{\top}\bbeta  -  \tilde{\bx}_i^{\top}\balpha\right),
\end{eqnarray*}
where $n_o = \sum_{i=1}^{n}\delta_i$. Consider the reparameterisation $\tau = 1/\sigma$, $\theta_0 =  \mu/\sigma$, $\btheta = -{\balpha}/\sigma$, and $\bmeta = -{\bbeta}$. After some algebra we can write down the log likelihood function as follows
\begin{eqnarray*}
\ell_n(\bPsi) &=&  n_o \nu - \sum_{i=1}^{n} \delta_i \log(t_i) + \sum_{i=1}^{n} {\delta_i}\left( \tilde{\bx}_i^{\top}\btheta  e^{-\nu} - \bx_i^{\top}\bmeta \right) \nonumber\\
&+& \sum_{i=1}^{n} {\delta_i} \log f\left(  e^{\nu}\log(t_i) - \tilde{\bx}_i^{\top}\btheta -\theta_0 \right)\nonumber\\
&+& \sum_{i=1}^{n} \log F\left( -e^{\nu}\log(t_i) + \tilde{\bx}_i^{\top}\btheta +\theta_0 \right) \left[\exp\left(  { \tilde{\bx}_i^{\top}\btheta e^{-\nu} - \bx_i^{\top}\bmeta }\right) -\delta_i\right],
\end{eqnarray*}
where $\tau = \exp(\nu)$, $\bPsi= (\nu,\theta_0,\btheta^{\top}, \bmeta^{\top})^{\top}$.

\subsection{The Gradient function}

The gradient function is:
\begin{eqnarray*}
g(\bPsi) =
\begin{pmatrix}
\dfrac{\partial}{\partial \nu}  \ell(\bPsi)  \\
\dfrac{\partial}{\partial \theta_0} \ell(\bPsi) \\
\nabla_{\btheta}  \ell(\bPsi)  \\
\nabla_{\bmeta}  \ell(\bPsi)
\end{pmatrix},
\end{eqnarray*}
where,
\begin{eqnarray*}
\dfrac{\partial}{\partial \nu}  \ell(\bPsi) &=& n_o
- \sum_{i=1}^{n} {\delta_i} \tilde{\bx}_i^{\top}\btheta e^{-\nu} +  \sum_{i=1}^{n} {\delta_i} \dfrac{ f'\left(u_i\right)}{ f\left(u_i\right)} \log(t_i)e^{\nu}  -  \sum_{i=1}^{n}  \dfrac{ f\left(u_i \right)}{ F\left(-u_i \right)} \log(t_i) \left[ v_i - \delta_i\right]e^{\nu}\\
&-&  \sum_{i=1}^{n} \log F\left(-u_i \right) v_i \tilde{\bx}_i^{\top}\btheta e^{-\nu} ,\\
\dfrac{\partial}{\partial \theta_0}\ell(\bPsi) &=& -  \sum_{i=1}^{n} {\delta_i} \dfrac{ f'\left(u_i\right)}{ f\left(u_i\right)}
+ \sum_{i=1}^{n}  \dfrac{ f\left(u_i \right)}{ F\left(-u_i \right)}
\left[v_i -\delta_i\right],\\
\nabla_{\btheta}  \ell(\bPsi) &=& \sum_{i=1}^{n} {\delta_i} \tilde{\bx}_i e^{-\nu}  -  \sum_{i=1}^{n} {\delta_i} \tilde{\bx}_i \dfrac{ f'\left(u_i\right)}{ f\left(u_i\right)}
+ \sum_{i=1}^{n}  \dfrac{ f\left(u_i \right)}{ F\left(-u_i \right)}  \tilde{\bx}_i\left[v_i -\delta_i\right]
+ \sum_{i=1}^{n} \log F\left( -u_i\right)v_i\tilde{\bx}_i e^{-\nu}, \\
\nabla_{\bmeta}  \ell(\bPsi) &=& -\sum_{i=1}^{n} {\delta_i} \bx_i  - \sum_{i=1}^{n} \log F\left( -u_i\right)v_i\bx_i,
\end{eqnarray*}
with $u_i =  e^{\nu}\log(t_i) - \tilde{\bx}_i^{\top}\btheta -\theta_0$ and $v_i = \exp\left({ e^{-\nu}\tilde{\bx}_i^{\top}\btheta - \bx_i^{\top}\bmeta } \right)$.

\subsection{The Hessian matrix}\label{sec:hesslik}
The Hessian matrix is
\begin{eqnarray*}
H(\bPsi) = \begin{pmatrix}
\dfrac{\partial^2}{\partial \nu^2}  \ell(\bPsi)  & \nabla_{\nu,\theta_0}^2  \ell(\bPsi)  & \nabla_{\nu,\btheta}^2  \ell(\bPsi) & \nabla_{\nu,\bmeta}^2  \ell(\bPsi) \\
 \nabla_{\nu,\theta_0}^2\ell(\bPsi)  &\dfrac{\partial^2}{\partial \theta_0^2} \ell(\bPsi) & \nabla_{\btheta,\theta_0}^2  \ell(\bPsi) & \nabla_{\bmeta,\theta_0}^2  \ell(\bPsi) \\
\nabla_{\nu,\btheta}^2  \ell(\bPsi)  &  \nabla_{\btheta,\theta_0}^2  \ell(\bPsi) & \nabla_{\btheta}^2  \ell(\bPsi) & \nabla_{\bmeta,\btheta}^2  \ell(\bPsi) \\
\nabla_{\nu,\bmeta}^2  \ell(\bPsi) &  \nabla_{\bmeta,\theta_0}^2  \ell(\bPsi) &  \nabla_{\bmeta,\btheta}^2  \ell(\bPsi) & \nabla_{\bmeta}^2  \ell(\bPsi)
\end{pmatrix}.
\end{eqnarray*}

The entries of the hessian matrix are:
\begin{eqnarray*}
\dfrac{\partial^2}{\partial \nu^2}  \ell(\bPsi) &=&  \sum_{i=1}^{n} {\delta_i} \tilde{\bx}_i^{\top}\btheta e^{-\nu}
+ \sum_{i=1}^{n} {\delta_i} \left[ \dfrac{ f{''}\left(u_i\right)}{ f\left(u_i\right)}  - \left(  \dfrac{ f{'}\left(u_i\right)}{ f\left(u_i\right)} \right)^2 \right] \log(t_i)^2 e^{2\nu} + \sum_{i=1}^n \delta_i \dfrac{ f{'}\left(u_i\right)}{ f\left(u_i\right)} \log(t_i) e^{\nu}\\
&-& \sum_{i=1}^n \left[ \dfrac{ f{'}\left(u_i\right)}{ F\left(-u_i\right)}  + \left(  \dfrac{ f\left(u_i\right)}{F\left(-u_i\right)} \right)^2 \right] \log(t_i)^2 \left[ v_i - \delta_i \right]e^{2\nu} - \sum_{i=1}^{n} \dfrac{ f\left(u_i \right)}{ F\left(-u_i \right)} \log(t_i) \left[ v_i - \delta_i \right] e^{\nu}\\
&+& \sum_{i=1}^{n} \log F(-u_i) \tilde{\bx}_i^{\top}\btheta e^{-\nu} v_i \left[ 1+\tilde{\bx}_i^{\top}\btheta e^{-\nu} \right] + 2 \sum_{i=1}^{n} \dfrac{ f\left(u_i \right)}{ F\left(-u_i \right)} \log(t_i) v_i \tilde{\bx}_i^{\top}\btheta , \\
\dfrac{\partial^2}{\partial \theta_0^2} \ell(\bPsi) &=& \sum_{i=1}^{n} {\delta_i} \left[ \dfrac{f{''}(u_i)}{f(u_i)} - \left( \dfrac{f'(u_i)}{f(u_i)} \right)^2\right]
- \sum_{i=1}^{n} \left[   \dfrac{f{'}(u_i)}{F(-u_i)} + \left( \dfrac{f(u_i)}{F(-u_i)} \right)^2\right]\left[ v_i - \delta_i\right],\\
\nabla_{\btheta}^2  \ell(\bPsi) &=& \sum_{i=1}^{n} {\delta_i} \left[ \dfrac{f{''}(u_i)}{f(u_i)} - \left( \dfrac{f'(u_i)}{f(u_i)} \right)^2 \right] \tilde{\bx}_i \tilde{\bx}_i^{\top}
- \sum_{i=1}^{n} \left[ \dfrac{f{'}(u_i)}{F(-u_i)} + \left( \dfrac{f(u_i)}{F(-u_i)} \right)^2 \right] \tilde{\bx}_i \tilde{\bx}_i^{\top} \left[ v_i - \delta_i\right] \\
&+&  2\sum_{i=1}^{n}  \dfrac{f(u_i)}{F(-u_i)}  v_i \tilde{\bx}_i \tilde{\bx}_i^{\top} e^{-\nu} + \sum_{i=1}^n\log F(-u_i) v_i {\tilde{\bx}_i \tilde{\bx}_i^{\top}} e^{-2\nu},\\
\nabla_{\bmeta}^2  \ell(\bPsi) &=&  \sum_{i=1}^{n} \log F\left( -u_i\right)v_i\bx_i\bx_i^{\top},\\
\nabla_{\btheta,\theta_0}^2  \ell(\bPsi) &=& \sum_{i=1}^{n} {\delta_i} \left[  \dfrac{f{''}(u_i)}{f(u_i)} - \left( \dfrac{f{'}(u_i)}{f(u_i)}  \right)^2 \right] \tilde{\bx}_i^{\top} +  \sum_{i=1}^{n}  \dfrac{f(u_i)}{F(-u_i)} v_i \tilde{\bx}_i^{\top}e^{-\nu} \\
&-& \sum_{i=1}^{n}  \left[  \dfrac{f{'}(u_i)}{F(-u_i)} + \left( \dfrac{f(u_i)}{F(-u_i)}  \right)^2 \right] \left[ v_i - \delta_i \right] \tilde{\bx}_i^{\top}\\
\nabla_{\bmeta,\theta_0}^2  \ell(\bPsi) &=& -\sum_{i=1}^{n}  \dfrac{ f\left(u_i \right)}{ F\left(-u_i \right)} v_i \bx_i,\\
\nabla_{\nu,\theta_0}^2  \ell(\bPsi) &=& -\sum_{i=1}^{n} {\delta_i} \left[  \dfrac{f{''}(u_i)}{f(u_i)} - \left( \dfrac{f{'}(u_i)}{f(u_i)}  \right)^2 \right] \log(t_i) e^{\nu} - \sum_{i=1}^{n}    \dfrac{f(u_i)}{F(-u_i)} v_i \tilde{\bx}_i^{\top}\btheta e^{-\nu}\\
&+& \sum_{i=1}^{n}  \left[  \dfrac{f{'}(u_i)}{F(-u_i)} + \left( \dfrac{f(u_i)}{F(-u_i)}  \right)^2 \right] \log(t_i) e^{\nu} \left[ v_i - \delta_i \right],\\
\nabla_{\bmeta,\btheta}^2  \ell(\bPsi) &=&  -\sum_{i=1}^{n}  \dfrac{f(u_i)}{F(-u_i)}  v_i \tilde{\bx}_i {\bx}_i^{\top} - \sum_{i=1}^n\log F(-u_i) v_i {\tilde{\bx}_i {\bx}_i^{\top}}e^{-\nu},\\
\nabla_{\nu,\btheta}^2  \ell(\bPsi) &=& -\sum_{i=1}^{n} {\delta_i} \tilde{\bx}_i e^{-\nu} - \sum_{i=1}^{n} {\delta_i} \left[  \dfrac{f{''}(u_i)}{f(u_i)} - \left( \dfrac{f{'}(u_i)}{f(u_i)}  \right)^2 \right] \log(t_i) \tilde{\bx}_i e^{\nu} - \sum_{i=1}^{n} \dfrac{f(u_i)}{F(-u_i)} v_i \tilde{\bx}_i \tilde{\bx}_i^{\top} \btheta e^{-\nu}  \\
&+& \sum_{i=1}^{n}  \left[  \dfrac{f{'}(u_i)}{F(-u_i)} + \left( \dfrac{f(u_i)}{F(-u_i)}  \right)^2 \right] \log(t_i) \left[ v_i - \delta_i \right]\tilde{\bx}_i e^{\nu}  -   \sum_{i=1}^{n} \log F(-u_i) v_i \tilde{\bx}_i e^{-\nu} \left[ \tilde{\bx}_i^{\top} \btheta e^{-\nu} + 1 \right] \\
&-&  \sum_{i=1}^{n}    \dfrac{f(u_i)}{F(-u_i)} \log(t_i)  v_i \tilde{\bx}_i,\\
\nabla_{\nu,\bmeta}^2  \ell(\bPsi) &=&  \sum_{i=1}^{n} \dfrac{f(u_i)}{F(-u_i)} \log(t_i)  v_i {\bx}_i e^{\nu} + \sum_{i=1}^{n} \log F(-u_i) v_i {\bx}_i \tilde{\bx}_i^{\top}\btheta e^{-\nu}
\end{eqnarray*}

\subsection{Special tractable cases}\label{sec:specialcase}
\begin{itemize}
\item For the case $f=\phi$ is the Normal pdf, we have that
\begin{eqnarray*}
\dfrac{ f\left(u_i \right)}{ F\left(-u_i \right)} &=& r(u_i),\\
\dfrac{ f'\left(u_i\right)}{ f\left(u_i\right)} &=& -u_i,\\
\dfrac{ f'\left(u_i \right)}{ F\left(-u_i \right)} &=& -u_i r(u_i),\\
\dfrac{ f{''}\left(u_i\right)}{ f\left(u_i\right)} &=& -1+u_i^2,
\end{eqnarray*}
where $r(x) = \text{IMR}(-x)$ is the complementary Normal Inverse Mills ratio.

\item For the case where $f(u_i)=\frac{e^{-u_i}}{\left(e^{-u_i}+1\right)^2}$ is the Logistic pdf, we have that
\begin{eqnarray*}
\dfrac{ f\left(u_i \right)}{ F\left(-u_i \right)} &=& F(u_i),\\
\dfrac{ f'\left(u_i\right)}{ f\left(u_i\right)} &=& - \tanh\left( \dfrac{u_i}{2} \right),\\
\dfrac{ f'\left(u_i \right)}{ F\left(-u_i \right)} &=& -f(u_i)(e^{u_i}-1),\\
\dfrac{ f{''}\left(u_i\right)}{ f\left(u_i\right)} &=& 1-\frac{3}{\cosh (u_i)+1}.
\end{eqnarray*}

\item For the case where $f(u_i)=\dfrac{1}{2}\operatorname{sech}\left(\dfrac{\pi}{2}u_i\right)$ is the hyperbolic secant pdf, we have that
\begin{eqnarray*}
F(u_i) &=&  \frac{2 \arctan\left(e^{\frac{\pi  u_i}{2}}\right)}{\pi },\\
\dfrac{ f\left(u_i \right)}{ F\left(-u_i \right)} &=& \frac{\pi  \text{sech}\left(\frac{\pi  u_i}{2}\right)}{4 \arctan\left(e^{-\frac{\pi  u_i}{2}}\right)},\\
\dfrac{ f'\left(u_i\right)}{ f\left(u_i\right)} &=& -\frac{1}{2} \pi  \tanh \left(\frac{\pi  u_i}{2}\right),\\
\dfrac{ f'\left(u_i \right)}{ F\left(-u_i \right)} &=& -\frac{\pi ^2 \tanh \left(\frac{\pi  u_i}{2}\right)
   \text{sech}\left(\frac{\pi  u_i}{2}\right)}{8 \arctan\left(e^{-\frac{\pi  u_i}{2}}\right)},\\
\dfrac{ f{''}\left(u_i\right)}{ f\left(u_i\right)} &=& \frac{1}{8} \pi ^2 (\cosh (\pi  u_i)-3) \text{sech}^2\left(\frac{\pi   u_i}{2}\right).
\end{eqnarray*}

\item For the case where $f(u_i)= \left(\frac{1}{u_i^2+2}\right)^{3/2}$ is the Student-$t$ with 2 degrees of freedom pdf, we have that
\begin{eqnarray*}
F(u_i) &=& \frac{u_i}{2 \sqrt{2} \sqrt{\frac{u_i^2}{2}+1}}+\frac{1}{2} ,\\
\dfrac{ f\left(u_i \right)}{ F\left(-u_i \right)} &=& \frac{\sqrt{\frac{1}{u_i^2+2}} \left(\sqrt{u_i^2+2}+u_i\right)}{\sqrt{u_i^2+2}},\\
\dfrac{ f'\left(u_i\right)}{ f\left(u_i\right)} &=&  -\frac{3 u_i}{u_i^2+2},\\
\dfrac{ f'\left(u_i \right)}{ F\left(-u_i \right)} &=& -3 u_i \left(\frac{1}{u_i^2+2}\right)^{3/2}
   \left(\frac{u_i}{\sqrt{u_i^2+2}}+1\right) ,\\
\dfrac{ f{''}\left(u_i\right)}{ f\left(u_i\right)} &=&  \frac{6 \left(2 u_i^2-1\right)}{\left(u_i^2+2\right)^2}.
\end{eqnarray*}
\end{itemize}

\section{Log-Prior, Gradient and Hessian functions}
\subsection{Product Prior}
The vague prior $\theta_0\sim N(0,K)$, $K=10^6$, is included in the computations, but it is effectively constant and is omitted here. The Product prior on $(\btheta_{\bgamma},\bmeta_{\bgamma},\nu)$ has the form of
\begin{equation*}
\pi_P(\btheta_{\bgamma},\bmeta_{\bgamma},\nu)
= \left[ N\left(\btheta_{\bgamma}; 0,  g_{C_t} n \left(\tbX_{\bgamma}^\top \tbX_{\bgamma}\right)^{-1}\right)
       N\left(\bmeta_{\bgamma}; 0, g_{C_h} n\left(\bX_{\bgamma}^\top \bX_{\bgamma}\right)^{-1}\right) \right]\pi(\nu),
\end{equation*}
where
\begin{equation*}
\pi(\nu)
= \frac{\beta_\nu^{\alpha_\nu} e^{-\beta_\nu e^\nu}}{\Gamma(\alpha_\nu)} e^{\alpha_\nu \nu}.
\end{equation*}
The logarithm of this prior is
\begin{align*}
\ell_P(\btheta_{\bgamma},\bmeta_{\bgamma},\nu) &= - \frac{1}{2n} \left[ \btheta_{\bgamma}^\top \left(\tbX_{\bgamma}^\top \tbX_{\bgamma}\right) \btheta_{\bgamma}/g_{C_t}  + \bmeta_{\bgamma}^\top \left(\bX_{\bgamma}^\top \bX_{\bgamma}\right) \bmeta_{\bgamma}/g_{C_h}\right] - \beta_\nu e^\nu + \alpha_\nu \nu  \\
&-\frac{d_{\bgamma}}{2}\log(2\pi) - \frac{1}{2}\left( \log \left|g_{C_t} n \left(\tbX_{\bgamma}^\top \tbX_{\bgamma}\right)^{-1}\right| + \log\left|g_{C_h} n \left(\bX_{\bgamma}^\top \bX_{\bgamma}\right)^{-1}\right| \right) \\
&- \log\Gamma(\alpha_\nu) + \alpha_\nu \log\beta_\nu,
\end{align*}
where $\vert \cdot \vert$ denotes the determinant.
The gradient function of log prior is:
\begin{eqnarray*}
g(\btheta_{\bgamma},\bmeta_{\bgamma},\nu) =
\begin{pmatrix}
-\frac{1}{g_{C_t} n} \left(\tbX_{\bgamma}^\top \tbX_{\bgamma}\right) \btheta_{\bgamma} \\
 -\frac{1}{g_{C_h} n} \left(\bX_{\bgamma}^\top \bX_{\bgamma}\right) \bmeta_{\bgamma}  \\
\alpha_\nu - \beta_\nu e^\nu
\end{pmatrix},
\end{eqnarray*}
and the Hessian matrix is :
\begin{equation*}
H(\btheta_{\bgamma},\bmeta_{\bgamma},\nu) =
\begin{pmatrix}
-\frac{1}{g_{C_t} n} \left(\tbX_{\bgamma}^\top \tbX_{\bgamma}\right) & \mathbf{0} & \mathbf{0} \\
\mathbf{0}  & -\frac{1}{g_{C_h} n} \left(\bX_{\bgamma}^\top \bX_{\bgamma} \right) & \mathbf{0}\\
\mathbf{0} & \mathbf{0} & -\beta_\nu e^\nu
\end{pmatrix}.
\end{equation*}

\section{Derivation of Proposition 1: Expected Fisher information Matrix for GH model}

Let $\bk = (\btheta_{\bgamma},\bmeta_{\bgamma})^\top$. Following the decomposition of the expected Fisher information
in \cite{DESANTIS2001}, we could write it as
\begin{equation*}
    \begin{aligned}
I({\bk}) &= -\mathbb E \left[ \nabla^2_{\bk} \ell({\bk}) \right] \\
&= \sum_{k=1}^n \mathbb E \left[ -\nabla^2_{\bk} \ell^{cen}({\bk}) \mid {\bk} \right]_k \cdot P(o_k > c_k) + \sum_{k=1}^n \mathbb E \left[ -\nabla^2_{\bk} \ell^{unc}({\bk})\mid {\bk} \right]_k \cdot P(o_k \leq c_k),
\end{aligned}
\end{equation*}

where $v_k = \exp\left({ e^{-\nu}\tilde{\bx}_k^{\top}\btheta - \bx_k^{\top}\bmeta } \right)$, $P(o_k \leq c_k) = 1- S(c_k) = 1- \exp[-H_0(c_k)v_k]$, and $P(o_k > c_k) = S(c_k) = \exp[-H_0(c_k)v_k]$. From Section~\ref{sec:hesslik} we know the Hessians are in the form of
$$
\begin{aligned}
\nabla_{\boldsymbol{\theta}}^2 \ell(\boldsymbol{\Psi}) & =\sum_{i=1}^n \delta_i\left[\frac{f^{\prime \prime}\left(u_i\right)}{f\left(u_i\right)}-\left(\frac{f^{\prime}\left(u_i\right)}{f\left(u_i\right)}\right)^2\right] \tilde{\mathbf{x}}_i \tilde{\mathbf{x}}_i^{\top}-\sum_{i=1}^n\left[\frac{f^{\prime}\left(u_i\right)}{F\left(-u_i\right)}+\left(\frac{f\left(u_i\right)}{F\left(-u_i\right)}\right)^2\right] \tilde{\mathbf{x}}_i \tilde{\mathbf{x}}_i^{\top}\left[v_i-\delta_i\right] \\
& +2 \sum_{i=1}^n \frac{f\left(u_i\right)}{F\left(-u_i\right)} v_i \tilde{\mathbf{x}}_i \tilde{\mathbf{x}}_i^{\top} e^{-\nu}+\sum_{i=1}^n \log F\left(-u_i\right) v_i \tilde{\mathbf{x}}_i \tilde{\mathbf{x}}_i^{\top} e^{-2 \nu}, \\
\nabla_{\boldsymbol{\eta}}^2 \ell(\boldsymbol{\Psi}) & =\sum_{i=1}^n \log F\left(-u_i\right) v_i \mathbf{x}_i \mathbf{x}_i^{\top}.
\end{aligned}
$$
with $u_i=e^\nu \log \left(t_i\right)-\tilde{\mathbf{x}}_i^{\top} \boldsymbol{\theta}-\theta_0$ and $ v_i=\exp \left(e^{-\nu} \tilde{\mathbf{x}}_i^{\top} \boldsymbol{\theta}-\mathbf{x}_i^{\top} \boldsymbol{\eta}\right)$. If we rearrange the above expressions, we have the following expression
$$
\begin{aligned}
& \nabla_{\boldsymbol{\theta}}^2 \ell(\mathbf{\Psi})=-{\boldsymbol{\tilde X}}^{\top} \boldsymbol{W}_{\boldsymbol{\theta}} {\boldsymbol{\tilde X}}, \\
& \nabla_{\boldsymbol{\eta}}^2 \ell(\mathbf{\Psi})=-\boldsymbol{X}^{\top} \boldsymbol{W}_{\boldsymbol{\eta}} \boldsymbol{X},
\end{aligned}
$$
where $\boldsymbol{W}_{\boldsymbol{\theta}}$ and $\boldsymbol{W}_{\boldsymbol{\eta}}$ are diagonal matrices with the following elements on the diagonals:
$$
\begin{aligned}
w_{\theta, i}&=-\delta_i\left[\frac{f^{\prime \prime}\left(u_i\right)}{f\left(u_i\right)}-\left(\frac{f^{\prime}\left(u_i\right)}{f\left(u_i\right)}\right)^2+\frac{f^{\prime}\left(u_i\right)}{F\left(-u_i\right)}+\left(\frac{f\left(u_i\right)}{F\left(-u_i\right)}\right)^2\right],\\
&-v_i\left[-\frac{f^{\prime}\left(u_i\right)}{F\left(-u_i\right)}-\left(\frac{f\left(u_i\right)}{F\left(-u_i\right)}\right)^2+2 \frac{f\left(u_i\right)}{F\left(-u_i\right)} e^{-\nu}+\log F\left(-u_i\right) e^{-2 \nu}\right] \\
w_{\eta, i}&=-v_i\log F\left(-u_i\right).
\end{aligned}
$$

If we write the above equations in terms of hazard functions, and consider the censored and uncensored part separately, these diagonal elements become
$$
\begin{aligned}
w_{\theta, i}^{cen}&=v_i\left[h_0'(c_i)-2 \sigma h_0(c_i)+\sigma^2H_0(c_i) \right],\\
w_{\eta, i}^{cen}&= v_i H_0(c_i),
\end{aligned}
$$
for the censored part, and
$$
\begin{aligned}
w_{\theta, i}^{unc}&=-\frac{f^{\prime \prime}\left(u_i\right)}{f\left(u_i\right)}+\left(\frac{f^{\prime}\left(u_i\right)}{f\left(u_i\right)}\right)^2-h_0'(o_i)+v_i\left[h_0'(o_i)-2 \sigma h_0(o_i)+\sigma^2H_0(o_i) \right], \\
&= 1 + h_0'(o_i)\left[v_i - 1 \right] - v_i\sigma \left[ 2h_0(o_i) - \sigma H_0(o_i) \right],\\
w_{\eta, i}^{unc}&= v_i H_0(o_i),
\end{aligned}
$$
for the uncensored part. Note that the simplification in the second line of $w_{\theta, i}^{unc}$ requires $f$ to be standard Gaussian.

Having the expressions for the negative Hessians, we could integrate $o_i$ and $c_i$ out. The integral of cumulative hazard have closed form solution, i.e. $\int v_i H_0(o_i)dF_t(o_i)=1$, where $F_t$ is the lifetime distribution derivable from the baseline $F_0$. A similar derivation can be found in the supplementary material of \cite{castellanos:2021} and will not be expanded here. However, the integrals of the baseline hazard and derivative of baseline hazard have no closed-form solution even under the assumption of $f$ being Gaussian. For example, the term $\int h_0(o_i)dF_t(o_i)$ could be written in the following form:
$$
\int h_0(o_i)dF_t(o_i) = (v_i e^{\nu})^2 \int^\infty_0 o_i^{-2}\frac{\phi\left(e^{\nu}\log(o_i) - \theta_0\right) \cdot \phi\left(e^{\nu}\log(o_i) - \tilde{\mathbf{x}}_i^{\top} \boldsymbol{\theta} - \theta_0\right)}{1-\Phi\left(e^{\nu}\log(o_i) - \theta_0\right)}do_i,
$$
where $\phi$ is the normal PDF and $\Phi$ is the normal CDF. This integral obviously does not have closed form solution, and the same issue exists for $\int h_0'(o_i)dF_t(o_i)$.

\begin{table}[htbp]
\centering
\caption{Expressions of diagonal elements of $\boldsymbol{M}^{\delta_k}_{\cdot}$ in Proposition 1 assuming $f$ is standard normal, where $i=1,2,...,n$ denotes the $i$-th diagonal element, $r_i = \exp\left({ e^{-\nu}\tilde{\bx}_{\bgamma,i}^{\top}\btheta_{\bgamma} } \right)$, and $v_i = \exp\left({ e^{-\nu}\tilde{\bx}_{\bgamma,i}^{\top}\btheta_{\bgamma} - \bx_{\bgamma,i}^{\top}\bmeta_{\bgamma} } \right)$. In addition, $\boldsymbol{M}^{\delta_k=1}_{\btheta_{\bgamma}\bmeta_{\bgamma}}=\boldsymbol{M}^{\delta_k=1}_{\bmeta_{\bgamma}\btheta_{\bgamma}}=-\sigma+\int v_i h_0\left(o_k\right) d F_t\left(o_k\right)$ and $\boldsymbol{M}^{\delta_k=0}_{\btheta_{\bgamma}\bmeta_{\bgamma}}=\boldsymbol{M}^{\delta_k=0}_{\bmeta_{\bgamma}\btheta_{\bgamma}}=-\sigma+\int v_i h_0\left(c_k\right) d F_t\left(c_k\right)$.}
\label{tab:eftab}
{\renewcommand{\arraystretch}{1.5}
\begin{tabular}{>{\centering\arraybackslash}m{1.5cm} >{\centering\arraybackslash}m{4.5cm} >{\centering\arraybackslash}m{1.5cm} >{\centering\arraybackslash}m{4.5cm} >{\centering\arraybackslash}m{1.5cm}}
\toprule
$(\boldsymbol{M}^{\delta_k}_{\cdot})_{i,i}$ & \multicolumn{2}{c}{\textbf{Uncensored}} & \multicolumn{2}{c}{\textbf{Censored}} \\
\cmidrule(lr){2-3} \cmidrule(lr){4-5}
\textbf{Model} & $\boldsymbol{M}^{\delta_k=1}_{\btheta_{\bgamma}}$ & $\boldsymbol{M}^{\delta_k=1}_{\bmeta_{\bgamma}}$ &  $\boldsymbol{M}^{\delta_k=0}_{\btheta_{\bgamma}}$ &  $\boldsymbol{M}^{\delta_k=0}_{\bmeta_{\bgamma}}$ \\
\midrule
\text{PH} & $-$ & $1$ & $-$ & $1$ \\
\addlinespace[0.3em]
\vspace{1em}\text{AH} & $\begin{aligned}
\sigma^2 + 1 &{}- \int 2\sigma r_i h_0(o_k) \, dF_t(o_k) \\
&+ \int (r_i-1) h_0'(o_k) \, dF_t(o_k)
\end{aligned}$ & \vspace{1em}$-$ & $\begin{aligned}
\sigma^2 &{}- \int 2\sigma r_i h_0(c_k) \, dF_t(c_k) \\
&+ \int r_i h_0'(c_k) \, dF_t(c_k)
\end{aligned}$ & \vspace{1em}$-$ \\
\addlinespace[0.3em]
\text{AFT} & $1$ & $-$ & $\int h_0'(c_k) \, dF_t(c_k)$ & $-$ \\
\addlinespace[0.3em]
\vspace{1.2em}\text{GH} & $\begin{aligned}
\sigma^2 + 1 &{}- \int 2\sigma v_i h_0(o_k) \, dF_t(o_k) \\
&+ \int (v_i-1) h_0'(o_k) \, dF_t(o_k)
\end{aligned}$ & \vspace{1.2em}$1$ & $\begin{aligned}
\sigma^2 &{}- \int 2\sigma v_i h_0(c_k) \, dF_t(c_k) \\
&+ \int v_i h_0'(c_k) \, dF_t(c_k)
\end{aligned}$ & \vspace{1.2em}$1$ \\
\bottomrule
\end{tabular}}
\end{table}

\section{Full expression for prior on model space}

Combining the factors in (8)--(10) in the main text, we derive the following full expression of $\pi(\bgamma)$:
\begin{equation}\label{eqn:modprior}
\pi(\bgamma) = \frac{\text{B}(a_\Lambda+|\boldsymbol{\Lambda}|, b_\Lambda+p_0)}{\text{B}(a_\Lambda, b_\Lambda)}\cdot
\begin{cases}
1, & \bgamma = \mathbf{0}\\
h_{l}/H, & \hz(\bgamma) = l \in \{\text{AH},\text{PH},\text{AFT}\}\\
\dfrac{h_{GH}}{H} \cdot \dfrac{w_\omega}{\sum_{\omega'=|\boldsymbol{\Lambda}|}^{2|\boldsymbol{\Lambda}|} N_{\omega'} w_{\omega'}}, &  \hz(\bgamma) = \text{GH},
\end{cases}
\end{equation}
where $H = h_{AH}+h_{PH}+h_{AFT}+h_{GH}$ and $N_\omega$ and $w_\omega$ as defined in Section 3.3 of the main text. For the default $q=1/3$, the prior of a GH model equals $\text{B}(a_\Lambda+|\boldsymbol{\Lambda}|, b_\Lambda+p_0)/\text{B}(a_\Lambda, b_\Lambda)\cdot (h_{GH}/H)/[(|\boldsymbol{\Lambda}|+1)N_\omega]$.

Since selection is performed with respect to a prior on the model space, one could impose additional conditions, such as guaranteeing positive-definite Hessians, by letting $\pi(\bgamma)$ assign zero mass to any $\bgamma$ for which $n_o<d_{\bgamma}+2$ \citep{rossell:2018,rossell:2023}. 

\section{Derivation of marginal likelihood under LCM prior under integrated Laplace approximation}

The marginal likelihood is given by
\begin{equation}\label{eq:ebmarglik}
    p(\bt \mid \bgamma)= \int \exp [\ell(\bPsi_{\bgamma}) ]\pi(\bPsi_{\bgamma}) \,d\bPsi_{\bgamma},
\end{equation}

where $\pi(\bPsi_{\bgamma}) = \pi(\btheta_{\bgamma},\bmeta_{\bgamma}) \pi(\nu) \pi(\theta_0)$ is the prior of the parameters. The following Gaussian priors are assigned to the parameters:
\begin{equation}\label{eq:ebprior}
    \left[\begin{array}{c}\boldsymbol{\theta}_{\bgamma}\\ \boldsymbol{\eta}_{\bgamma} \end{array} \right] \mid g_E \sim N(0, ng_E\mathcal J_{\boldsymbol{\theta},\boldsymbol{\eta}}(\widehat\bPsi_{\bgamma})^{-1}),\quad \theta_0 \sim N(0,K), \quad \nu \sim N(\mu_\nu,\sigma^2_\nu),
\end{equation}
where $\mj_{\boldsymbol{\theta},\boldsymbol{\eta}}(\widehat\bPsi_{\bgamma}) = -\nabla^2_{\boldsymbol{\theta},\boldsymbol{\eta}}(\widehat\bPsi_{\bgamma})$ is the $(\btheta_{\bgamma},\bmeta_{\bgamma})^{\top}$ block of observed Fisher information evaluated at $\widehat\bPsi_{\bgamma}$, the MLE of $\bPsi_{\bgamma}$. $K$ is a large constant, $n$ is the sample size, and $\mu_\nu$, $g_E$, and $\sigma^2_\nu$ are hyperparameters.

Following the idea of integrated Laplace approximation in \cite{WangGeorge:2007}, we start by expanding the integrated log-likelihood $\ell(\bPsi_{\bgamma})$ at $\widehat \bPsi_{\bgamma}$, the MLE of $\bPsi_{\bgamma}$:
\begin{equation}\label{eq:ebintexp}
    \ell(\bPsi_{\bgamma}) \approx \ell(\widehat\bPsi_{\bgamma}) - \frac12 \left( \bPsi_{\bgamma}-\widehat\bPsi_{\bgamma} \right)^{\top} \mj (\widehat \bPsi_{\bgamma}) \left( \bPsi_{\bgamma}-\widehat\bPsi_{\bgamma} \right),
\end{equation}
where $\mj (\widehat \bPsi_{\bgamma}) = -\nabla^2 \ell(\widehat\bPsi_{\bgamma})$ is the observed Fisher information evaluated at $\widehat\bPsi_{\bgamma}$. Plugging this approximation back to the integral in \eqref{eq:ebmarglik} and substituting in the prior, we have
\begin{equation*}
    \begin{aligned}
p(\bt \mid \bgamma) &\approx \exp[\ell(\widehat\bPsi_{\bgamma})]\int \exp \left\{-\frac12 \left( \bPsi_{\bgamma}-\widehat\bPsi_{\bgamma} \right)^{\top} \mj (\widehat \bPsi_{\bgamma}) \left( \bPsi_{\bgamma}-\widehat\bPsi_{\bgamma} \right)\right\} \pi(\bPsi_{\bgamma}) \,d\bPsi_{\bgamma}\\
&\propto \exp[\ell(\widehat\bPsi_{\bgamma})] (2 \pi)^{-\frac{d_{\bgamma}+2}{2}}\left|ng_E \mathcal{J}_{\boldsymbol{\theta}, \boldsymbol{\eta}}(\widehat\bPsi_{\bgamma})^{-1}\right|^{-\frac{1}{2}}
\\
\times&\int \exp \left\{-\frac{1}{2}\left( \bPsi_{\bgamma}-\widehat\bPsi_{\bgamma} \right)^{\top} \mj (\widehat \bPsi_{\bgamma}) \left( \bPsi_{\bgamma}-\widehat\bPsi_{\bgamma} \right)-\frac{1}{2 ng_E}\left[\begin{array}{c}\boldsymbol{\theta}_{\bgamma}\\ \boldsymbol{\eta}_{\bgamma} \end{array} \right]^{\top} \mathcal{J}_{\theta, \eta}(\widehat\bPsi_{\bgamma}) \left[\begin{array}{c}\boldsymbol{\theta}_{\bgamma}\\ \boldsymbol{\eta}_{\bgamma} \end{array} \right]\right.\\
&\qquad \qquad \left. -\frac{1}{2 \sigma_\nu^2}\left(\nu-\mu_\nu\right)^2-\frac{1}{2 K} \theta_0^2 \right\}\,d\bPsi_{\bgamma}.
\end{aligned}
\end{equation*}

For cleaner presentation, we omit the $(\widehat\bPsi_{\bgamma})$ in $\mj_{\cdot \cdot}$, as the Fisher information will all be evaluated at $\widehat\bPsi_{\bgamma}$. For the rest of this section only, denote $\bk_{\bgamma} = [\boldsymbol{\theta}_{\bgamma},\boldsymbol{\eta}_{\bgamma}]^{\top}$, then the exponent in the integral could be written as
\begin{equation*}
    \begin{aligned}
&-\frac{1}{2}\left( \bPsi_{\bgamma}-\widehat\bPsi_{\bgamma} \right)^{\top} \mj (\widehat \bPsi_{\bgamma}) \left( \bPsi_{\bgamma}-\widehat\bPsi_{\bgamma} \right)-\frac{1}{2ng_E}\left[\begin{array}{c}\boldsymbol{\theta}_{\bgamma}\\ \boldsymbol{\eta}_{\bgamma} \end{array} \right]^{\top} \mathcal{J}_{\theta, \eta}(\widehat\bPsi_{\bgamma}) \left[\begin{array}{c}\boldsymbol{\theta}_{\bgamma}\\ \boldsymbol{\eta}_{\bgamma} \end{array} \right]-\frac{1}{2 \sigma_\nu^2}\left(\nu-\mu_\nu\right)^2-\frac{1}{2 K} \theta_0^2\\
=& -\frac12[(\bk_{\bgamma}-\widehat{\bk}_{\bgamma})^{\top} \mathcal{J}_{\theta, \eta}(\bk_{\bgamma}-\widehat{\bk}_{\bgamma})+2(\bk_{\bgamma}-\widehat{\bk}_{\bgamma})^{\top} \mathcal{J}_{\theta, \eta, \nu}(\nu-\widehat{\nu})\\
&+2(\bk_{\bgamma}-\widehat{\bk}_{\bgamma})^{\top} \mathcal{J}_{\theta, \eta, \theta_0}\left(\theta_0-\widehat{\theta}_0\right)+(\nu-\widehat{\nu})^2 \mathcal{J}_{\nu \nu}+2(\nu-\widehat{\nu})\left(\theta_0-\widehat{\theta}_0\right) \mathcal{J}_{\nu, \theta_0}+\left(\theta_0-\widehat{\theta}_0\right)^2 \mathcal{J}_{\theta_0 \theta_0} ]\\
&-\frac{1}{2ng_E} \bk_{\bgamma}^{\top} \mathcal{J}_{\theta, \eta}\bk_{\bgamma}-\frac{1}{2 \sigma_\nu^2}\left(\nu-\mu_\nu\right)^2-\frac{1}{2 K} \theta_0^2.
\end{aligned}
\end{equation*}

We will now discuss the non-null model case and null model case separately.

\subsection{Non-null model case}

\paragraph{$\bk_{\bgamma} = [\boldsymbol{\theta}_{\bgamma},\boldsymbol{\eta}_{\bgamma}]^{\top}$ part.}
Consider first the terms containing $\bk_{\bgamma}$ or $\widehat{\bk}_{\bgamma}$. Expanding the centred quadratic and the cross terms around $\widehat{\bk}_{\bgamma}$, we obtain
\begin{equation*}
    \begin{aligned}
&-\frac12 (\bk_{\bgamma}-\widehat{\bk}_{\bgamma})^{\top} \mathcal{J}_{\theta, \eta}(\bk_{\bgamma}-\widehat{\bk}_{\bgamma})
-(\bk_{\bgamma}-\widehat{\bk}_{\bgamma})^{\top} \mathcal{J}_{\theta, \eta, \nu}(\nu-\widehat{\nu})
-(\bk_{\bgamma}-\widehat{\bk}_{\bgamma})^{\top} \mathcal{J}_{\theta, \eta, \theta_0}\left(\theta_0-\widehat{\theta}_0\right)
-\frac{1}{2ng_E} \bk_{\bgamma}^{\top} \mathcal{J}_{\theta, \eta}\bk_{\bgamma}\\
=& -\frac{1}{2} \bk_{\bgamma}^{\top} \mathcal{J}_{\boldsymbol{\theta}, \boldsymbol{\eta}}\left(1+\frac{1}{ng_E}\right) \bk_{\bgamma}+\bk_{\bgamma}^{\top} \mathcal{J}_{\boldsymbol{\theta}, \boldsymbol{\eta}} \widehat{\bk}_{\bgamma}-\bk_{\bgamma}^{\top} \mathcal{J}_{\boldsymbol{\theta}, \boldsymbol{\eta}, \nu}(\nu-\widehat{\nu})-\bk_{\bgamma}^{\top} \mathcal{J}_{\boldsymbol{\theta}, \boldsymbol{\eta}, \theta_0}\left(\theta_0-\widehat{\theta}_0\right)\\
&-\frac{1}{2} \widehat{\bk}_{\bgamma}^{\top} \mathcal{J}_{\theta, \eta} \widehat{\bk}_{\bgamma}
+\widehat{\bk}_{\bgamma}^{\top} \mathcal{J}_{\theta, \eta, \nu}(\nu-\widehat{\nu})
+\widehat{\bk}_{\bgamma}^{\top} \mathcal{J}_{\theta, \eta, \theta_0}\left(\theta_0-\widehat{\theta}_0\right)\\
=& -\frac{1+ng_E}{2ng_E} \bk_{\bgamma}^{\top} \mathcal{J}_{\theta, \eta} \bk_{\bgamma}+\bk_{\bgamma}^{\top} \mathbf{b}
-\frac{1}{2} \widehat{\bk}_{\bgamma}^{\top} \mathcal{J}_{\theta, \eta} \widehat{\bk}_{\bgamma}
+\widehat{\bk}_{\bgamma}^{\top} \mathcal{J}_{\theta, \eta, \nu}(\nu-\widehat{\nu})
+\widehat{\bk}_{\bgamma}^{\top} \mathcal{J}_{\theta, \eta, \theta_0}\left(\theta_0-\widehat{\theta}_0\right)\\
=& -\frac{1+ng_E}{2ng_E}\left(\bk_{\bgamma}-\frac{ng_E}{1+ng_E} \mathcal{J}_{\theta, \eta}^{-1} \mathbf{b}\right)^{\top} \mathcal{J}_{\theta, \eta}\left(\bk_{\bgamma}-\frac{ng_E}{1+ng_E} \mathcal{J}_{\theta, \eta}^{-1} \mathbf{b}\right)+\frac{ng_E}{2(1+ng_E)} \mathbf{b}^{\top} \mathcal{J}_{\theta, \eta}^{-1} \mathbf{b}\\
&-\frac{1}{2} \widehat{\bk}_{\bgamma}^{\top} \mathcal{J}_{\theta, \eta} \widehat{\bk}_{\bgamma}
+\widehat{\bk}_{\bgamma}^{\top} \mathcal{J}_{\theta, \eta, \nu}(\nu-\widehat{\nu})
+\widehat{\bk}_{\bgamma}^{\top} \mathcal{J}_{\theta, \eta, \theta_0}\left(\theta_0-\widehat{\theta}_0\right),
\end{aligned}
\end{equation*}
where $\mathbf{b}=\mathcal{J}_{\boldsymbol{\theta}, \eta} \widehat{\bk}_{\bgamma}-\mathcal{J}_{\boldsymbol{\theta}, \eta, \nu}(\nu-\widehat{\nu})-\mathcal{J}_{\boldsymbol{\theta}, \eta, \theta_0}\left(\theta_0-\widehat{\theta}_0\right)$. The three terms in the last line do not involve $\bk_{\bgamma}$ and will be carried forward to the $(\nu,\theta_0)$ part below. After the completion of square, the integral with respect to $\bk_{\bgamma}$ becomes
\begin{equation}\label{eq:ebbeta}
    \int \exp \left\{-\frac{1+ng_E}{2ng_E}\left(\bk_{\bgamma}-\frac{ng_E}{1+ng_E} \mathcal{J}_{\theta, \eta}^{-1} \mathbf{b}\right)^{\top} \mathcal{J}_{\theta, \eta}\left(\bk_{\bgamma}-\frac{ng_E}{1+ng_E} \mathcal{J}_{\theta, \eta}^{-1} \mathbf{b}\right)\right\} \,d \bk_{\bgamma}=\left(\frac{2 \pi ng_E}{1+ng_E}\right)^{\frac{d_{\bgamma}}{2}}\left|\mathcal{J}_{\theta, \eta}\right|^{-\frac{1}{2}}.
\end{equation}

\paragraph{$\nu, \theta_0$ part} The remaining part consists of the $(\nu,\theta_0)$ quadratic from the expansion \eqref{eq:ebintexp}, the priors on $\nu$ and $\theta_0$, the residual term from the completion of the square, and the terms carried over from the $\bk_{\bgamma}$ part:
\begin{equation*}
\begin{aligned}
&-\frac{1}{2 \sigma_\nu^2}\left(\nu-\mu_\nu\right)^2-\frac{1}{2 K} \theta_0^2 -\frac{1}{2}(\nu-\widehat{\nu})^2 \mathcal{J}_{\nu \nu}-(\nu-\widehat{\nu})\left(\theta_0-\widehat{\theta}_0\right) \mathcal{J}_{\nu, \theta_0}-\frac{1}{2}\left(\theta_0-\widehat{\theta}_0\right)^2 \mathcal{J}_{\theta_0 \theta_0} \\
&+\frac{ng_E}{2(1+ng_E)} \mathbf{b}^{\top} \mathcal{J}_{\theta, \eta}^{-1} \mathbf{b}
-\frac{1}{2} \widehat{\bk}_{\bgamma}^{\top} \mathcal{J}_{\theta, \eta} \widehat{\bk}_{\bgamma}
+\widehat{\bk}_{\bgamma}^{\top} \mathcal{J}_{\theta, \eta, \nu}(\nu-\widehat{\nu})
+\widehat{\bk}_{\bgamma}^{\top} \mathcal{J}_{\theta, \eta, \theta_0}\left(\theta_0-\widehat{\theta}_0\right).
\end{aligned}
\end{equation*}
Expanding $\mathbf{b}^{\top} \mathcal{J}_{\theta, \eta}^{-1} \mathbf{b}$, we have
\begin{equation*}
\begin{aligned}
\mathbf{b}^{\top} \mathcal{J}_{\theta, \eta}^{-1} \mathbf{b}
=&\ \widehat{\bk}_{\bgamma}^{\top} \mathcal{J}_{\theta, \eta} \widehat{\bk}_{\bgamma}
-2\, \widehat{\bk}_{\bgamma}^{\top} \mathcal{J}_{\theta, \eta, \nu}(\nu-\widehat{\nu})
-2\, \widehat{\bk}_{\bgamma}^{\top} \mathcal{J}_{\theta, \eta, \theta_0}\left(\theta_0-\widehat{\theta}_0\right)\\
&+(\nu-\widehat{\nu})^2 \mathcal{J}_{\boldsymbol{\theta}, \boldsymbol{\eta}, \nu}^{\top} \mathcal{J}_{\boldsymbol{\theta}, \boldsymbol{\eta}}^{-1} \mathcal{J}_{\boldsymbol{\theta}, \boldsymbol{\eta}, \nu}+2(\nu-\widehat{\nu})\left(\theta_0-\widehat{\theta}_0\right) \mathcal{J}_{\boldsymbol{\theta}, \boldsymbol{\eta}, \nu}^{\top} \mathcal{J}_{\boldsymbol{\theta}, \boldsymbol{\eta}}^{-1} \mathcal{J}_{\boldsymbol{\theta}, \boldsymbol{\eta}, \theta_0}+\left(\theta_0-\widehat{\theta}_0\right)^2 \mathcal{J}_{\boldsymbol{\theta}, \boldsymbol{\eta}, \theta_0}^{\top} \mathcal{J}_{\boldsymbol{\theta}, \boldsymbol{\eta}}^{-1} \mathcal{J}_{\boldsymbol{\theta}, \boldsymbol{\eta}, \theta_0}.
\end{aligned}
\end{equation*}
Multiplying by $\frac{ng_E}{2(1+ng_E)}$ and combining with the terms carried over from the $\bk_{\bgamma}$ part, the coefficient of the $\widehat{\bk}_{\bgamma}^{\top} \mathcal{J}_{\theta, \eta} \widehat{\bk}_{\bgamma}$ becomes
\begin{equation*}
\frac{ng_E}{2(1+ng_E)} - \frac12 = -\frac{1}{2(1+ng_E)},
\end{equation*}
and the coefficient of each of the linear terms $\widehat{\bk}_{\bgamma}^{\top} \mathcal{J}_{\theta, \eta, \nu}(\nu-\widehat{\nu})$ and $\widehat{\bk}_{\bgamma}^{\top} \mathcal{J}_{\theta, \eta, \theta_0}(\theta_0-\widehat{\theta}_0)$ becomes
\begin{equation*}
1 - \frac{ng_E}{1+ng_E} = \frac{1}{1+ng_E}.
\end{equation*}

Define the partial information matrix for $(\nu, \theta_0)$ as
\begin{equation*}
    \tilde{\mathcal{J}} = \begin{bmatrix} \tilde{\mathcal{J}}_{\nu\nu} & \tilde{\mathcal{J}}_{\nu,\theta_0} \\ \tilde{\mathcal{J}}_{\theta_0,\nu} & \tilde{\mathcal{J}}_{\theta_0\theta_0} \end{bmatrix},
\end{equation*}
where each entry is an adjusted Schur complement given by
\begin{equation*}
\begin{aligned}
&\tilde{\mathcal{J}}_{\nu \nu}=\mathcal{J}_{\nu \nu}-\frac{ng_E}{1+ng_E} \mathcal{J}_{\boldsymbol{\theta}, \eta, \nu}^{\top} \mathcal{J}_{\boldsymbol{\theta}, \eta}^{-1} \mathcal{J}_{\boldsymbol{\theta}, \eta, \nu}\\
&\tilde{\mathcal{J}}_{\nu, \theta_0}=\mathcal{J}_{\nu, \theta_0}-\frac{ng_E}{1+ng_E} \mathcal{J}_{\boldsymbol{\theta}, \eta, \nu}^{\top} \mathcal{J}_{\boldsymbol{\theta}, \eta}^{-1} \mathcal{J}_{\boldsymbol{\theta}, \eta, \theta_0}\\
&\tilde{\mathcal{J}}_{\theta_0 \theta_0}=\mathcal{J}_{\theta_0 \theta_0}-\frac{ng_E}{1+ng_E} \mathcal{J}_{\boldsymbol{\theta}, \boldsymbol{\eta}, \theta_0}^{\top} \mathcal{J}_{\boldsymbol{\theta}, \boldsymbol{\eta}}^{-1} \mathcal{J}_{\boldsymbol{\theta}, \boldsymbol{\eta}, \theta_0}.
\end{aligned}
\end{equation*}

Then we could rewrite the remaining part as
\begin{equation*}
\begin{aligned}
    &-\frac{1}{2}\left[\begin{array}{c}
\nu-\widehat{\nu} \\
\theta_0-\widehat{\theta}_0
\end{array}\right]^{\top} \tilde{\mathcal{J}}\left[\begin{array}{c}
\nu-\widehat{\nu} \\
\theta_0-\widehat{\theta}_0
\end{array}\right]
+\frac{1}{1+ng_E}\left[\widehat{\bk}_{\bgamma}^{\top} \mathcal{J}_{\theta, \eta, \nu}(\nu-\widehat{\nu})+\widehat{\bk}_{\bgamma}^{\top} \mathcal{J}_{\theta, \eta, \theta_0}\left(\theta_0-\widehat{\theta}_0\right)\right]\\
&-\frac{1}{2(1+ng_E)} \widehat{\bk}_{\bgamma}^{\top} \mathcal{J}_{\theta, \eta} \widehat{\bk}_{\bgamma}
-\frac{1}{2 \sigma_\nu^2}\left(\nu-\mu_\nu\right)^2-\frac{1}{2 K} \theta_0^2
= -\frac{1}{2} \mathbf{z}^{\top} \mathbf{P}_{\bgamma} \mathbf{z}+\mathbf{h}_{\bgamma}^{\top} \mathbf{z} + C_{\bgamma},
\end{aligned}
\end{equation*}
where
\begin{equation*}
\mathbf{z}=\left(\nu, \theta_0\right)^{\top}, \quad \mathbf{P}_{\bgamma}=\left[\begin{array}{cc}
\tilde{\mathcal{J}}_{\nu \nu}+\frac{1}{\sigma_\nu^2} & \tilde{\mathcal{J}}_{\nu, \theta_0} \\
\tilde{\mathcal{J}}_{\nu, \theta_0} & \tilde{\mathcal{J}}_{\theta_0 \theta_0}+\frac{1}{K}
\end{array}\right],\quad
\mathbf{h}_{\bgamma}=\left[\begin{array}{c}
\tilde{\mathcal{J}}_{\nu \nu} \widehat{\nu}+\tilde{\mathcal{J}}_{\nu, \theta_0} \widehat{\theta}_0+\frac{\mu_\nu}{\sigma_\nu^2}+\frac{1}{1+ng_E} \widehat{\bk}_{\bgamma}^{\top} \mathcal{J}_{\theta, \eta, \nu} \\
\tilde{\mathcal{J}}_{\nu, \theta_0} \widehat{\nu}+\tilde{\mathcal{J}}_{\theta_0 \theta_0} \widehat{\theta}_0+\frac{1}{1+ng_E} \widehat{\bk}_{\bgamma}^{\top} \mathcal{J}_{\theta, \eta, \theta_0}
\end{array}\right],
\end{equation*}
and
\begin{equation*}
\begin{aligned}
C_{\bgamma}=&-\frac{\mu_\nu^2}{2 \sigma_\nu^2}-\frac{1}{2} \widehat{\nu}^2 \mathcal{J}_{\nu \nu}-\widehat{\nu} \widehat{\theta}_0 \mathcal{J}_{\nu, \theta_0}-\frac{1}{2} \widehat{\theta}_0^2 \mathcal{J}_{\theta_0 \theta_0}+\frac{ng_E}{2(1+ng_E)} \widehat{\nu}^2 \mathcal{J}_{\theta, \eta, \nu}^{\top} \mathcal{J}_{\theta, \eta}^{-1} \mathcal{J}_{\theta, \eta, \nu}\\
&+\frac{ng_E}{(1+ng_E)} \widehat{\nu} \widehat{\theta}_0 \mathcal{J}_{\theta, \eta, \nu}^{\top} \mathcal{J}_{\theta, \eta}^{-1} \mathcal{J}_{\theta, \eta, \theta_0} +\frac{ng_E}{2(1+ng_E)} \widehat{\theta}_0^2 \mathcal{J}_{\theta, \eta, \theta_0}^{\top} \mathcal{J}_{\theta, \eta}^{-1} \mathcal{J}_{\theta, \eta, \theta_0}\\
&-\frac{1}{2(1+ng_E)} \widehat{\bk}_{\bgamma}^{\top} \mathcal{J}_{\theta, \eta} \widehat{\bk}_{\bgamma}
-\frac{1}{1+ng_E}\left(\widehat{\bk}_{\bgamma}^{\top} \mathcal{J}_{\theta, \eta, \nu}\, \widehat{\nu}+\widehat{\bk}_{\bgamma}^{\top} \mathcal{J}_{\theta, \eta, \theta_0}\, \widehat{\theta}_0\right).
\end{aligned}
\end{equation*}
Therefore, after completing the square, the integral with respect to $\nu$ and $\theta_0$ becomes
\begin{equation}\label{eq:ebv0}
    \int \exp \left\{-\frac{1}{2} \mathbf{z}^{\top} \mathbf{P}_{\bgamma} \mathbf{z}+\mathbf{h}_{\bgamma}^{\top} \mathbf{z} + C_{\bgamma}\right\} d \mathbf{z}=(2 \pi)|\mathbf{P}_{\bgamma}|^{-\frac{1}{2}} \exp \left\{\frac{1}{2} \mathbf{h}_{\bgamma}^{\top} \mathbf{P}_{\bgamma}^{-1} \mathbf{h}_{\bgamma}+C_{\bgamma}\right\}.
\end{equation}

Finally, combining \eqref{eq:ebbeta} and \eqref{eq:ebv0} and plugging back into \eqref{eq:ebintexp}, the marginal likelihood is approximately proportional to
\begin{equation}
p(\bt \mid \bgamma) \approx \widehat p(\bt \mid \bgamma) \propto \exp[\ell(\widehat\bPsi_{\bgamma})](1+ng_E)^{-\frac{d_{\bgamma}}{2}} |\mathbf{P}_{\bgamma}|^{-\frac{1}{2}} \exp \left\{\frac{1}{2} \mathbf{h}_{\bgamma}^{\top} \mathbf{P}_{\bgamma}^{-1} \mathbf{h}_{\bgamma}+C_{\bgamma}\right\}.
\end{equation}


These terms have the following interpretations: the first two terms, $\exp[\ell(\widehat\bPsi_{\bgamma})](1+ng_E)^{-\frac{d_{\bgamma}}{2}} $, resemble BIC: the likelihood term plus a dimension penalty term. The third term, $|\mathbf{P}_{\bgamma}|^{-\frac{1}{2}}$, is the Occam factor for the common parameters. When $g_E\to 0$, we keep the raw information $\mj_{\nu\theta_0}$; when $g_E\to \infty$, we have the full Schur complement, which is the profile information for $(\nu,\theta_0)$ after estimating $\bkappa_{\bgamma}$ from data. So $|\mathbf{P}_{\bgamma}|^{-\frac{1}{2}}$ is the volume of the posterior uncertainty in the common parameters after adjusting to the shared information with the coefficients.

The final exponent, $\frac{1}{2} \mathbf{h}_{\bgamma}^{\top} \mathbf{P}_{\bgamma}^{-1} \mathbf{h}_{\bgamma}+C_{\bgamma}$, could be expanded as
$$
\frac{1}{2} \mathbf{h}_\gamma^{\top} \mathbf{P}_\gamma^{-1} \mathbf{h}_\gamma+C_\gamma=-\frac{1}{2}\left(\hat{\mathbf {z}}_{\bgamma}-\mathbf {z}_0\right)^{\top}\left(\tilde{\mj}_{\nu\theta_0}^{-1}+\Sigma_0\right)^{-1}\left(\hat{\mathbf {z}}_{\gamma}-\mathbf {z}_0\right)-\frac{W_\gamma}{2\left(1+n g_E\right)}+ \frac12 \tilde{\mathbf{h}}_{\bgamma}^\top \tilde\mj^{-1} \tilde{\mathbf{h}}_{\bgamma},
$$
where
$$
\begin{aligned}
\hat{\mathbf {z}}_\gamma=&\left(\hat{\nu}, \hat{\theta}_0\right)^{\top}+\tilde\mj^{-1}\tilde{\mathbf{h}}_{\bgamma}, ~\tilde{\mathbf{h}}_{\bgamma} = \frac{1}{1+ng_E}\left( \hat\bkappa^\top_{\bgamma}\mj_{\btheta,\bmeta,\nu}, ~\hat\bkappa^\top_{\bgamma}\mj_{\btheta,\bmeta,\theta_0}\right)^\top,\\
&\mathbf {z}_0=\left(\mu_\nu, 0\right)^{\top}, ~
\Sigma_0=\text{diag}\left(\sigma_\nu^2, K\right), ~
W_{\bgamma}=\widehat{\bk}_{\bgamma}^{\top} \mathcal{J}_{\theta, \eta} \widehat{\bk}_{\bgamma}.
\end{aligned}
$$

In particular, the first quadratic term is the coefficient-adjusted distance between the MLE of common parameter and the prior centre. The second term, $-W_{\bgamma}/(2(1+ng_E))$, shrinks the contribution of the maximised likelihood, where $W_{\bgamma}$ resembles the Wald statistic for the model-specific coefficients. It is the analogue for the GH model of the shrinkage factor appearing in the ILA-approximated marginal likelihood under information based $g$-prior in generalised linear models \citep{WangGeorge:2007}. The last term is the small leftover term from the coefficient-adjustment.

\color{black}

\subsection{Null model case}

When $\bgamma = \mathbf{0}$, we only have the $(\nu, \theta_0)$ part, and the exponent in the integral becomes
$$
\begin{aligned}
&-\frac12[(\nu-\widehat{\nu})^2 \mathcal{J}_{\nu \nu}+2(\nu-\widehat{\nu})\left(\theta_0-\widehat{\theta}_0\right) \mathcal{J}_{\nu, \theta_0}+\left(\theta_0-\widehat{\theta}_0\right)^2 \mathcal{J}_{\theta_0 \theta_0} ]-\frac{1}{2 \sigma_\nu^2}\left(\nu-\mu_\nu\right)^2-\frac{1}{2 K} \theta_0^2\\
=& -\frac{1}{2} \mathbf{z}^{\top} \mathbf{P}_0 \mathbf{z}+\mathbf{h}_0^{\top} \mathbf{z} + C_0,
\end{aligned}
$$
where $\widehat{\nu}$ and $\widehat{\theta}_0$ denote the components of the MLE $\widehat{\bPsi}_{\mathbf{0}}$ under the null model,
and
$$\mathbf{z}=\left(\nu, \theta_0\right)^{\top}, \quad \mathbf{P}_0=\left[\begin{array}{cc}
\mathcal{J}_{\nu \nu}+\frac{1}{\sigma_\nu^2} & \mathcal{J}_{\nu, \theta_0} \\
\mathcal{J}_{\nu, \theta_0} & \mathcal{J}_{\theta_0 \theta_0}+\frac{1}{K}
\end{array}\right],\quad \mathbf{h}_0=\left[\begin{array}{c}
\mathcal{J}_{\nu \nu} \widehat{\nu}+\mathcal{J}_{\nu, \theta_0} \widehat{\theta}_0+\frac{\mu_\nu}{\sigma_\nu^2} \\
\mathcal{J}_{\nu, \theta_0} \widehat{\nu}+\mathcal{J}_{\theta_0 \theta_0} \widehat{\theta}_0
\end{array}\right],$$
and
$$
C_0=-\frac{\mu_\nu^2}{2 \sigma_\nu^2}-\frac{1}{2} \widehat{\nu}^2 \mathcal{J}_{\nu \nu}-\widehat{\nu} \widehat{\theta}_0 \mathcal{J}_{\nu, \theta_0}-\frac{1}{2} \widehat{\theta}_0^2 \mathcal{J}_{\theta_0 \theta_0}.
$$
Since the null model contains no model-specific coefficients, no terms involving $\widehat{\bk}_{\bgamma}$ arise in this case. We can see that the expressions above coincide with those of the non-null case upon deleting all terms involving $\widehat{\bk}_{\bgamma}$ and replacing $\tilde{\mathcal J}$ with the corresponding block of $\mathcal J$.

After completing the square, we have the following approximation for the marginal likelihood:
$$
\widehat p(\bt \mid \bgamma = \mathbf{0}) \propto\exp[\ell(\widehat{\bPsi}_{\mathbf{0}})] |\mathbf{P}_0|^{-\frac{1}{2}} \exp \left\{\frac{1}{2} \mathbf{h}_0^{\top} \mathbf{P}_0^{-1} \mathbf{h}_0+C_0\right\}.
$$

\section{Proof of Theoretical Results}
Let us define the following limit function. By the weak law of large numbers, we have that
\begin{equation*}
M_n\left({\bPsi}_{\bgamma}\right)= \dfrac{\ell_n({\bPsi}_{\bgamma})}{n} \stackrel{\Pr}{\to} M({\bPsi}_{\bgamma}),
\end{equation*}
where
\begin{eqnarray*}
M\left({\bPsi}_{\bgamma}\right) = P(\delta_1 = 1) E_{\varphi_0}\left[\log\left( h\left(o_1 \mid {\bPsi}_{\bgamma}\right)\right) \mid \delta_1 = 1 \right] - E_{\varphi_0}\left[H\left(t_1 \mid{\bPsi}_{\bgamma}\right)\right],
\end{eqnarray*}
and $\varphi_0$ is the true generating model.
Define also $\bv = (t_1, o_1, c_1, \delta_1)^{\top}$, and the function
\begin{eqnarray*}
m\left({\bPsi}_{\bgamma}, \bv \right) = \delta_1 \log\left( h\left(o_1 \mid {\bPsi}_{\bgamma} \right) \right) - H\left(t_1 \mid {\bPsi}_{\bgamma} \right),
\end{eqnarray*}
which is assumed to be finite for each ${\bPsi}_{\bgamma} \in \Gamma_{\bgamma}$.

Let us also denote by $\bX_{\bgamma}$ and $\tbX_{\bgamma}$, the design matrices associated with model $\bgamma$, and $\bX_{\bgamma,O}$ and $\tbX_{\bgamma,O}$, the design matrices associated with model $\bgamma$ and the uncensored observations.
Consider the following technical conditions.
\begin{itemize}
\item[C1.]  The parameter space is a compact subset $ \Gamma_{\bgamma} \subset {\mathbb R}^{d_{\bgamma}+2} $.

\item[C2.] There exists $\tilde{n}>0$ such that, for $n>\tilde{n}$, the matrices $\bX_{\bgamma}^{\top}\bX_{\bgamma}$, $\tbX_{\bgamma}^{\top}\tbX_{\bgamma}$, $\bX_{\bgamma,O}^{\top}\bX_{\bgamma,O}$ and $\tbX_{\bgamma,O}^{\top}\tbX_{\bgamma,O}$ are positive definite.

\item[C3.] Conditionally independent censoring. $O_i \bot C_i \mid \bz_i$.

\item[C4.] The maximum ${\bPsi}^*_{\bgamma} =\operatorname{argmax}_{\Gamma_{\bgamma}} M({\bPsi}_{\bgamma})$ is a unique interior point of $\Gamma_{\bgamma}$.

\item[C5.] The baseline hazard $h(t \mid \mu, \sigma)$ is continuous in $t>0$. Moreover, $M({\bPsi}_{\bgamma}) < \infty$ for all ${\bPsi}_{\bgamma}$, and $\vert m({\bPsi}_{\bgamma}; \bv) \vert \leq \psi_0 (\bv)$ for all ${\bPsi}_{\bgamma} \in \Gamma_{\bgamma}$, where
    \begin{equation*}
        \int \psi_0 (\bv) d\varphi(\bv) <\infty.
    \end{equation*}

\item[C6.] $\dfrac{ \partial^2 m(\bPsi_{\bgamma}; \bv)}{\partial \bPsi_{\bgamma_i}\partial \bPsi_{\bgamma_j}}$ is continuous in $\bv$, and there exist functions $\psi_1(\bv)$ and $\psi_2(\bv)$ such that
    \begin{equation*}
    \begin{split}
    \left \vert \dfrac{ \partial m(\bPsi_{\bgamma}; \bv)}{\partial \bPsi_{\bgamma_i}} \cdot \dfrac{ \partial m(\bPsi_{\bgamma}; \bv)}{\partial \bPsi_{\bgamma_j}}  \right\vert &\leq \psi_1(\bv), \\
    \left \vert \dfrac{ \partial^2 m(\bPsi_{\bgamma}; \bv)}{\partial \bPsi_{\bgamma_i}\partial \bPsi_{\bgamma_j}}  \right\vert &\leq \psi_2(\bv),
    \end{split}
    \end{equation*}
    where
    \begin{equation*}
        \begin{split}
        \int \psi_1 (\bv) d\varphi(\bv) &<\infty,\\
        \int \psi_2 (\bv) d\varphi(\bv) &<\infty.
            \end{split}
    \end{equation*}
 \item[C7.] The matrix $E_{\varphi}\left[ \nabla m(\bPsi^*_{\bgamma}; \bv) \nabla m(\bPsi^*_{\bgamma}; \bv)^{\top}\right]$ is non-singular and the Hessian matrix $\nabla^2 M(\bPsi_{\bgamma})$ has constant rank in a neighbourhood of $\bPsi_{\bgamma}^{*}$.

\end{itemize}

Condition C1 is imposed mainly for convenience, but the results extend to locally compact sets. The set $\Gamma_{\bgamma}$ may be taken to be any sufficiently large compact set. Condition C2 ensures the absence of collinearity in the design matrices for both the time-level and hazard-level effects for sufficiently large samples. Condition C3 restricts the censoring mechanism to be non-informative, since informative censoring would require modifications to the likelihood function. Conditions C4–C7 are standard regularity assumptions used in asymptotic theory \citep{white:1981,white:1982,vandervaart:2000}.

\begin{lemma}\label{lemma:consistency}
Suppose that conditions C1-C5 are satisfied. Then, $\widehat{\bPsi}_{\bgamma} \stackrel{\Pr}{\rightarrow} {\bPsi}_{\bgamma}^*$, as $n\rightarrow \infty$.
\end{lemma}

\begin{lemma}\label{lemma:asympnorm}
Suppose that conditions C1-C7 are satisfied. Then, $\sqrt{n}\left(\widehat{\bPsi}_{\bgamma}-\bPsi_{\bgamma}^*\right) \stackrel{D}{\longrightarrow} N\left(0, V\left({\bPsi_{\bgamma}^*}\right)^{-1} {\mathbb E}_{F_0}[ \nabla m(\bPsi_{\bgamma}^*) \nabla m(\bPsi_{\bgamma}^*)^{\top}] V\left({\bPsi_{\bgamma}^*}\right)^{-1}\right)$,
where $V\left({\bPsi_{\bgamma}^*}\right)$ is the Hessian matrix of $M(\bPsi_{\bgamma})$ evaluated at $\bPsi_{\bgamma}^*$.
\end{lemma}

\begin{lemma}\label{lemma:fisher}
Suppose that conditions C1-C7 are satisfied. Let $\mj(\widehat\bPsi)$ denote the observed Fisher information evaluated at the MLE $\widehat\bPsi$, then $$\mj(\widehat\bPsi_{\bgamma})/n \stackrel{\Pr}{\rightarrow} -\nabla^2M(\bPsi^*_{\bgamma}), \quad n \to \infty.$$
For each entry of the observed Fisher information $\mj_{a,b}(\widehat\bPsi_{\bgamma})$, $a,b \in (\nu,\theta_0,\theta_1,...,\theta_{p_{\gamma}},\eta_1,...,\eta_{q_{\gamma}})$, we have
$$\mj_{a,b}(\widehat\bPsi_{\bgamma})/n \stackrel{\Pr}{\rightarrow} -\nabla^2M_{a,b}(\bPsi^*_{\bgamma}) = c , \quad n \to \infty,$$

where $c$ is a constant.
\end{lemma}

\subsection*{Proof of Lemma 1}

First, note that the GH model is identifiable as we are using baseline hazards associated with distributions from the log-location scale family, and the GH model is identifiable provided that the baseline hazard is not a member of the Weibull family of distributions \citep{chen:2001}. Moreover, conditions C1-C5 are equivalent to Assumptions A1-A3 in \cite{white:1982}. Similar to A1 in \cite{white:1982}, the compactness assumption of C1 is made for theoretical convenience, but it could be relaxed with a locally compact assumption \cite{white:1981}. C2 indicates that for sufficiently large sample size, the design matrices have full rank, and C3 implies the censoring mechanism is independent of the survival time given the set of observed variables (non-informative censoring). C4 and C5 corresponds to A3 of \cite{white:1982}. C4 is the fundamental identification condition \citep{white:1982}, which assumes the existence and uniqueness of the maximum of the expected log-likelihood. C5 ensures that the model discrimination measure, the Kullback-Leibler Information Criterion, is well defined. Consequently, the conditions of Theorem 2.2 in \cite{white:1982} are satisfied, which implies that $\widehat{\bPsi}_{\bgamma} \stackrel{\Pr}{\rightarrow} {\bPsi}_{\bgamma}^*$, as $n\rightarrow \infty$. That is, the MLE is consistent under model misspecification.

\subsection*{Proof of Lemma 2}

The result follows by noticing that conditions C1-C7 are equivalent to Assumptions A1-A6 in \cite{white:1982}. C6 corresponds to A4 and A5, and it ensures that the first two derivatives of $m({\bPsi}_{\bgamma}; \bv)$ with respect to ${\bPsi}_{\bgamma}$ exists and dominated by functions integrable with respect to $\varphi$ for all $\bv$. These assumptions justifies the use of mean value theorem and law of large numbers for proving the asymptotic normality of Theorem 3.2 in \cite{white:1982}. Lastly, C7, corresponding to A6, is essential for deriving locally identifiable property of ${\bPsi}_{\bgamma}$ \citep{white:1982}, and guarantees the entries of the Fisher information matrix are well-defined. Moreover, it ensures that $\nabla^2 M(\bPsi_{\bgamma})$ is invertible and the covariance matrix of the asymptotic normal distribution in Lemma \ref{lemma:asympnorm} is well-defined. The result then follows from Theorem 3.2 in \cite{white:1982}, which establishes the asymptotic normality of the MLE under misspecification of the GH model.

\subsection{Proof of Lemma 3}

Appealing to Lemma \ref{lemma:consistency}, we have $\widehat{\bPsi}_{\bgamma} \stackrel{\Pr}{\to} \bPsi_{\bgamma}^*$ as $n \to \infty$. Applying continuous mapping theorem and have $n^{-1}H(\widehat\bPsi_{\bgamma}) \stackrel{\Pr}{\to} \nabla^2 M(\bPsi_{\bgamma}^*)$ as $n \to \infty$, where $\nabla^2 M\left({\bPsi}^*_{\bgamma} \right)$ is a negative definite matrix according to C4. And since $\mj(\widehat\bPsi_{\bgamma}) = -H(\widehat\bPsi_{\bgamma})$, we have $n^{-1}\mj(\widehat\bPsi_{\bgamma}) \stackrel{\Pr}{\to} -\nabla^2 M(\bPsi_{\bgamma}^*)$ as $n \to \infty$.

For each entry of $\nabla^2M({\bPsi}^*_{\bgamma})$, we could refer to the expressions in section \ref{sec:hesslik} after taking their expectations. By C4 we have ${\bPsi}^*_{\bgamma}$ is a unique interior maximum, therefore all elements in ${\bPsi}^*_{\bgamma} = (\nu^*,\theta_0^*,\theta_1^*,...,\theta_{p_{\gamma}}^*,\eta_1^*,...,\eta^*_{q_{\gamma}})$ are constants, which implies $v_i^*$ is constant. Assuming the observed times $t_1,...,t_n$ are finite, combining with the fact that $X$ and $\tilde X$ are assumed fixed, then $u_i^*$ is also constant. Since $f$ and $F$ are valid PDF and valid CDF and therefore continuous, $f(u_i^*)$ and $F(u_i^*)$ are also constants greater than $0$. Moreover, by C6, we have that $f''$ and $f'$ are continuous. Therefore, the ratios $\dfrac{f\left(u_i^*\right)}{F\left(u_i^*\right)}$, $\dfrac{f{''}(u_i^*)}{f(u_i^*)} - \left( \dfrac{f'(u_i^*)}{f(u_i^*)} \right)^2$, and $\dfrac{f^{\prime}\left(u_i^*\right)}{F\left(-u_i^*\right)}+\left(\dfrac{f\left(u_i^*\right)}{F\left(-u_i^*\right)}\right)^2$ are also constants. For example, when $f$ is the PDF of a normal or a logistic distribution, we could refer to section \ref{sec:specialcase} and have those ratios in nice forms. After characterizing all terms involved in $\nabla^2M({\bPsi}^*_{\bgamma})$, we have that for each element of $\nabla^2 M_{a^*,b^*}(\bPsi_{\bgamma}^*)$, $a^*,b^* \in (\nu^*,\theta_0^*,\theta_1^*,...,\theta_{p_{\gamma}}^*,\eta_1^*,...,\eta^*_{q_{\gamma}})$,
$$\nabla^2 M_{a^*,b^*}(\bPsi_{\bgamma}^*) = \lOp(1).$$

Combining with the previous result of $n^{-1}\mj(\widehat\bPsi_{\bgamma}) \stackrel{\Pr}{\to} -\nabla^2 M(\bPsi_{\bgamma}^*)$ as $n \to \infty$ entry by entry, we conclude the proof.

\subsection{Proof of Proposition 2}
\label{app:proof_bfrates_EB}
We characterise the asymptotic rates of:
\begin{equation*}
\widehat{B}_{\bgamma,\bgamma^*}= \frac{\widehat{p}(\bt \mid \bgamma)}{\widehat{p}(\bt \mid \bgamma^*)},
\end{equation*}
where $\widehat{p}(\bt \mid \bgamma)$ is obtained via an integrated Laplace approximation:
\begin{equation*}
\widehat p(\bt \mid \bgamma)\approx \exp[\ell(\widehat{\bPsi}_{\bgamma} )](1+ng_E)^{-\frac{d_{\bgamma}}{2}} |\mathbf{P}_{\bgamma}|^{-\frac{1}{2}} \exp \left\{\frac{1}{2} \mathbf{h}_{\bgamma}^{\top} \mathbf{P}_{\bgamma}^{-1} \mathbf{h}_{\bgamma}+C_{\bgamma} \right\}
\end{equation*}

where $\widehat{\bPsi}_{\bgamma}= \arg\max_{\bPsi_{\bgamma}}  \ell(\bPsi_{\bgamma}) $ is the maximum likelihood estimator (MLE). We consider the case of $\bgamma^* = \bf0$ and $\bgamma^* \not= \bf0$ separately. For convenience, we omit the subscript of ${\bgamma}$ under $\mathbf {h}_{\bgamma}$, $\mathbf {P}_{\bgamma}$, and $C_{\bgamma}$, and we use $*$ to denote the subscript ${\bgamma}^*$.

\paragraph{Case $\bgamma^* \not= \bf0$} Let us factorise the Bayes factor as:
\begin{eqnarray*}
\widehat{B}_{\bgamma, \bgamma^*} = (1+ng_E)^{\frac{d_{\bgamma} - d_{\bgamma^*}}{2}} \exp\{Z_1\} Z_2 \exp \{Z_3\},
\end{eqnarray*}
where
\begin{eqnarray*}
    Z_1 &=& \ell(\widehat{\bPsi}_{\bgamma} ) - \ell(\widehat{\bPsi}_{\bgamma^*} ),\\
    Z_2 &=& \dfrac{|\mathbf{P}|^{-\frac{1}{2}}}{|\mathbf{P}_*|^{-\frac{1}{2}}},\\
    Z_3 &=& \left(\frac{1}{2} \mathbf{h}^{\top} \mathbf{P}^{-1} \mathbf{h}+C\right) - \left(\frac{1}{2} \mathbf{h}_*^{\top} \mathbf{P}_*^{-1} \mathbf{h}_*+C_*\right).
\end{eqnarray*}

For $Z_1$, we again consider the case of $\bgamma^* \subset \bgamma$ and $\bgamma^* \not\subset \bgamma$ separately. For $\bgamma^* \subset \bgamma$, we perform a second order Taylor expansion
\begin{align}\label{eqn:Z1-taylor-p1}
Z_1 &= \ell(\widehat{\bPsi}_{\bgamma} ) - \ell(\widehat{\bPsi}_{\bgamma^*} ) + \mathcal{O}_{\mathrm{p}}(1) \nonumber\\
&=-\frac{1}{2}\left(\widehat{\bPsi}_{\bgamma} - \widehat{\bPsi}_{\bgamma^*} \right)^{\top} \nabla^2  \ell(\widehat{\bPsi}_{\bgamma} ) \left(\widehat{\bPsi}_{\bgamma} - \widehat{\bPsi}_{\bgamma^*} \right) + \mathcal{O}_{\mathrm{p}}(1)\nonumber\\
&=-\frac{1}{2}\sqrt{n}\left(\widehat{\bPsi}_{\bgamma} - \widehat{\bPsi}_{\bgamma^*} \right)^{\top} \nabla^2  M({\bPsi}^*_{\bgamma} ) \sqrt{n} \left(\widehat{\bPsi}_{\bgamma} - \widehat{\bPsi}_{\bgamma^*} \right) + \mathcal{O}_{\mathrm{p}}(1)\nonumber\\
&= -\frac{1}{2}\boldsymbol\phi_n ^{\top} \nabla^2  M({\bPsi}^*_{\bgamma} ) \boldsymbol\phi_n + \mathcal{O}_{\mathrm{p}}(1),
\end{align}

where $\boldsymbol\phi_n = \sqrt{n}\left(\widehat{\bPsi}_{\bgamma} - \widehat{\bPsi}_{\bgamma^*} \right)$ and the third line follows from $\dfrac1n\nabla^2  \ell(\widehat{\bPsi}_{\bgamma} ) \xrightarrow{\Pr} \nabla^2  M({\bPsi}^*_{\bgamma})$. In this case of $\bgamma^* \subset \bgamma$, we have that ${\bPsi}^*_{\bgamma}={\bPsi}^*_{\bgamma^*}$. So we could rewrite $\boldsymbol\phi$ as
$$
\boldsymbol\phi_n = \sqrt{n}\left(\widehat{\bPsi}_{\bgamma} - \widehat{\bPsi}_{\bgamma^*} \right) = \sqrt{n}\left(\widehat{\bPsi}_{\bgamma} - {\bPsi}^*_{\bgamma} \right)
-\sqrt{n}\left({\bPsi}^*_{\bgamma^*} - \widehat{\bPsi}_{\bgamma^*} \right).
$$

Applying Lemma \ref{lemma:asympnorm} on the final terms, we have that $\sqrt{n}\left(\widehat{\bPsi}_{\bgamma} - {\bPsi}^*_{\bgamma} \right) \stackrel{D}{\longrightarrow} N(\mathbf0, \Sigma_{\bgamma})$ and $\sqrt{n}\left({\bPsi}^*_{\bgamma^*} - \widehat{\bPsi}_{\bgamma^*} \right)\stackrel{D}{\longrightarrow} N(\mathbf0, \Sigma_{\bgamma^*})$. Therefore,
$$
\boldsymbol\phi_n \stackrel{D}{\longrightarrow} N(\mathbf0, \Sigma), \quad n \to \infty,
$$
where $\Sigma$ is the asymptotic covariance matrix of $\boldsymbol\phi_n$, which exists by the joint asymptotic normality of $(\widehat{\bPsi}_{\bgamma},\widehat{\bPsi}_{\bgamma^*})$ \citep{white:1982}. Since $\Sigma$ is a positive semi-definite covariance matrix and $\nabla^2  M({\bPsi}^*_{\bgamma})$ is a real symmetric matrix, applying Lemma 1 and Theorem 1 in \cite{baldessari:1967}, the first term in (\ref{eqn:Z1-taylor-p1}) could be written asymptotically as a linear combination of independent Chi-squared random variables (see also Proposition 1 in \citealp{iqbal:2025}):
$$
\boldsymbol\phi^{\top}_n  \nabla^2  M({\bPsi}^*_{\bgamma}) \boldsymbol\phi_n \stackrel{D}{\longrightarrow} \sum_j  \lambda_j \chi_{r_j}^2,
$$
where $\lambda_j$ is the $j$-th unique eigenvalue of $\nabla^2  M({\bPsi}^*_{\bgamma})\Sigma$ and $r_j$ is its corresponding order of multiplicity. Combining this result with (\ref{eqn:Z1-taylor-p1}), we have that $Z_1$ is stochastically bounded
\begin{align*}
    Z_1 \stackrel{D}{\longrightarrow} -\frac{1}{2}\sum_j  \lambda_j \chi_{r_j}^2 + \mathcal{O}_{\mathrm{p}}(1) = \mathcal{O}_{\mathrm{p}}(1).
\end{align*}

For $\bgamma^* \not\subset \bgamma$, by Lemma \ref{lemma:consistency} and a uniform law of large numbers (C1, C5), we have $\frac{1}{n}\ell(\widehat{\bPsi}_{\bgamma}) \xrightarrow{\Pr} M(\bPsi^*_{\bgamma})$ and $\frac{1}{n}\ell(\widehat{\bPsi}_{\bgamma^*}) \xrightarrow{\Pr} M(\bPsi^*_{\bgamma^*})$. Therefore,
\begin{align}\label{eqn:case2-z2-p1}
\dfrac 1nZ_1 = \dfrac 1n \left(\ell(\widehat{\bPsi}_{\bgamma} ) - \ell(\widehat{\bPsi}_{\bgamma^*} ) \right)
\xrightarrow{\Pr}
M(\bPsi^*_{\bgamma}) - M(\bPsi^*_{\bgamma^*}) <0.
\end{align}

Although $M(\bPsi^*_{\bgamma}) - M(\bPsi^*_{\bgamma^*})$ could not be further bounded, it is driven by the KL-optimal errors of $\bgamma$ and $\bgamma^*$ \citep{rossell:2023}.

For $Z_2$, recall the expression for the
$2\times2$ matrix $\mathbf P$ is
\begin{equation*}
\mathbf{P}=\left[\begin{array}{cc}
\mathcal{J}_{\nu \nu}-\frac{ng_E}{1+ng_E} \mathcal{J}_{\boldsymbol{\theta}, \eta, \nu}^{\top} \mathcal{J}_{\boldsymbol{\theta}, \eta}^{-1} \mathcal{J}_{\boldsymbol{\theta}, \eta, \nu}+\frac{1}{\sigma_\nu^2} & \mathcal{J}_{\nu, \theta_0}-\frac{ng_E}{1+ng_E} \mathcal{J}_{\boldsymbol{\theta}, \eta, \nu}^{\top} \mathcal{J}_{\boldsymbol{\theta}, \eta}^{-1} \mathcal{J}_{\boldsymbol{\theta}, \eta, \theta_0} \\
\mathcal{J}_{\nu, \theta_0}-\frac{ng_E}{1+ng_E} \mathcal{J}_{\boldsymbol{\theta}, \eta, \nu}^{\top} \mathcal{J}_{\boldsymbol{\theta}, \eta}^{-1} \mathcal{J}_{\boldsymbol{\theta}, \eta, \theta_0} & \mathcal{J}_{\theta_0 \theta_0}-\frac{ng_E}{1+ng_E} \mathcal{J}_{\boldsymbol{\theta}, \boldsymbol{\eta}, \theta_0}^{\top} \mathcal{J}_{\boldsymbol{\theta}, \boldsymbol{\eta}}^{-1} \mathcal{J}_{\boldsymbol{\theta}, \boldsymbol{\eta}, \theta_0}+\frac{1}{K}
\end{array}\right].
\end{equation*}

Note that all entries in $\bf P$ is composed of entries in the observed Fisher information matrix.
Therefore, by Lemma \ref{lemma:fisher}, all of the four entries are of order $\lOp(n)$, and the determinant ratio is given by
\begin{equation}
    Z_2 = \dfrac{|\mathbf{P}|^{-\frac{1}{2}}}{|\mathbf{P}_*|^{-\frac{1}{2}}} = \dfrac{\lOp(n^{-1})}{\lOp(n^{-1})} = \lOp(1),
\end{equation}
where the orders of the numerator and denominator follow from the eigen decomposition of $\bf P$ and $\bf P^*$.

For $Z_3$, we use the decomposition derived in Section 6 of this script,
\begin{equation*}
\tfrac{1}{2}\mathbf{h}^{\top}\mathbf{P}^{-1}\mathbf{h}+C=-\tfrac{1}{2}(\hat{\mathbf z}_{\bgamma}-\mathbf z_0)^{\top}(\tilde{\mj}^{-1}+\Sigma_0)^{-1}(\hat{\mathbf z}_{\bgamma}-\mathbf z_0)-\frac{W_{\bgamma}}{2(1+ng_E)}+\tfrac{1}{2}\tilde{\mathbf h}_{\bgamma}^{\top}\tilde{\mj}^{-1}\tilde{\mathbf h}_{\bgamma}.
\end{equation*}
Since $\tilde{\mj}/n$ converges to a positive definite matrix by Lemma~\ref{lemma:fisher}, $(\tilde{\mj}^{-1}+\Sigma_0)^{-1}\to\Sigma_0^{-1}$. Moreover, $\hat{\mathbf z}_{\bgamma}=\lOp(1)$ by Lemma~\ref{lemma:consistency} and C1, $W_{\bgamma}=\lOp(n)$ and $\tilde{\mathbf h}_{\bgamma}=\lOp(1)$. The three terms are therefore $\lOp(1)$, $\lOp(1)$ and $\lOp(n^{-1})$, so $\frac{1}{2}\mathbf{h}^{\top}\mathbf{P}^{-1}\mathbf{h}+C=\lOp(1)$. The same argument applies to $\bgamma^*$, and therefore
\begin{equation}
    Z_3 = \left(\frac{1}{2} \mathbf{h}^{\top} \mathbf{P}^{-1} \mathbf{h}+C\right) - \left(\frac{1}{2} \mathbf{h}_*^{\top} \mathbf{P}_*^{-1} \mathbf{h}_*+C_*\right) = \lOp(1).
\end{equation}

Combining the three exponents together, for $\bgamma^* \subset \bgamma$ we have
\begin{eqnarray*}
\widehat{B}_{\bgamma, \bgamma^*} =(1+ng_E)^{\frac{d_{\bgamma} - d_{\bgamma^*}}{2}} \exp\{\lOp(1)\} \lOp(1) \exp \{\lOp(1)\} = (1+ng_E)^{\frac{d_{\bgamma} - d_{\bgamma^*}}{2}}\exp\{\lOp(1)\},
\end{eqnarray*}
and for $\bgamma^* \not\subset \bgamma$ we have
\begin{eqnarray*}
\widehat{B}_{\bgamma, \bgamma^*} = (1+ng_E)^{\frac{d_{\bgamma} - d_{\bgamma^*}}{2}} \exp\left\{n\left(M(\bPsi^*_{\bgamma}) - M(\bPsi^*_{\bgamma^*})\right) +\lOp(1)\right\} .
\end{eqnarray*}

\paragraph{Case $\bgamma^* = \bf0$} When the true model is null, the Bayes factor becomes:
\begin{eqnarray*}
\widehat{B}_{\bgamma, \bgamma^*} = (1+ng_E)^{\frac{d_{\bgamma} }{2}} \exp\{Z_1\} Z_2 \exp \{Z_3\},
\end{eqnarray*}
where
\begin{eqnarray*}
    Z_1 &=& \ell(\widehat{\bPsi}_{\bgamma} ) - \ell(\widehat{\bPsi}_{\mathbf{0}} ),\\
    Z_2 &=& \dfrac{|\mathbf{P}|^{-\frac{1}{2}}}{|\mathbf{P}_0|^{-\frac{1}{2}}},\\
    Z_3 &=& \left(\frac{1}{2} \mathbf{h}^{\top} \mathbf{P}^{-1} \mathbf{h}+C\right) - \left(\frac{1}{2} \mathbf{h}_0^{\top} \mathbf{P}_0^{-1} \mathbf{h}_0+C_0\right).
\end{eqnarray*}

Recall the expression for $\mathbf{h}_0, \mathbf{P}_0, C_0$ is
$$\mathbf{P}_0=\left[\begin{array}{cc}
\mathcal{J}_{\nu \nu}+\frac{1}{\sigma_\nu^2} & \mathcal{J}_{\nu, \theta_0} \\
\mathcal{J}_{\nu, \theta_0} & \mathcal{J}_{\theta_0 \theta_0}+\frac{1}{K}
\end{array}\right],\quad \mathbf{h}_0=\left[\begin{array}{c}
\mathcal{J}_{\nu \nu} \widehat{\nu}+\mathcal{J}_{\nu, \theta_0} \widehat{\theta}_0+\frac{\mu_\nu}{\sigma_\nu^2} \\
\mathcal{J}_{\nu, \theta_0} \widehat{\nu}+\mathcal{J}_{\theta_0 \theta_0} \widehat{\theta}_0
\end{array}\right],$$
and
$$
C_0=-\frac{\mu_\nu^2}{2 \sigma_\nu^2}-\frac{1}{2} \widehat{\nu}^2 \mathcal{J}_{\nu \nu}-\widehat{\nu} \widehat{\theta}_0 \mathcal{J}_{\nu, \theta_0}-\frac{1}{2} \widehat{\theta}_0^2 \mathcal{J}_{\theta_0 \theta_0}.
$$

We could observe that each of the entries in $\mathbf{h}_0, \mathbf{P}_0$ are of $\lOp(n)$, and $C_0$ is also of $\lOp(n)$. Note that, since the null model contains no model-specific coefficients, no additional terms involving $\widehat{\bk}_{\bgamma}$ arise in $\mathbf{h}_0$ or $C_0$. Therefore, $\mathbf{P}_0$ has the same rate as $\mathbf{P}$, the decomposition above holds for the null model with the terms involving $\widehat{\bk}_{\bgamma}$ removed, and $Z_2$ and $Z_3$ are then of $\lOp(1)$. For $Z_1$, the proof is identical to the previous case, which leads to $Z_1 = \lOp(1), n\to \infty$ for $\bgamma^* \subset \bgamma$, and $Z_1=n\left(M(\bPsi^*_{\bgamma}) - M(\bPsi^*_{\mathbf{0}})\right), n\to \infty$ for $\bgamma^* \not\subset \bgamma$. Combining these together, we have that for $\bgamma^* \subset \bgamma$
\begin{eqnarray*}
\widehat{B}_{\bgamma, \bgamma^*} = (1+ng_E)^{\frac{d_{\bgamma}}{2}} \exp\{\lOp(1)\} \lOp(1) \exp \{\lOp(1)\} = (1+ng_E)^{\frac{d_{\bgamma}}{2}}\exp\{\lOp(1)\},
\end{eqnarray*}
and for $\bgamma^* \not\subset \bgamma$ we have
\begin{eqnarray*}
\widehat{B}_{\bgamma, \bgamma^*} =(1+ng_E)^{\frac{d_{\bgamma}}{2}} \exp\left\{n\left(M(\bPsi^*_{\bgamma}) - M(\bPsi^*_{\bgamma^*})\right) +\lOp(1)\right\}.
\end{eqnarray*}

Note that the Bayes factor rates are the same for both $\bgamma^* \not = \bf0$ and $\bgamma^* = \bf0$, and conclude the proof.

\subsection{Proof of Corollary 1}

We could write the statement as

\begin{equation*}
\widehat{\pi}\left(\boldsymbol{\bgamma}^* \mid \boldsymbol{t}, \boldsymbol{\delta}\right) = \left(1+ \sum_{\bgamma \not = \bgamma^*} \widehat B_{\bgamma,\bgamma^*}\frac{\pi(\bgamma)}{\pi(\bgamma^*)} \right)^{-1} \xrightarrow{\Pr} 1.
\end{equation*}

By Proposition 2, for case $\bgamma^*\subset \bgamma$, we have $d_{\bgamma}>d_{\bgamma^*}$, so $\widehat B_{\bgamma,\bgamma^*} = (1+ng_E)^{-(d_{\bgamma}-d_{\bgamma^*})/2}\exp\{\lOp(1)\}$; for case $\bgamma^* \not \subset \bgamma$, $M(\bPsi^*_{\bgamma^*}) - M(\bPsi^*_{\bgamma}) > 0$, so the factor $\exp\{-n[M(\bPsi^*_{\bgamma^*}) - M(\bPsi^*_{\bgamma})]\}$ dominates the polynomial term. Therefore, $\widehat B_{\bgamma,\bgamma^*} \xrightarrow{\Pr} 0$ for either of the cases. And hence, we have
\begin{equation*}
\sum_{\bgamma \not = \bgamma^*} \widehat B_{\bgamma,\bgamma^*}\frac{\pi(\bgamma)}{\pi(\bgamma^*)} \xrightarrow{\Pr} 0,
\end{equation*}
and we conclude that
\begin{equation*}
\left(1+ \sum_{\bgamma \not = \bgamma^*} \widehat B_{\bgamma,\bgamma^*}\frac{\pi(\bgamma)}{\pi(\bgamma^*)} \right)^{-1} \xrightarrow{\Pr} 1.
\end{equation*}

\subsection{Proof of Proposition 3}
\label{app:proof_bfrates}

The proof of this result represents an extension of the Bayes factor rates characterisation in \cite{rossell:2023} to the context where a variable plays more than one role in the model. We emphasise that this proof presents a characterisation of the asymptotic behaviour of Bayes factors obtained with the Laplace approximation. That is, we characterise the asymptotic rates of:
\begin{equation*}
\tilde{B}_{\bgamma,\bgamma^*}= \frac{\tilde{p}(\bt \mid \bgamma)}{\tilde{p}(\bt \mid \bgamma^*)},
\end{equation*}
where $\tilde{p}(\bt \mid \bgamma)$ is obtained via a Laplace approximation:
\begin{equation*}
\tilde{p}(\bt \mid \bgamma)= \exp\{\ell(\tilde{\bPsi}_{\bgamma}) + \log \pi(\tilde{\bPsi}_{\bgamma}) \}
(2\pi)^{d_{\bgamma}/2} \left|H(\tilde{\bPsi}_{\bgamma}) + \nabla^2 \log\pi(\tilde{\bPsi}_{\bgamma}) \right|^{-1/2},
\end{equation*}
where $\tilde{\bPsi}_{\bgamma}= \arg\max_{\bPsi_{\bgamma}} \left\{ \ell(\bPsi_{\bgamma}) + \log \pi(\bPsi_{\bgamma})\right\}$ is the maximum a posteriori (MAP) under prior $\pi(\bPsi_{\bgamma})$.

First, let us factorise the Bayes factor as:
\begin{eqnarray*}
\tilde{B}_{\bgamma, \bgamma^*} = (2\pi)^{\frac{d_{\bgamma} - d_{\bgamma^*}}{2}} \exp\{Z_1\} Z_2 Z_3,
\end{eqnarray*}
where
\begin{eqnarray*}
    Z_1 &=& \ell(\tilde{\bPsi}_{\bgamma} ) - \ell(\tilde{\bPsi}_{\bgamma^*} ),\\
    Z_2 &=& \dfrac{\pi(\tilde{\bPsi}_{\bgamma} \mid \bgamma)}{\pi(\tilde{\bPsi}_{\bgamma^*} \mid \bgamma)},\\
    Z_3 &=& \dfrac{\left|H(\tilde{\bPsi}_{\bgamma^*}) + \nabla^2 \log\pi(\tilde{\bPsi}_{\bgamma^*}) \right|^{-1/2}}{\left|H(\tilde{\bPsi}_{\bgamma}) + \nabla^2 \log\pi(\tilde{\bPsi}_{\bgamma}) \right|^{-1/2}}.
\end{eqnarray*}

We start with $Z_3$ by rewriting it as
\begin{align*}
    Z_3=n^{\frac{d_{\bgamma^*}-d_{\bgamma}}{2}} \frac{\left|n^{-1}\left[H\left(\tilde{\bPsi}_{\bgamma^*}\right)+\nabla^2 \log \pi\left(\tilde{\bPsi}_{\bgamma^*}\right)\right]\right|^{\frac{1}{2}}}{\left|n^{-1}\left[H\left(\tilde{\bPsi}_{\bgamma}\right)+\nabla^2 \log \pi\left(\tilde{\bPsi}_{\bgamma}\right)\right]\right|^{\frac{1}{2}}}.
\end{align*}

From Lemma \ref{lemma:consistency}, the MAP estimators $\tilde{\bPsi}_{\bgamma^*} \xrightarrow{\Pr} {\bPsi}^*_{\bgamma^*}$ and $\tilde{\bPsi}_{\bgamma} \xrightarrow{\Pr} {\bPsi}^*_{\bgamma}$ as $n \to \infty$. Therefore, applying the continuous mapping theorem, we have $n^{-1}H\left({\tbPsi}^*_{\bgamma}\right) \xrightarrow{\Pr} \nabla^2 M\left({\bPsi}^*_{\bgamma} \right)$, as $n\to\infty$. The same result follows for $\bgamma^*$: $n^{-1}H\left({\tbPsi}^*_{\bgamma^*}\right)  \xrightarrow{\Pr} \nabla^2 M\left({\bPsi}^*_{\bgamma^*} \right)$, as $n\to\infty$. According to C4, ${\bPsi}^*_{\bgamma}$ and ${\bPsi}^*_{\bgamma^*}$ are unique interior maximums, which implies the negative definiteness of $\nabla^2 M\left({\bPsi}^*_{\bgamma} \right)$ and $\nabla^2 M\left({\bPsi}^*_{\bgamma^*} \right)$. We also have $n^{-1}\bX_{\bgamma}^{\top}\bX_{\bgamma}$ and $n^{-1}\tbX_{\bgamma}^{\top}\tbX_{\bgamma}$ converges in probability to positive definite matrices. Therefore, the matrices $n^{-1}\nabla^2 \log \pi\left(\tilde{\bPsi}_{\bgamma}\right)= \mathcal{O}_{\mathrm{p}}(1)$ and $n^{-1}\nabla^2 \log \pi\left(\tilde{\bPsi}_{\bgamma^*}\right)=\mathcal{O}_{\mathrm{p}}(1)$ . Combining these together, we have
\begin{align*}
\frac{\left|n^{-1}\left[H\left(\tilde{\bPsi}_{\bgamma^*}\right)+\nabla^2 \log \pi\left(\tilde{\bPsi}_{\bgamma^*}\right)\right]\right|^{\frac{1}{2}}}{\left|n^{-1}\left[H\left(\tilde{\bPsi}_{\bgamma}\right)+\nabla^2 \log \pi\left(\tilde{\bPsi}_{\bgamma}\right)\right]\right|^{\frac{1}{2}}} \xrightarrow{\Pr} \frac{\left|\nabla^2 M\left({\bPsi}^*_{\bgamma^*} \right)\right|^{\frac{1}{2}}}{\left|\nabla^2 M\left({\bPsi}^*_{\bgamma} \right)\right|^{\frac{1}{2}}}, \quad n\to\infty.
\end{align*}

And since ${\bPsi}^*_{\bgamma}$ and ${\bPsi}^*_{\bgamma^*}$ are fixed by assumption, then $\left|\nabla^2 M\left({\bPsi}^*_{\bgamma^*} \right)\right|^{\frac{1}{2}}$ and $\left|\nabla^2 M\left({\bPsi}^*_{\bgamma} \right)\right|^{\frac{1}{2}}$ would be constants, so we have
$$
Z_3 n^{\frac{d_{\bgamma}-d_{\bgamma^*}}{2}} \xrightarrow{\Pr} \frac{\left|\nabla^2 M\left({\bPsi}^*_{\bgamma^*} \right)\right|^{\frac{1}{2}}}{\left|\nabla^2 M\left({\bPsi}^*_{\bgamma} \right)\right|^{\frac{1}{2}}} \in(0, \infty) .
$$

Therefore, since $Z_3 n^{\frac{d_{\bgamma}-d_{\bgamma^*}}{2}}$ converges in probability into a constant, we have
\begin{align}\label{eqn:z3}
Z_3 =  \mathcal{O}_{\mathrm{p}}\left(n^{\frac{d_{\bgamma^*}-d_{\bgamma}}{2}}\right).
\end{align}

Another pair of results that follows from applying continuous mapping theorem on Lemma \ref{lemma:consistency} is $\pi(\tilde{\bPsi}_{\bgamma} \mid \bgamma) \xrightarrow{\Pr} \pi(\tilde{\bPsi}^*_{\bgamma} \mid \bgamma)$ and $\pi(\tilde{\bPsi}_{\bgamma^*} \mid \bgamma) \xrightarrow{\Pr} \pi(\tilde{\bPsi}^*_{\bgamma^*} \mid \bgamma)$. Then we can construct the following asymptotic equation for $Z_2$
\begin{align}\label{eqn:z2_ratio}
Z_2 \dfrac{\pi(\tilde{\bPsi}^*_{\bgamma^*} \mid \bgamma)}{\pi(\tilde{\bPsi}^*_{\bgamma} \mid \bgamma)} \xrightarrow{\Pr} 1, \quad n\to\infty,
\end{align}
with any continuous non-vanishing prior $\pi$. We start by characterizing $\tilde\nu$ and $\tilde\theta_0$. Recall that $e^{\nu} \sim \text{Gamma}(a_\nu, b_\nu)$ and $\theta_0 \sim N(0,K)$, so that
\begin{align*}
\dfrac{\pi(\tilde\nu_{\bgamma}) \pi(\tilde\theta_{0_{\bgamma}})}{\pi(\tilde\nu_{\bgamma^*}) \pi(\tilde\theta_{0_{\bgamma^*}})}
&= \exp\Big\{ a_\nu(\tilde{\nu}_{\bgamma}-\tilde{\nu}_{\bgamma^*}) - b_\nu\big(e^{\tilde{\nu}_{\bgamma}}-e^{\tilde{\nu}_{\bgamma^*}}\big) - \frac{\tilde\theta_{0_{\bgamma}}^2-\tilde\theta_{0_{\bgamma^*}}^2}{2K} \Big\}  \\
& = \lOp(1),
\end{align*}
as, by C1, $(\tilde{\nu}_{\bgamma},\tilde\theta_{0_{\bgamma}})$ and $(\tilde{\nu}_{\bgamma^*},\tilde\theta_{0_{\bgamma^*}})$ lie in compact sets, so the exponent is bounded.

Therefore, the ratio in the left-hand side of (\ref{eqn:z2_ratio}) could be expanded as
\begin{align}
\dfrac{\pi(\tilde{\bPsi}^*_{\bgamma^*} \mid \bgamma)}{\pi(\tilde{\bPsi}^*_{\bgamma} \mid \bgamma)}
&\propto \dfrac{\pi(\tilde{\btheta}^*_{\bgamma^*} \mid \bgamma) \pi(\tilde{\bmeta}^*_{\bgamma^*} \mid \bgamma)}{\pi(\tilde{\btheta}^*_{\bgamma} \mid \bgamma) \pi(\tilde{\bmeta}^*_{\bgamma} \mid \bgamma) } \nonumber \\ \nonumber
&= g_{C_t}^{\frac{q_{\bgamma}-q_{\bgamma^*}}{2}}g_{C_h}^{\frac{p_{\bgamma}-p_{\bgamma^*}}{2}}
\frac{
\left|n^{-1} \tilde \bX_{\bgamma^*}^\top \tilde \bX_{\bgamma^*}\right|^{\frac{1}{2}}
\left|n^{-1} \bX_{\bgamma^*}^\top \bX_{\bgamma^*}\right|^{\frac{1}{2}}
}{\left|n^{-1} \tilde \bX_{\bgamma}^\top \tilde \bX_{\bgamma}\right|^{\frac{1}{2}}
\left|n^{-1} \bX_{\bgamma}^\top \bX_{\bgamma}\right|^{\frac{1}{2}}}\\ \nonumber
& ~ \times \exp \left(
-\dfrac{1}{2n g_{C_t}} \left[ \tilde{\btheta}^{*T}_{\bgamma} \tilde\bX_{\bgamma}^\top \tilde\bX_{\bgamma} \tilde{\btheta}^{*}_{\bgamma} -
\tilde{\btheta}^{*T}_{\bgamma^*} \tilde\bX_{\bgamma^*}^\top \tilde\bX_{\bgamma^*} \tilde{\btheta}^{*}_{\bgamma^*}
\right]\right)\\ \nonumber
& ~ \times \exp \left(
-\dfrac{1}{2n g_{C_h}} \left[ \tilde{\bmeta}^{*T}_{\bgamma} \bX_{\bgamma}^\top \bX_{\bgamma} \tilde{\bmeta}^{*}_{\bgamma} -
\tilde{\bmeta}^{*T}_{\bgamma^*} \bX_{\bgamma^*}^\top \bX_{\bgamma^*} \tilde{\bmeta}^{*}_{\bgamma^*}
\right]\right)
\end{align}

And since $n^{-1}\tilde\bX_{\bgamma^*}^\top \tilde\bX_{\bgamma^*} $, $n^{-1}\tilde\bX_{\bgamma}^\top \tilde\bX_{\bgamma} $, $n^{-1}\bX_{\bgamma^*}^\top \bX_{\bgamma^*}$, and $n^{-1}\bX_{\bgamma}^\top \bX_{\bgamma}$ converges in probability to positive definite matrices, and $\tilde{\bmeta}^*_{\bgamma}$, $\tilde{\bmeta}^*_{\bgamma^*}$, $\tilde{\btheta}^*_{\bgamma}$, and $\tilde{\btheta}^*_{\bgamma^*}$ are fixed by assumption, the product of the determinants and exponential terms converges into a constant $k\in (0,\infty)$. So we have
\begin{align*}
\dfrac{\pi(\tilde{\bPsi}^*_{\bgamma^*} \mid \bgamma)}{\pi(\tilde{\bPsi}^*_{\bgamma} \mid \bgamma)}  \xrightarrow{\Pr} k\cdot g_{C_t}^{\frac{q_{\bgamma}-q_{\bgamma^*}}{2}}g_{C_h}^{\frac{p_{\bgamma}-p_{\bgamma^*}}{2}}.
\end{align*}

Now, appealing to Slutsky's theorem, we have
\begin{align*}
Z_2 \dfrac{\pi(\tilde{\bPsi}^*_{\bgamma^*} \mid \bgamma)}{\pi(\tilde{\bPsi}^*_{\bgamma} \mid \bgamma)} \xrightarrow{\Pr} 1 \implies Z_2 \cdot k g_{C_t}^{\frac{q_{\bgamma}-q_{\boldsymbol{\gamma}^*}}{2}}g_{C_h}^{\frac{p_{\bgamma}-p_{\bgamma^*}}{2}} \xrightarrow{\Pr} 1, \quad n\to\infty.
\end{align*}

And since $k$ is a constant, we have
\begin{align}\label{eqn:z2}
    Z_2 = \mathcal{O}_{\mathrm{p}}\left(g_{C_t}^{\frac{q_{\bgamma^*}-q_{\bgamma}}{2}}g_{C_h}^{\frac{p_{\bgamma^*}-p_{\bgamma}}{2}}\right).
\end{align}

For the characterization of $Z_1$, we need to consider the case of $\bgamma^* \not \subset \bgamma$ and $\bgamma^* \subset \bgamma$ separately.

\paragraph{Case $\bgamma^* \subset \bgamma$} This represents the situation where the optimal model is within the specified model family. We start by establishing the relationship between the MAP $\tilde{\bPsi}_{\bgamma}$ and MLE $\widehat{\bPsi}_{\bgamma}$. Performing a second order Taylor expansion with remainder around the MLE estimate $\widehat{\bPsi}_{\bgamma}$ (Chapter 16, \cite{vandervaart:2000}), we have that there exists $\bar{\bPsi}_{\bgamma}$ in the line joining the MAP $\tilde{\bPsi}_{\bgamma}$ and MLE $\widehat{\bPsi}_{\bgamma}$ such that,
\begin{align*}
\ell(\tilde{\bPsi}_{\bgamma} ) - \ell(\widehat{\bPsi}_{\bgamma} ) &= \left(\tilde{\bPsi}_{\bgamma} - \widehat{\bPsi}_{\bgamma} \right)^{\top} \nabla \ell(\widehat{\bPsi}_{\bgamma} ) + \frac{1}{2}\left(\tilde{\bPsi}_{\bgamma} - \widehat{\bPsi}_{\bgamma} \right)^{\top} \nabla^2  \ell(\bar{\bPsi}_{\bgamma} ) \left(\tilde{\bPsi}_{\bgamma} - \widehat{\bPsi}_{\bgamma} \right)\\
&= 0+\frac{1}{2}\sqrt{n} \left(\tilde{\bPsi}_{\bgamma} - \widehat{\bPsi}_{\bgamma} \right)^{\top} \dfrac{\nabla^2  \ell(\bar{\bPsi}_{\bgamma} )}{n} \sqrt{n}\left(\tilde{\bPsi}_{\bgamma} - \widehat{\bPsi}_{\bgamma} \right) \\
&= \frac{1}{2}\sqrt{n} \left(\tilde{\bPsi}_{\bgamma} - \widehat{\bPsi}_{\bgamma} \right)^{\top} \nabla^2  M(\bar{\bPsi}_{\bgamma} ) \sqrt{n}\left(\tilde{\bPsi}_{\bgamma} - \widehat{\bPsi}_{\bgamma} \right) + \mathcal{O}_{\mathrm{p}}(1) \\
&= \mathcal{O}_{\mathrm{p}}(1),
\end{align*}
where the last line follows from the fact $\dfrac{\nabla^2  \ell(\bar{\bPsi}_{\bgamma} )}{n} \xrightarrow{\Pr} \nabla^2  M(\bar{\bPsi}_{\bgamma} )$, $\nabla^2  M({\bPsi}_{\bgamma} ) $ is a symmetric negative-definite matrix (C7), and appealing to Proposition 2 in \cite{RT17}. The same results follows for ${\tbPsi}_{\bgamma^*}$. Therefore, we would have
$$
Z_1 = \ell(\tilde{\bPsi}_{\bgamma} ) - \ell(\tilde{\bPsi}_{\bgamma^*} ) = \ell(\widehat{\bPsi}_{\bgamma} ) - \ell(\widehat{\bPsi}_{\bgamma^*} ) + \mathcal{O}_{\mathrm{p}}(1).
$$

We now perform a second order Taylor expansion on $Z_1$
\begin{align}\label{eqn:Z1-taylor}
Z_1 &= \ell(\widehat{\bPsi}_{\bgamma} ) - \ell(\widehat{\bPsi}_{\bgamma^*} ) + \mathcal{O}_{\mathrm{p}}(1) \nonumber\\
&=-\frac{1}{2}\left(\widehat{\bPsi}_{\bgamma} - \widehat{\bPsi}_{\bgamma^*} \right)^{\top} \nabla^2  \ell(\widehat{\bPsi}_{\bgamma} ) \left(\widehat{\bPsi}_{\bgamma} - \widehat{\bPsi}_{\bgamma^*} \right) + \mathcal{O}_{\mathrm{p}}(1)\nonumber\\
&=-\frac{1}{2}\sqrt{n}\left(\widehat{\bPsi}_{\bgamma} - \widehat{\bPsi}_{\bgamma^*} \right)^{\top} \nabla^2  M({\bPsi}^*_{\bgamma} ) \sqrt{n} \left(\widehat{\bPsi}_{\bgamma} - \widehat{\bPsi}_{\bgamma^*} \right) + \mathcal{O}_{\mathrm{p}}(1)\nonumber\\
&= -\frac{1}{2}\boldsymbol\phi_n ^{\top} \nabla^2  M({\bPsi}^*_{\bgamma} ) \boldsymbol\phi_n + \mathcal{O}_{\mathrm{p}}(1),
\end{align}
where $\boldsymbol\phi_n = \sqrt{n}\left(\widehat{\bPsi}_{\bgamma} - \widehat{\bPsi}_{\bgamma^*} \right)$ and the third line follows from $\dfrac1n\nabla^2  \ell(\widehat{\bPsi}_{\bgamma} ) \xrightarrow{\Pr} \nabla^2  M({\bPsi}^*_{\bgamma})$. In this case of $\bgamma^* \subset \bgamma$, we have that ${\bPsi}^*_{\bgamma}={\bPsi}^*_{\bgamma^*}$. So we could rewrite $\boldsymbol\phi$ as
$$
\boldsymbol\phi_n = \sqrt{n}\left(\widehat{\bPsi}_{\bgamma} - \widehat{\bPsi}_{\bgamma^*} \right) = \sqrt{n}\left(\widehat{\bPsi}_{\bgamma} - {\bPsi}^*_{\bgamma} \right)
-\sqrt{n}\left({\bPsi}^*_{\bgamma^*} - \widehat{\bPsi}_{\bgamma^*} \right).
$$

Applying Lemma \ref{lemma:asympnorm} on the final terms, we have that $\sqrt{n}\left(\widehat{\bPsi}_{\bgamma} - {\bPsi}^*_{\bgamma} \right) \stackrel{D}{\longrightarrow} N(\mathbf0, \Sigma_{\bgamma})$ and $\sqrt{n}\left({\bPsi}^*_{\bgamma^*} - \widehat{\bPsi}_{\bgamma^*} \right)\stackrel{D}{\longrightarrow} N(\mathbf0, \Sigma_{\bgamma^*})$. Therefore,
$$
\boldsymbol\phi_n \stackrel{D}{\longrightarrow} N(\mathbf0, \Sigma), \quad n \to \infty,
$$
where $\Sigma$ is the asymptotic covariance matrix of $\boldsymbol\phi_n$, which exists by the joint asymptotic normality of $(\widehat{\bPsi}_{\bgamma},\widehat{\bPsi}_{\bgamma^*})$ \citep{white:1982}. Since $\Sigma$ is a positive semi-definite covariance matrix and $\nabla^2  M({\bPsi}^*_{\bgamma})$ is a real symmetric matrix, applying Lemma 1 and Theorem 1 in \cite{baldessari:1967}, the first term in (\ref{eqn:Z1-taylor}) could be written asymptotically as a linear combination of independent Chi-squared random variables \citep{rossell:2018,iqbal:2025}:
$$
\boldsymbol\phi^{\top}_n  \nabla^2  M({\bPsi}^*_{\bgamma}) \boldsymbol\phi_n \stackrel{D}{\longrightarrow} \sum_j  \lambda_j \chi_{r_j}^2,
$$
where $\lambda_j$ is the $j$-th unique eigenvalue of $\nabla^2  M({\bPsi}^*_{\bgamma})\Sigma$ and $r_j$ is its corresponding order of multiplicity. Combining this result with (\ref{eqn:Z1-taylor}), we have that $Z_1$ is stochastically bounded
\begin{align}\label{eqn:case1-z1}
    Z_1 \stackrel{D}{\longrightarrow} -\frac{1}{2}\sum_j  \lambda_j \chi_{r_j}^2 + \mathcal{O}_{\mathrm{p}}(1) = \mathcal{O}_{\mathrm{p}}(1).
\end{align}

Lastly, combining (\ref{eqn:z3}), (\ref{eqn:z2}), and (\ref{eqn:case1-z1}), we obtain the desired result for case $\bgamma^* \subset \bgamma$
\begin{align*}
    \tilde{B}_{\bgamma, \bgamma^*}= (2\pi)^{\frac{d_{\bgamma} - d_{\bgamma^*}}{2}} \exp (Z_1)  {Z_2 Z_3} = \mathcal{O}_{\mathrm{p}}\left(1 \cdot n^{\frac{d_{\bgamma^*}-d_{\bgamma}}{2}} \cdot g_{C_t}^{\frac{q_{\bgamma^*}-q_{\bgamma}}{2}} g_{C_h}^{\frac{p_{\bgamma^*}-p_{\bgamma}}{2}} \right).
\end{align*}

\paragraph{Case $\bgamma^* \not\subset \bgamma$} By the consistency of the MAP estimators (see the beginning of this proof) and a uniform law of large numbers (C1, C5), we have $\frac{1}{n}\ell(\tilde{\bPsi}_{\bgamma}) \xrightarrow{\Pr} M(\bPsi^*_{\bgamma})$ and $\frac{1}{n}\ell(\tilde{\bPsi}_{\bgamma^*}) \xrightarrow{\Pr} M(\bPsi^*_{\bgamma^*})$. Therefore,
\begin{align}\label{eqn:case2-z2}
\dfrac 1nZ_1 = \dfrac 1n \left(\ell(\tilde{\bPsi}_{\bgamma} ) - \ell(\tilde{\bPsi}_{\bgamma^*} ) \right)
\xrightarrow{\Pr}
M(\bPsi^*_{\bgamma}) - M(\bPsi^*_{\bgamma^*}) <0.
\end{align}

Although $M(\bPsi^*_{\bgamma}) - M(\bPsi^*_{\bgamma^*})$ could not be further bounded, it is driven by the KL-optimal errors of $\bgamma$ and $\bgamma^*$ \citep{rossell:2023}. Combining (\ref{eqn:case2-z2}) with (\ref{eqn:z3}) and (\ref{eqn:z2}), we derive the desired result for case $\bgamma^* \not\subset \bgamma$:
\begin{align*}
    \tilde{B}_{\bgamma, \bgamma^*}= \mathcal{O}_{\mathrm{p}}\left( 1\cdot
    \exp\left\{n\left(M(\bPsi^*_{\bgamma}) - M(\bPsi^*_{\bgamma^*})\right)\right\}\cdot n^{\frac{d_{\bgamma^*}-d_{\bgamma}}{2}} \cdot g_{C_t}^{\frac{q_{\bgamma^*}-q_{\bgamma}}{2}} g_{C_h}^{\frac{p_{\bgamma^*}-p_{\bgamma}}{2}} \right),
\end{align*}
and conclude the proof of Proposition 3.

\subsection{Proof of Corollary 2}
\label{app:proof_coro}

We could write the statement as
\begin{equation*}
\tilde{\pi}\left(\boldsymbol{\bgamma}^* \mid \boldsymbol{t}, \boldsymbol{\delta}\right) = \left(1+ \sum_{\bgamma \not = \bgamma^*} \tilde B_{\bgamma,\bgamma^*}\frac{\pi(\bgamma)}{\pi(\bgamma^*)} \right)^{-1} \xrightarrow{\Pr} 1.
\end{equation*}

By Proposition 3, for case $\bgamma^*\subset \bgamma$, we have $\tilde B_{\bgamma,\bgamma^*} = \lOp\big((ng_{C_h})^{(p_{\bgamma^*}-p_{\bgamma})/2}(ng_{C_t})^{(q_{\bgamma^*}-q_{\bgamma})/2}\big)$, where $p_{\bgamma^*}\le p_{\bgamma}$ and $q_{\bgamma^*}\le q_{\bgamma}$ with at least one strict inequality; for case $\bgamma^* \not \subset \bgamma$, $M(\bPsi^*_{\bgamma^*}) - M(\bPsi^*_{\bgamma}) > 0$, so the factor $\exp\{-n[M(\bPsi^*_{\bgamma^*}) - M(\bPsi^*_{\bgamma})]\}$ dominates the polynomial term. Therefore, $\tilde B_{\bgamma,\bgamma^*} \xrightarrow{\Pr} 0$ for either of the cases. And hence, we have
\begin{equation*}
\sum_{\bgamma \not = \bgamma^*} \tilde B_{\bgamma,\bgamma^*}\frac{\pi(\bgamma)}{\pi(\bgamma^*)} \xrightarrow{\Pr} 0,
\end{equation*}
and we conclude that
\begin{equation*}
\left(1+ \sum_{\bgamma \not = \bgamma^*} \tilde B_{\bgamma,\bgamma^*}\frac{\pi(\bgamma)}{\pi(\bgamma^*)} \right)^{-1} \xrightarrow{\Pr} 1.
\end{equation*}

\subsection{Proof of Corollary 3}
\label{app:proof_coro3}
{
For any deterministic map $g$, the event $\{g(\bgamma)=g(\bgamma^*)\}$ contains $\{\bgamma=\bgamma^*\}$, so $\widehat\pi(g(\bgamma)=g(\bgamma^*)\mid\bt,\boldsymbol{\delta})\ge
\widehat\pi(\bgamma^*\mid\bt,\boldsymbol{\delta})\xrightarrow{\Pr}1$ by Corollary 1; the displayed limits follow by taking $g=\mathcal{S}$, $g=\mathrm{Hz}$ and
$g(\bgamma)=I(\gamma_j\neq 0)$. Under correct specification, the Kullback--Leibler Information Criterion of \citet{white:1982} minimised by $\bPsi^*_{\bgamma}$ (well defined by C5) attains the value zero at the data-generating parameter $\bPsi_0$; identifiability of the GH model with a non-Weibull log-location-scale baseline \citep{chen:2001}, together with C2 and C4, makes the minimiser unique, so  $\bPsi^*_{\bgamma_0}=\bPsi_0$. The parsimony in the definition of $\bgamma^*$ then forces $\bgamma^*=\bgamma_0$.
}

\section{Propriety of the posterior under improper prior for common parameters}

In this section, we justify the propriety of the posterior $\pi(\bgamma \mid \bt)$ under the improper prior of $\pi(\mu,\sigma)\propto \sigma^{-k}, k\geq 1$.
Recall $\bPsi=(\nu,\theta_0,\btheta^\top,\bmeta^\top)^\top$ with $e^{\nu}=\tau=1/\sigma$, $\theta_0=\mu/\sigma$, $\btheta=-\balpha/\sigma$, $\bmeta=-\bbeta$, and the standard log--location--scale density $f$ and survival $S=1-F$. For observation $i$ we have
\begin{align*}
y_i \;=\; \frac{\log t_i-\mu}{\sigma}\;=\;e^{\nu}\log t_i-\theta_0,&\qquad
z_i \;=\; \frac{\log t_i+\tbx_i^\top\balpha-\mu}{\sigma}\;=\;e^{\nu}\log t_i-\tbx_i^\top\btheta-\theta_0\;=\;y_i-\tbx_i^\top\btheta,\\
\zeta_i \;=&\; e^{\,\bx_i^\top\bbeta-\tbx_i^\top\balpha}\;=\;e^{\,\tbx_i^\top\btheta e^{-\nu}-\bx_i^\top\bmeta}>0 .
\end{align*}
Here $y_i$ is the standardised log-time and $z_i=y_i-\tbx_i^\top\btheta$ is the time-level adjusted (covariate-corrected) standardised residual; $z_i$ is the argument of the density $f$, while the hazard ``tilt'' $\zeta_i$ is free of the common parameters $(\mu,\sigma)$. Let $E=\{i:\delta_i=1\}$ index the $n_o=\sum_i\delta_i$ uncensored observations.
For $\zeta>0$ define
\begin{equation}\label{eq:gc}
g_\zeta(z)\;=\;\zeta\,f(z)\,S(z)^{\,\zeta-1}\;=\;\zeta \,h(z)\,e^{-\zeta H(z)}, \qquad z\in\R ,
\end{equation}
where $h=f/S$ and $H=-\log S$ are the hazard and cumulative hazard of the standard log--location--scale law.
\begin{lemma}[Properties $g_\zeta$]\label{lem:gc}
For every $\zeta>0$ and $z\in\R$:
\begin{enumerate}
\item[(i)] $g_\zeta$ is a probability density: $\int_\R g_\zeta(z)\,dz=1$.
\item[(ii)] $g_\zeta(z)\le \zeta\,h(z)$.
\item[(iii)] $\displaystyle\sup_{\zeta>0}g_\zeta(z)=\frac{h(z)}{e\,H(z)}$, attained at $\zeta^\star(z)=1/H(z)$.
\end{enumerate}
\end{lemma}

\begin{proof}
    \begin{enumerate}
        \item[(i)] Since $S'=-f$, $\frac{d}{dz}S(z)^{\zeta}=-\zeta\,f(z)S(z)^{\zeta-1}=-g_\zeta(z)$, hence $\int_\R g_\zeta=-[S^\zeta]_{-\infty}^{+\infty}=1$.
        \item[(ii)] Immediate from $g_\zeta(z)=\zeta \,h(z)\,e^{-\zeta H(z)}$ since $e^{-\zeta H}\leq1$.
        \item[(iii)] For fixed $z$, $g_\zeta(z)>0$ on $(0,\infty)$ with $\lim_{\zeta\to 0^+} g_\zeta=0$ and $\lim_{\zeta \to +\infty}g_\zeta=0$, so $g_\zeta$ attains a positive maximum in the interior, where $\frac{d}{d\zeta}g_\zeta=h(z)\,(1-\zeta H(z))e^{-\zeta H(z)}=0$. The only zero is $\zeta^\star=1/H(z)$, which is therefore the global maximiser; evaluating gives $\sup_{\zeta>0}g_\zeta(z)=h(z)/(eH(z))$.
    \end{enumerate}
\end{proof}

Using the per-observation likelihood, an uncensored observation contributes $h(t_i)S(t_i)=\frac{\zeta_i}{\sigma t_i}\,\frac{f(z_i)}{S(z_i)}\,S(z_i)^{\zeta_i}
=\frac{1}{\sigma t_i}\,\zeta_i\,f(z_i)\,S(z_i)^{\zeta_i-1}
=\frac{1}{\sigma t_i}\,g_{\zeta_i}(z_i)$, while a censored observation contributes $S(t_i)=S(z_i)^{\zeta_i}\le1$. Therefore, we only need to consider the uncensored part, as the censored part is already bounded. The full likelihood factorises as
\begin{equation}\label{eq:Lik}
L(\balpha,\bbeta,\mu,\sigma)
=\sigma^{-n_o}\,\Big(\prod_{i\in E}\tfrac{1}{t_i}\Big)\,
\underbrace{\prod_{i\in E} g_{\zeta_i}(z_i)}_{\text{uncensored}}\;
\underbrace{\prod_{i\notin E}S(z_i)^{\zeta_i}}_{\text{censored,}\,\le\,1}.
\end{equation}

For common parameters, we consider the following improper prior:
\begin{equation}\label{eq:improper}
\pi(\mu,\sigma)\;\propto\;\sigma^{-k},\qquad k\geq 1,
\end{equation}
and the model-specific coefficients $\bkappa_\gamma=(\btheta_\gamma,\bmeta_\gamma)$ keep the proper LCM $g$-prior,
\begin{equation}\label{eq:re_lcm}
\bkappa_\gamma\mid g_E\;\sim\;N\big(\mathbf 0,\;n\,g_E\,\bm{\mathcal J}_\gamma^{-1}\big),\qquad
\bm{\mathcal J}_\gamma\;:=\;\bm{\mathcal J}_{\btheta_\gamma,\bmeta_\gamma}(\widehat\bPsi_\gamma),
\end{equation}
where $\bm{\mathcal J}_\gamma$ is the $(\btheta_\gamma,\bmeta_\gamma)$ block of the observed Fisher
information evaluated at the model-specific MLE $\widehat\bPsi_\gamma$.
It is useful to note that: (i) because $\bm{\mathcal J}_\gamma$ is evaluated at the fixed point $\widehat\bPsi_\gamma$, the covariance is a constant positive-definite matrix given the data---it does not depend on the running parameters $(\mu,\sigma,\btheta_\gamma,\bmeta_\gamma)$; (ii) $g_E$ is assumed fixed in this section. While it may be possible to justify the propriety of posterior under a proper hyper-prior $g_E\sim \pi(g_E)$, as the main text focuses on fixed-$g$ scenarios, we will not look into it here. The same argument applies to the Product prior of the main text, which is also a proper Gaussian prior on $\bkappa_\gamma$ that does not depend on $(\mu,\sigma)$, provided $\bX_\gamma^\top\bX_\gamma$ and $\tbX_\gamma^\top\tbX_\gamma$ are positive definite.

It is also worth noting the connection between $\pi(\mu,\sigma)$ in \eqref{eq:improper} with $\pi(\nu,\theta_0)$, the reparametrisation:
\begin{align*}
\pi(\mu,\sigma)\;\propto\;\sigma^{-1} \iff& \pi(\nu,\theta_0)\propto e^{-\nu}\ \text{(reference prior)},\\
\pi(\mu,\sigma)\;\propto\;\sigma^{-2} \iff& \pi(\nu,\theta_0)\propto 1\ \text{(flat)} .
\end{align*}
The second choice is exactly the improper limit for the vague specification used in the main text, where $\sigma^{-1}=e^{\nu} \sim \text{Gamma}(a_{\nu},b_\nu)$ and $\theta_0\sim N(0,K)$ when $a_{\nu},b_\nu \to 0$ and $K\to \infty$.

The posterior for model $\bgamma$ is proper iff the marginal likelihood $m_\gamma=\int L(\mu,\sigma,\balpha,\bbeta)\,\pi(\mu,\sigma,\balpha,\bbeta) d\mu d\sigma d\balpha d\bbeta$ is finite. We give conditions under which $0<m_\gamma<\infty$ for every $\gamma$ in the model space, so that all pairwise Bayes factors exist. Throughout, $d_\gamma=p_\gamma+q_\gamma$ is the number of model-specific coefficients ($p_\gamma$ hazard-level, $q_\gamma$ time-level).

\begin{lemma}\label{lem:reduce}
With $\pi(\mu,\sigma)=\sigma^{-k}$ and the prior \eqref{eq:re_lcm}, the marginal likelihood satisfies
\begin{equation}
\label{eq:marg}
m_\gamma\ \le\ \Big(\prod_{i\in E}t_i^{-1}\Big)\int_{\R^{d_\gamma}} \Nrm(\bkappa_\gamma;\mathbf0,\bSigma_\gamma)\,G(\bkappa_\gamma)\,d\bkappa_\gamma,
\end{equation}
where
\begin{equation}\label{eq:G_k}
    G(\bkappa_\gamma):=\int_\R\!\int_\R e^{(k+n_o-2)\nu} \prod_{i\in E} g_{\zeta_i}(z_i)\,d\theta_0\,d\nu.
\end{equation}
\end{lemma}
The bound follows from a change of variable from $(\mu,\sigma)$ to $(\nu,\theta_0)$ (Jacobian $e^{-2\nu}$) in the integrated likelihood $L$, after dropping the censored factors $\prod_{i\notin E}S(z_i)^{\zeta_i}\le1$ from \eqref{eq:Lik}. With this factorisation and the fact that the LCM prior is a proper distribution free of $(\nu,\theta_0)$, the proof will be constructed as follows: (i) Proposition~\ref{prop:LS} shows that for each fixed $\bkappa_\gamma$, $G(\bkappa_\gamma)$ is finite; (ii) Proposition~\ref{prop:poly} shows that $G$ grows at most polynomially in $\bkappa_\gamma$, hence is integrable against the Gaussian; (iii) Proposition~\ref{prop:propriety} establishes the finiteness of $m_\gamma$; (iv) Theorem \ref{theo:1} concludes the propriety of the posterior $\pi(\bgamma \mid \bt)$. Consider the following technical conditions:
\begin{itemize}
    \item[(B1)] $f$ is continuous, symmetric, strictly positive, and bounded. The corresponding $g_\zeta(z)$ satisfies, for some peak exponent $r<1$ and constant $V$,
    \begin{equation}
    \label{eq:peak}
    M_\zeta:=\sup_{z}g_\zeta(z)\ \le\ V\,\big(1+|\log \zeta|\big)^{r}\qquad(\zeta>0),
    \end{equation}
    together with the light--tail property that, on compact $\zeta,\zeta'$--ranges, $\int_\R g_{\zeta}(u)\,g_{\zeta'}(u-\delta)\,du\to0$ faster than any polynomial in $|\delta|$.
    \item[(B2)] Censoring is noninformative.
    \item[(B3)] $\mj_\gamma\succ0$, so the prior \eqref{eq:re_lcm} is proper.
    \item[(B4)] $n_o\geq2$, and at least two uncensored units have distinct event times, $\log t_a\ne\log t_b$.
\end{itemize}
(B1) ensures the tails decay super-polynomially and can be peak-bounded. The three mentioned log-location-scale distributions in the main text---log-normal, log-logistic, and log-Laplace---satisfy this condition with $r<1$:
\begin{itemize}
\item \emph{Log--normal}: From Lemma~\ref{lem:gc}(iii) the peak is attained at $\zeta^\star=1/H(z)$. Then a direct estimate gives $M_\zeta\asymp\sqrt{2\log \zeta}/e$, \textit{i.e.} $r=1/2$.
\item \emph{Log--logistic \& log--Laplace}: $h$ is bounded and the reverse hazard is bounded, so $M_\zeta=\sup_z g_\zeta(z)$ is bounded uniformly in $\zeta$ (e.g.\ for the standard logistic $M_\zeta=\big(\tfrac{\zeta}{1+\zeta}\big)^{\zeta+1}\le e^{-1}$) and hence $r=0$.
\end{itemize}
(B2) ensures the censoring mechanism contributes no parameters and is captured by $\prod_{\notin E}S(z_i)^{\zeta_i}\le1$. (B3) requires the model-specific MLE $\widehat\bPsi_\gamma$ to exist and the observed-information block $\mj_\gamma$ to be positive definite; this is the only place the hazard-level design re-enters the conditions. Concretely, $\mj_\gamma\succ0$ presupposes (i) $\widehat\bPsi_\gamma$ is finite; (ii) that the combined design has full column rank; and (iii) enough uncensored observations to identify all $d_\gamma+2$ parameters, \textit{i.e.} $n_o\ge d_\gamma+2$. Thus, although the propriety argument in Propositions~\ref{prop:LS}--\ref{prop:poly} places no rank or count condition on $\btX_{\gamma,O}$ or $\bX_{\gamma,O}$ beyond (B4), the well-definedness of the prior \eqref{eq:re_lcm} carries the rank, non-separation, and $n_o\ge d_\gamma+2$ requirements through (B3). Lastly, (B4) is the location-scale requirement of a null model: it ensures the location is pinned by an event and the scale is pinned by two distinct event times.

\setcounter{proposition}{3}

\begin{proposition}\label{prop:LS}
Under \textup{(B1)--(B4)} and $k\geq1$, $G(\bkappa_\gamma)<\infty$ for every $\bkappa_\gamma\in\R^{d_\gamma}$.
\end{proposition}

\begin{proposition}\label{prop:poly}
Under \textup{(B1)--(B4)}, there are $V<\infty$ and an integer $m_0\geq 0$ such that for every $\bkappa_\gamma\in\R^{d_\gamma}$, $G(\bkappa_\gamma) \leq V\left(1+\left\|\bkappa_\gamma\right\|\right)^{m_0}$.
\end{proposition}

\begin{proposition}\label{prop:propriety}
Under \textup{(B1)--(B4)}, $\displaystyle\int_{\R^{d_\gamma}}\Nrm(\bkappa_\gamma;\mathbf0,\bSigma_\gamma)\,G(\bkappa_\gamma)\,d\bkappa_\gamma<\infty$.
\end{proposition}

\begin{theorem}\label{theo:1}
    Under \textup{(B1)--(B4)}, the improper prior $\pi(\mu,\sigma)\propto \sigma^{-k}$, $k\geq1$, and the fixed-$g$ LCM prior in \eqref{eq:re_lcm}, all Bayes factors are well-defined and the posterior $\pi(\bgamma \mid \bt)$ is proper under proper $\pi(\bgamma)$ for every sample.
\end{theorem}

\subsection{Proof of Proposition \ref{prop:LS}}
Since the censored factors are already dropped in the definition \eqref{eq:G_k}, $G(\bkappa_\gamma)=\int_\R e^{(k+n_o-2)\nu}\,A(\nu)\,d\nu$ with
\[
A(\nu):=\int_\R \prod_{i\in E} g_{\zeta_i(\nu)}\big(r_i(\nu)-\theta_0\big)\,d\theta_0 , \qquad r_i(\nu)=e^{\nu}\log t_i-\tbx_i^\top\btheta_\gamma ,
\]
so the integrand of $A$ is $\prod_{i\in E}g_{\zeta_i}(z_i)$ with $z_i=r_i(\nu)-\theta_0$. To bound $A(\nu)$, picking two distinct events $1,2\in E$ with $\log t_1\ne\log t_2$, their residual gap is
\[
\delta(\nu):=r_1(\nu)-r_2(\nu)=e^{\nu}(\log t_1-\log t_2)-(\tbx_1-\tbx_2)^\top\btheta_\gamma .
\]
Bounding the remaining $n_o-2$ events by their suprema $M_{\zeta_i}=\sup_z g_{\zeta_i}(z)$, with $u=r_1(\nu)-\theta_0$ we have
\begin{equation}
\label{eq:Anu}
A(\nu)\ \le\ \Big(\prod_{i\ge3}M_{\zeta_i(\nu)}\Big)\,\Psi(\nu),
\qquad
\Psi(\nu):=\int_\R g_{\zeta_1(\nu)}(u)\,g_{\zeta_2(\nu)}\big(u-\delta(\nu)\big)\,du .
\end{equation}
We consider splitting the $\nu$-integral into three regions: $\nu \to +\infty$, $\nu \to -\infty$, and the remaining middle part, and discuss how the integrand in $G$ is finite and integrable at each of them.

\paragraph{When $\nu \to +\infty$} In this case, $e^{(k+n_o-2)\nu}$ goes to infinity, so we need the other term, $A(\nu)$, goes to 0 faster. Then we have $\zeta_i(\nu)\to e^{-\bx_i^\top\bmeta_\gamma}$, which is a fixed finite value, so the $M_{\zeta_i(\nu)}$ converge to finite constants and $g_{\zeta_1},g_{\zeta_2}$ converge to fixed densities. Meanwhile $|\delta(\nu)|\sim e^{\nu}|\log t_1-\log t_2|\to\infty$, consider splitting $\Psi(\nu)$ at $\delta/2$:
$$
\Psi(\nu)=\int_{-\infty}^{\delta / 2} g_{\zeta_1}(u) g_{\zeta_2}(u-\delta) d u+\int_{\delta / 2}^{\infty} g_{\zeta_1}(u) g_{\zeta_2}(u-\delta) d u .
$$
In the left piece $u \leq \delta / 2 \Rightarrow u-\delta \leq-\delta / 2$, so $g_{\zeta_2}(u-\delta)$ sits at the deep left tail of $g_{\zeta_2}$. In the right piece $u>\delta / 2$, so $g_{\zeta_1}(u)$ sits at the deep right tail of $g_{\zeta_1}$. Bounding $g_{\zeta_1} \leq M_{\zeta_1}$ in the left piece and bounding $g_{\zeta_2} \leq M_{\zeta_2}$ in the right piece, the integrals give
$$
\Psi(\nu) \leq M_{\zeta_1} \bar{F}_{\zeta_2}^{-}\left(\frac{|\delta|}{2}\right)+M_{\zeta_2} \bar{F}_{\zeta_1}^{+}\left(\frac{|\delta|}{2}\right),
$$
where $\bar{F}_{\zeta}^{\pm}$ are the tail probabilities of $g_\zeta$:
$$
\bar{F}_\zeta^{+}(s)=\int_s^{\infty} g_\zeta(z) d z=\left[-S(z)^\zeta\right]_s^{\infty}=S(s)^\zeta, \quad \bar{F}_\zeta^{-}(s)=\int_{-\infty}^{-s} g_\zeta(z) d z=\left[-S(z)^\zeta\right]_{-\infty}^{-s}=1-S(-s)^\zeta .
$$
Therefore,
$$
\Psi(\nu) \leq M_{\zeta_1} \bar{F}_{\zeta_2}^{-}\left(\frac{|\delta|}{2}\right)+M_{\zeta_2} \bar{F}_{\zeta_1}^{+}\left(\frac{|\delta|}{2}\right)=M_{\zeta_1}\left(1-S\left(-\frac{|\delta|}{2}\right)^{\zeta_2}\right)+M_{\zeta_2} S\left(\frac{|\delta|}{2}\right)^{\zeta_1}.
$$
By the light tail assumption of (B1), as $s\to \infty$, $S(s)$ and (by symmetry) $F(-s)=1-S(-s)=S(s)$ decay faster than any polynomial in $s$. As $\zeta>0$, $S(s)^\zeta$ inherits the super-polynomial decay. For $1-S(-s)^\zeta$, we have
$$
1-S(-s)^\zeta=1-e^{-\zeta H(-s)} \leq \zeta\, H(-s),
$$
where as $s \to \infty$, by Taylor series expansion,
$$
H(-s)=-\log \left(1-F(-s)\right) \sim F(-s) \rightarrow 0
$$
super-polynomially according to (B1). With $M_{\zeta_i}$ bounded, given $|\delta(\nu)| \sim e^\nu$, $\Psi(\nu)$ decays super-polynomially in $e^\nu$.
Since $e^{(k+n_o-2)\nu}$ is only a fixed power of $e^{\nu}$, the integrand $e^{(k+n_o-2)\nu}A(\nu)$ is finite and integrable at $+\infty$.

\paragraph{When $\nu \to -\infty$} Now $e^{(k+n_o-2)\nu}\to0$ exponentially because $n_o\geq2$. Therefore, we need to show that the speed at which $A(\nu)$ increases is slower than this exponential rate. Bounding all but one event by their peaks, we have
\begin{equation}
\label{eq:Anu2}
A(\nu)\ \le\ \Big(\prod_{i\neq i^*}M_{\zeta_i(\nu)}\Big) \int_\R g_{\zeta_{i^*}(\nu)}\big(r_{i^*}(\nu)-\theta_0\big)\,d\theta_0 =\Big(\prod_{i\neq i^*}M_{\zeta_i(\nu)}\Big),
\end{equation}
as $\int_{\mathbb{R}} g_{\zeta_{i^*}}\left(r_{i^*}(\nu)-\theta_0\right) d \theta_0=\int_{\mathbb{R}} g_{\zeta_{i^*}}(u) d u=1$. Given that $\log \zeta_i(\nu)=e^{-\nu}\tbx_i^\top\btheta_\gamma-\bx_i^\top\bmeta_\gamma$, we have $|\log \zeta_i(\nu)|\le e^{-\nu}|\tbx_i^\top\btheta_\gamma|+|\bx_i^\top\bmeta_\gamma|$. Since $\bkappa_\gamma$ is fixed and $e^{-\nu}\to\infty$ as $\nu\to-\infty$, the first term dominates and
$$
1+|\log \zeta_i(\nu)|\ \le\ V''\,e^{|\nu|},
$$
where $V'':=\max_{i\in E}\big(1+|\tbx_i^\top\btheta_\gamma|+|\bx_i^\top\bmeta_\gamma|\big)$ is a constant depending only on the fixed $\bkappa_\gamma$ and sample of size $n$. Hence, for fixed $\bkappa_\gamma$,
$$
M_{\zeta_i(\nu)}\le V\big(1+|\log \zeta_i(\nu)|\big)^{r}\le V\,(V'')^{r}\,e^{\,r|\nu|}\ =:\ V'\,e^{\,r|\nu|},\qquad\nu\to-\infty,
$$
with $V'=V\,(V'')^{r}$. As there are $n_o-1$ retained events, $A(\nu)\le\prod_{i\neq i^*}M_{\zeta_i(\nu)}\le (V')^{n_o-1}e^{(n_o-1)r|\nu|}$; combined with $e^{(k+n_o-2)\nu}=e^{-(k+n_o-2)|\nu|}$ for $\nu<0$, and absorbing the constant $(V')^{n_o-1}$ into $\lesssim$,
$$
e^{\left(k+n_o-2\right) \nu} A(\nu) \lesssim e^{-\left(k+n_o-2\right)|\nu|} e^{\left(n_o-1\right) r|\nu|}=e^{-\left[\left(k+n_o-2\right)-\left(n_o-1\right) r\right]|\nu|},
$$
and $\left(k+n_o-2\right)-\left(n_o-1\right) r>0$ because $r<1$, $k\geq 1$, $n_o\geq2$ gives $\left(n_o-1\right) r<n_o-1 \leq n_o+k-2$. Therefore the integrand in $G(\bkappa_\gamma)$ is finite and integrable at $-\infty$.

\paragraph{When $-\infty<\nu< \infty$} Since $\zeta_i(\nu)$ lives in a compact subset of $(0,\infty)$, the $g_{\zeta_i}$'s are uniformly bounded and continuous, with $A(\nu)\leq \prod_{i \neq i^*} M_{\zeta_i(\nu)}< \infty$.

Combining the three regimes, we have $G(\bkappa_\gamma)<\infty$ for each fixed $\bkappa_\gamma$.

\subsection{Proof of Proposition \ref{prop:poly}}
Consider the change of variable $w = e^\nu$, so that $e^{(k+n_o-2)\nu}\,d\nu = w^{k+n_o-3}\,dw$ and $\delta(w)=w\,(\log t_1-\log t_2)-(\tbx_1-\tbx_2)^\top\btheta_\gamma$. Then \eqref{eq:Anu} becomes
\begin{equation}
\label{eq:Gw}
G(\bkappa_\gamma)\ \le\ \int_0^\infty w^{\,k+n_o-3}
\Big(\prod_{i\ge3}M_{\zeta_i}\Big)\,\widetilde\Psi(w)\,dw,
\qquad
\widetilde\Psi(w):=\int_\R g_{\zeta_1}(u)\,g_{\zeta_2}(u-\delta)\,du ,
\end{equation}
with all $\zeta_i$ evaluated at $\nu=\log w$. By (B1) and Proposition~\ref{prop:LS}, $\widetilde\Psi(w)$ decays super-polynomially when $|\delta|$ is large, so the integrand is appreciable only on a $w$-window of width $\lO(1)$, where $|\delta(w)|=\lO(1)$, centred at
$$
w^\star := \frac{(\tbx_1-\tbx_2)^\top\btheta_\gamma}{\log t_1-\log t_2},
\qquad
|w^\star|\ \le\ \frac{\|\tbx_1-\tbx_2\|}{|\log t_1-\log t_2|}\,\|\btheta_\gamma\|\ =\ \lO\!\big(\|\btheta_\gamma\|\big) .
$$
When $w^\star\le0$ the window sits at bounded $w$ and $G$ is bounded in $\btheta_\gamma$; we therefore take $w^\star>0$. Changing $\btheta_\gamma$ relocates the window but not its $\lO(1)$ width. We bound the integrand on this window through its three ingredients.

\paragraph{Prefactor} On the window $w=w^\star+\lO(1)$, and since $k+n_o-3\ge0$ (as $k\ge1,\,n_o\ge2$) the map $w\mapsto w^{\,k+n_o-3}$ is nondecreasing; combined with $w^\star=\lO(\|\btheta_\gamma\|)$ this gives
$$
w^{\,k+n_o-3}\ =\ \lO\!\big((1+\|\btheta_\gamma\|)^{\,k+n_o-3}\big)
$$
on the window.

\paragraph{Peaks} On the window $w\asymp w^\star$, so using $e^{-\nu}=1/w$ and $w^\star=(\tbx_1-\tbx_2)^\top\btheta_\gamma/(\log t_1-\log t_2)$,
$$
e^{-\nu}\,\tbx_i^\top\btheta_\gamma=\frac{\tbx_i^\top\btheta_\gamma}{w}\ \asymp\ \frac{\tbx_i^\top\btheta_\gamma}{w^\star}
=(\log t_1-\log t_2)\,\frac{\tbx_i^\top\btheta_\gamma}{(\tbx_1-\tbx_2)^\top\btheta_\gamma},
$$
a ratio of linear forms in $\btheta_\gamma$ and hence homogeneous of degree $0$ in $\btheta_\gamma$; in particular it is $\lO(1)$ as $\|\btheta_\gamma\|\to\infty$. Hence, using $|\log \zeta_i|\le|e^{-\nu}\tbx_i^\top\btheta_\gamma|+|\bx_i^\top\bmeta_\gamma|$, the Cauchy--Schwarz bound $|\bx_i^\top\bmeta_\gamma|\le(\max_i\|\bx_i\|)\|\bmeta_\gamma\|$, and (B1),
$$
M_{\zeta_i}\le V\big(1+|\log \zeta_i|\big)^{r}\le V'''\,(1+\|\bmeta_\gamma\|)^{r}\qquad\text{on the window,}
$$
so the product $\prod_{i\ge3}M_{\zeta_i}\le V\,(1+\|\bmeta_\gamma\|)^{(n_o-2)r}$ carries no $\btheta_\gamma$-growth (as $e^{-\nu}\tbx_i^\top\btheta_\gamma = \lO(1)$).

\paragraph{Overlap} We bound $\widetilde\Psi$ by length $\times$ height. For the height, since $\int g_{\zeta_2}=1$,
$$
\widetilde\Psi(w)=\int_\R g_{\zeta_1}(u)\,g_{\zeta_2}(u-\delta)\,du\ \le\ \Big(\sup_u g_{\zeta_1}(u)\Big)\int_\R g_{\zeta_2}(u-\delta)\,du\ =\ M_{\zeta_1}.
$$
For the length, $\delta(w)$ is affine in $w$ with fixed slope $\log t_1-\log t_2$, so the window $\{w:|\delta(w)|=\lO(1)\}$ has width $\lO(1)$. Hence
$$
\int_{\text{window}}\widetilde\Psi(w)\,dw\ \le\ |\text{window}|\cdot\sup_{w\in\text{window}}M_{\zeta_1}\ \le\ V'\,(1+\|\bmeta_\gamma\|)^{r}.
$$
For the last inequality, note that $w$ enters $M_{\zeta_1}$ only through $\zeta_1(w)$ in the peak bound in (B1), and we have
$$
\log \zeta_1(w)=e^{-\nu}\tbx_1^\top\btheta_\gamma-\bx_1^\top\bmeta_\gamma=\frac{\tbx_1^\top\btheta_\gamma}{w}-\bx_1^\top\bmeta_\gamma ,
$$
where the only $w$-dependent term is $\tbx_1^\top\btheta_\gamma/w$, and on the window $w\asymp w^\star$ this term equals $\tbx_1^\top\btheta_\gamma/w^\star$ up to $\lO(1)$, which is $\lO(1)$ by the Peaks bound (the ratio is homogeneous of degree $0$ in $\btheta_\gamma$). Hence
$$
\sup_{w\in\text{window}}|\log \zeta_1(w)|\ \le\ \lO(1)+(\max_i\|\bx_i\|)\|\bmeta_\gamma\| ,
$$
and (B1) gives $M_{\zeta_1(w)}\le V'(1+\|\bmeta_\gamma\|)^{r}$ uniformly over the window. This contributes one further factor $(1+\|\bmeta_\gamma\|)^{r}$ and no $\btheta_\gamma$-growth.

Combining the three (the off-window region is subordinated by the super-polynomial decay of $\widetilde\Psi$),
$$
G(\bkappa_\gamma)\ \lesssim(1+\|\btheta_\gamma\|)^{\,k+n_o-3}\,(1+\|\bmeta_\gamma\|)^{(n_o-1)r}\ \le\ V\,(1+\|\bkappa_\gamma\|)^{m_0},
$$
where the last step uses $\|\btheta_\gamma\|,\|\bmeta_\gamma\|\le\|\bkappa_\gamma\|$ together with $(1+\|\btheta_\gamma\|)^a(1+\|\bmeta_\gamma\|)^b\le(1+\|\bkappa_\gamma\|)^{a+b}$, and rounding the power up gives
$$
m_0=\big\lceil (k+n_o-3)+(n_o-1)\,r\big\rceil .
$$

\subsection{Proof of Proposition \ref{prop:propriety}}
By Proposition~\ref{prop:poly}, we have $G(\bkappa_\gamma)\le V(1+\|\bkappa_\gamma\|)^{m_0}$, and
$$
\int_{\mathbb R^{d_\gamma}}\Nrm(\bkappa_\gamma)\,G(\bkappa_\gamma)\,d\bkappa_\gamma \leq V\int_{\mathbb R^{d_\gamma}}\Nrm(\bkappa_\gamma)\,(1+\|\bkappa_\gamma\|)^{m_0}\,d\bkappa_\gamma = V\,\E_{\Nrm}\big[(1+\|\bkappa_\gamma\|)^{m_0}\big].
$$
Since $V$ is a finite constant, a Gaussian centred at $\bm0$ with finite covariance $\bSigma_\gamma$ has finite moments of every order, so $V\,\E_{\Nrm}\big[(1+\|\bkappa_\gamma\|)^{m_0}\big]$ is finite. Concretely, with $\lambda=\lambda_{\max}(\bSigma_\gamma)<\infty$, it is known that $\|\bkappa_\gamma\|\leq \sqrt{\lambda}\|\bZ\|$, where $\bZ \sim \Nrm(\bm0,\bm{I}_{d_\gamma})$ and hence $\|\bZ\| \sim \chi_{d_\gamma}$. Therefore, using $\E[\|\bZ\|^{m_0}]=2^{m_0/2}\,\Gamma\!\big(\tfrac{d_\gamma+m_0}{2}\big)/\Gamma\!\big(\tfrac{d_\gamma}{2}\big)$,
$$
\E_{\Nrm}[\|\bkappa_\gamma\|^{m_0}]\leq \lambda^{m_0 / 2}\, \E[\|\mathbf{Z}\|^{m_0}] = (2\lambda)^{m_0 / 2}\, \frac{\Gamma\left(\frac{d_\gamma+m_0}{2}\right)}{\Gamma\left(\frac{d_\gamma}{2}\right)}<\infty.
$$
As $\left(1+\| \bkappa_\gamma \|\right)^{m_0} \leq 2^{m_0-1}\left(1+\| \bkappa_\gamma \|^{m_0}\right)$, we have $\E_{\Nrm}\big[(1+\| \bkappa_\gamma \|)^{m_0}\big] \leq 2^{m_0-1}\left(1+\E_{\Nrm}[\| \bkappa_\gamma \|^{m_0}]\right) < \infty$. Hence the integral is finite.

\subsection{Proof of Theorem \ref{theo:1}}
By Lemma~\ref{lem:reduce} and Proposition~\ref{prop:propriety}, $m_\gamma\le(\prod_{i\in E} t_i^{-1})\int\Nrm(\bkappa_\gamma)\,G(\bkappa_\gamma)\,d\bkappa_\gamma<\infty$. Moreover $m_\gamma>0$: writing $m_\gamma=\int L(\balpha,\bbeta,\mu,\sigma)\,\pi(\mu,\sigma)\,\Nrm(\bkappa_\gamma)\,d\mu\,d\sigma\,d\bkappa_\gamma$, the integrand is continuous and strictly positive on $\{\sigma>0\}\times\R\times\R^{d_\gamma}$ (the standard density $f$ and survival $S$ are positive, and $\pi(\mu,\sigma)=\sigma^{-k}>0$), so the integral over this set of positive measure is strictly positive. Hence $0<m_\gamma<\infty$. Therefore, under the improper prior $\pi(\mu,\sigma)\propto \sigma^{-k}$, $k\geq1$, fix-$g$ LCM prior, and the conditions (B1)-(B4), the Bayes factors are well-defined and the posterior $\pi(\bgamma \mid \bt)$ is proper under proper $\pi(\bgamma)$ for every sample.

\color{black}

\newpage

\section{Detail of the extended ADS sampler}

\begin{table}[h]
\centering
\caption{Move selection probabilities by hazard class and state pattern. ``All equal'' means that all $p$ covariates are included with the same $\gamma_j$. From the null model, the only move is A\_null, which adds one variable with $\gamma_j$ drawn uniformly from $\{1,2,3,4\}$.}\label{tab:moves}
\begin{tabular}{llccccc}
\toprule
Class & Condition & A/D & A/D\_GH & S & C & $C^A$ \\
\midrule
Null & — & — & — & — & — & — \\
AH/PH & All equal & 0.50 & — & — & 0.25 & 0.25 \\
AH/PH & Mixed & 0.50 & — & 0.10 & 0.20 & 0.20 \\
AFT & All 4s & 0.60 & — & — & — & 0.40 \\
AFT & Mixed & 0.60 & — & 0.20 & — & 0.20 \\
GH & All in $\{0,3\}$, all 3s & — & 0.50 & — & 0.25 & 0.25 \\
GH & All in $\{0,3\}$, mixed & — & 0.50 & 0.15 & 0.15 & 0.20 \\
GH & Not all in $\{0,3\}$ & — & 0.50 & 0.25 & 0.25 & — \\
\bottomrule
\end{tabular}
\end{table}

\begin{algorithm}[H]
\caption{GetValidIdx: Filter out valid indices for A/D\_GH}
\begin{algorithmic}[1]
\Require $\bgamma$ with $\text{Haz}(\bgamma) = \text{GH}$
\State $p_k \gets |\{j : \gamma_j = k\}|$ for $k = 0, 1, 2, 3$
\If{$p_3 = 1$ and remaining forms AH/PH} \Return $\{j : \gamma_j \neq 3\}$
\ElsIf{$p_3 = 0, p_1 > 1, p_2 = 1$} \Return $\{j : \gamma_j \neq 2\}$
\ElsIf{$p_3 = 0, p_2 > 1, p_1 = 1$} \Return $\{j : \gamma_j \neq 1\}$
\ElsIf{$p_3 = 0, p_1 = 1, p_2 = 1$} \Return $\{j : \gamma_j = 0\}$
\Else{} \Return $\{1, \ldots, p\}$
\EndIf
\end{algorithmic}
\end{algorithm}

\begin{algorithm}[H]
\caption{Extended ADS sampler — Part 1: A\_null, A/D, A/D\_GH}
\begin{algorithmic}[1]
\Require Current state $\bgamma$
\Ensure Proposed state $\bgamma'$ and $\log Q = \log q_{\text{rev}} - \log q_{\text{fwd}}$
\State $\bgamma' \gets \bgamma$; $H \gets \text{Haz}(\bgamma)$; select move $M$ per Table~\ref{tab:moves}
\If{$M = $ \textbf{A\_null}}
    \State $j \sim \text{Unif}\{1,\ldots,p\}$; $v \sim \text{Unif}\{1,2,3,4\}$; $\gamma'_j \gets v$
    \State $q_{\text{fwd}} \gets 1/p/4$
    \State $q_{\text{rev}} \gets \begin{cases} P(\text{A/D\_GH}|\bgamma')/|\text{GetValidIdx}(\bgamma')| & \text{if GH} \\ P(\text{A/D}|\bgamma')/p & \text{o.w.} \end{cases}$
\ElsIf{$M = $ \textbf{A/D}}
    \State $j \sim \text{Unif}\{1,\ldots,p\}$
    \If{$\gamma_j > 0$} $\gamma'_j \gets 0$ \Comment{Delete}
    \Else{} $\gamma'_j \gets \mathbf{1}[H{=}\text{AH}] + 2\cdot\mathbf{1}[H{=}\text{PH}] + 4\cdot\mathbf{1}[H{=}\text{AFT}]$ \Comment{Add}
    \EndIf
    \State $q_{\text{fwd}} \gets P(\text{A/D}|\bgamma)/p$
    \State $q_{\text{rev}} \gets \begin{cases} 1/p/4 & \text{if } \bgamma' = \mathbf{0} \\ P(\text{A/D}|\bgamma')/p & \text{o.w.} \end{cases}$
\ElsIf{$M = $ \textbf{A/D\_GH}}
    \State $\mathcal{V} \gets \text{GetValidIdx}(\bgamma)$; $j \sim \text{Unif}(\mathcal{V})$
    \If{$\gamma_j > 0$} $\gamma'_j \gets 0$; \, $q_{\text{fwd}} \gets P(\text{A/D\_GH}|\bgamma)/|\mathcal{V}|$ \Comment{Delete}
        \State $q_{\text{rev}} \gets \begin{cases} 1/p/4 & \text{if } \bgamma' = \mathbf{0} \\ P(\text{A/D\_GH}|\bgamma')/|\mathcal{V}'|/3 & \text{o.w.} \end{cases}$
    \Else{} $\gamma'_j \sim \text{Unif}\{1,2,3\}$; \, $q_{\text{fwd}} \gets P(\text{A/D\_GH}|\bgamma)/|\mathcal{V}|/3$ \Comment{Add}
        \State $q_{\text{rev}} \gets P(\text{A/D\_GH}|\bgamma')/|\mathcal{V}'|$ where $\mathcal{V}' = \text{GetValidIdx}(\bgamma')$
    \EndIf
\EndIf
\end{algorithmic}
\end{algorithm}

\begin{algorithm}[H]
\caption{Extended ADS sampler — Part 2: C, S, $C^A$}
\begin{algorithmic}[1]
\If{$M = $ \textbf{C}}
    \State $\mathcal{N} \gets \{j : \gamma_j \neq 0\}$; $j \sim \text{Unif}(\mathcal{N})$; $\gamma'_j \sim \text{Unif}(\{1,2,3\}\setminus\{\gamma_j\})$
    \State $q_{\text{fwd}} \gets P(\text{C}|\bgamma)/|\mathcal{N}|/2$; \, $q_{\text{rev}} \gets P(\text{C}|\bgamma')/|\mathcal{N}|/2$
\ElsIf{$M = $ \textbf{S}}
    \State $j_1 \sim \text{Unif}\{1,\ldots,p\}$; $v_1 \gets \gamma_{j_1}$; $\mathcal{O} \gets \{j:\gamma_j \neq v_1\}$; $j_2 \sim \text{Unif}(\mathcal{O})$; $v_2 \gets \gamma_{j_2}$
    \If{$v_1 + v_2 \neq 3$} $(\gamma'_{j_1}, \gamma'_{j_2}) \gets (v_2, v_1)$ \Comment{Standard swap}
    \ElsIf{$v_1 \in \{1,2\}$} $(\gamma'_{j_1}, \gamma'_{j_2}) \sim \text{Unif}\{(0,3),(3,0)\}$ \Comment{$(1,2) \to (0,3)/(3,0)$}
    \Else{} $(\gamma'_{j_1}, \gamma'_{j_2}) \sim \text{Unif}\{(1,2),(2,1)\}$ \Comment{$(0,3) \to (1,2)/(2,1)$}
    \EndIf
    \State $p'_{\mathcal{O}} \gets |\{j : \gamma'_j \neq \gamma'_{j_1}\}|$
    \State $q_{\text{fwd}} \gets P(\text{S}|\bgamma)/p/|\mathcal{O}|$; if $v_1+v_2=3$: $q_{\text{fwd}} \gets q_{\text{fwd}}/2$
    \State $q_{\text{rev}} \gets P(\text{S}|\bgamma')/p/p'_{\mathcal{O}}$; if $\gamma'_{j_1}+\gamma'_{j_2}=3$: $q_{\text{rev}} \gets q_{\text{rev}}/2$
\ElsIf{$M = C^A$}
    \If{$H = \text{AFT}$}
        \State $v \sim \text{Unif}\{1,2,3\}$; replace all 4s with $v$
        \State $q_{\text{fwd}} \gets P(C^A|\bgamma)/3$; $q_{\text{rev}} \gets \begin{cases} P(C^A|\bgamma') & \text{if GH} \\ P(C^A|\bgamma')/2 & \text{if AH/PH} \end{cases}$
    \ElsIf{$H = \text{GH}$}
        \State Replace all 3s with 4
        \State $q_{\text{fwd}} \gets P(C^A|\bgamma)$; $q_{\text{rev}} \gets P(C^A|\bgamma')/3$
    \Else{} \Comment{AH or PH}
        \State $v \sim \text{Unif}(\{1,2,4\}\setminus\{v_{\text{old}}\})$; replace all nonzeros with $v$
        \State $q_{\text{fwd}} \gets P(C^A|\bgamma)/2$; $q_{\text{rev}} \gets \begin{cases} P(C^A|\bgamma')/3 & \text{if AFT} \\ P(C^A|\bgamma')/2 & \text{o.w.} \end{cases}$
    \EndIf
\EndIf
\State \Return $\bgamma'$, $\log Q \gets \log q_{\text{rev}} - \log q_{\text{fwd}}$
\end{algorithmic}
\end{algorithm}

Once the MCMC sample has been obtained, two standard approaches can be used to estimate the posterior model probabilities. The first, and most common, is to calculate the proportion of times each model appears in the MCMC sample.
This provides an asymptotically unbiased estimate of the model posterior probabilities, but it may require a large number of iterations to stabilise. Let $M_{mc}$ be the set of models evaluated by the MCMC chain(s), that is, all proposed models, accepted or not, including those evaluated during burn-in, and $\pi(\bt,\bgamma)$ be the un-normalised posterior distribution of models under either ILA or LA. The second method estimates the model posterior probabilities by normalising over $M_{mc}$:
$$
\widehat p(\bgamma_m \mid \bt) = \dfrac{\pi(\bt,\bgamma_m)}{\sum_{m\in M_{mc}} \pi(\bt,\bgamma_m)}.
$$
Since most proposals are rejected, $M_{mc}$ covers considerably more of the posterior mass than the set of visited models, at no extra cost, as the sampler stores the un-normalised posterior of every evaluated model. Although the second method depends on accurate approximation of the marginal likelihood and MCMC convergence, it offers greater stability and lower variance. Note that this approach is not asymptotically unbiased with respect to the number of samples, since it requires evaluating all models with non-negligible posterior probability, a condition that can only be met with a sufficiently large number of MCMC iterations. Both estimates are available in our implementation, but in this paper we adopt the second method and evaluate its performance. Similarly, the posterior distribution of the hazard structures could be computed in the same manner:
$$
\widehat p(\hz(\bgamma)=l\mid \bt) = \dfrac{\sum_{\bgamma_k}\pi(\bt,\bgamma_k, \hz(\bgamma_k)=l)}{\sum_{m\in M_{mc}} \pi(\bt,\bgamma_m)}.
$$

\clearpage

\section{Additional information for the simulation study}\label{sec:add_sim}

In addition to the simulation setting mentioned in Section 6 of the main text, for the underlying baseline hazard, we consider using either a $\text{Lognormal}(1.55,0.7)$ distribution or a $\text{PGW}(1,1,2)$ distribution. The parameters for the baseline hazards are chosen such that the corresponding survival functions are around $0.05$ at $t=15$.
The active variables are positioned at $0.2p, 0.4p, 0.6p, 0.8p$. Their coefficients under the true models are: $(1, -1, 0.25, -0.25)$ for PH/AFT models and $\balpha_{active} =(1, 0.25, -1, -0.25)$ and $\bbeta_{active} =(0.25, -1, -0.25, 1)$ for GH model. The survival times are generated using the \verb|simGH| function from \verb|HazReg| R package, followed by administrative censoring to create corresponding level of right-censoring on average.
The setting for MCMC across all runs is as follows: $1$ chain with $20{,}000$ total iterations, $10{,}000$ burn-in, and thinning period of 2, which results in $5{,}000$ posterior samples of models. The initial model is chosen by the following procedure: we fit a Cox-LASSO model (using \verb|cv.glmnet| from \verb|glmnet| package, $10$-fold cross-validation, restricted to at most $n_o/5$ variables, i.e. at least five events per variable), followed by a step-wise selection on the selected PH model on the time-dependent part, based on BIC.
This initialisation is very cheap to compute and would yield a GH model that is better than a random initialisation. In case where $p$ is larger than the values considered here, the initialisation process could be simplified to only using the variables selected by Cox-LASSO model and start with the GH model containing them with all $\bgamma=3$. No optimisation failed to converge and only three of more than $1.5\times10^{8}$ evaluated models had a non-positive-definite Hessian. And as reported in Figure~\ref{fig:sim_GR}, chains started from dispersed models give mean Gelman--Rubin statistics of at most $1.011$ after $20{,}000$ iterations.

The model sensitivity and specificity results in Figure 2 of the main text are the true positive rate and true negative rate, respectively. The number of true positives of a model $\bgamma$ given the true model $\bgamma_0$ is calculated by $\sum^p_{j=1} \mathbf{1}(\gamma_j\neq 0 \text{ and } \gamma_{0j} \neq 0)$, and the number of true negatives of $\bgamma$ given the true model $\bgamma_0$ is calculated by $\sum^p_{j=1} \mathbf{1}(\gamma_j= 0 \text{ and } \gamma_{0j} = 0)$.

\subsection{Figures for large-scale simulation}

\begin{figure}[h]
    \centering
    \begin{subfigure}[b]{0.495\textwidth}
        \centering
        \includegraphics[width=\textwidth]{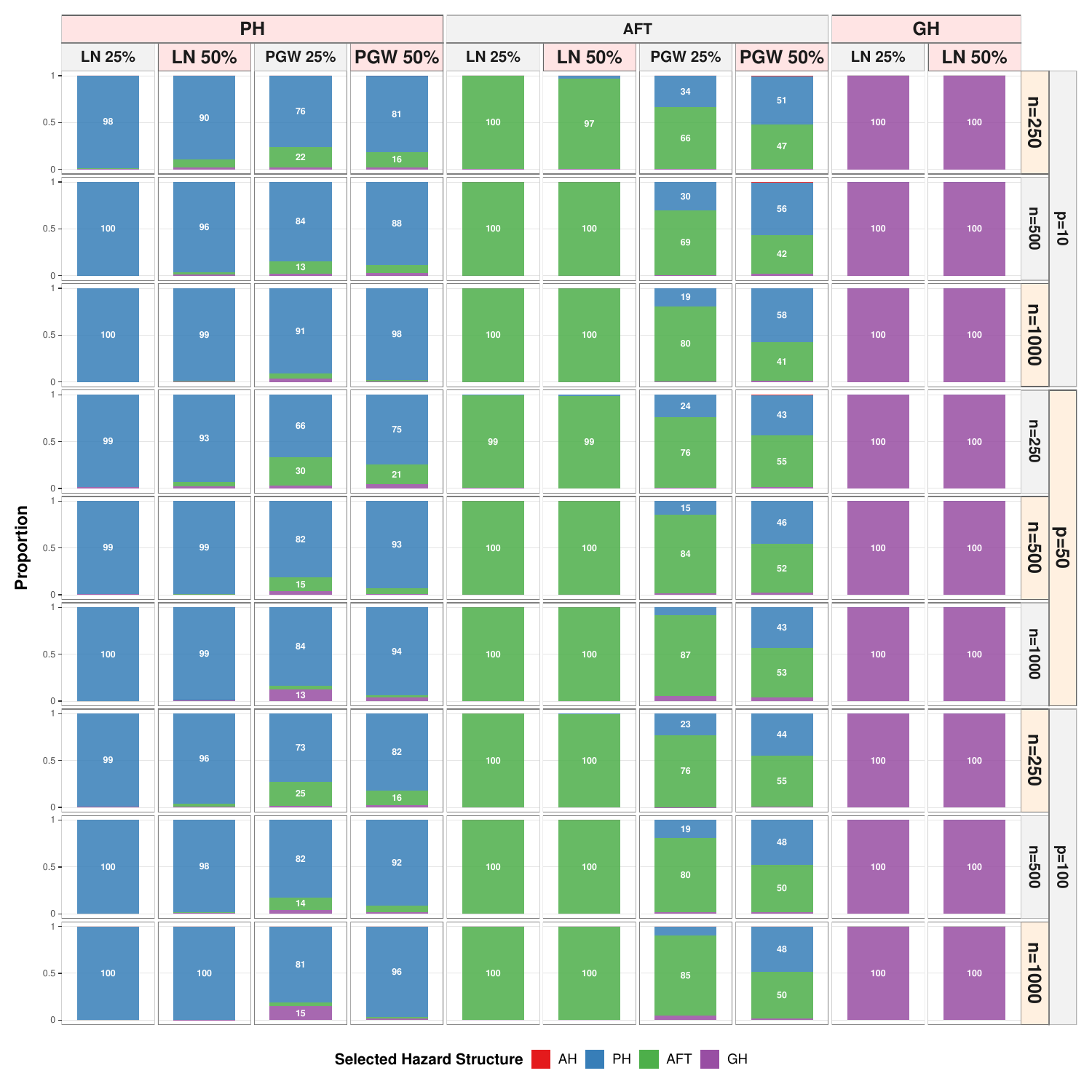}
        \caption{Hazard selection (LCM)}
    \end{subfigure}
    \hfill
    \begin{subfigure}[b]{0.495\textwidth}
        \centering
        \includegraphics[width=\textwidth]{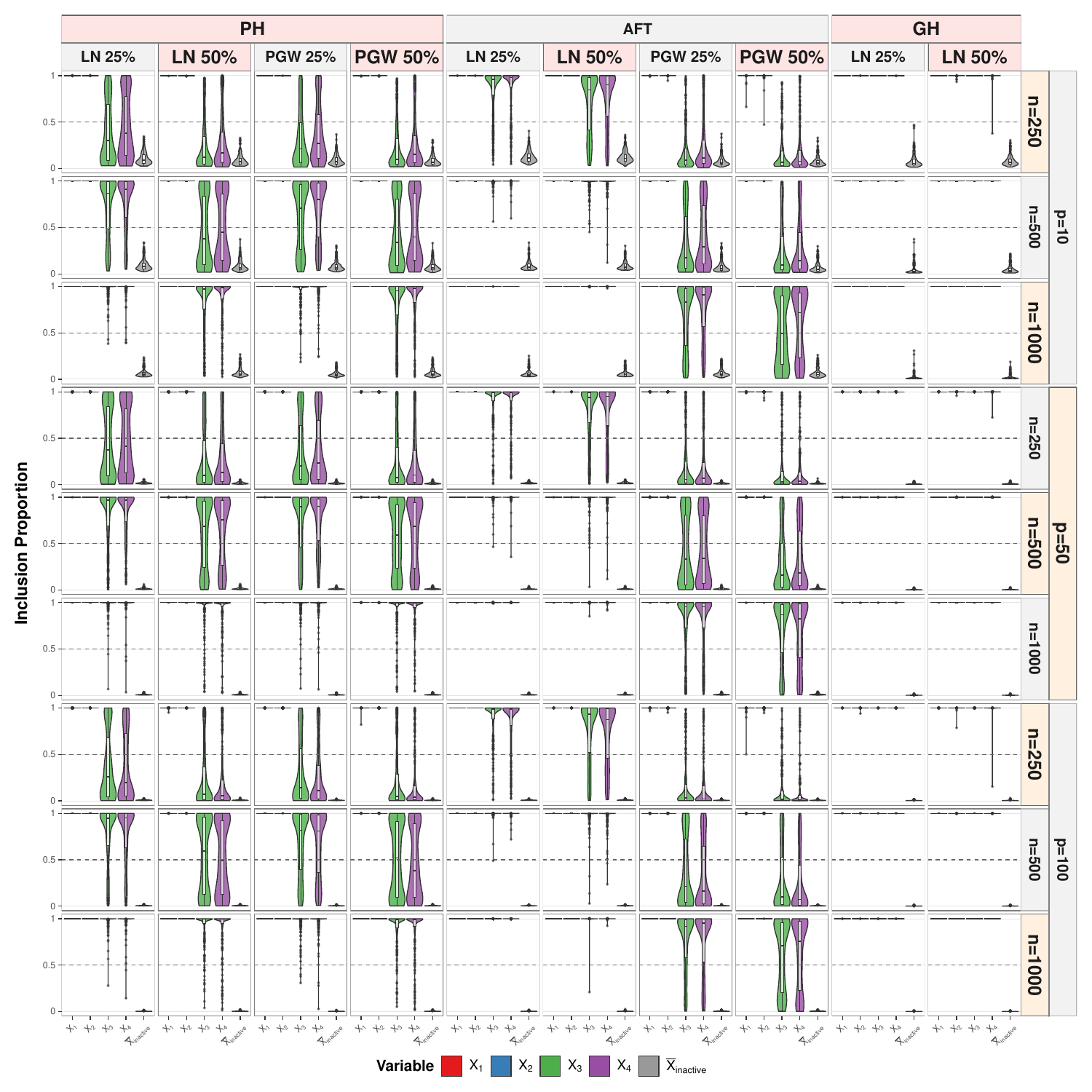}
        \caption{Variable inclusion (LCM)}
    \end{subfigure}
    \vspace{1em}
    \begin{subfigure}[b]{0.495\textwidth}
        \centering
        \includegraphics[width=\textwidth]{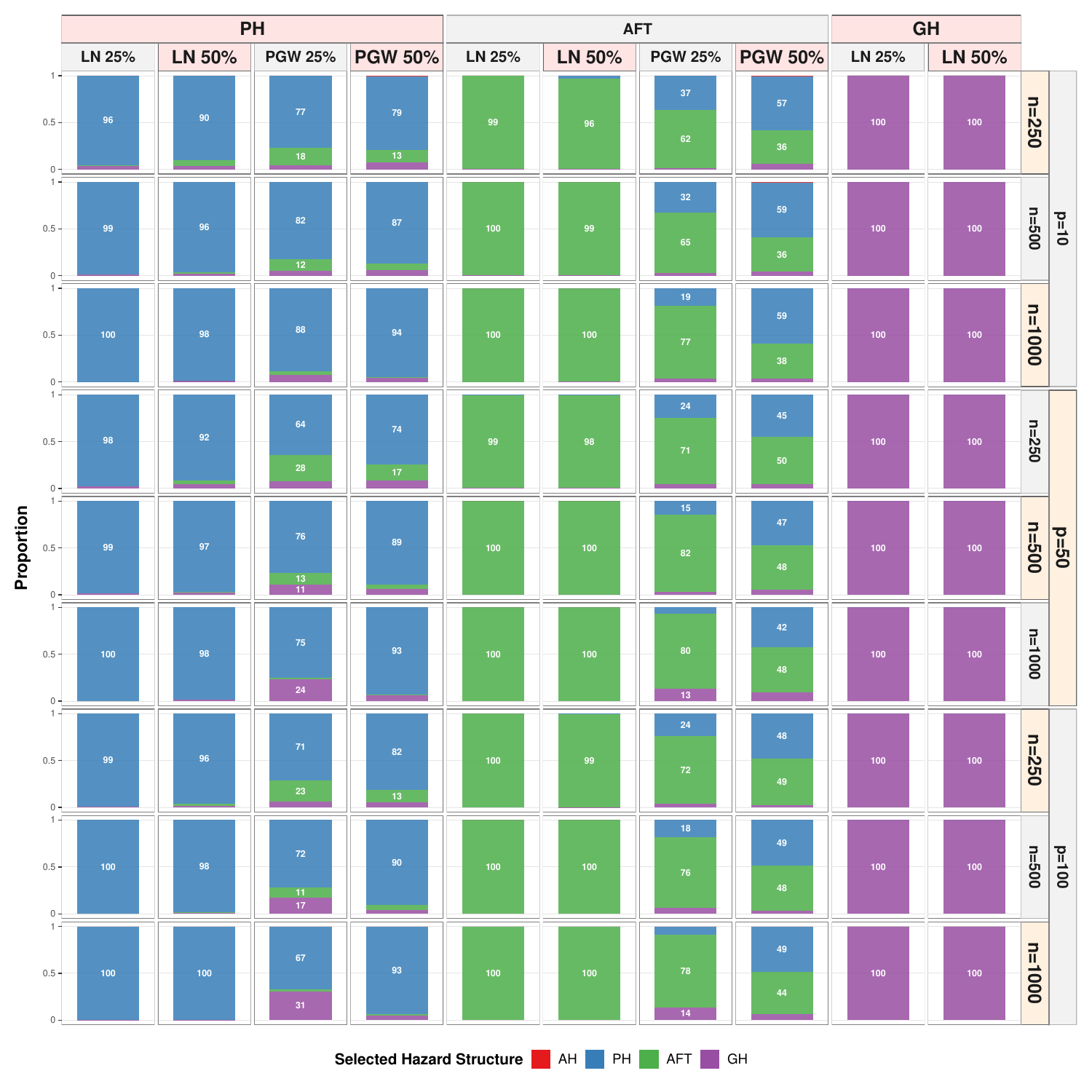}
        \caption{Hazard selection (Prod)}
    \end{subfigure}
    \hfill
    \begin{subfigure}[b]{0.495\textwidth}
        \centering
        \includegraphics[width=\textwidth]{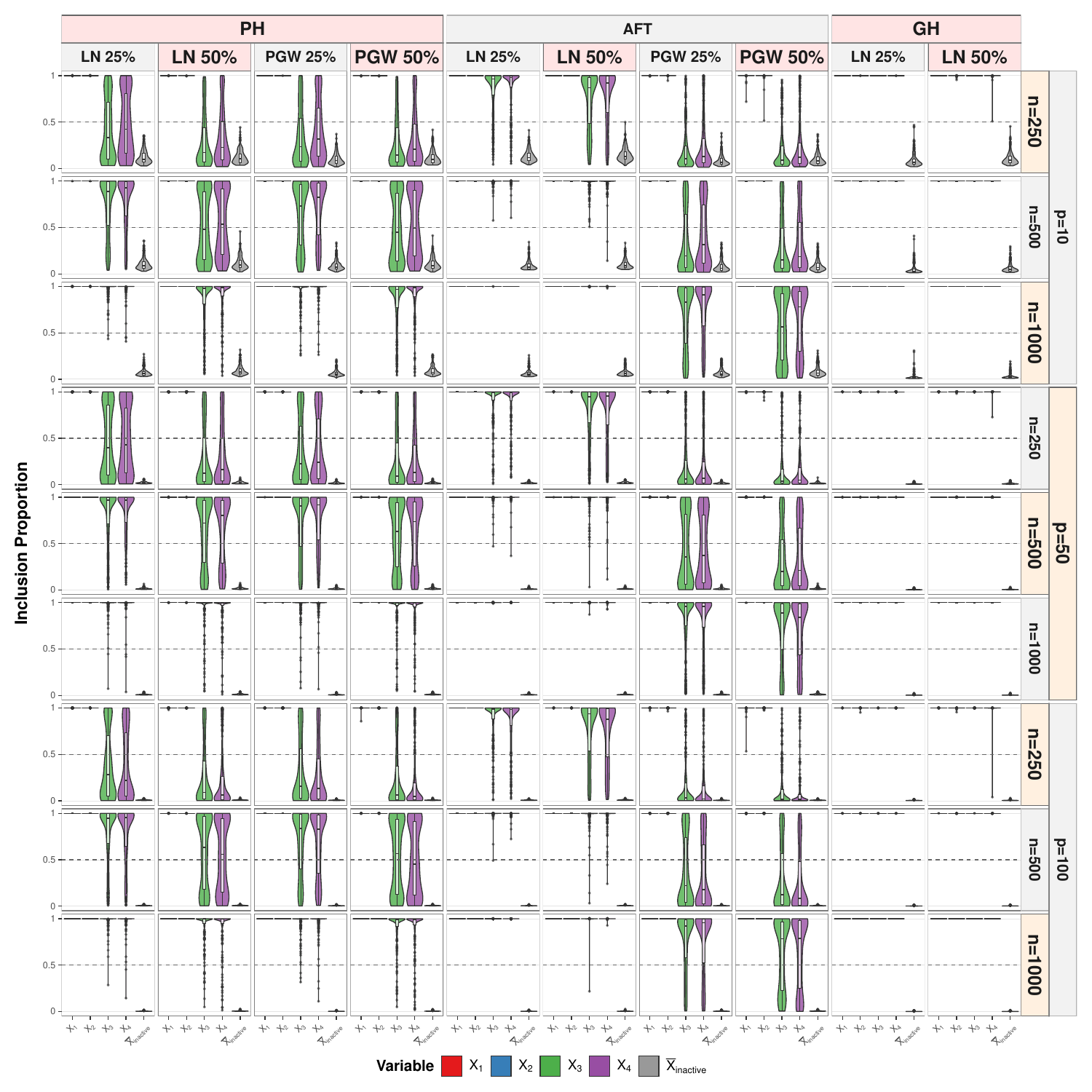}
        \caption{Variable inclusion (Prod)}
    \end{subfigure}
    \caption{Detailed hazard structure selection and variable inclusion results across the $250$ MC simulations for all scenarios and prior settings. }
    \label{fig:sim_vio_hstr}
\end{figure}

\begin{figure}[h]
    \centering
    \begin{subfigure}[b]{0.495\textwidth}
        \centering
        \includegraphics[width=\textwidth]{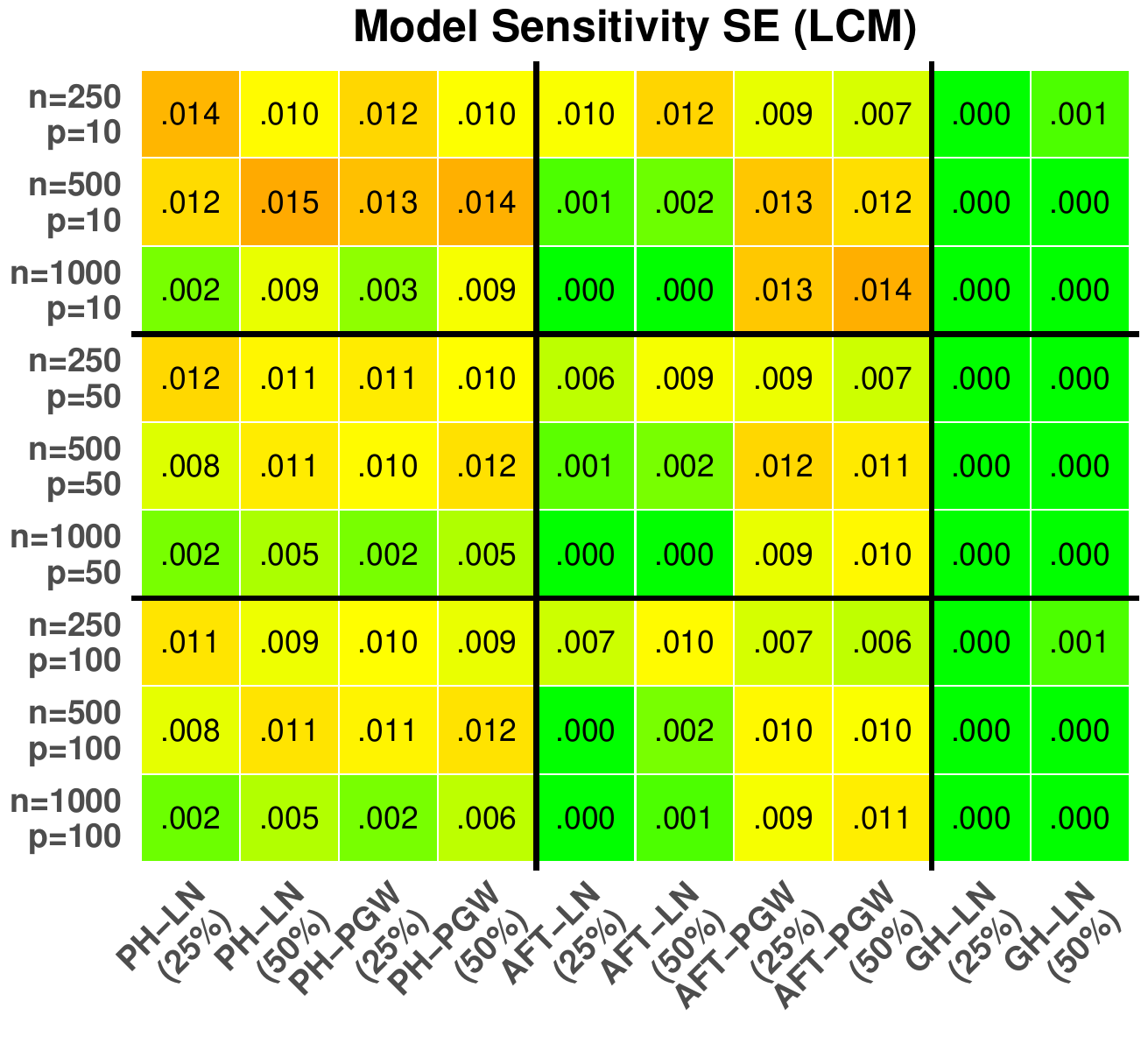}
        \caption{}
    \end{subfigure}
    \hfill
    \begin{subfigure}[b]{0.495\textwidth}
        \centering
        \includegraphics[width=\textwidth]{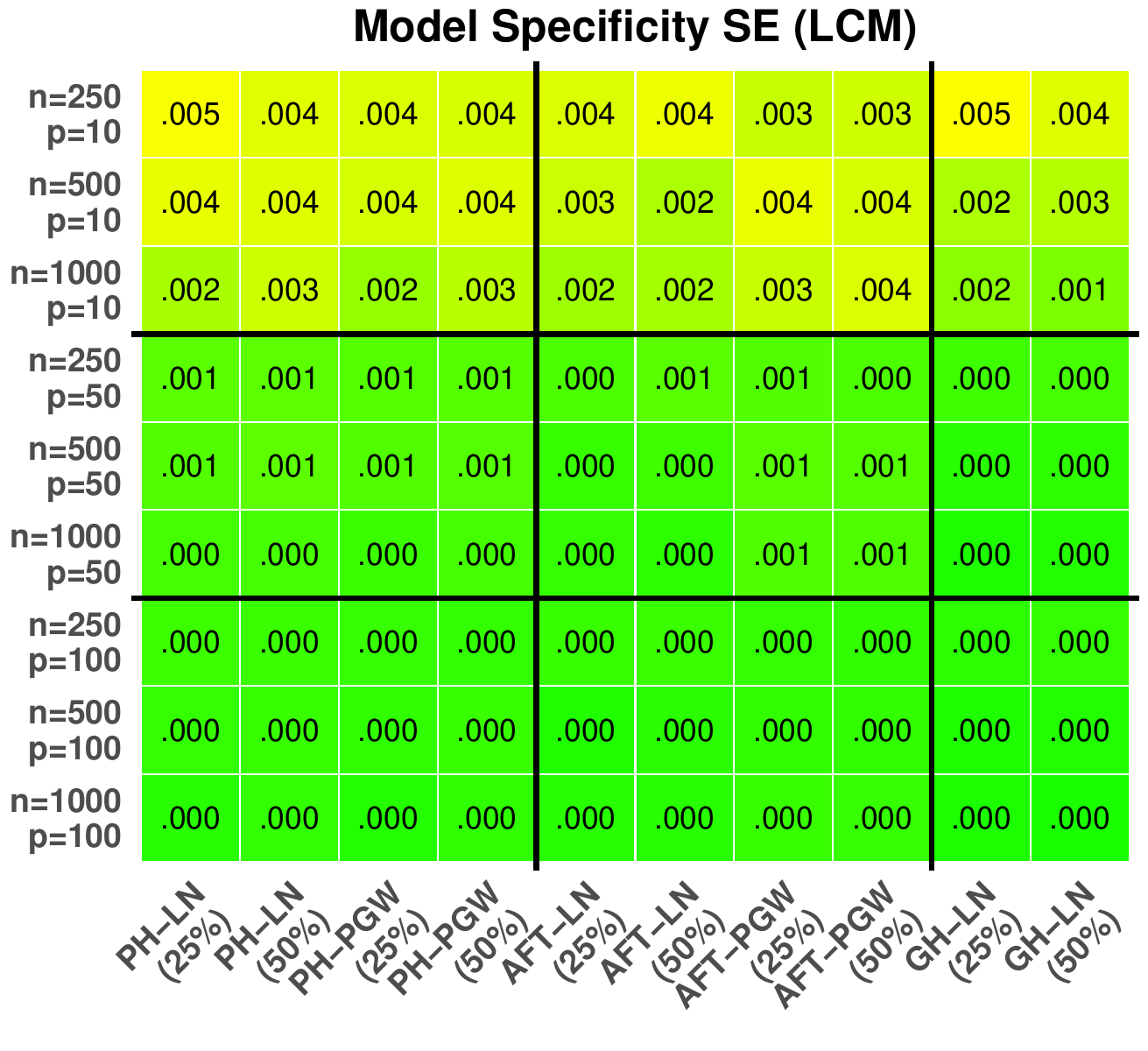}
        \caption{}
    \end{subfigure}
    \vspace{1em}
    \begin{subfigure}[b]{0.495\textwidth}
        \centering
        \includegraphics[width=\textwidth]{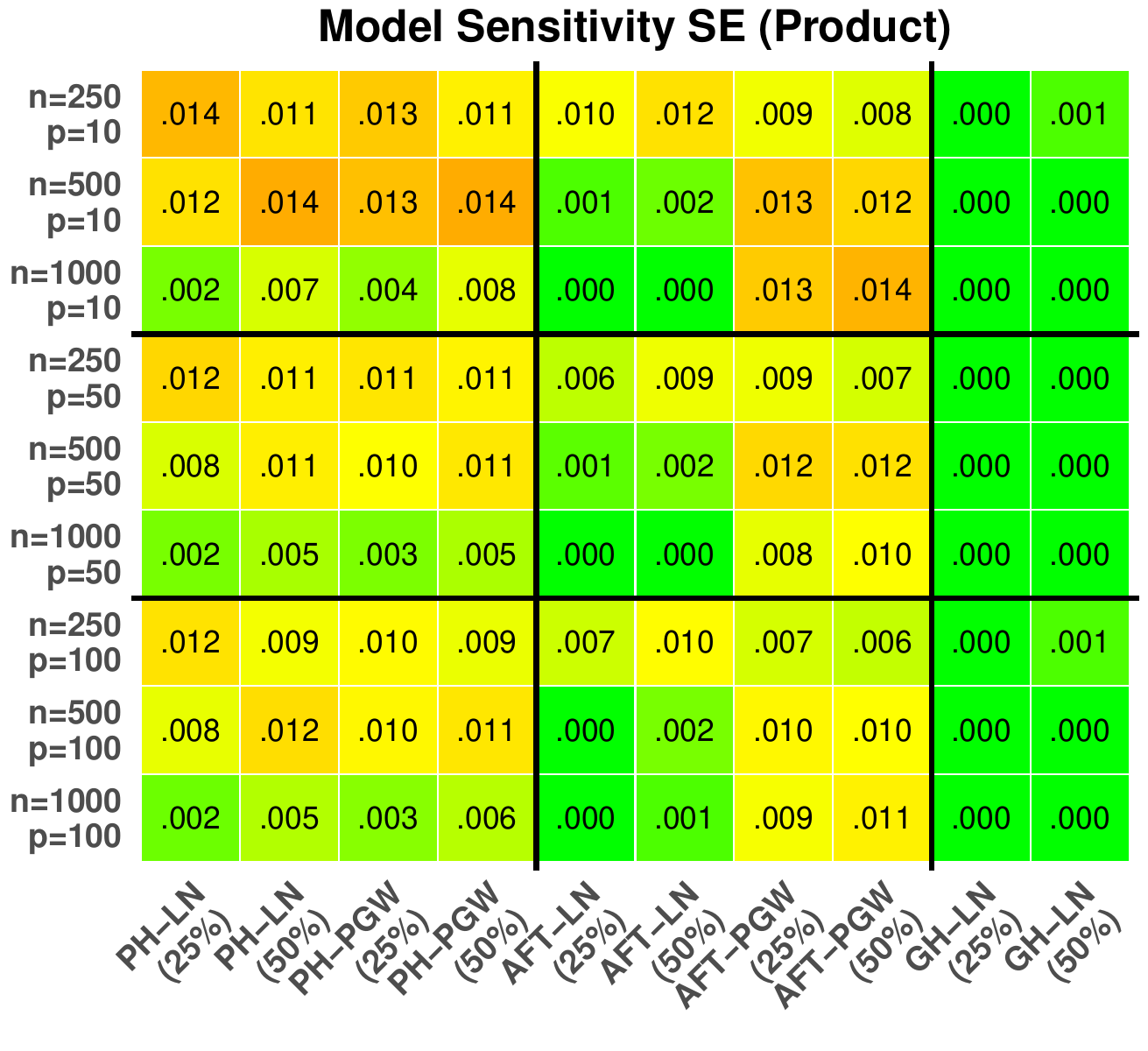}
        \caption{}
    \end{subfigure}
    \hfill
    \begin{subfigure}[b]{0.495\textwidth}
        \centering
        \includegraphics[width=\textwidth]{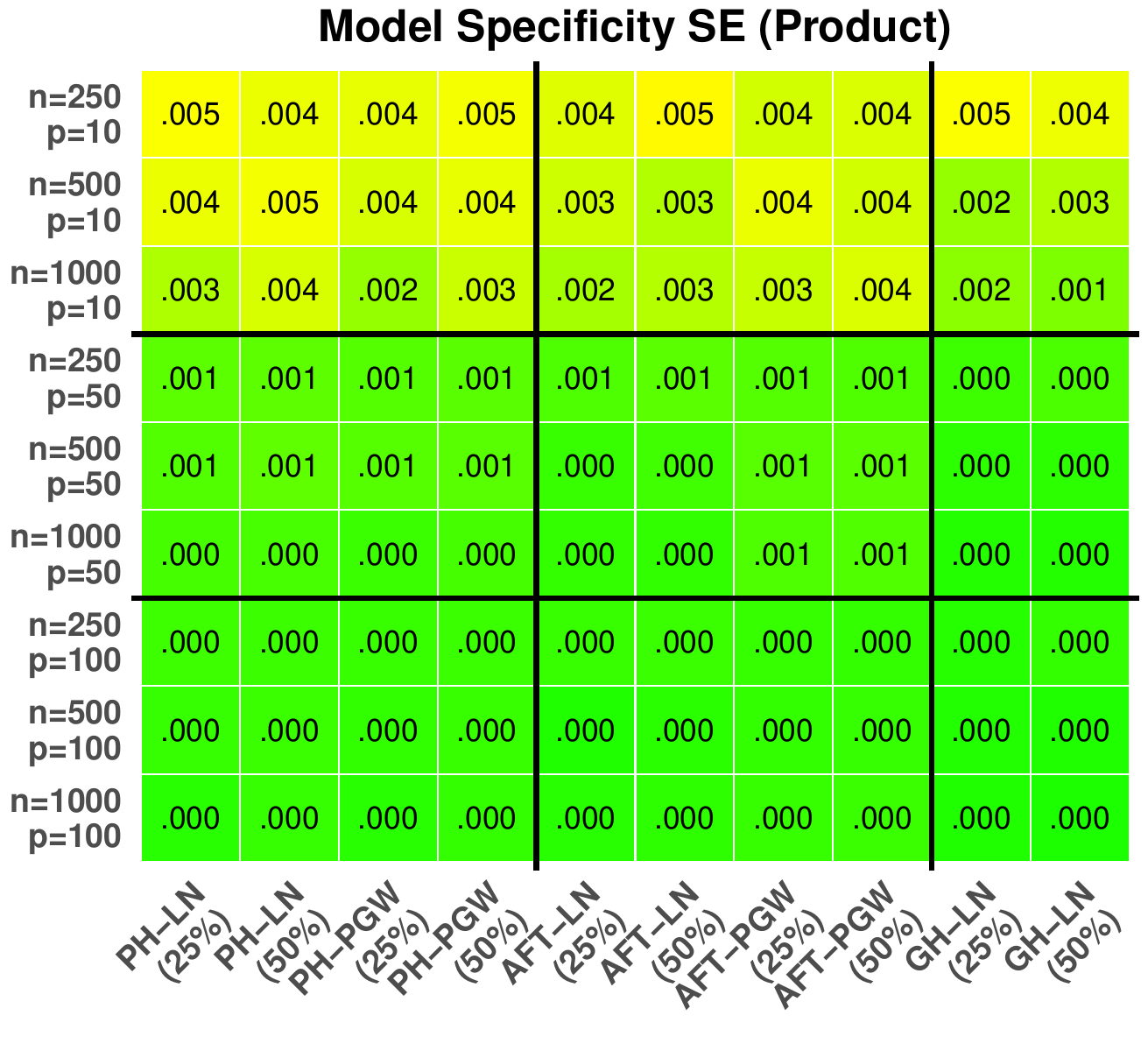}
        \caption{}
    \end{subfigure}
    \caption{Standard error of sensitivity and specificity of variable selection across the $250$ MC simulations for all scenarios and prior settings, based on the highest estimated posterior probability models. (a) SE of sensitivity for all simulation scenarios under LCM prior; (b) SE of specificity under LCM prior; (c) SE of sensitivity under Product prior; (d) SE of specificity under Product prior.}
    \label{fig:sim_varsel_SE}
\end{figure}

\begin{figure}[h]
    \centering
    \begin{subfigure}[b]{0.3\textwidth}
        \centering
        \includegraphics[width=\textwidth]{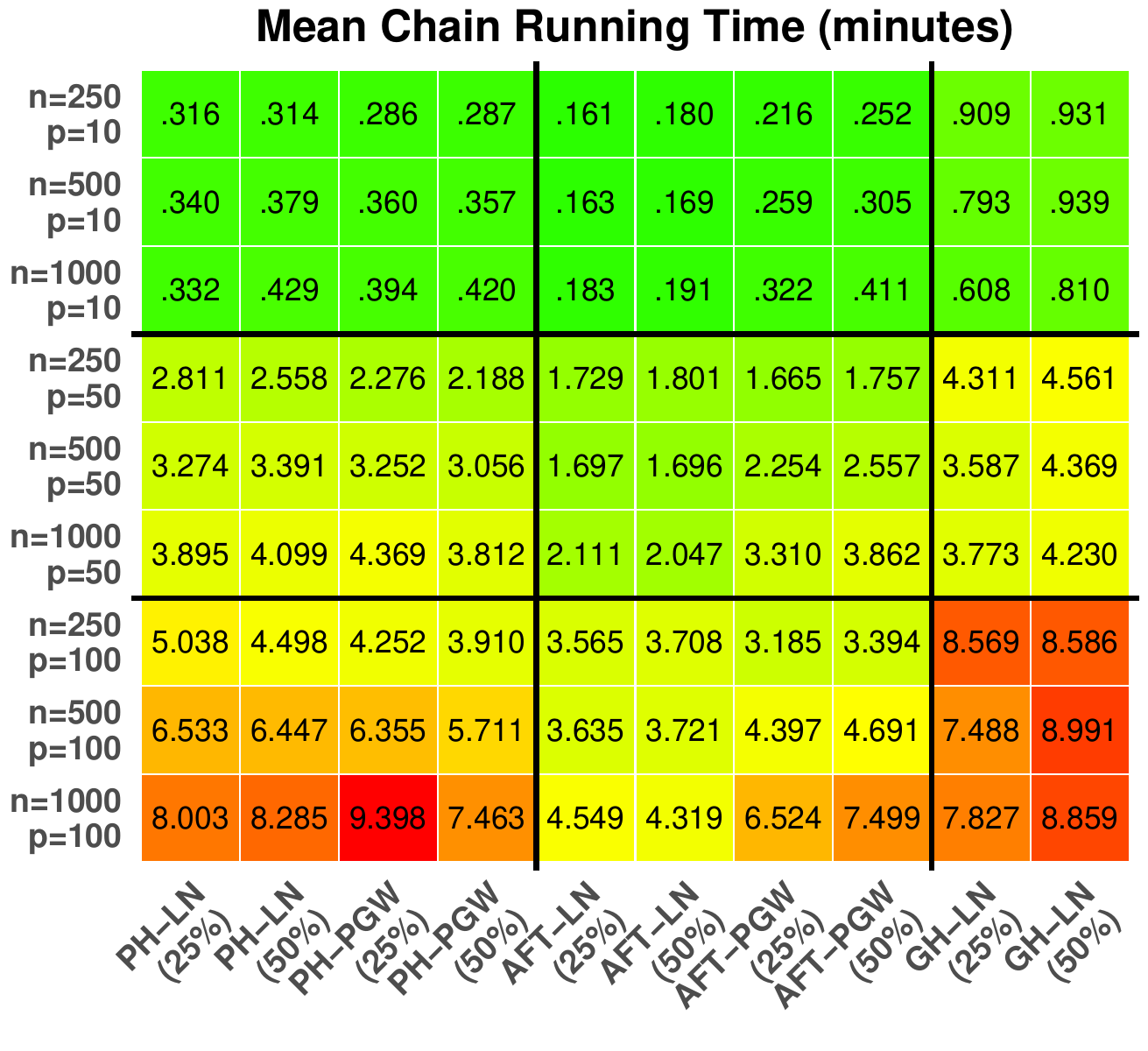}
        \caption{}
    \end{subfigure}
    \hfill
    \begin{subfigure}[b]{0.3\textwidth}
        \centering
        \includegraphics[width=\textwidth]{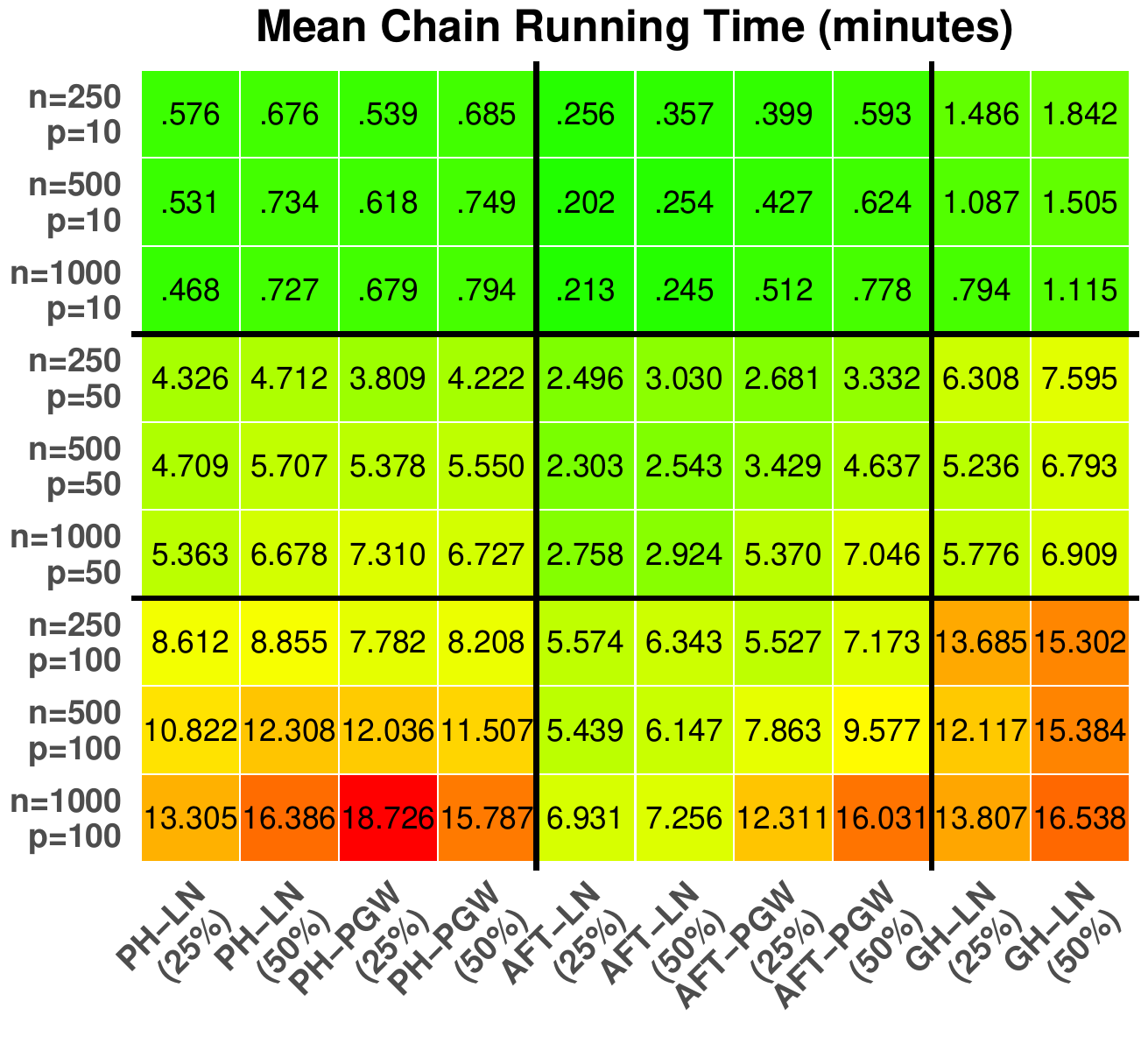}
        \caption{}
    \end{subfigure}
    \hfill
    \begin{subfigure}[b]{0.3\textwidth}
        \centering
        \includegraphics[width=\textwidth]{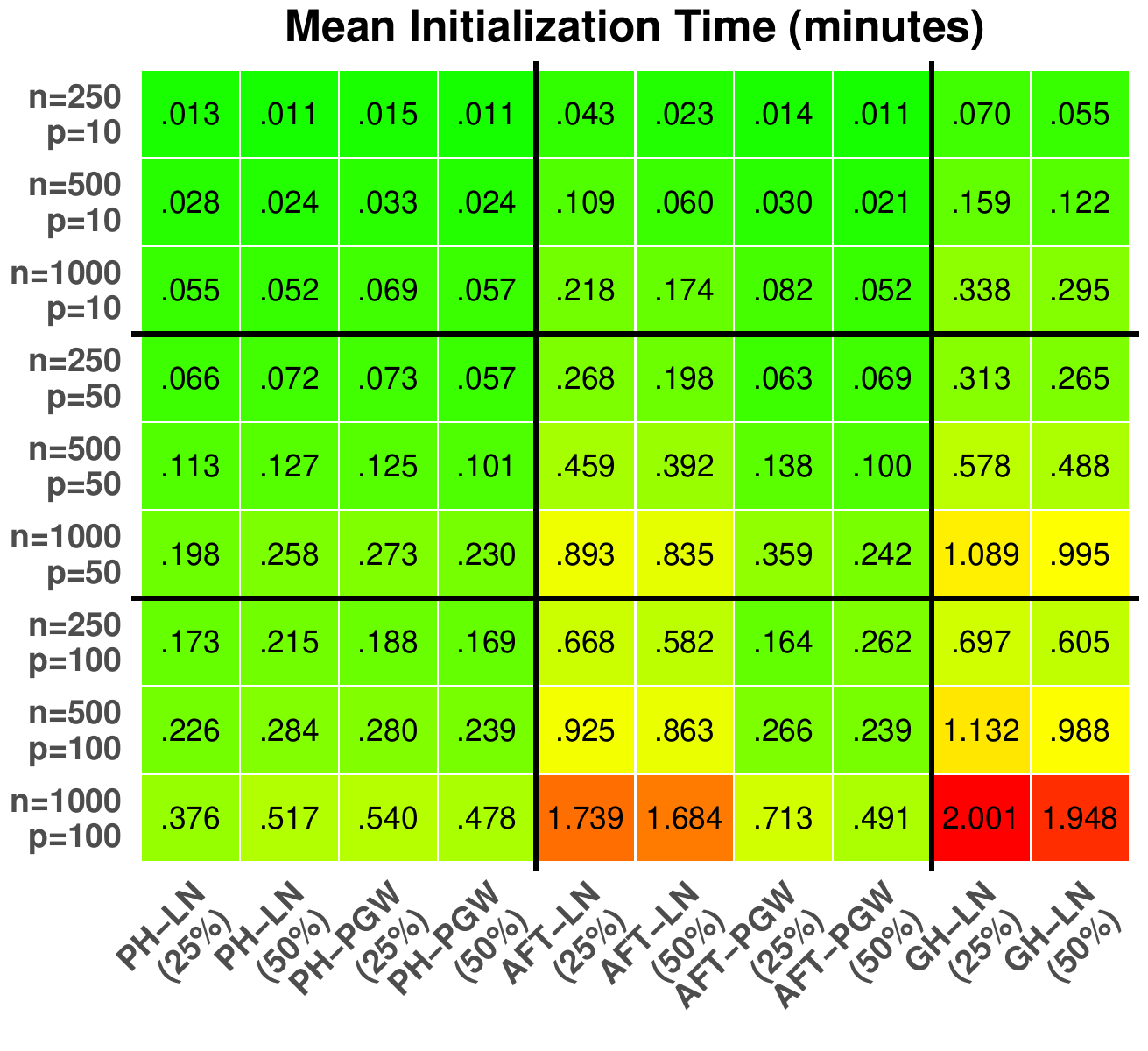}
        \caption{}
    \end{subfigure}
    \caption{Mean time (in minutes) of initialisation and running the chains across the $250$ MC simulations for all scenarios and prior settings. Each chain runs on a single core of an Intel(R) Xeon(R) Gold 6240 CPU @2.60GHz for $p=10$ and of an Intel(R) Xeon(R) Gold 6248 CPU @2.50GHz for $p\in\{50,100\}$. (a) Mean running time for $20{,}000$ iterations of MCMC under LCM prior; (b) Mean running time for $20{,}000$ iterations of MCMC under Product prior; (c) Initialisation time using the method discussed in Section~\ref{sec:add_sim}.}
    \label{fig:sim_varsel_time}
\end{figure}

\clearpage

\begin{figure}[h]
    \centering
    \begin{subfigure}[b]{0.49\textwidth}
        \centering
        \includegraphics[width=\textwidth]{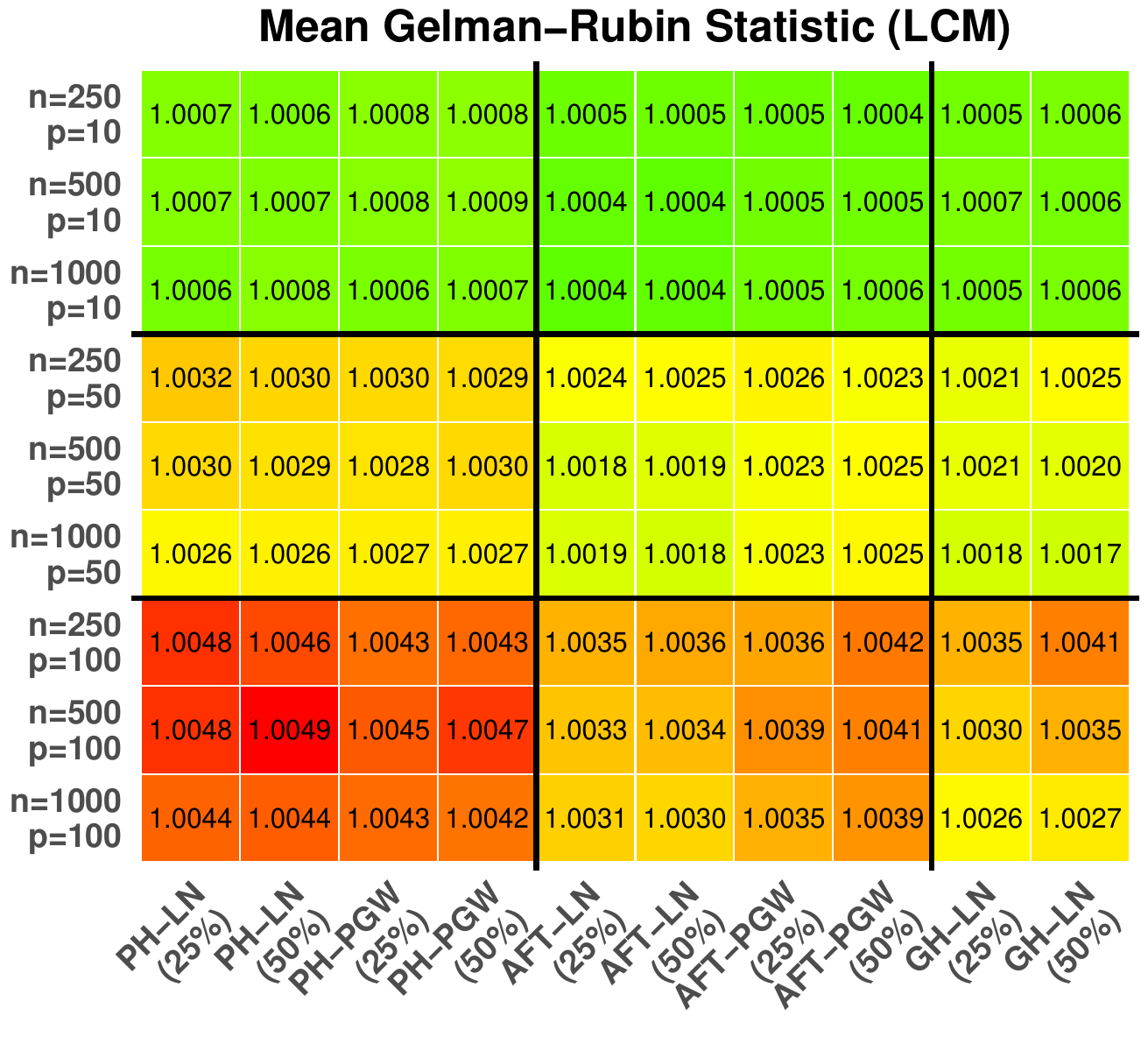}
        \caption{}
    \end{subfigure}
    \hfill
    \begin{subfigure}[b]{0.49\textwidth}
        \centering
        \includegraphics[width=\textwidth]{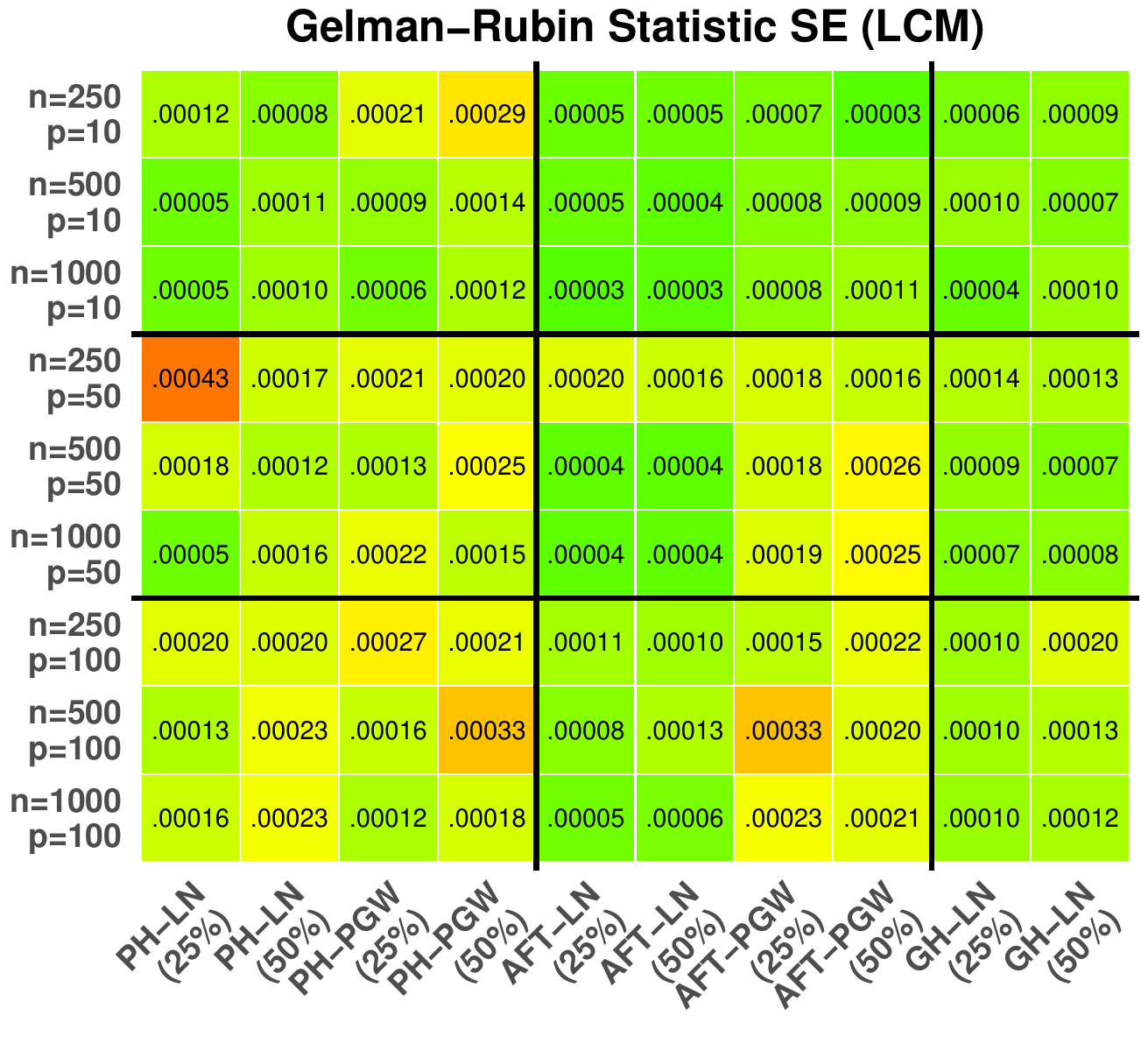}
        \caption{}
    \end{subfigure}
    \vfill
    \begin{subfigure}[b]{0.49\textwidth}
        \centering
        \includegraphics[width=\textwidth]{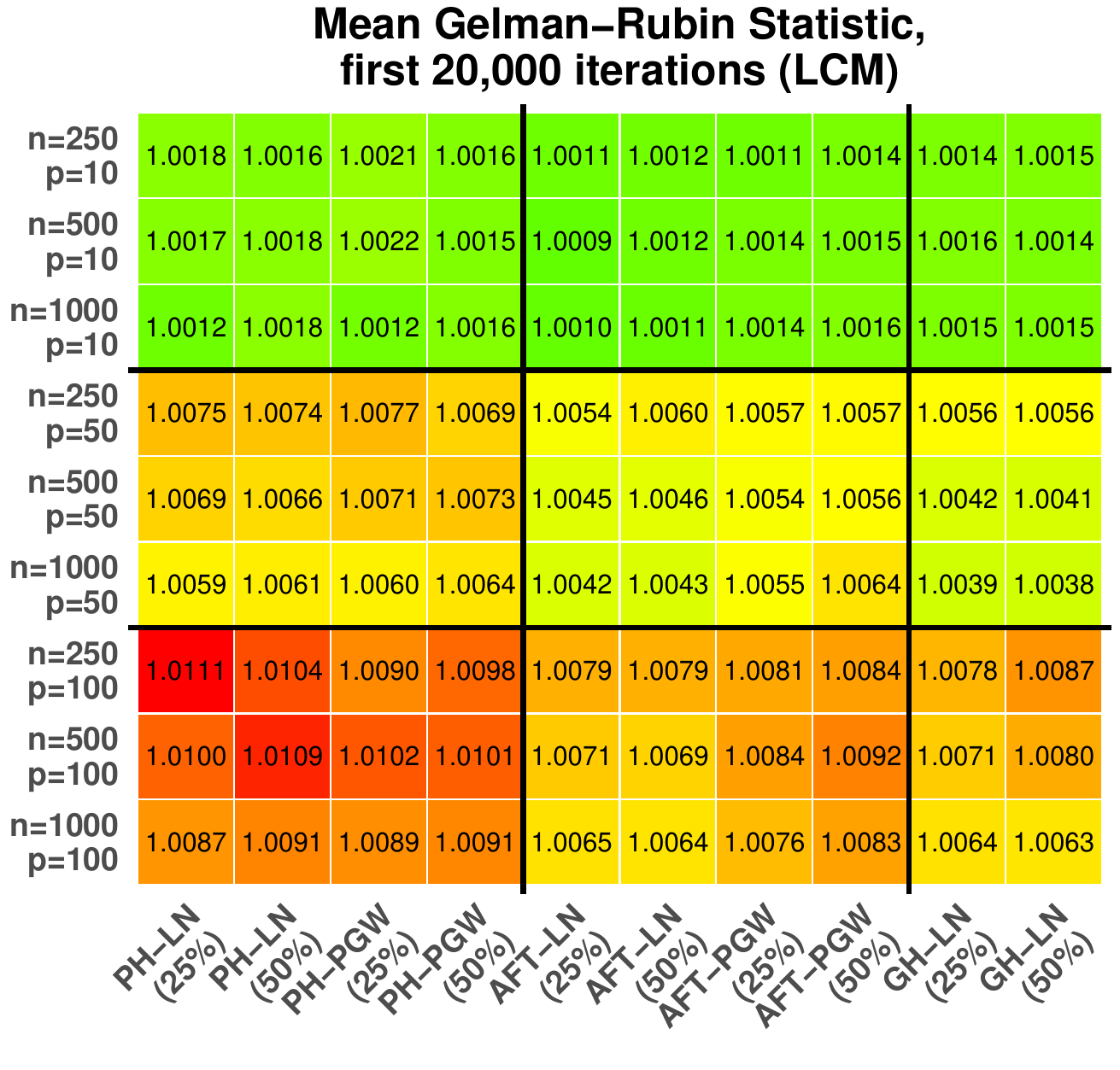}
        \caption{}
    \end{subfigure}
    \hfill
    \begin{subfigure}[b]{0.49\textwidth}
        \centering
        \includegraphics[width=\textwidth]{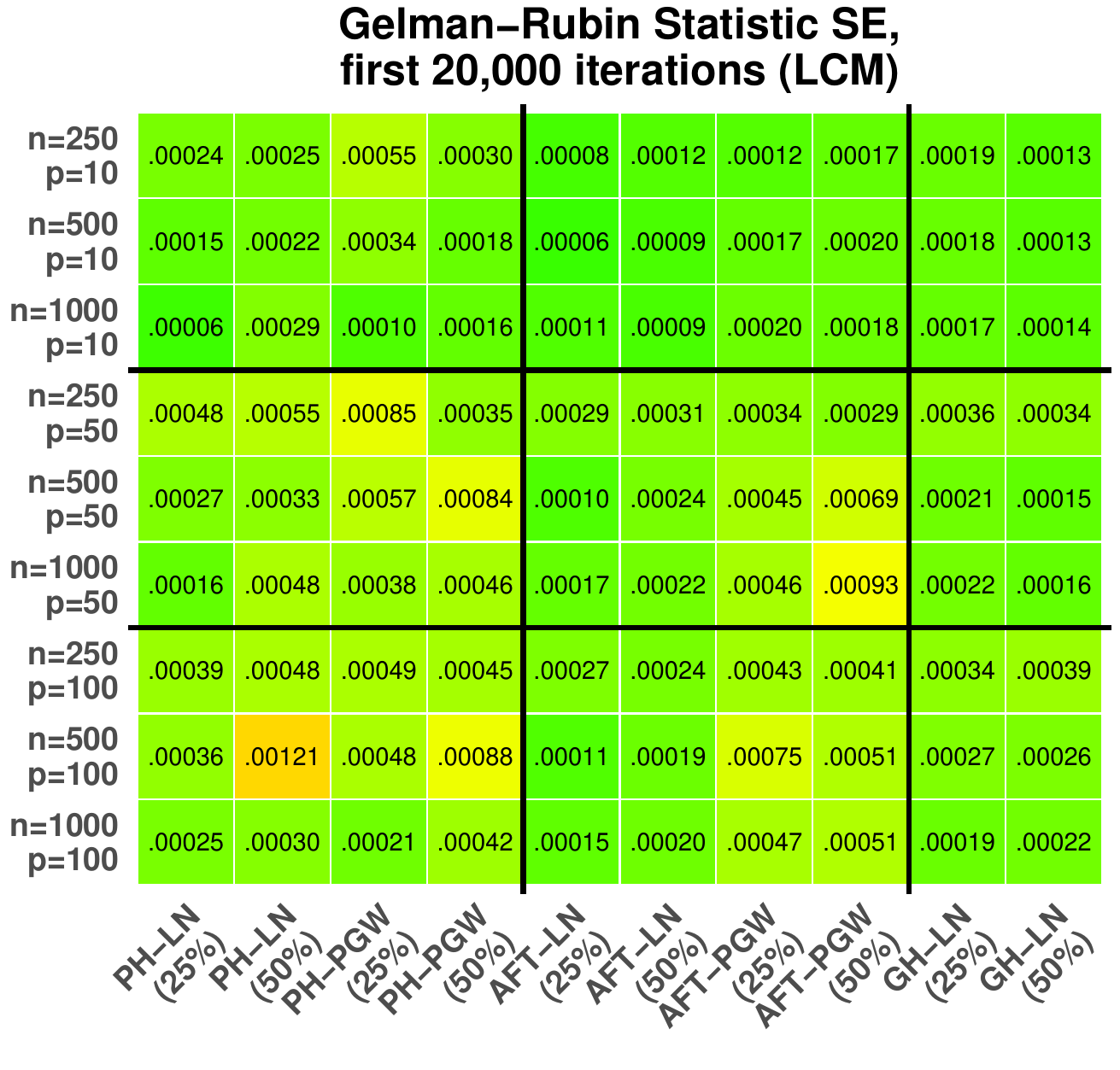}
        \caption{}
    \end{subfigure}
    \caption{Mean Gelman-Rubin statistics and standard errors across $10$ MC runs under LCM prior. Each run consists of 8 chains: two of them are the same initial model using the default strategy mentioned in Section \ref{sec:add_sim}, four are random models with the same model size as the default strategy, one under-dispersed with only one active variable (picked randomly), and one over-dispersed with two times the model size of the default strategy. (a)-(b) Mean and SE of Gelman-Rubin statistics under $50,000$ iterations with $25,000$ burn-in and a thinning period of $2$; (c)-(d) Mean and SE of Gelman-Rubin statistics under $20,000$ iterations with $10,000$ burn-in and a thinning period of $2$.}
    \label{fig:sim_GR}
\end{figure}

\clearpage

\subsection{Results for additional study on influence of hazard selection on variable selection}

\begin{figure}[h]
    \centering
    \begin{subfigure}[b]{\textwidth}
        \centering
        \includegraphics[width=\textwidth]{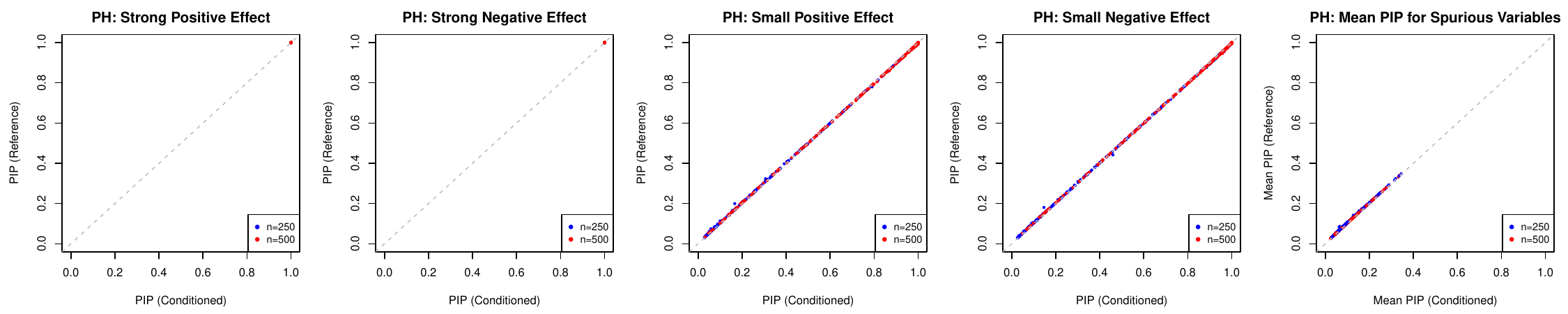}
        \caption{PH models}
    \end{subfigure}
    \vfill
    \begin{subfigure}[b]{\textwidth}
        \centering
        \includegraphics[width=\textwidth]{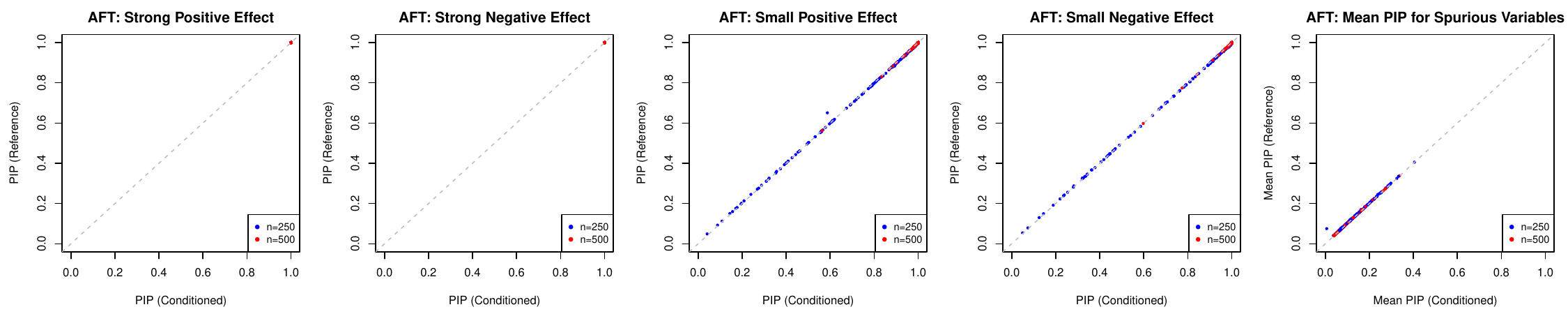}
        \caption{AFT models}
    \end{subfigure}
    \caption{Demonstration of the influence of conducting variable selection with the more complex GH model under different simpler true hazard structures, for $n=250,500$ with LCM prior. The censoring rate is $25\%$ with log-normal baseline hazard. The $x$-axis is the PIP, conditioned on the given hazard structure, from the large-scale simulation, and $y$-axis is the reference PIP from direct fitting of the simpler structures. The first four columns correspond to the four active variables, and the last column is the average PIP of all the spurious variables. }
    \label{fig:sim_comp}
\end{figure}

\begin{figure}[h]
    \centering
    \begin{subfigure}[b]{\textwidth}
        \centering
        \includegraphics[width=\textwidth]{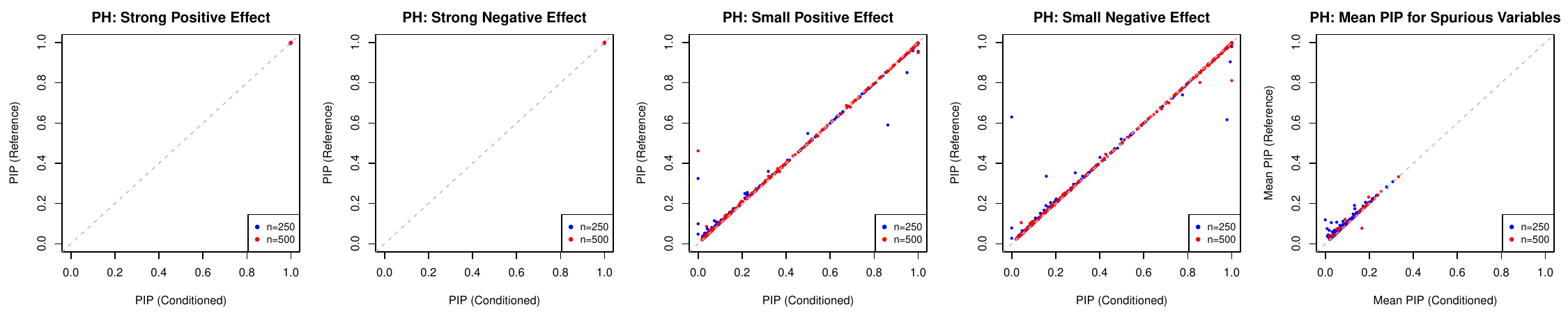}
        \caption{PH models}
    \end{subfigure}
    \vfill
    \begin{subfigure}[b]{\textwidth}
        \centering
        \includegraphics[width=\textwidth]{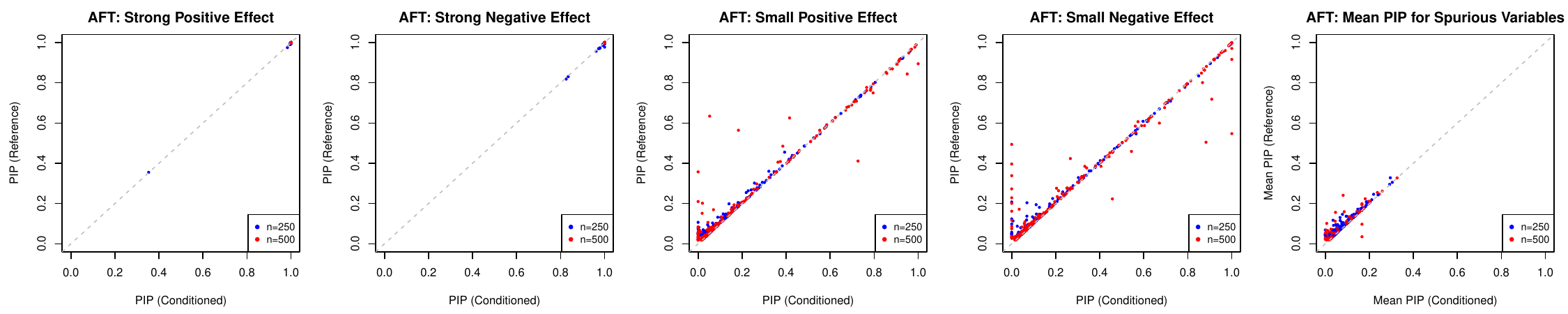}
        \caption{AFT models}
    \end{subfigure}
    \caption{Demonstration of the influence of conducting variable selection with the more complex GH model under different simpler true hazard structures, for $n=250,500$ with LCM prior. The censoring rate is $50\%$ with PGW baseline hazard. The $x$-axis is the PIP, conditioned on the given hazard structure, from the large-scale simulation, and $y$-axis is the reference PIP from direct fitting of the simpler structures. The first four columns correspond to the four active variables, and the last column is the average PIP of all the spurious variables. }
    \label{fig:sim_comp2}
\end{figure}

Here we assess the interplay of performing simultaneous variable and hazard-structure selection. For each configuration with $p = 10$, $n \in {250, 500}$, a censoring rate of $25\%$, a log-normal baseline hazard, and true hazard structures of PH and AFT, we conduct an additional $250$ Monte Carlo simulations. In each simulation, we modify the prior on the model space so that only models corresponding to the true hazard structure receive positive prior probability, effectively preventing all other structures from being visited. We then compare the resulting PIPs with those obtained in the earlier large-scale simulation under the full model space.
This allows us to determine whether variable-selection performance changes when using a more complex hazard structure than the true one. Figure \ref{fig:sim_comp} displays the relationship between the PIPs for each of the four active variables and the mean PIP of the spurious variables under the LCM prior. The points lie closely along the diagonal reference line in all panels, indicating that variable-selection performance is hardly affected by adopting a richer hazard structure when the true hazard structure receives a high posterior probability.

\begin{figure}
    \centering
    \includegraphics[width=\textwidth]{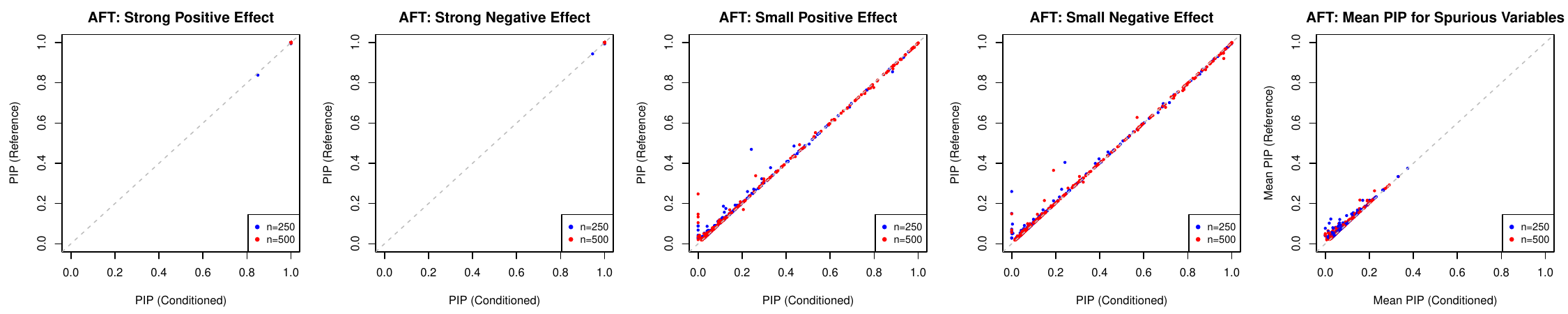}
    \caption{Demonstration of the influence of conducting variable selection with the more complex GH model under true hazard structure being AFT, for $n=250,500$ with LCM prior. The censoring rate is $25\%$ with PGW baseline hazard.}
    \label{fig:sim_comp3}
\end{figure}

We additionally conducted a similar set of simulations with a censoring rate of $50\%$ and a PGW baseline hazard to investigate whether poor hazard-structure selection could influence variable selection. The corresponding results are provided in Figure \ref{fig:sim_comp2}. For PH models, the variable-selection performance again appears stable, with most points remaining tightly clustered around the diagonal. This is because the hazard selection accuracy under this scenario is still very high, around $81\%$ for $n=250$. However, for the AFT model, the true positive rates for the active variables are affected by the substantially poorer hazard-structure selection accuracy,
as can be seen from the decrease in hazard selection accuracy in Figure 2 of the main text, from $0.996$ for AFT-LN ($25\%$) to $0.472$ for AFT-PGW ($50\%$) under $n=250$, $p=10$. Moreover, as suggested by Figure \ref{fig:sim_vio_hstr}, the PH model is often selected instead, as time-level effects are harder to detect for a monotonic baseline with a high censoring rate. However, this effect is limited: in the plots for the small effects, most points remain on the diagonal line, and only in $5\%$ ($n=250$) to $9\%$ ($n=500$) of the replicates do the two PIPs differ by more than $0.05$, mostly with a higher PIP when the model space is restricted to the true structure. The specificity is also hardly affected. In addition, as implied by Figure \ref{fig:sim_comp3}, the difference is a synergy of both higher censoring rate and misspecified baseline hazard. These findings suggest that hazard-structure selection would usually not affect variable selection, but under poor hazard selection due to misspecified baseline hazard and higher censoring (shorter follow-ups), the difference in selection of smaller time-level effects, comparing to the true hazard, may have larger variability.

\subsection{Comparison with BIC-based strategies}

In this section we compare the proposed method under default setting with the following alternative strategies:
\begin{enumerate}
    \item \textit{Frequentist stepwise}: conduct forward selection for each of the four structures individually based on BIC, and select the best one out of the four final models. Run on all scenarios.
    \item \textit{BIC-flat}: replacing $\widehat p(\bt,\bgamma)$ with BIC, and derive posterior samples using the proposed MCMC algorithm. Run on $p=10$, $n\in\{250,500,1000\}$.
    \item \textit{BIC-$\gamma$}: replacing only the approximated marginal likelihood $\widehat p(\bt\mid \bgamma)$ with BIC, and derive posterior samples using the proposed MCMC algorithm. Run on $p\in\{10,50\}$, $n\in\{250,500,1000\}$.
    \item \textit{LCM-unif}: replacing only the Beta-binomial$(1,1)$ prior for $\Lambda$, the binary vector of variable inclusion, with a uniform prior. The rest remains the same as the default strategy in the main text. Run on $p\in\{10,50\}$, $n\in\{250,500,1000\}$.
\end{enumerate}
The frequentist approach serves as a baseline. The middle two approaches provide insights to the effect of the proposed $\pi(\bgamma)$ and the extra terms, comparing to BIC, in the approximated marginal likelihood. Lastly, the LCM-unif comparison is used to assess how the proposed methodology performs under different specification of $\pi(\Lambda)$. While there are differences in the scenarios being run for the four strategies due to computational budget, they are adequate to convey the most important messages.

\begin{figure}[h]
    \centering
    \begin{subfigure}[b]{0.32\textwidth}
        \centering
        \includegraphics[width=\textwidth]{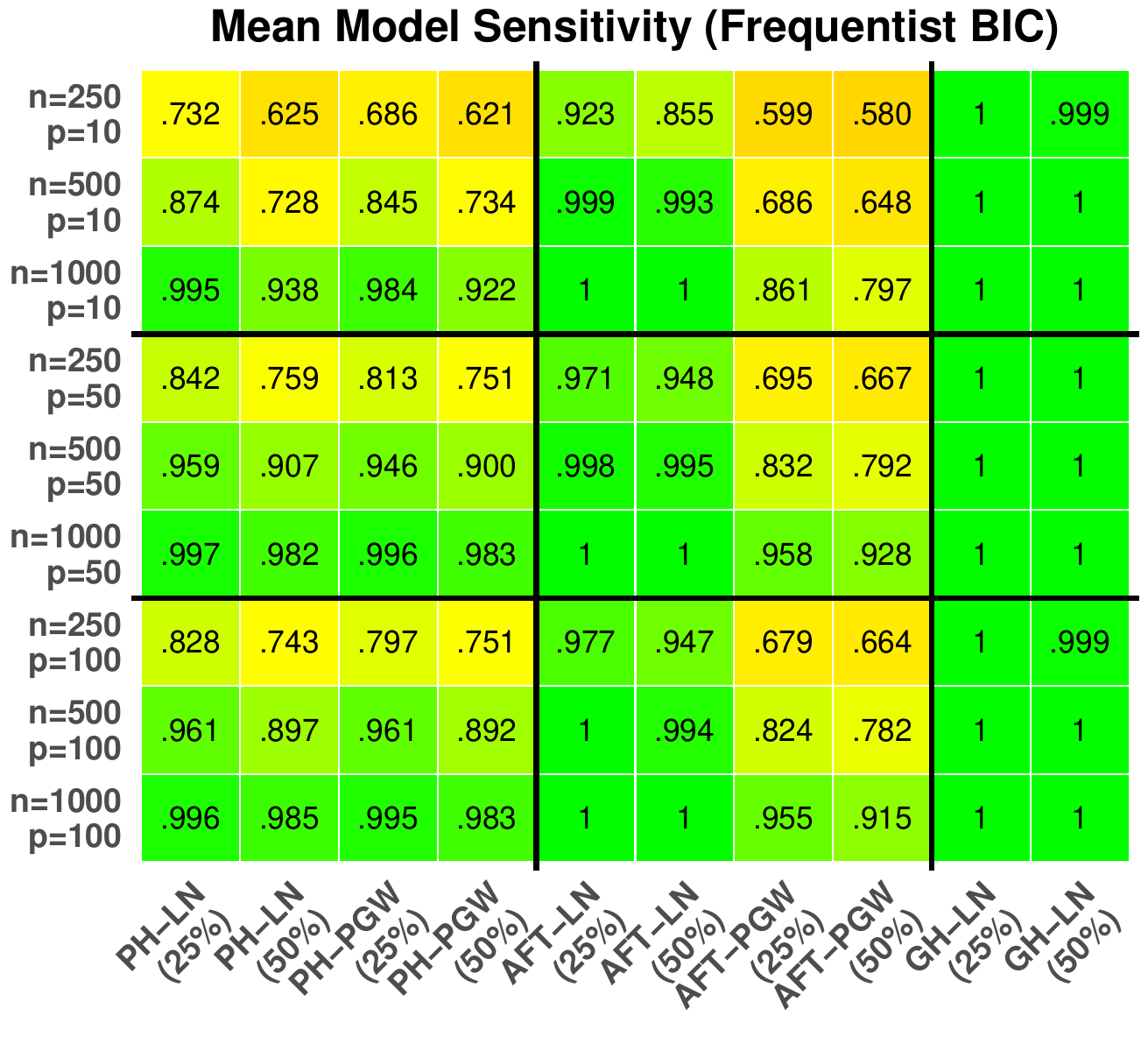}
        \caption{Frequentist approach}
    \end{subfigure}
    \hfill
    \begin{subfigure}[b]{0.32\textwidth}
        \centering
        \includegraphics[width=\textwidth]{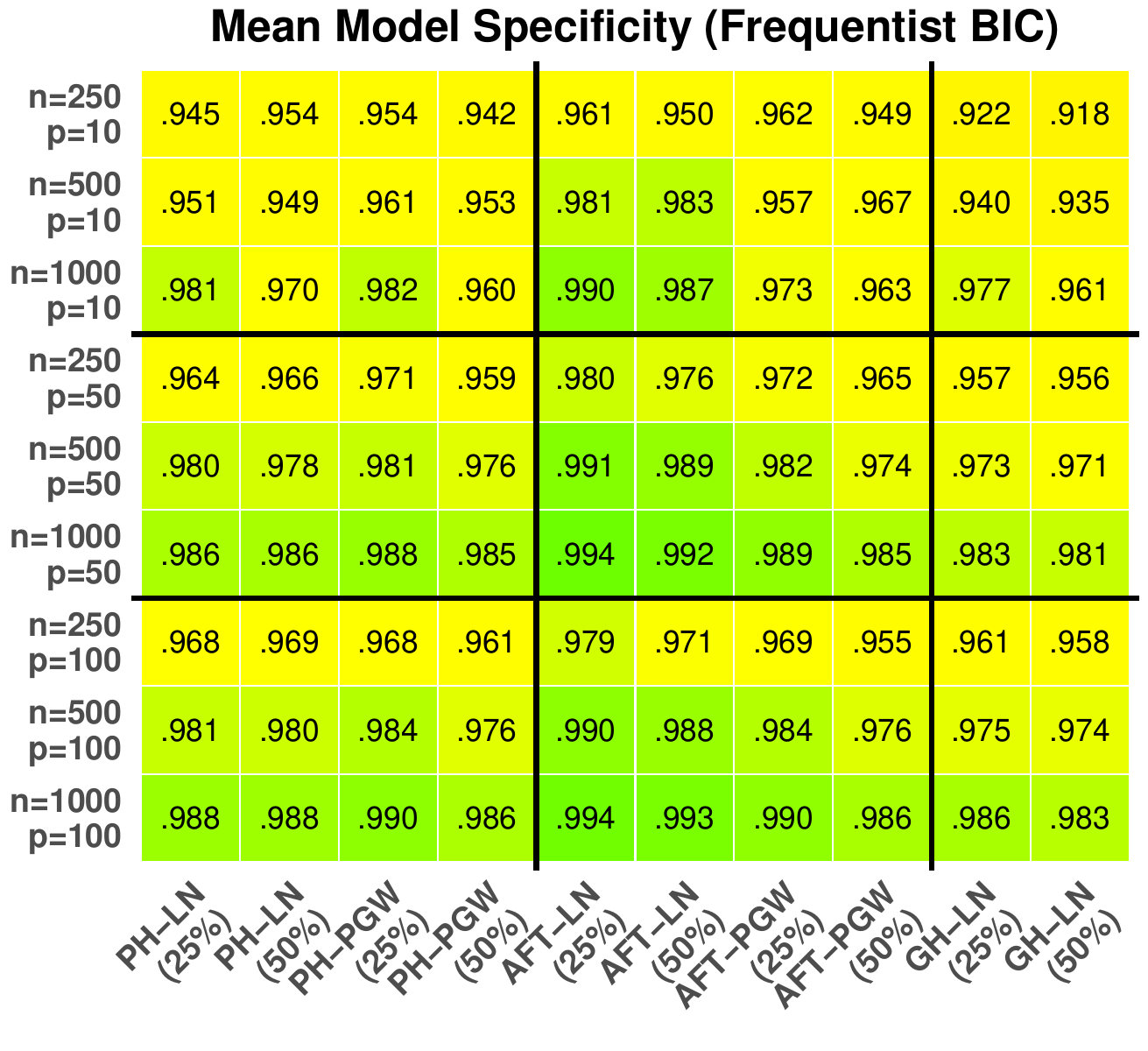}
        \caption{Frequentist approach}
    \end{subfigure}
    \hfill
    \begin{subfigure}[b]{0.32\textwidth}
        \centering
        \includegraphics[width=\textwidth]{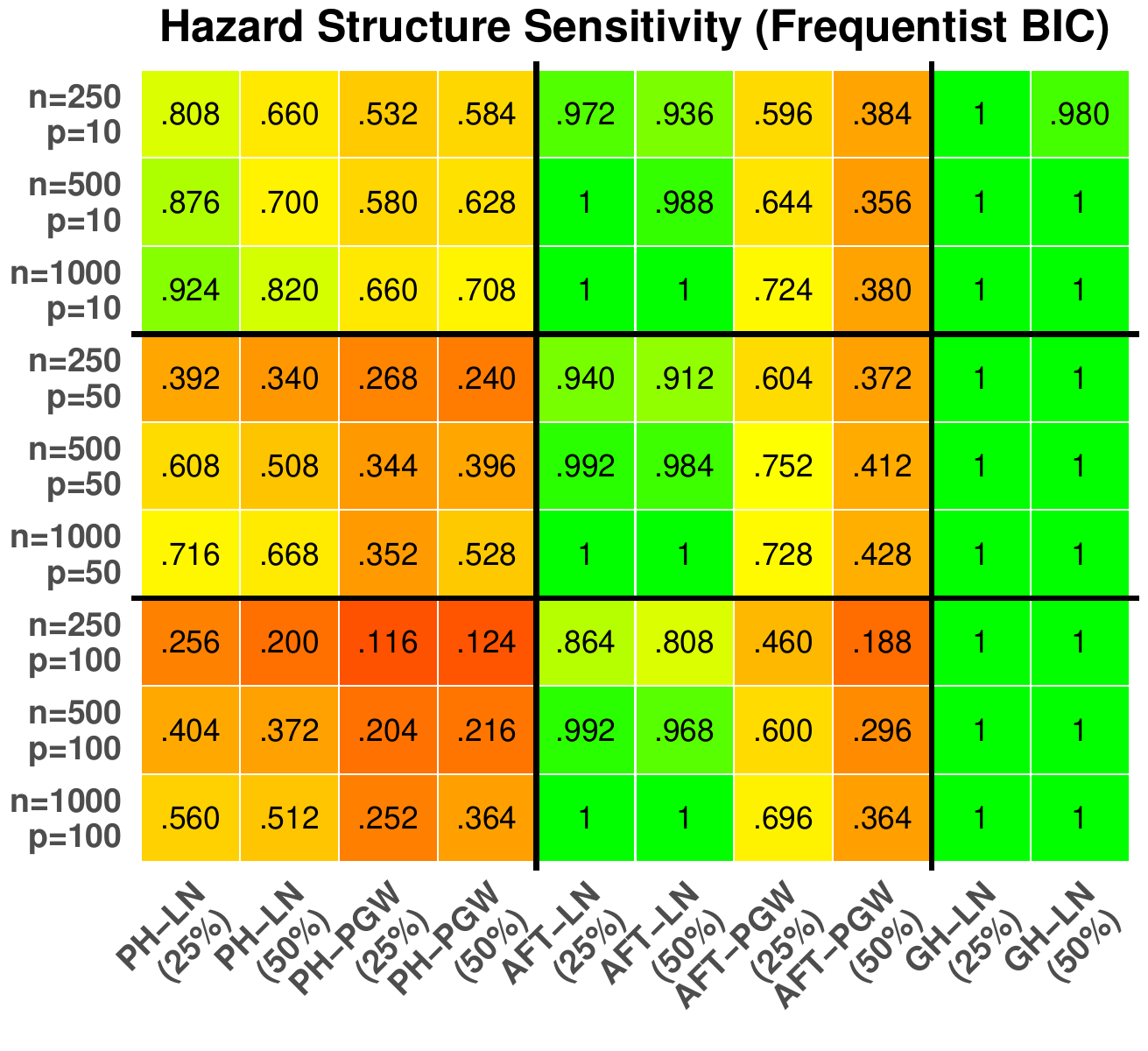}
        \caption{Frequentist approach}
    \end{subfigure}
    \vfill
    \begin{subfigure}[b]{0.32\textwidth}
        \centering
        \includegraphics[width=\textwidth]{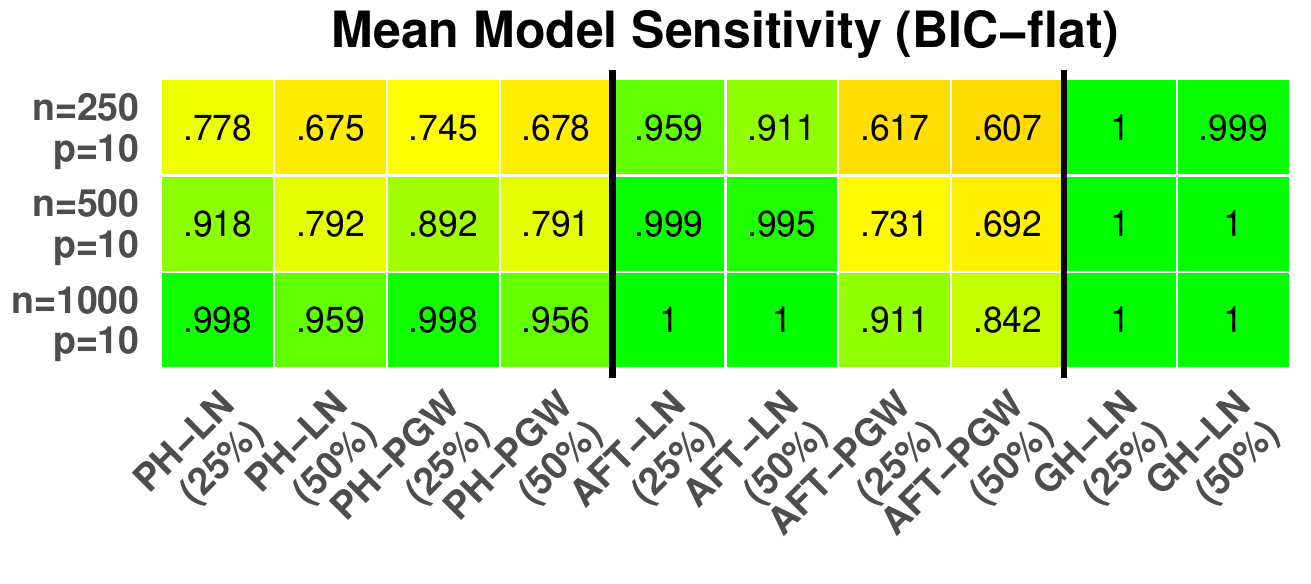}
        \caption{BIC-flat}
    \end{subfigure}
    \hfill
    \begin{subfigure}[b]{0.32\textwidth}
        \centering
        \includegraphics[width=\textwidth]{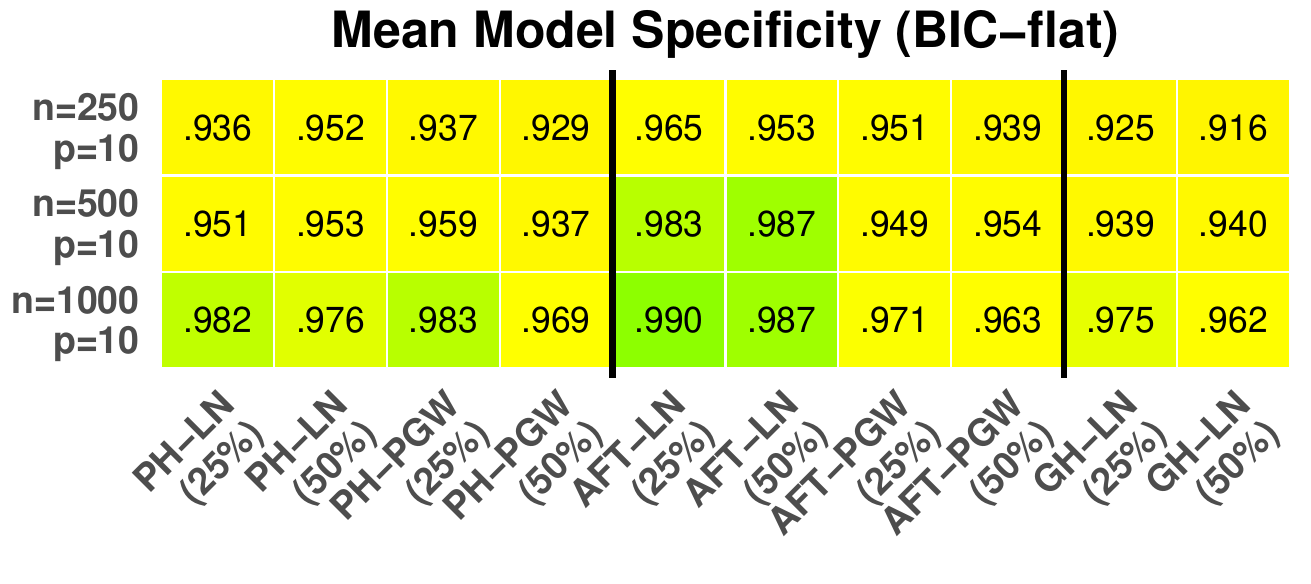}
        \caption{BIC-flat}
    \end{subfigure}
    \hfill
    \begin{subfigure}[b]{0.32\textwidth}
        \centering
        \includegraphics[width=\textwidth]{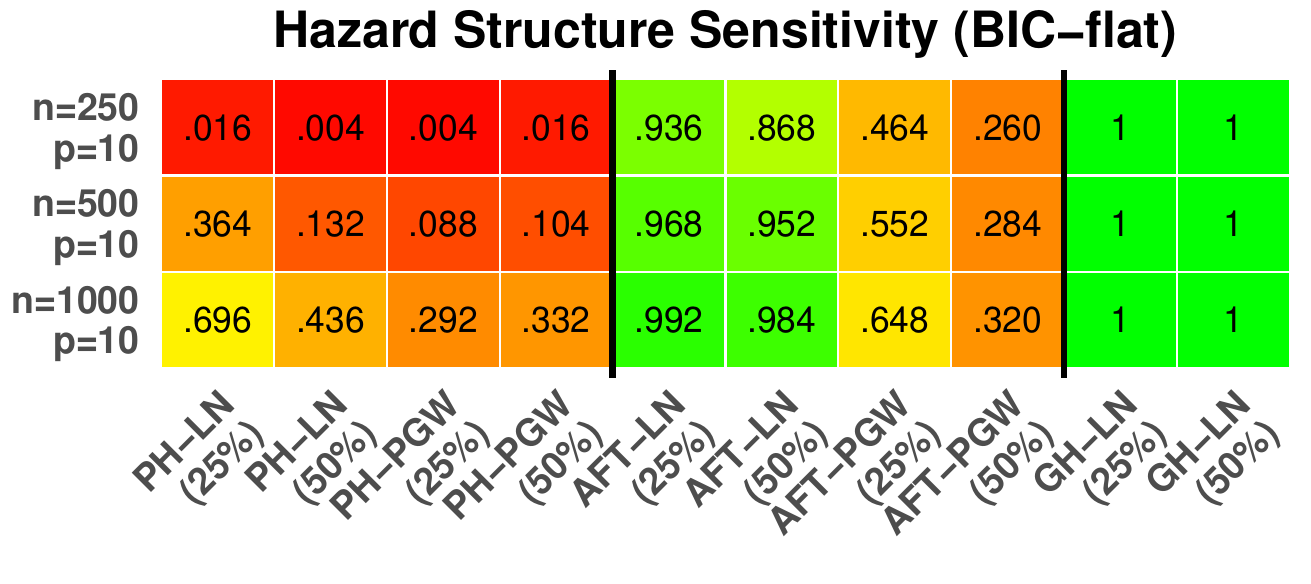}
        \caption{BIC-flat}
    \end{subfigure}
    \vfill
    \begin{subfigure}[b]{0.32\textwidth}
        \centering
        \includegraphics[width=\textwidth]{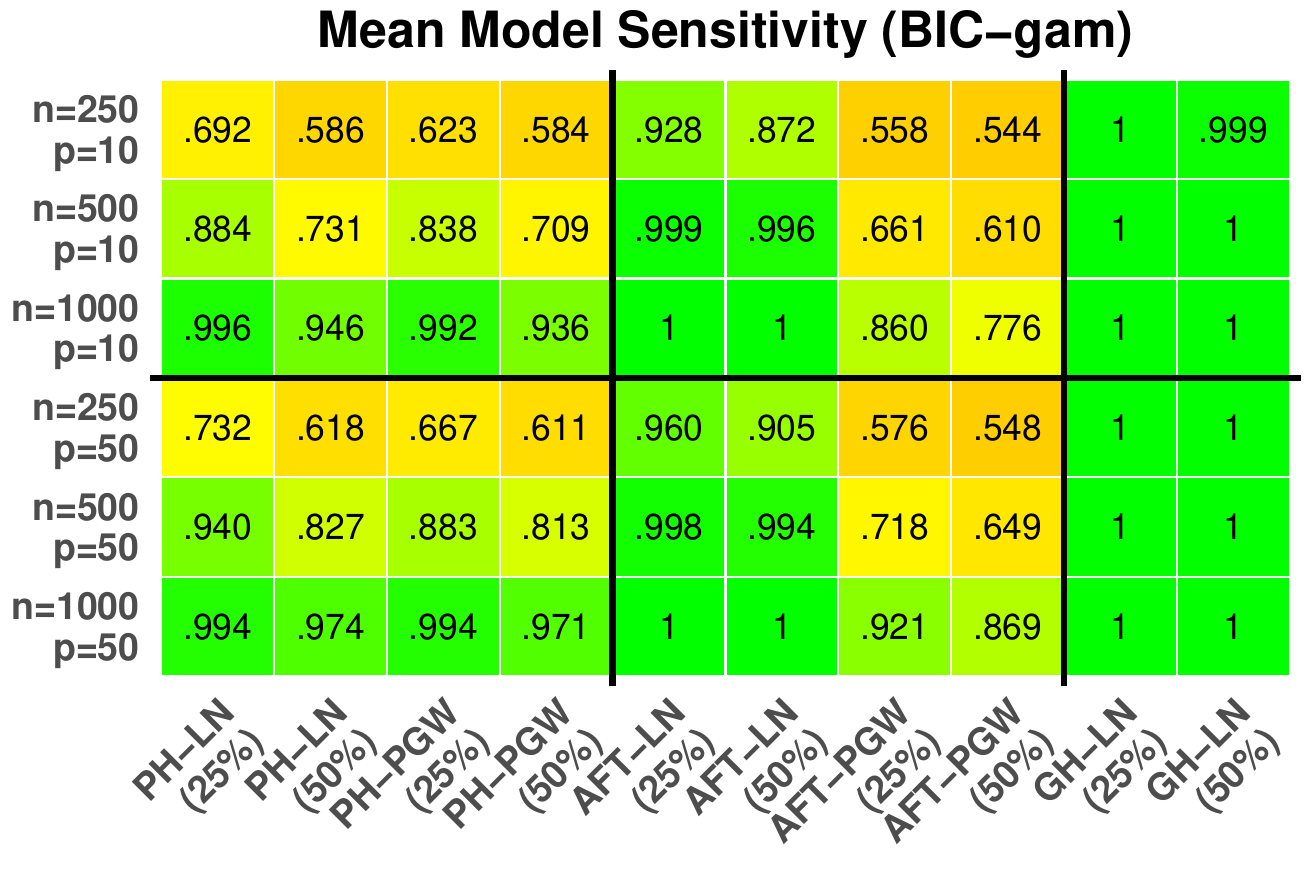}
        \caption{BIC-$\gamma$}
    \end{subfigure}
    \hfill
    \begin{subfigure}[b]{0.32\textwidth}
        \centering
        \includegraphics[width=\textwidth]{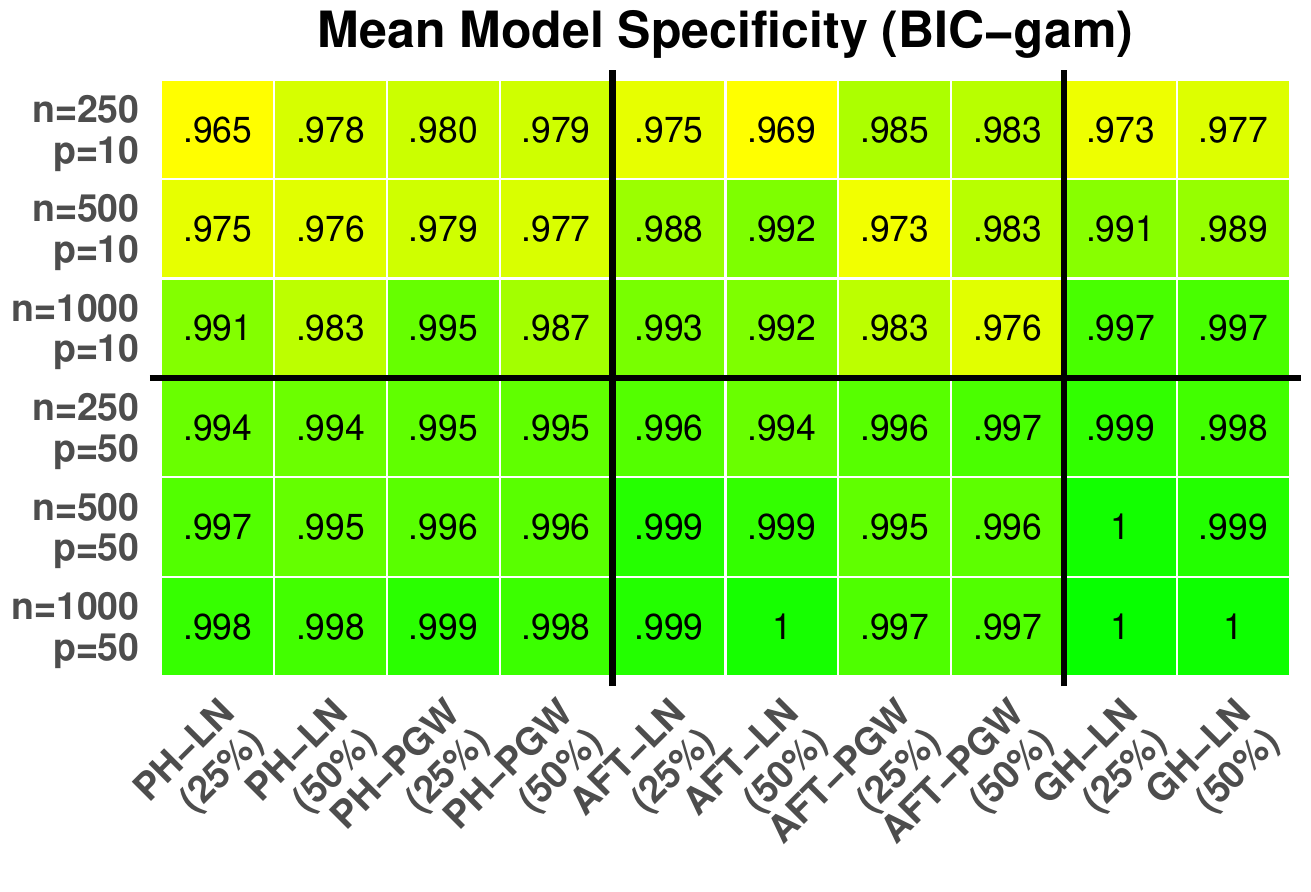}
        \caption{BIC-$\gamma$}
    \end{subfigure}
    \hfill
    \begin{subfigure}[b]{0.32\textwidth}
        \centering
        \includegraphics[width=\textwidth]{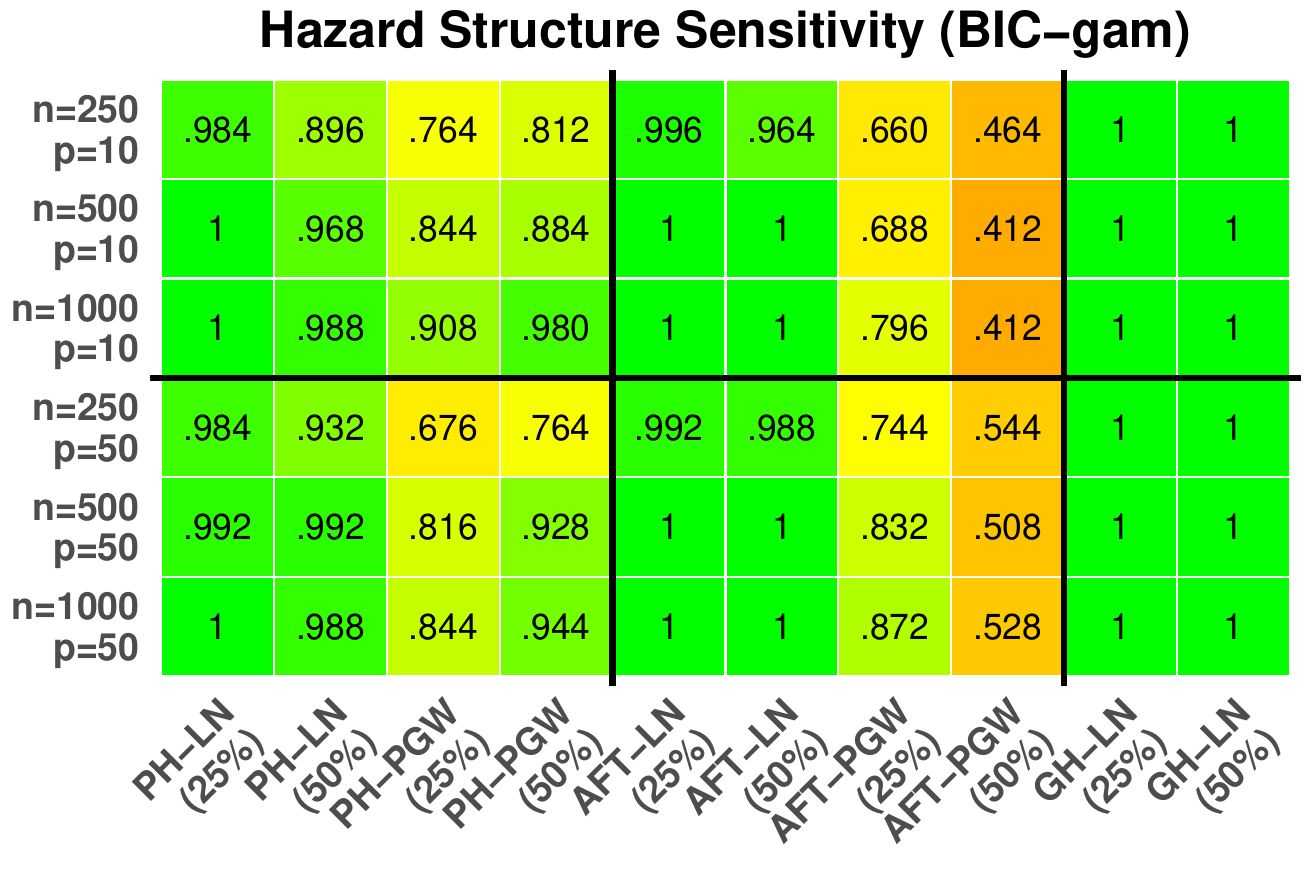}
        \caption{BIC-$\gamma$}
    \end{subfigure}
    \vfill
        \begin{subfigure}[b]{0.32\textwidth}
        \centering
        \includegraphics[width=\textwidth]{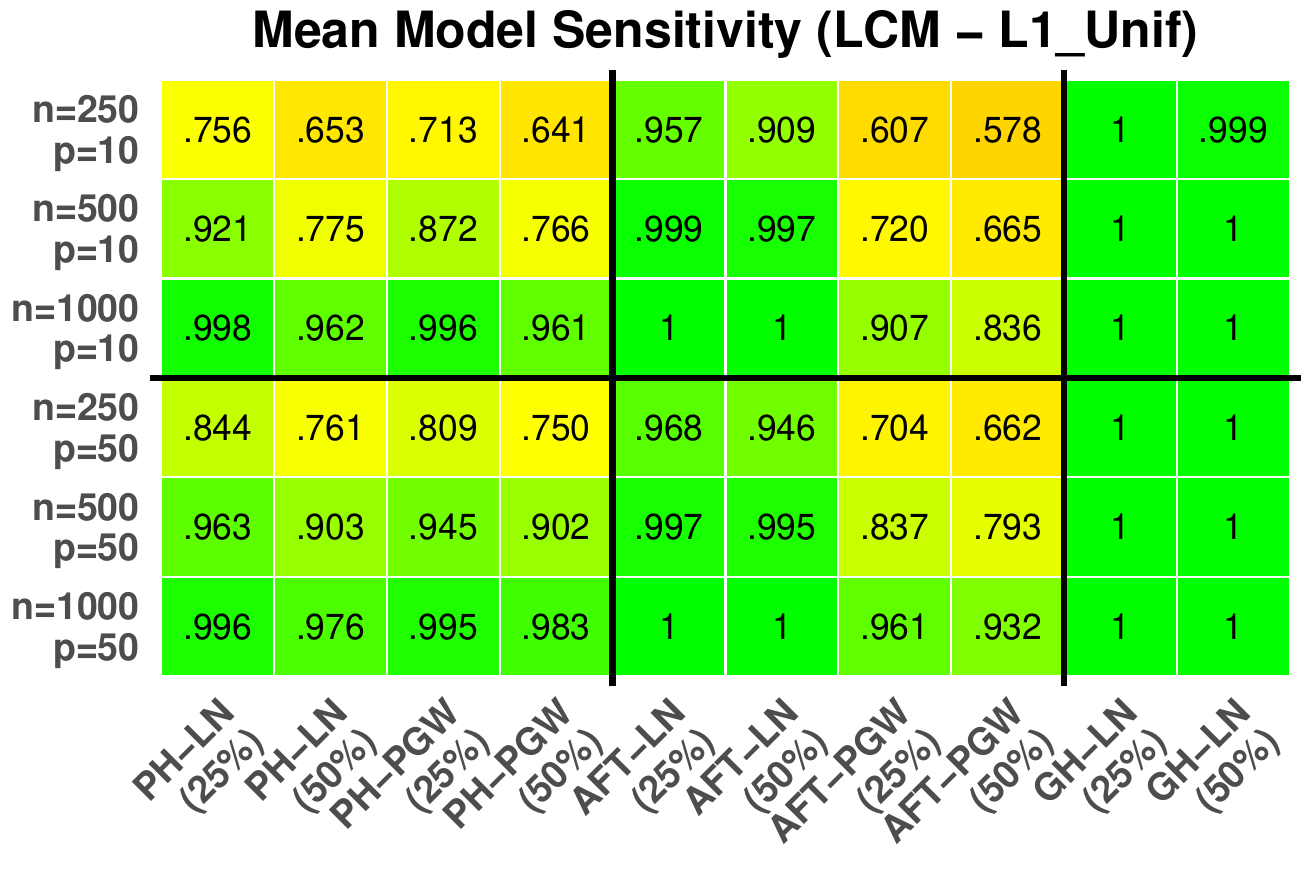}
        \caption{LCM-unif}
    \end{subfigure}
    \hfill
    \begin{subfigure}[b]{0.32\textwidth}
        \centering
        \includegraphics[width=\textwidth]{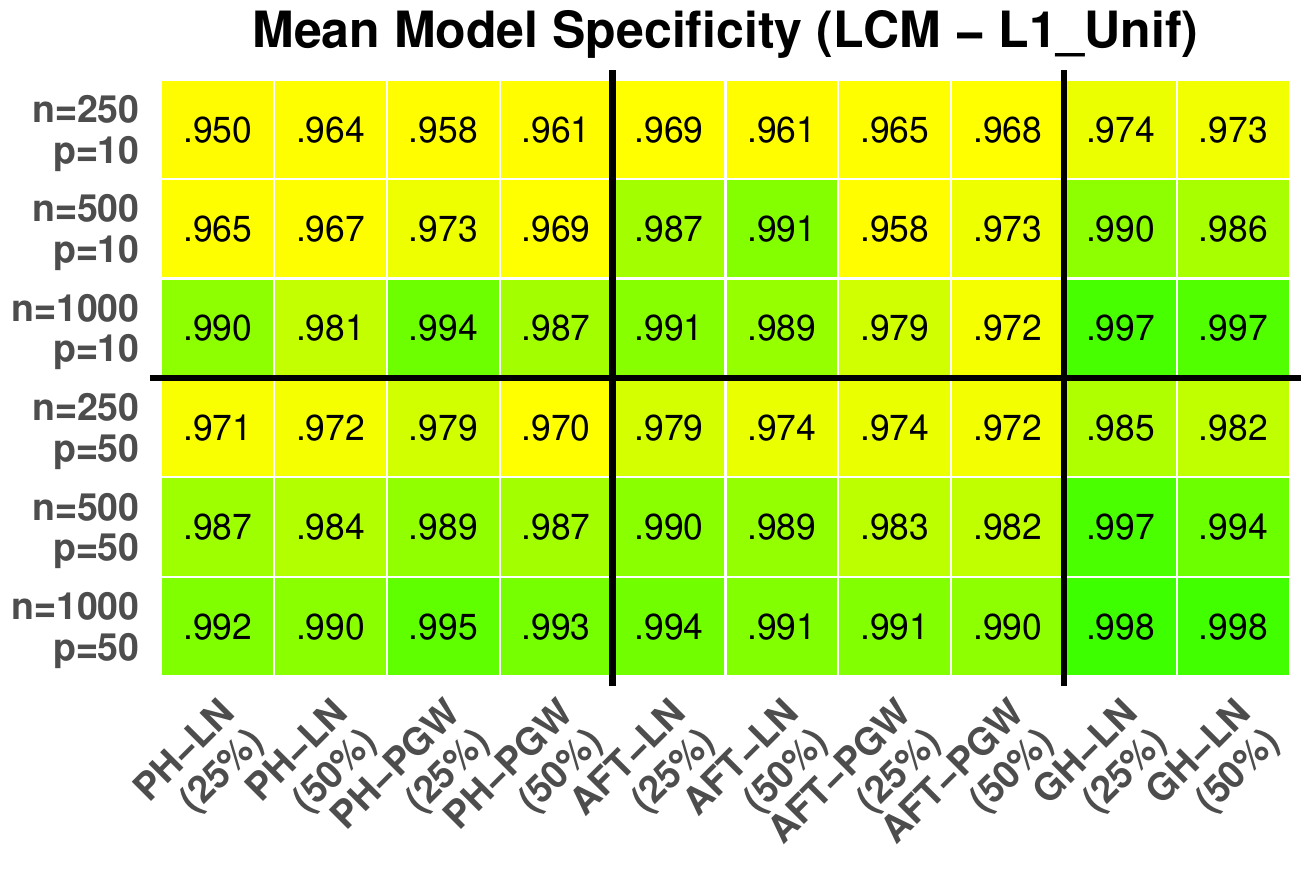}
        \caption{LCM-unif}
    \end{subfigure}
    \hfill
    \begin{subfigure}[b]{0.32\textwidth}
        \centering
        \includegraphics[width=\textwidth]{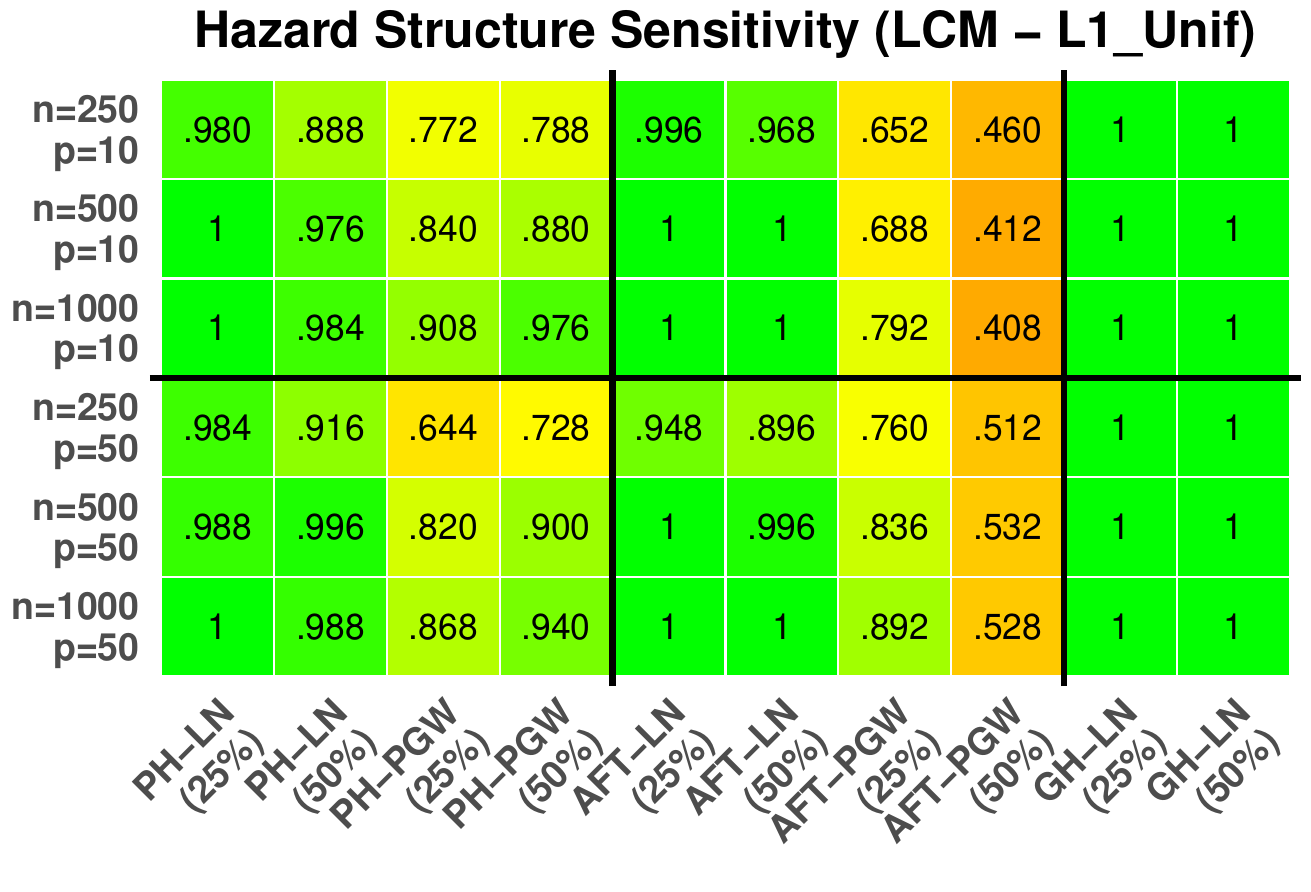}
        \caption{LCM-unif}
    \end{subfigure}
    \caption{Results for the comparison study. The MCMC settings for (d)-(l) are the same as the simulation study described in Section 10.}
    \label{fig:alt_sim}
\end{figure}

Figure~\ref{fig:alt_sim} shows the results of the comparison. Comparing the frequentist approach with the results in the main text, the hazard structure selection accuracy of the proposed approach is significantly higher. The frequentist approach often mistakenly selects GH while the true structure is PH --- this is because the former fails to account for the multiplicity difference between GH and other model types, whereas the proposed $\pi(\bgamma)$ does. On the other hand, although the specificity of the variable selection is higher for the proposed method, the sensitivity is lower in many scenarios. However, this is not a failure of the proposed methodology, instead it is directly related to the specification of $\pi(\mathbf{\Lambda})$.
The comparison between (i) LCM-unif with the frequentist approach and between (ii) LCM-unif with the manuscript's result sheds more light on this. From (i) we can find that the specificity of variable selection is higher for LCM-unif (in $55$ of the $60$ scenarios), its sensitivity is higher for $p=10$ and similar for $p=50$, and its hazard structure selection accuracy is significantly higher. From (ii) we see that while the hazard structure selection accuracy is almost identical, LCM-unif has higher sensitivity but lower specificity than the manuscript's result, which uses $\Lambda \sim \text{BetaBin}(1,1)$.
The direct reason is that, the beta-binomial's dimension penalty happens to be similar to the difference in likelihood between the true model and the model without one or both of the smaller effects. The comparison between BIC-flat and BIC-$\gamma$ also confirms this: while the former has higher sensitivity, the latter has higher specificity and much higher hazard selection accuracy.
Therefore, the phenomenon that the proposed approach behaves very differently under different choice of $\pi(\mathbf{\Lambda})$ illustrates a highlight of the proposed structure of $\pi(\bgamma)$: by choosing different $\pi(\mathbf{\Lambda})$, one is essentially balancing the trade-off between variable selection sensitivity and specificity, \emph{independent} of the accuracy of hazard structure selection, and the remaining two parts of $\pi(\bgamma)$ largely improve the hazard selection accuracy upon the baseline frequentist approach.

Next, comparing BIC-$\gamma$ with the results in the main text, one can see that the difference in selection performance is very small: the mean sensitivity, specificity and hazard selection accuracy of the two agree to within $0.002$. Therefore, this comparison suggests that, under the default hyperparameter choice of $g_E=1$, the proposed methodology's selection performance is consistent with BIC, under the same model space prior. This similarity is expected, since, for $g_E=1$, the dimension penalty $(1+n)^{-d_{\bgamma}/2}$ of the approximated marginal likelihood matches that of BIC to first order (Proposition 2). The expression of the approximated marginal likelihood under LCM prior with $g_E=1$ is approximately BIC plus several additional penalty terms on how surprising the MLE is with respect to the prior. However, under the log-normal baseline and vague common priors, these additional terms do not contribute as much as the likelihood. With a different value or a hyper prior on $g_E$, or a more complex baseline distribution, the difference between the proposed approximation and BIC is expected to be larger.

\color{black}

\clearpage

\section{Variable dictionary for real application}

\begin{table}[h]
\centering
\caption{Variables from the flchain (FLC) data set in \texttt{R} package \texttt{survival}, adopted from \cite{survival-package}.}
\label{tab:flchain}
\begin{tabular}{@{}lp{0.65\textwidth}@{}}
\toprule
\textbf{Variable} & \textbf{Description} \\
\midrule
age & Age in years \\
\addlinespace
sex & Sex of the subject (F = female, M = male) \\
\addlinespace
kappa & Serum free light chain, kappa portion \\
\addlinespace
lambda & Serum free light chain, lambda portion \\
\addlinespace
creatinine & Serum creatinine level \\
\addlinespace
mgus & Indicator for monoclonal gammopathy diagnosis (1 = diagnosed with MGUS, 0 = not diagnosed) \\
\addlinespace
futime & Days from enrollment until death (note: 3 subjects had sample obtained on death date) \\
\addlinespace
death & Vital status (0 = alive at last contact, 1 = dead)\\
\bottomrule
\end{tabular}
\end{table}

\begin{table}[h]
\centering
\caption{Variables from the nki70 data set in \texttt{R} package \texttt{penalized}, adopted from \cite{penalized-package}. The details of the $70$ genes are omitted.}
\label{tab:nki70}
\begin{tabular}{@{}lp{0.65\textwidth}@{}}
\toprule
\textbf{Variable} & \textbf{Description} \\
\midrule
Diam & Diameter of the tumor (two levels) \\
\addlinespace
N & Number of affected lymph nodes (two levels) \\
\addlinespace
ER & Estrogen receptor status (two levels) \\
\addlinespace
Grade & Grade of the tumor (three ordered levels) \\
\addlinespace
Age & Patient age at diagnosis (years) \\
\addlinespace
TSPYL5--DKFZP586A0522 & Gene expression measurements of 70 prognostic genes (70 gene columns) \\
\bottomrule
\end{tabular}
\end{table}

\clearpage

\section{MCMC detail for study of (\texttt{NKI70})}

For each of the two proposed priors, we run $10$ chains of $110{,}000$ iterations with $10{,}000$ burn-in, and renormalise the model posterior probabilities over all models evaluated by the $10$ chains (Section 9). Across chains, the estimated posterior probabilities of the ten most probable models differ by at most $0.001$ under the LCM prior and $0.011$ under the Product prior. Running multiple long chains ensures adequate exploration of high-probability models and reduces the variability of the estimated model posterior probabilities.

For the AFT model with the Zellner-type prior, we run $10$ chains of $110{,}000$ Gibbs sweeps (each updating all $75$ inclusion indicators) with $10{,}000$ burn-in, and renormalise over all models visited after burn-in; across chains, the probabilities of the ten most probable models differ by less than $0.001$. For the non-local piMOM prior partial-Cox model, \texttt{BVSNLP} implements a variant of the Shotgun Stochastic Search (SSS) algorithm called S5, initially proposed by \cite{S5:2018}.
We perform $10$ independent runs, each with $11{,}000$ iterations for each of the $10$ temperatures ($110{,}000$ in total), and renormalise over the $1{,}000$ most probable models returned by each run.

\section{Experiment on polarised censoring time}

One particular scenario where the assumption of equal contribution of all samples to the prior covariance could be problematic is when the censoring times are concentrated at the extremes of the follow-up period, as discussed in \cite{castellanos:2021}, where this situation is referred to as ``polarised censoring time''. They considered a toy data set to demonstrate the bias induced by the regular $g$-prior under polarised censoring times in their Table 1. We generate the data set following the same procedure as in their Section 3, and conduct variable selection using both of LCM and Product priors. We restrict the model space within the AFT models and run the chain for $20{,}000$ iterations with $10{,}000$ burn-in's. Table \ref{tab:motive-ex} shows the PIP for the three variables under the two priors. We could observe a big difference between PIP of the two priors. Comparing this result with Table 2 of \cite{castellanos:2021}, we could observe that the variable selection performance of the LCM prior is not impacted by the polarised censoring time, since the observed Fisher information matrix automatically adjust the contribution of these abnormal observations to the prior covariance matrix. On the other hand, Product prior gives equal weights to all observations, which leads to underestimation of the prior covariance and therefore a poorer variable selection performance.

\begin{table}[ht]
\centering
\caption{PIP for the polarised censoring time experiment.}
\label{tab:motive-ex}
\begin{tabular}{@{}cccc@{}}
\toprule
   Method    & X1    & X2    & X3    \\ \midrule
AFT-LCM & 0.994 & 0.997 & 0.424 \\
AFT-Product & 0.611 & 0.704 & 0.437 \\ \bottomrule
\end{tabular}
\end{table}

\section{Strategies for tuning hyperparameter for the priors on coefficients}

The choice of hyperparameters, $g_E$ for the LCM prior and $(g_{C_t},g_{C_h})$ for the product prior, is critical to selection performance. As mentioned in the main text, the default strategy we use is setting them to $1$. When $n \to \infty$ but the likelihood vanishes (\textit{e.g.} censoring rate $\to 1$), the Bayes factor between an active model and the null model, say for the LCM prior, is $(1+ng_E)^{-d_\gamma/2} \to 0$. In other words, this choice encodes a mild preference for the null model that strengthens with the nominal sample size.
\cite{castellanos:2021} considers an alternative strategy: they replace $n$ with the effective sample size $n_{eff}$, which is the expected Fisher information for the intercept ($\theta_0$ in the current context) under the null model, under the assumption that the censoring times are fixed and known. Concretely, under the lognormal baseline,
\begin{equation}\label{eq:ess}
n_{eff}=\sum_{i=1}^n w\left(z_i\right),\quad w(z)=\Phi(z)+\phi(z)\{h(z \mid \hat \nu, \hat \theta_0)-z\}, \quad z_i=e^\nu \log c_i-\theta_0.
\end{equation}
One can view this as specifying the hyperparameters as $n_{eff}/n$, and it coincides with the previous strategy when there is no censoring.
They show that this choice satisfies the predictive matching criteria \citep{Bayarri2012CriteriaSelection}: when the information vanishes, the BF between any active model and the null model goes to $1$, \textit{i.e.} no decision on variable inclusion should be made.

\begin{table}[h]
\centering
\caption{Value of $n_{eff}/n$ for different censoring levels.}
\label{tab:cevs_ess}
\begin{tabular}{@{}cccccc@{}}
\toprule
Censoring Level & 0.818 & 0.924 & 0.971 & 0.989 & 0.996 \\ \midrule
$n_{eff}/n$     & 0.506 & 0.300 & 0.158 & 0.078 & 0.037 \\ \bottomrule
\end{tabular}%
\end{table}

\begin{figure}[h]
    \centering
    \begin{subfigure}[b]{\textwidth}
        \centering
        \includegraphics[width=0.85\textwidth]{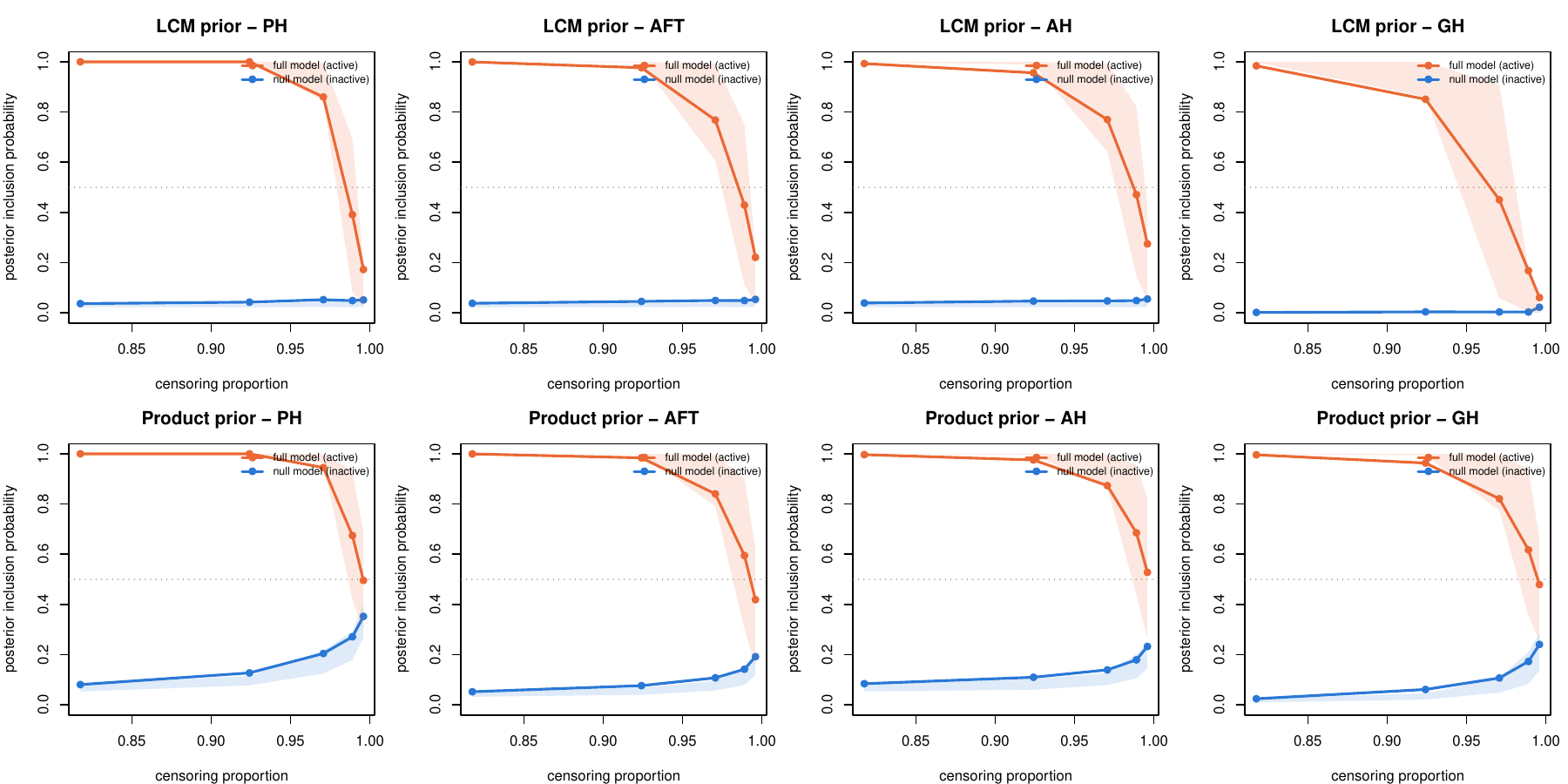}
        \caption{The default setup with $(g_E,g_{C_t},g_{C_h})=(1,1,1)$}.
    \end{subfigure}
    \vfill
    \begin{subfigure}[b]{\textwidth}
        \centering
        \includegraphics[width=0.85\textwidth]{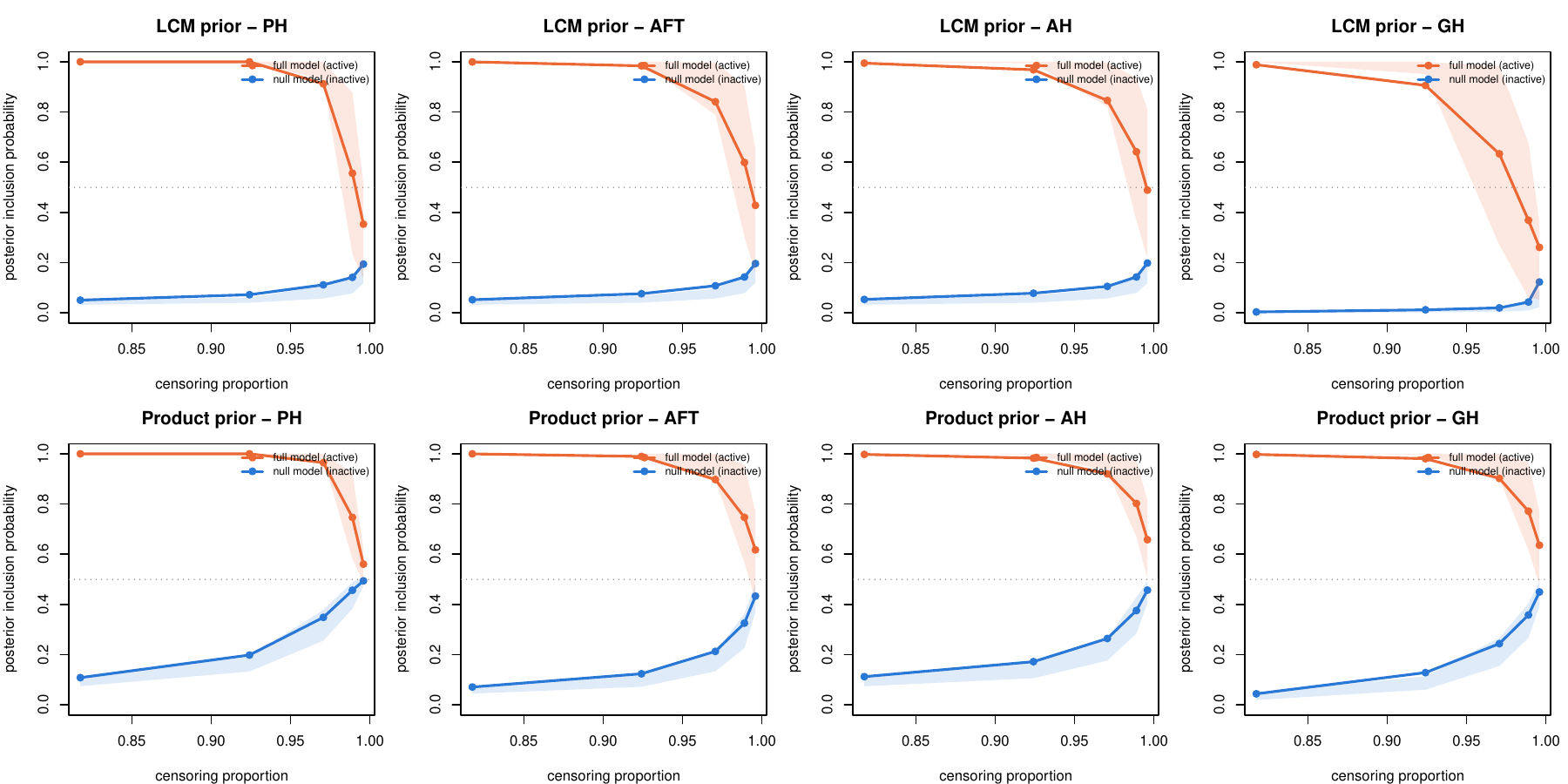}
        \caption{Applying effective sample size correction in \eqref{eq:ess}.}
    \end{subfigure}
    \caption{Average PIP and its middle $50\%$ band of the full model and null model under LCM and product prior over $100$ Monte Carlo replicates, with (a) the default set of hyperparameter and (b) the effective sample size correction.}
    \label{fig:ess}
\end{figure}

When the information contained in the likelihood function is not too weak compared to the prior (\textit{i.e.} when the censoring rate is not extremely high), the difference in selection performance between the two strategies is negligible, as the likelihood and the model space prior will dominate. To illustrate the difference between the two tuning strategies, we conduct a simple simulation under extreme censoring rates and weak signal.
The setup is as follows: a single covariate is considered for each of the four hazard structures, with the baseline of $\text{LogNormal}(1.55,0.7)$ (same as the large-scale simulation). Five administrative censoring levels are considered, $(0.818, 0.924, 0.971, 0.989, 0.996)$, with sample size of $n=2000$ to enable the estimation of all parameters. The corresponding value of $n_{eff}/n$ is presented in Table~\ref{tab:cevs_ess}, where the MLE of the null model is used to evaluate the expected information.
The coefficients for PH, AFT, AH, and GH are $0.5$, $0.142$, $0.209$, and $(\alpha,\beta)=(0.168, 0.084)$. They produce weak enough likelihood and are calibrated such that the four structures have the same cut-off to produce the $0.99$ administrative censoring level. The effect of the prior on model space is removed. The results are presented in Figure~\ref{fig:ess}. They are just as expected: when the censoring rate is not extreme enough (before $0.971$), the two strategies are very similar; after $0.971$ censoring rate, with the $n$-scaling, under the LCM prior BF gets closer to 0 (PIPs get closer to 0), whereas under the Product prior the PIPs of the inactive covariate increase (to $0.19$--$0.35$ at the highest censoring level) and those of the active covariate decrease to about $1/2$; and with the $n_{eff}$-scaling, BF gets closer to 1 (PIPs get closer to 1/2). Under the LCM prior, the PIP of the GH model stays below $1/2$, as there are two effects so it pays double dimension penalty (in the $(1+ng_E)^{-d_\gamma/2}$ part); under the Product prior, it is still above $1/2$ at the highest censoring level considered.

While it is not reasonable to claim the superiority of one over the other, they do reflect the different beliefs that people may want to encode into the hyperparameter in the presence of vanishing information: the default strategy is more conservative and prefers sparsity, while the effective sample size strategy prefers to make no decision in the absence of information. Under the LCM prior, the former acts as an Occam-like penalty: it drives the PIP close to $0$ but not exactly, while a large enough $n$ drives it closer when censoring rate gets closer to 1. But given the difference between the two strategies requires high enough censoring rate, we are already assuming that the sample size is large enough (precisely, $n_o \ge d_\gamma+2$) so that we can estimate the parameters (and the null MLE for $n_{eff}$). On the other hand, an advantage of the latter is that the sparsity is being fully controlled by the prior on model space, while the former blends in extra sparsity inclination as information vanishes. In summary, we do recommend running both strategies under such scenarios as a sensitivity check before drawing any conclusions on selection result.

\color{black}
\clearpage

\bibliography{bibliography.bib}